%% file: main.tex
\documentclass[manuscript,trackchanges]{aastex631} 
\usepackage{color,float,amsmath,amssymb,booktabs,footnote,makecell,multirow,geometry,graphicx,subfigure,newtxtext,newtxmath,threeparttable} 
\usepackage[T1]{fontenc}
\usepackage[section]{placeins}
\usepackage[version=4]{mhchem}
\usepackage[figuresright]{rotating}

\received{XXX}
\revised{XXX}
\accepted{XXX}

\submitjournal{ApJS}

\begin{document}

\title[GECAM Localization]{GECAM Localization of High Energy Transients and the Systematic Error}

\shorttitle{GECAM Localization}
\shortauthors{Zhao et al.}

\input{authors}

\begin{abstract}
    Gravitational wave high-energy Electromagnetic Counterpart All-sky Monitor (GECAM) is a pair of microsatellites (i.e. GECAM-A and GECAM-B) dedicated to monitoring gamma-ray transients including gravitational waves high-energy electromagnetic counterparts, Gamma-ray Bursts, Soft Gamma-ray Repeaters, Solar Flares and Terrestrial Gamma-ray Flashes. Since launch in December 2020, GECAM-B has detected hundreds of astronomical and terrestrial events. For these bursts, localization is the key for burst identification and classification as well as follow-up observations in multi-wavelength. Here, we propose a Bayesian localization method with Poisson data with Gaussian background profile likelihood to localize GECAM bursts based on the burst counts distribution in detectors with different orientations. We demonstrate that this method can work well for all kinds of bursts, especially for extremely short ones. In addition, we propose a new method to estimate the systematic error of localization based on a confidence level test, which can overcome some problems of the existing method in literature. We validate this method by Monte Carlo simulations, and then apply it to a burst sample with accurate location and find that the mean value of the systematic error of GECAM-B localization is $\sim 2.5^{\circ}$. By considering this systematic error, we can obtain a reliable localization probability map for GECAM bursts. Our methods can be applied to other gamma-ray monitors.
\end{abstract}


\section{Introduction}

    Gravitational wave high-energy Electromagnetic Counterpart All-sky Monitor (GECAM) \citep{Ins_GECAM_Li2022} is a space-based instrument dedicated to the detection of gamma-ray electromagnetic counterparts (EMs) of the Gravitational Waves (GWs) \citep{common_GWGRB_Abbott2017a, common_GWGRB_Abbott2017b, common_GWGRB_Goldstein2017, common_GWGRB_Savchenko2017, common_GWGRB_Li2018} and Fast Radio Bursts (FRBs) \citep{FRB_Find_Lorimer2007, FRB_HXMT_Li2021}, as well as other high-energy astrophysical and terrestrial transient sources, such as Gamma-ray Bursts (GRBs) \citep{common_GRB_K1973}, Soft Gamma-ray Repeaters (SGRs) \citep{common_SGR_Woods2004}, Solar Flares (SFLs) \citep{common_SFL_Parks1969}, Terrestrial Gamma-ray Flashes (TGFs) \citep{TGF_BATSE_Fishman1994} and Terrestrial Electron Beams (TEBs) \citep{TEB_Find_Dwyer2008, TEB_GBM_Xiong2012}.

    Launched in December 2020, GECAM has been operating in low earth orbit (600 km altitude and 29$^{\circ}$ inclination angle) \citep{Ins_GECAM_Han2020}. GECAM consists of twin micro-satellites (i.e. GECAM-A and GECAM-B) and each of them comprises 25 Gamma-ray Detectors (GRDs) \citep{Ins_GECAM_Lv2018, Ins_GECAM_Zhang2019, Ins_GECAM_An2022} and 8 Charged Particle Detectors (CPDs) \citep{Ins_GECAM_Xv2021, Ins_GECAM_LiCY2022, Ins_GECAM_Zhang2022}. With the LaBr$_3$ crystal read out by a silicon photomultiplier (SiPM) array, GRDs could monitor the entire un-occulted sky in the energy range of $\sim 15$ keV to $\sim 5$ MeV \citep{Ins_GECAM_SiPM_Zhang2022a, Ins_GECAM_SiPM_Zhang2022b}. The CPD is designed to detect the charged particles in the orbit, which is powerful to observe TEBs and help the GRDs to distinguish between gamma-ray burst events and charged particle burst events \citep{Ins_GECAM_Yun2021}.

    Since the launch, GECAM-B has detected many GRBs \citep{GRB_Study_Wang2021, GRB_Study_Song2022}, SGRs \citep{SGR_Study_Xie2022}, SFLs and TGFs, while GECAM-A has not been able to observe yet due to power supply issues. The main scientific goal of GECAM is to detect high-energy electromagnetic counterpart associated with GWs and FRBs: such association analysis usually requires the coincidence of temporal and localization. For all kinds of bursts, the localization is the crucial information for joint observations with other instruments. Therefore, localization capability is of fundamental importance for the GECAM mission.

    \begin{table*}
        \centering
        \caption{ Different localizations provided by GECAM in different stages after burst trigger. }
        \label{TABLE_GECAMLOCSTA}
        \begin{tabular}{cccccccccccc}
            Time After Burst & Localization Algorithm & Method \\
            \hline
            $\sim$ 1 min   & Flight Location                 & $\chi^{2}$ minimization      \\
            $\sim$ 10 min  & Ground Location with alert data & $\chi^{2}$ minimization      \\
            $\sim$ 10 hour & Ground Location with full data  & Bayesian with PGSTAT         \\
            $\sim$ 1 day   & Triangulation Location          & Time-delay with Li-CCF       \\
            \hline
        \end{tabular}
    \end{table*}

    As shown in Table 1, a series of localization algorithms have been developed by the GECAM team for different stages after the trigger of a burst (\citealt{Loc_MCMC_Liao2020}; \citealt{Ins_GECAM_Yun2021}; Cai et al. 2022 (in preparation); Huang et al. 2022 (in preparation)):

    \begin{itemize}

        \item For onboard triggered bursts, the flight software computes their locations and provides a preliminary classification of the bursts within $\sim$ 20 seconds \citep{Ins_GECAM_Yun2021}. The flight software employs a $\chi^{2}$ minimization localization method considering the very limited onboard resources of computing and storage.

        \item To allow rapid follow-up observations with GECAM triggers, the GECAM alert data are downlinked in real-time by the global short message communication service \citep{Ins_GECAM_LiG2020} of BeiDou navigation satellite System \cite[BDS,][]{Ins_GECAM_Yang2019}. With the alert data, an automatic on-ground analysis pipeline (Huang et al. 2022, submitted for publication) is used to provide a refined localization result, which is based on the $\chi^{2}$ minimization localization method. Moreover, a one-time spectrum-location iteration is used to optimize the localization spectral template. Thus this localization is expected to be more reliable than the in-flight ones.

        \item After the full science data (including time tagged events data and binned data) are downloaded to the ground, both the automatic and human-in-the-loop analyses for the burst will be initiated, including the detailed temporal, positional, and spectral analyses. Among them, the localization is based on a Bayesian localization method with Poisson data with Gaussian background (PGSTAT) profile likelihood, which is the main topic of the present paper.


        \item If the burst is jointly observed by other instruments, such as \textit{Insight}-HXMT/HE, \textit{Fermi}/GBM and \textit{Swift}/BAT, the triangulation (time-delay) localization within our ETJASMIN pipeline would also be implemented to improve the location \citep{Loc_TimeDelay_Xiao2021, Ins_GECAM_ETJASMIN_Xiao2022}.

    \end{itemize}

    In the present paper, we focus on the ground localization with full science data, rather than flight location or ground location with alert data. The Bayesian localization methods with fixed spectral templates have been discussed in previous work \citep{YiZhao_LOC_Principle}. Here we extend this method to real observation data by replacing the Poisson likelihood with the Poisson data with Gaussian background (PGSTAT) profile likelihood. We also propose a new method to estimate the systematic error of location.

    We note that in this work only GECAM-B data are used since GECAM-A has not been able to observe yet \citep{Ins_GECAM_Li2022}.

    This paper is structured as follows: The GECAM instruments, detectors, and data used in localization analysis are presented in Section 2. In Section 3, a Bayesian localization method with PGSTAT profile likelihood is proposed for GECAM. In Section 4, we present the localization analysis for 23 bright bursts (including GRBs, SGRs, and SFLs) and 3 extremely short bursts (TGFs). We discuss the systematic error and demonstrate the systematic uncertainty estimation using GECAM localizations in Section 5. Finally, a summary is given in Section 6.

\clearpage
\section{The GECAM Instrument}

    \begin{figure*}
        \centering
        \includegraphics[height=6cm]{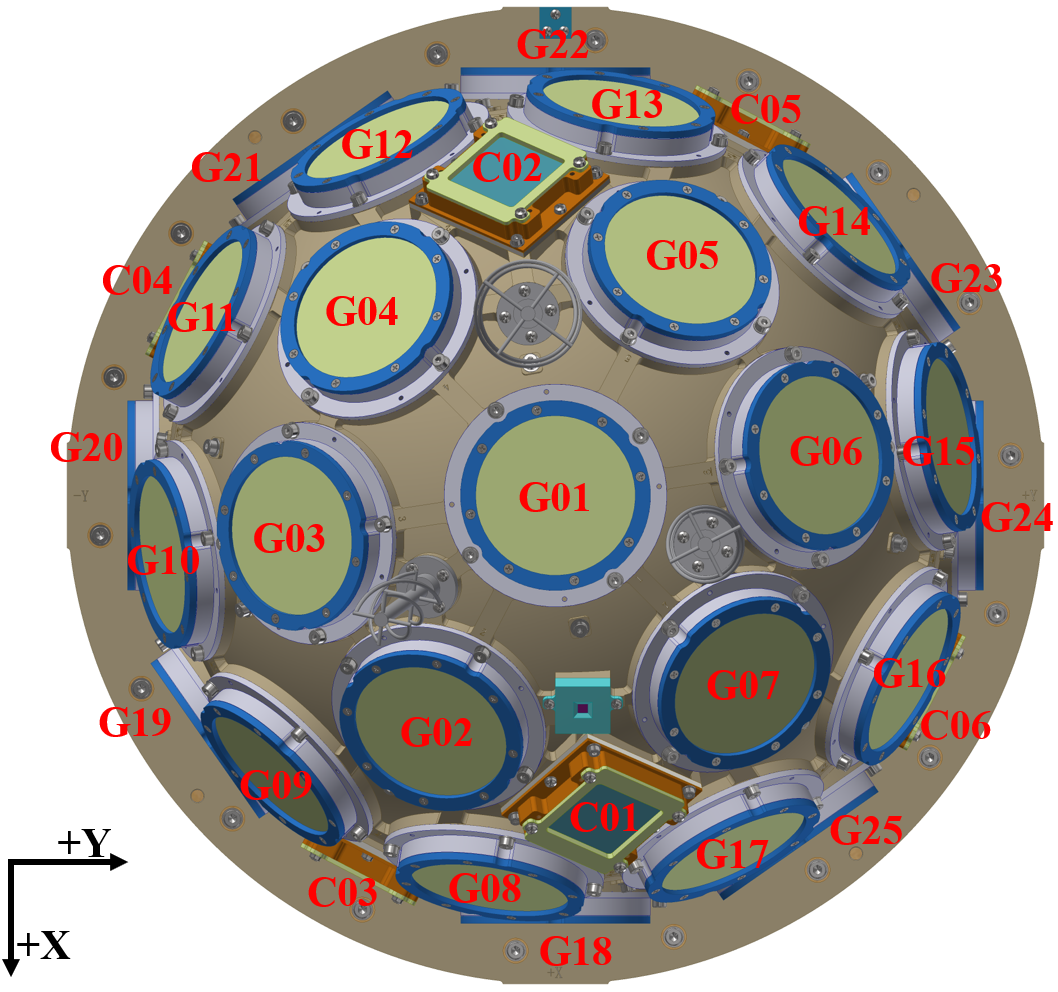}
        \includegraphics[height=6cm]{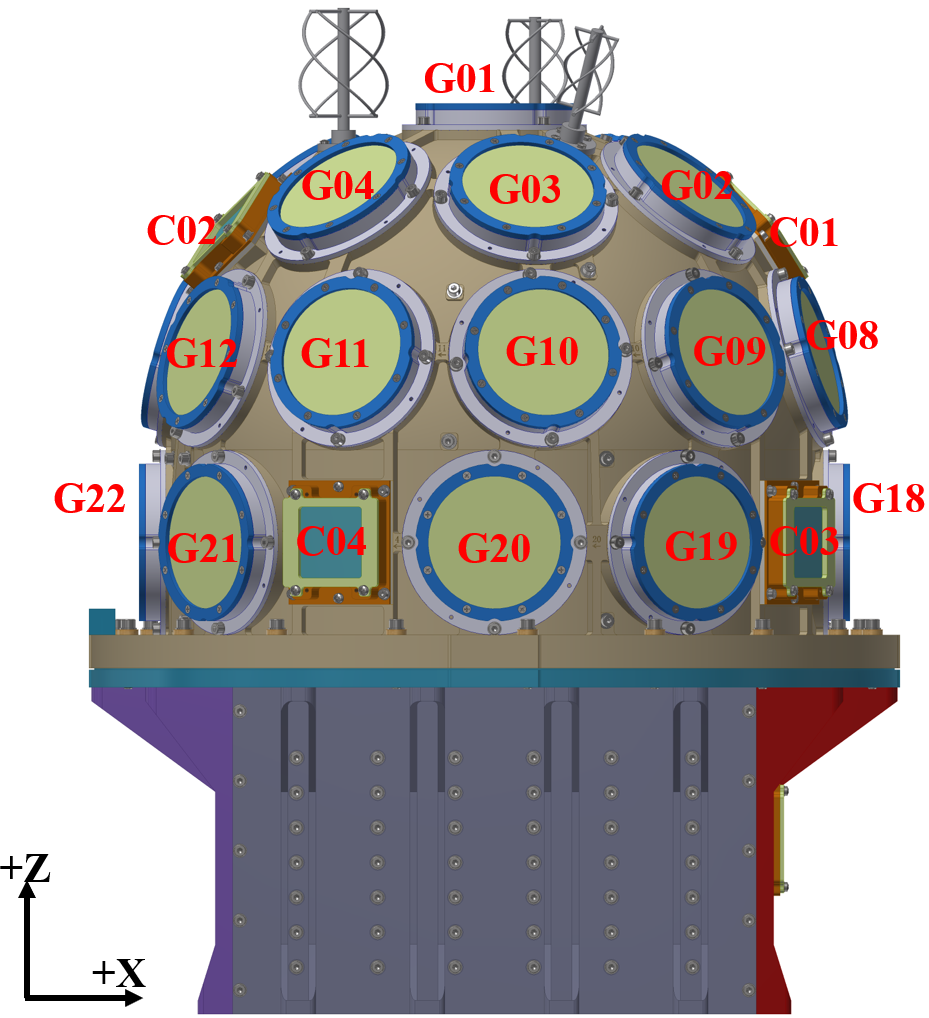}
        \includegraphics[height=6cm]{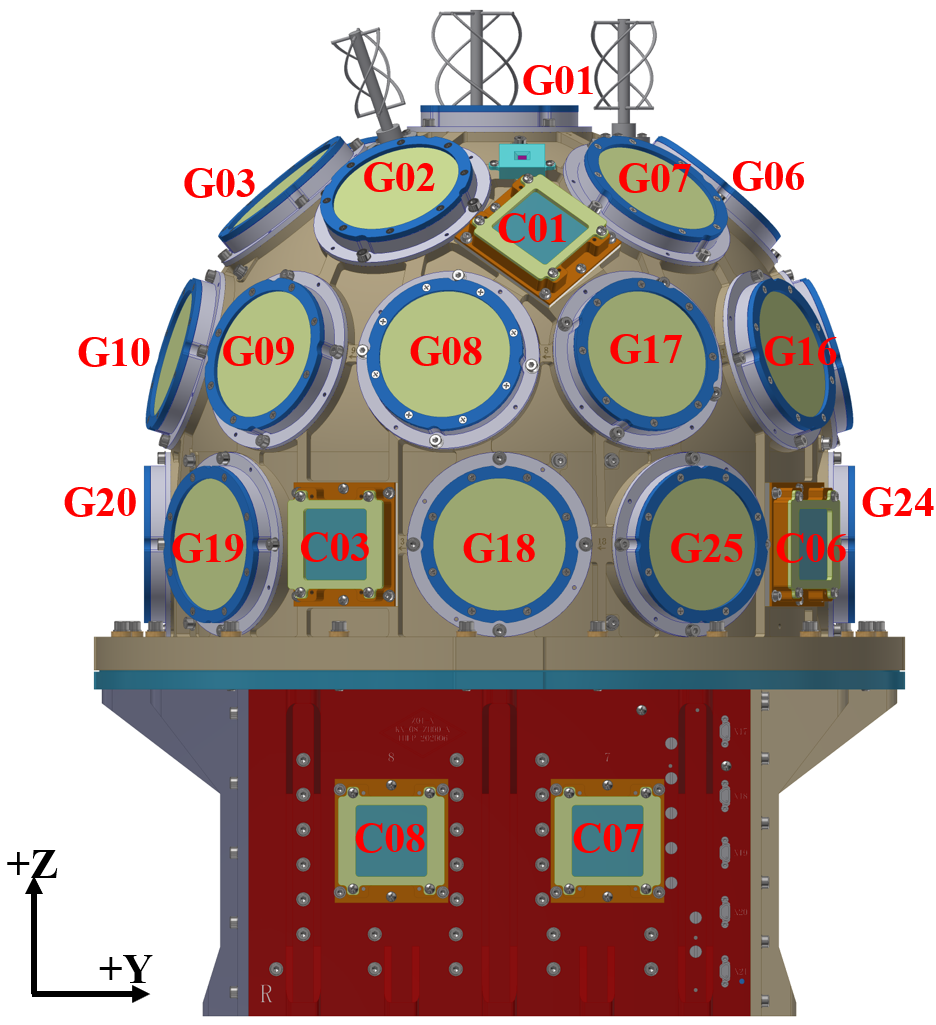}
        \includegraphics[height=6cm]{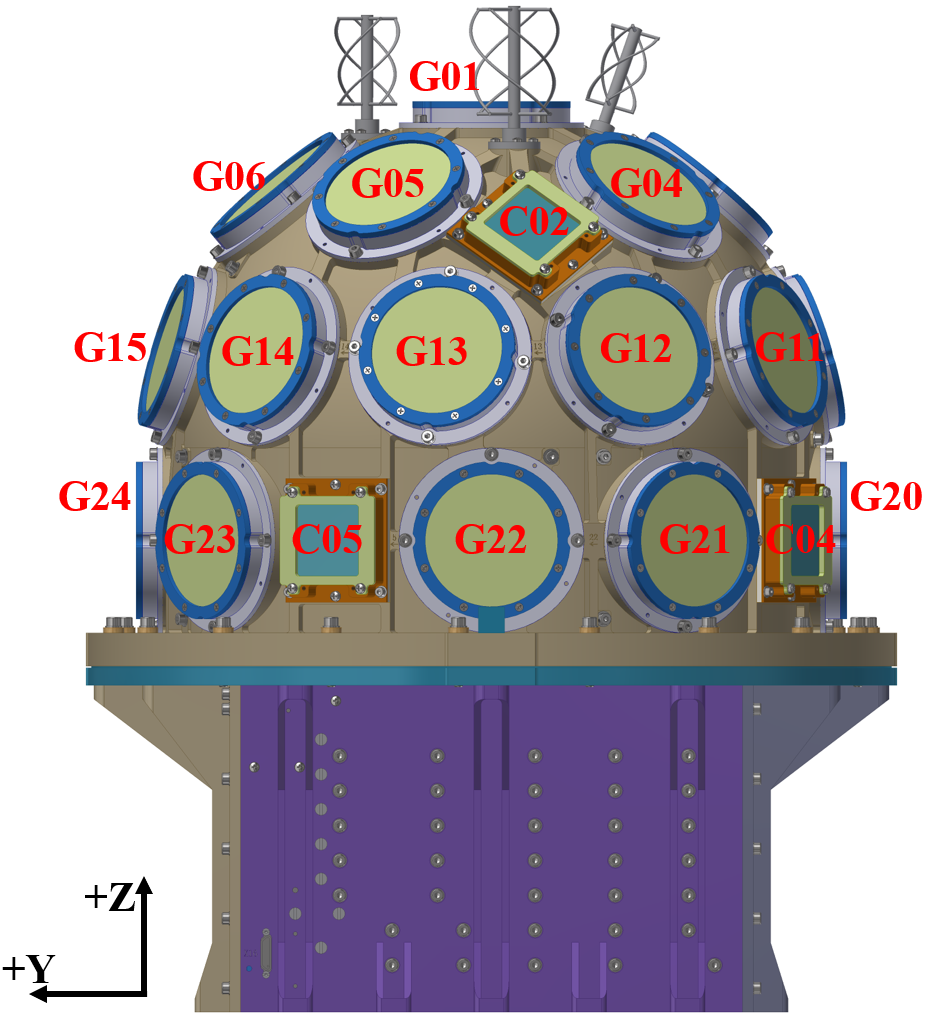}
        \caption{ The schematic diagram illustrates the GECAM payload from different views. 25 GRDs are labeled with an index from G01 to G25 and 8 CPDs are labeled from C01 to C08. It comprises a detector dome and an Electronic Box (EBOX). Most of the detectors (all 25 GRDs and 6 CPDs) are assembled in the detector dome, and 2 CPDs are installed on the +X side of the payload. In the payload coordinate system, the +X[+Y][+Z] axis is aligned with the pointing direction of GRD18[GRD24][GRD01]. The 25 GRDs pointing directions are listed in Table \ref{TABLE_GRD_Point}. }
        \label{fig1}
    \end{figure*}

    \begin{table}
        \centering
        \caption{ Pointing directions of the 25 GRDs in the GECAM payload coordinate system. $\theta$ and $\phi$ are the zenith and azimuth angles in the payload coordinate (see Figure 1), respectively. }
        \label{TABLE_GRD_Point}
        \begin{tabular}{cccccccccccc}
            GRD \# & $\theta$ (${}^{\circ}$) & $\phi$ (${}^{\circ}$) & GRD \# & $\theta$ (${}^{\circ}$) & $\phi$ (${}^{\circ}$) \\
            \hline
            01 &  0.00 &   0.00 & 14 & 73.50 & 134.50 \\
            02 & 40.00 & 332.17 & 15 & 73.50 &  98.50 \\
            03 & 40.00 & 278.67 & 16 & 73.50 &  62.50 \\
            04 & 40.00 & 225.17 & 17 & 73.50 &  26.50 \\
            05 & 40.00 & 152.17 & 18 & 90.00 &   0.00 \\
            06 & 40.00 &  98.67 & 19 & 90.00 & 305.00 \\
            07 & 40.00 &  45.17 & 20 & 90.00 & 270.00 \\
            08 & 40.00 & 350.50 & 21 & 90.00 & 215.00 \\
            09 & 73.50 & 314.50 & 22 & 90.00 & 180.00 \\
            10 & 73.50 & 278.50 & 23 & 90.00 & 125.00 \\
            11 & 73.50 & 242.50 & 24 & 90.00 &  90.00 \\
            12 & 73.50 & 206.50 & 25 & 90.00 &  35.00 \\
            13 & 73.50 & 170.50 &    &       &        \\
            \hline
        \end{tabular}
    \end{table}

    The GECAM payload mainly consists of 25 GRDs (for 15 keV-5 MeV X/$\gamma$ rays), 8 CPDs (for 150 keV-5 MeV charged particles) and the Electronic Box (EBOX) \citep{Ins_GECAM_Li2022, Ins_GECAM_An2022, Ins_GECAM_Xv2021}. As illustrated in Figure \ref{fig1}, 25 GRDs and 6 CPDs are placed with different orientations in the detector dome of the GECAM satellite to monitor most of the sky region, while 2 CPDs are installed on the +X side of the payload EBOX. The flight localization and alert data are generated by the in-flight trigger and localization software which is executed on the Data Management board of the EBOX \citep{Ins_GECAM_Li2022}. Since the CPDs have low detection efficiency to gamma-rays \citep{Ins_GECAM_Xv2021, Ins_GECAM_LiCY2022, Ins_GECAM_Zhang2022}, we only use 25 GRDs to compute burst localization. The pointing directions of GRDs in GECAM payload coordinates are listed in Table \ref{TABLE_GRD_Point}.

    The event-by-event (EVT) data of GRDs are employed to conduct localization analysis. GECAM features the highest time resolution among instruments of its kind, i.e. $0.1~\mu$s \citep{Ins_GECAM_TimeCalibration_Xiao2022}. The dead time is $4~\mu$s for a normal event and $> 69~\mu$s for an overflow event (i.e. event with higher energy deposition than the maximum measurable energy) \citep{Ins_GECAM_SiPM_Liu2021}.

    Each GRD detector has two read-out channels: high-gain channel and low-gain channel \citep{Ins_GECAM_SiPM_Liu2021}. The normal events of GRDs (i.e. within the detection energy range) are registered in 4096 raw ADC channels by data acquisition (DAQ) electronics and converted to 447 PI energy channels for high gain ($\sim 15$ keV to $\sim 300$ keV) and low gain ($\sim 300$ keV to $\sim 5$ MeV). For simplicity, only the counts from $\sim$30 keV to $\sim$200 keV (i.e. PI channels from 45 to 150) of the high-gain channel are adopted in the localization of the present work, because most counts from detected bursts are recorded in this energy range. The detector response is constructed based on a series of ground tests \citep{Ins_GECAM_Zhang2019, Ins_GECAM_An2022, Ins_GECAM_Li2022} before launch and comprehensive simulations incorporating the GECAM spacecraft mass model into GEANT4 \citep{Ins_GECAM_Sim_Guo2020}, and has been further calibrated in flight with characteristic lines and known bursts (\citealt{Ins_GECAM_Calibration_Zhang2022, Ins_GECAM_Calibration_Zheng2022}; Qiao et al. 2022 (in preparation); Zhang et al. 2022 (in preparation)).

\clearpage
\section{Localization Methodology}

\subsection{Localization Method}

    \citealt{YiZhao_LOC_Principle} proposed a Bayesian localization method with Poisson likelihood for the fixed spectral template, which was validated through comprehensive tests. But that method is designed for the ideal case where the expected background is known.

    For the real observation, the expected background is unknown and we can only use the estimated background and have to take into account its uncertainty. Therefore, here we propose the Bayesian localization method with Poisson data with Gaussian background (PGSTAT) profile likelihood to account for the uncertainties of background estimation. The derivation of PGSTAT profile likelihood and localization methodology is detailed in Appendix \ref{SECTION_LocMethod}.

    To quantitatively validate the Bayesian localization method with the PGSTAT profile likelihood mentioned above, we conduct a Monte Carlo (MC) simulation tests with GECAM detector response, following the procedures in \citep{YiZhao_LOC_Principle}. For simplicity, some assumptions are made: (1) The background is known precisely, thus the PGSTAT profile likelihood is reduced to the simple Poisson likelihood (Equation \ref{EQ_Pois_P}). (2) The utilized localization template is derived from the input spectrum of the simulated bursts. The simulated counts in each detector are derived from the Poisson fluctuation of the expectation of total counts which is the sum of the burst counts (the input burst spectrum convolved with the detector response) and the background. The true position for simulation and more detailed setting can be found in Table \ref{TABLE_SimulationSetting}.

    The localization result for simulated bursts is shown in Figure \ref{fig5} (a). To validate the location probability map and credible region, we check the distribution of the Highest Posterior Density (HPD) cumulative probability of the true position\footnote{ The cumulative of the probability for each HEALPix pixel starting from the highest pixel \citep{YiZhao_LOC_Principle}. } in the location maps for simulated bursts. These results indicate that this localization method could give reliable and correct location results for simulated bursts and no systematic uncertainties exist in the samples.

\subsection{Localization Procedures}

    To present the procedure of localization analysis in detail, here we take the GRB 220511A as an example:

    \begin{itemize}

        \item[STEP 1] The source and background intervals are manually selected, and the polynomial fitting for the background is implemented, as shown in Figure \ref{fig2a} (a). \textcolor[rgb]{1,0,0}{ We utilize the GECAMtools to implement the polynomial fitting of background, which uses a weighted least-squares ($\chi^{2}$) statistic to fit the polynomial coefficients through a two-pass approach, as used in the GBM data analysis in RMFIT\footnote{RMFit: \href{http://fermi.gsfc.nasa.gov/ssc/data/analysis/rmfit/}{http://fermi.gsfc.nasa.gov/ssc/data/analysis/rmfit/}} and GBM Data Tools\footnote{GBM Data Tools: \href{https://fermi.gsfc.nasa.gov/ssc/data/analysis/rmfit/gbm\_data\_tools/gdt-docs/}{https://fermi.gsfc.nasa.gov/ssc/data/analysis/rmfit/gbm\_data\_tools/gdt-docs/}}. For the time interval selection of source, we tried several different ranges, and choose the best one by the localization results, i.e. the size of statistical error. } Then, for the source interval, the estimated background and its uncertainty, the exposure time, and total observed counts corrected for dead time are obtained.

        \item[STEP 2] As shown in Figure \ref{fig2a} (b), a \textbf{fixed} templates localization with all 25 GRDs is performed. Use three fixed spectral templates (Table \ref{TABLE_FixedTemplate}) \citep[see also New Spectral Templates of Table 1,][]{Loc_GBM_Goldstein2020} through the Bayesian localization method with PGSTAT profile likelihood. In analogy to GBM DoL \citep{Loc_GBM_Connaughton2015}, the maxima of the maximum PGSTAT profile likelihood of three spectral template's localization is regarded as the location result. This localization is termed FIX hereafter.

        \item[STEP 3] To derive a \textbf{refined} spectral template for localization, we implement a one-time spectrum-location iteration. A spectral fitting to the location center of FIX localization is first implemented. After the consistency check of the observational and expected counts, we redo the localization for which the spectral template is re-constructed by the spectral fitting result. This localization is termed RFD hereafter.

        \begin{itemize}

            \item[(a)] As shown in Figure \ref{fig2a} (c), to derive a refined spectrum for localization, a spectral fitting to the location center of FIX localization is implemented. These three GRDs (GRD \# 11, 12, and 21) within $60^{\circ}$ from the location center and with detection significance $>5\sigma$ are used for this spectral fitting.

            \item[(b)] Check the consistency of the expected counts (the sum of estimated background and expected source contribution) and measured counts registered in detectors, as shown in Figure \ref{fig2a} (d). Although excess counts in some detectors are significant, their expected counts are inconsistent with the measured data due to the imperfection of the off-axis response. Therefore, only detectors within 3 $\sigma$ deviation between the measured and expected counts are adopted for localization in the following refined template localization. The source contribution is derived from the refined spectrum.

            \item[(c)] As shown in Figure \ref{fig2a} (e), using the refined spectral fitting results and the selected detectors, we construct the spectral template and localize the burst again.

        \end{itemize}

        \item[STEP 4] To check the best scenario of the one-time iteration localization, similar to the RFD localization, we do a one-time spectrum-location iteration, but the \textbf{accurate position based on the reference location}\footnote{ Reference location: The accurate position provided by instruments such as \textit{Swift} and IPN, etc. } is adopted for spectral fitting. This localization is termed APR hereafter.

        \begin{itemize}

            \item[(a)] First conduct the spectral fitting to the reference location, as shown in Figure \ref{fig2b} (a).

            \item[(b)] As shown in Figure \ref{fig2b} (b), do the consistency check for detectors.

            \item[(c)] Then re-construct the spectral template and redo localization with the the spectral fitting parameters and selected detectors. The location's credible region is shown in Figure \ref{fig2b} (c). 

        \end{itemize}

    \end{itemize}

    For the burst localization of \textit{Fermi}/GBM, \citet{Loc_BALROG_Michael2017} noted that localization with all detectors could introduce systematic deviation, which is also seen in GECAM localization. As shown in Figure \ref{fig2b} (d), compared to the case that only uses the selected detectors, APR localization with all 25 GRDs deviates further from the reference location.

    We should note that the RFD localization is what we do for normal observations of bursts. For those bursts with accurate position, we just do the APR localization for the localization methodology study. 

    For RFD localization, it should be noted that the spectrum-location iteration localization should be used. However, the automatic iteration localization procedure is challenging for the observation data, because it includes many steps, such as the background and source intervals selection, spectral model selection, the evaluation of spectral fitting results, consistency check, detector selection, etc. These steps may influence the final localization result. The automatic iteration localization procedure is still under development and testing.

\clearpage
\section{Localization Results}

\subsection{GRBs, SGRs and SFLs}

    By the end of May of 2022, hundreds of bursts have been detected by GECAM-B (\citealt{Ins_GECAM_Yun2021}; Cai et al. 2022 (in preparation); Huang et al. 2022 (in preparation)), including 23 bright bursts (9 GRBs, 12 SGRs, and 2 SFLs) which are accurately localized by external instruments (e.g. \textit{Swift}, IPN and INTEGRAL-IBAS). The reference locations of the 23 bright bursts and their location results of the flight and the ground with alert data (see Table \ref{TABLE_GECAMLOCSTA}) are listed in Table \ref{TABLE_LocRes_Part1}. These bursts could be further used to estimate the localization systematic error (see the next section). The FIX, RFD and APR location results of these bursts can be found in Table \ref{TABLE_LocRes_Part2} and Appendix \ref{SECTION_LocRes_GRB}. The spectral parameters for RFD and APR localization are listed in Table \ref{TABLE_SpeRes}.

    The incident angles of these bursts in GECAM payload coordinates are shown in Figure \ref{fig3a}. The distributions of the angular separation between the location centers and reference locations for different localization (FIX, RFD, and APR) are shown in Figure \ref{fig3b}. As expected, the RFD localization (all bursts have offset $<$ 8 $^{\circ}$) is closer to the reference location than that of the FIX localization (all bursts have offset $<$ 13 $^{\circ}$), while the APR one (all bursts have offset $<$ 8 $^{\circ}$) is closer than that of the RFD. The influence of spectral template and detector selection will be further discussed in Section 5. We note that, the $1\sigma$ statistical error of GRBs with fluence $\sim 1 \times 10^{-5} ~\rm erg/cm^{2} ~(10-1000  keV)$ (e.g. GRB 210927B, GRB 220511A and GRB 220514A) is $\sim 2^{\circ}$, which is generally consistent with the design of the GECAM mission. 

\subsection{TGFs}

    TGFs are very short events with a few counts \citep{TGF_BATSE_Fishman1994, TGF_BeppoSAX_Ursi2017, TGF_RHESSI_Grefenstette2009, TGF_RHESSI_Ostgaard2015, TGF_AGILE_Lindanger2020, TGF_AGILE_Maiorana2020, TGF_GBM_Roberts2018, TGF_ASIM_Ostgaard2019}. 
    Generally, a luminous TGF contains $\sim$ 100 observed counts in several hundred microseconds, which means there are usually only $\sim$ 10 counts registered in a detector. Locating these sub-millisecond-duration weak bursts is very challenging, for the following reasons:

    \begin{itemize}

        \item The limited counting statistics of measured counts registered in each detector challenge the accuracy and precision of localization.

        \item Most of the time, there are less than 10 GRDs that are facing the Earth simultaneously. A constrained location cannot be obtained for these cases since the imperfect response of off-axis detectors.

        \item The Compton scattering between the TGFs and the atmosphere may occur. Thus the incident direction of TGF photons is likely dispersed, rather than a single direction for the point source.

    \end{itemize}

    For TGFs detected by the Burst and Transient Source Experiment (BATSE) aboard the Compton Gamma Ray Observatory (CGRO) \citep{TGF_BATSE_Fishman1994}, they are localized with the BATSE Burst Location Algorithm (LOCBURST), which belongs to $\chi^{2}$ method \citep{Loc_BATSE_Pendleton1998}. However, \citep{YiZhao_LOC_Principle} showed that the $\chi^{2}$ method is likely problematic for this kind of weak burst. Besides, the background uncertainties are not considered in LOCBURST. 
    
    Here, we use TGFs to further test the performance of our localization algorithm in the case of limited counting statistics.

    We adopt three luminous TGFs detected by GECAM-B (Zhao et al. 2022, in preparation) with $\sim$ 100 counts detected by $\sim$ 11 GRDs facing to the Earth. The geographic positions of the GECAM-B nadir when these three TGFs were detected are shown in Figure \ref{fig4a}. 

    Since a reasonably good spectral fitting for such weak bursts is nearly impossible due to the very limited number of counts, we choose a typical TGF photon spectrum \citep{TGF_ASIM_Sarria2021} to generate the localization template:

        \begin{align}
            f(E) = A \cdot E^{-1} \cdot \exp( -\frac{ E }{ E_{\rm cut} } ) ,
            \label{EQ_TGFspec}
        \end{align}
    where $A$ is amplitude, $E$ is energy, and $E_{\rm cut}$ is cut-off energy which is set to 7 MeV.

    As shown in Figure \ref{fig4b} (a), there are $\sim$ 100 counts measured in a short duration of $\sim$ 200 $\mu$s for an observed TGF by GECAM, which is remarkably bright, thus the dead time effect should be considered. For dead time correction, we do it for every 30 $\mu$s considering the extremely short duration.

    The detailed GECAM localization results of TGFs are shown in Figure \ref{fig4b} and Appendix \ref{SECTION_LocRes_TGF}. It should be noted that there are only about 5 to 10 counts registered in a single detector (see Figure \ref{fig4b} (b)), thus the statistical error of localization is usually very large. We found that using all detectors could result in the best localization result for TGFs.

    If there is lightning associated with a TGF, then the position of the lightning could be used as the source location of the TGF. However, most TGFs have no associated lightning for various reasons \citep{TGF_Review_Dwyer2012}, thus their precise locations are unknown. But a good estimation of TGFs location is that they should come from the region within $\sim$ 800 km from the GECAM-B nadir point when the TGF is detected, which is usually narrow on the sky map (see Figure \ref{fig4b} (c)) and thus considered to be the most likely area of TGF origin \citep{TGF_RHESSI_Grefenstette2009, TGF_AGILE_Lindanger2020, TGF_GBM_Roberts2018}. Therefore, the geocentric HPD cumulative probability could be used to evaluate the correctness of TGF localization.

    As shown in Figure \ref{fig4b} (c) and Appendix \ref{SECTION_LocRes_TGF}, although the error region is large on the sky map, the location error region could cover most of the region nearby the nadir, and the geocentric HPD cumulative probabilities for all of these TGFs are $<$ 95.45\% (corresponding to 2 $\sigma$), which is consistent with the terrestrial origin of these bursts. These results also demonstrate that the TGF localization given by our Bayesian localization method with PGSTAT profile likelihood is reasonable.

\clearpage
\section{Systematic Error of Localization}

    In this Section, we investigate the systematic error estimation approach for localization and analyze the location systematic error for GECAM. Many factors could lead to the systematic error in localization:

    \begin{itemize}

        \item The inaccuracies of detector responses. Generally, the source counts comes from three components: (a) The direct response towards a burst. (b) The response of Compton scattering of the spacecraft and other payloads. (c) The response of atmospheric scattering \citep{Ins_GECAM_Sim_Guo2020}. Due to the complexity of Compton scattering and atmospheric scattering which relies heavily on the assumption and the Monte Carlo simulation process, the response of the detectors with large off-axis angles is usually imperfect. These effects are still under study for GECAM.


        \item The localization method and parameter settings, i.e. the spectral parameters as well as detector and energy channel selection. As noted by \citep{YiZhao_LOC_Principle}, the $\chi^{2}$ localization method for weak bursts could give an incorrect probability error region. While \citet{Loc_BALROG_Michael2017} and \citet{Loc_BALROG_Berlato2019} reported that the fixed spectral templates localization will introduce systematic uncertainties. As we will discuss in the following, the detector selection will also affect the results.


    \end{itemize}

\subsection{Estimation Method for Localization Systematic Error}

    In the previous works, a Bayesian approach was proposed and adopted to estimate the localization systematic uncertainties \citep{Loc_BATSE_Briggs1999, Loc_GBM_Connaughton2015, Loc_POLAR_Wang2021}, which is summarized in Appendix \ref{SECTION_GBMsysERR}. This method assumes that the localization error region is symmetric, and the uncertainty regions are circular. However, these two assumptions are usually not satisfied in the localization of bursts. As shown in Figure \ref{fig2a} and \ref{fig2b}, the localization error region is irregular. To illustrate the limitation of this estimation approach, we derive the systematic error with this method for simulated bursts which only contain statistical error, as shown in the black line of Figure \ref{fig5} (b). A non-zero systematic error could be seen with the peak around $\sim 0.5$ deg, due to the shape of the non-circular error region. We find that this systematic uncertainty will increase as the non-circular tendency of the error region increases.

    Here, we propose to derive the systematic error based on confidence level (CL) tests, described as follows:

    \begin{itemize}

        \item[\textbf{Step 1:}] Total localization map: (1) For a given assumed systematic error $\sigma_{\rm SYS}$, calculate the total localization error map by convolving the statistical probability map with $\sigma_{\rm SYS}$ for each burst; (2) And then derive the HPD cumulative probability for the true position of the simulated burst (see Appendix \ref{SECTION_LocMethod}); (3) Divide the confidence level evenly into $M_{s}$ bins\footnote{ For example, if we take 10 bins, then the confidence levels for each bin are 0-10\%, 10-20\%, ..., 90-100\%. }, and calculate the burst number $U(\rm t)$ in each confidence level bin $t$.

        \item[\textbf{Step 2:}] Confidence level test: For a reasonable $\sigma_{\rm SYS}$, the expectation of the burst number in each confidence level bin should be $\frac{ N_{\rm s} }{ M_{\rm s} }$, where $N_{\rm s}$ is the total number of simulated bursts. Thus, we can construct the likelihood that the distribution of burst number $U(\rm t)$ in a confidence level bin $t$ is observed for each assumed systematic error $\sigma_{\rm SYS}$:

            \begin{align}
                P_{\rm CL}^{'} = \prod_{t=1}^{M_{s}} \frac{ (\frac{ N_{\rm s} }{ M_{\rm s} })^{ U(\rm t) } \cdot \exp( -\frac{ N_{\rm s} }{ M_{\rm s} } ) }{ U(\rm t)! } .
                \label{EQ_SysErr_CL_1}
            \end{align}

        Then we can normalize this likelihood to obtain the probability of systematic error:

            \begin{align}
                P_{\rm CL} = \frac{ P_{\rm CL}^{'} }{ \sum_{t} P_{\rm CL}^{'} } .
                \label{EQ_SysErr_CL_2}
            \end{align}

    \end{itemize}

    As shown in the red line of Figure \ref{fig5} (b), for the sample of bursts with statistical error only, our systematic error estimation method could give the correct zero systematic error (i.e. peaks at $\sim 0$ deg), compared to the non-zero results given by the existing method in the literature.

    To further test our method, the true position of the burst sample is artificially moved to a fake position which deviates by 6 degrees away from the true one. As shown in Figure \ref{fig5} (c), our method can correctly give the systematic error whose probability distribution peaks at 6 deg. The simulated burst distribution in each confidence level bin for typical systematic uncertainties is shown in Figure \ref{fig5} (d). An individual inspection of simulated bursts is shown in Figure \ref{fig5} (e).

\subsection{Systematic Error of GECAM Location}

    To quantitatively investigate what factors contribute to systematic error, we conduct the systematic estimation for (1) FIX localization, (2) RFD localization, and (3) APR localization. Due to the limited number of bursts, we choose five confidence level bins with a bin width of $20\%$ for our systematic error estimation. As shown in Figure \ref{fig6}, the maximum probability of systematic uncertainties of FIX[RFD][APR] peaks at 4.0[2.5][2.5] deg. For FIX and RFD locations, the Gaussian fitting to their distribution results in the peaks at 4.18 deg and 2.62 deg, respectively.

    From the comparison between the FIX and RFD locations, we note that an extra systematic error of $\sim$ 1.5 deg is introduced in the FIX location. The reason for this could be: (1) FIX location makes use of the fixed spectral parameters; (2) FIX location has all detectors used. The systematic error of RFD and APR locations are similar since the absolute offset between the location (which is given by the FIX localization) used by RFD is quite close to the accurate location used by APR location, thus the spectral fitting results and the localization templates are very similar between RFD and APR localization.

    In the previous work, the detectors $< \sim$ 80 deg offset between the location center and detector pointing and the satisfied significance are generally used for localization and spectrum analysis. It is because the off-axis detectors' response is imperfect in most cases. However, all detectors should be used if the data obeys Poisson fluctuation \citep{Loc_BALROG_Michael2017}. Although no significant counts are registered in some detectors, they still help to reject some location region. The tendency could be found in the simulated result. Therefore, the previous detector selection scheme ($< \sim$ 80 deg offset between the location center and detector pointing and the satisfied significance) may introduce extra deviation and reduce the statistics.

\clearpage
\section{Summary}

    In this work, we propose a Bayesian localization method with PGSTAT profile likelihood for GECAM bursts. Compared to the method in \citep{YiZhao_LOC_Principle}, this method takes the background estimation and uncertainties into account, making it applicable for real observation data.

    This localization method is applied to a sample of 23 bright sources (including 9 GRBs, 12 SGRs and 2 SFLs) with external accurate locations. We found that the $1\sigma$ statistical error of the three bright GRBs (fluence $\sim 1 \times 10^{-5} \rm erg/cm^{2})$ is $\sim 2^{\circ}$, which is generally consistent with the original design of the GECAM mission. We also find that GECAM can reasonably locate millisecond-duration weak bursts (TGFs) with our method. 

    Regarding the estimation of systematic error, we find that the existing Bayesian systematic uncertainties estimation approach in the literature will give an incorrect result because it assumes a circular statistical error region. We propose an approach based on confidence level validation which is validated with simulations. We applied our method to the burst sample with accurate location and found that the mean value of the systematic error of GECAM localization is $\sim 2.5^{\circ}$. We notice that by considering this systematic error, we can obtain a final localization probability map which can pass the confidence level test. 
    
    Since our methods of the burst localization and systematic error estimation are universal for gamma-ray monitors, they can be applied to other missions, such as \textit{Fermi}/GBM \citep{Ins_GBM_Review_Meegan2009}, SVOM/GRM \citep{Ins_SVOM_Review_Shun2020}, POLAR \citep{Ins_POLAR_Review_Produit2018} and POLAR-2 \citep{Ins_POLAR2_Review_Hulsman2020} and GRID \citep{Ins_GRID_Found_Wen2019}.




\clearpage
\section*{Acknowledgements}
    The GECAM (HuaiRou-1) mission is supported by the Strategic Priority Research Program on Space Science of the Chinese Academy of Sciences, China. This work is supported by the National Key R\&D Program of China (2021YFA0718500). We thank the support from the Strategic Priority Research Program on Space Science, the Chinese Academy of Sciences (Grant No. XDA15360102, XDA15360300, XDA15052700), the National Natural Science Foundation of China (Grant No. 12173038, U2038106) and the National HEP Data Center (Grant No. E029S2S1). We thank Xi Long (Harvard University) for helpful discussions.

\clearpage

    \begin{figure*}[htb]
        \centering
        \subfigure[]{\includegraphics[height=3.0cm]{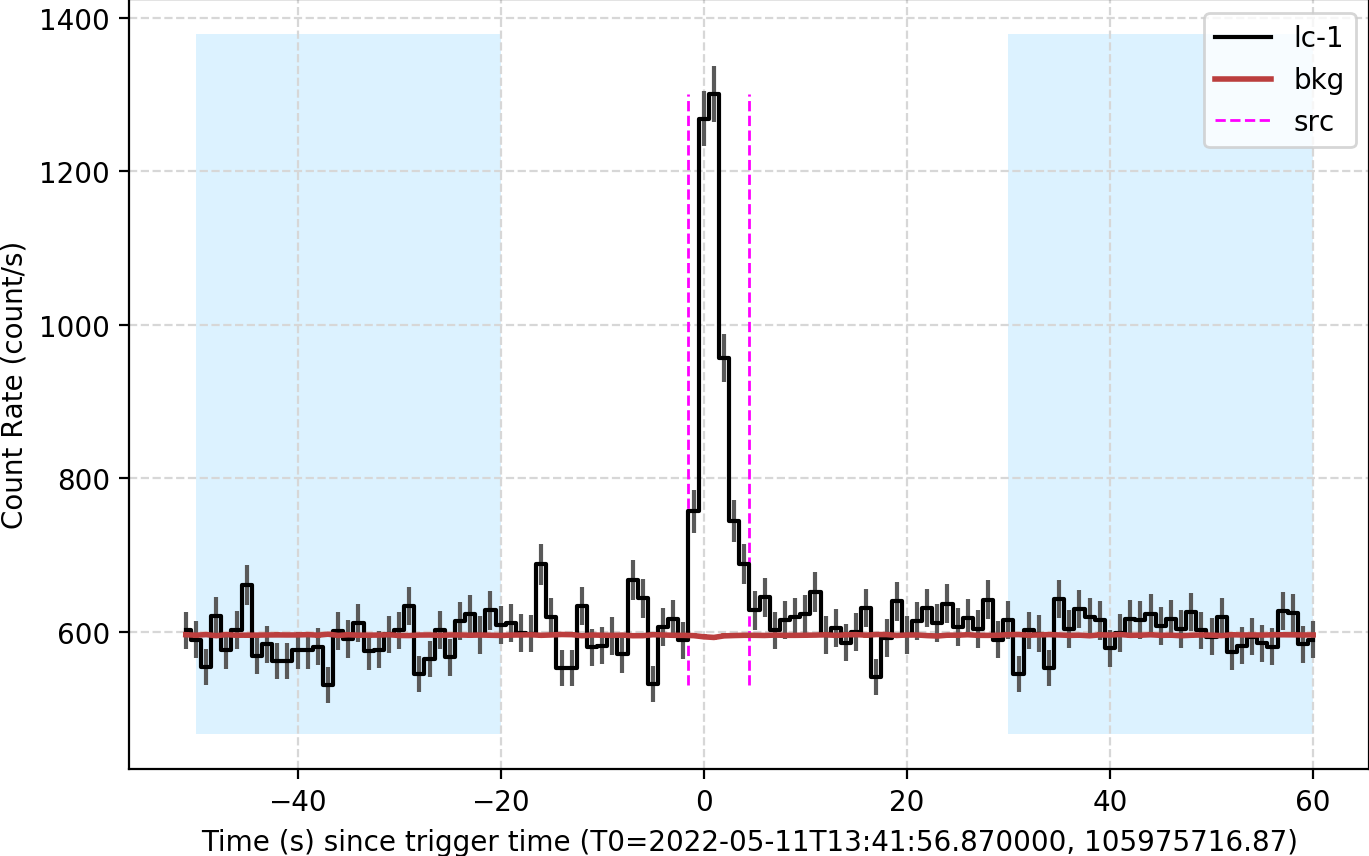}}
        \quad
        \subfigure[]{\includegraphics[height=3.0cm]{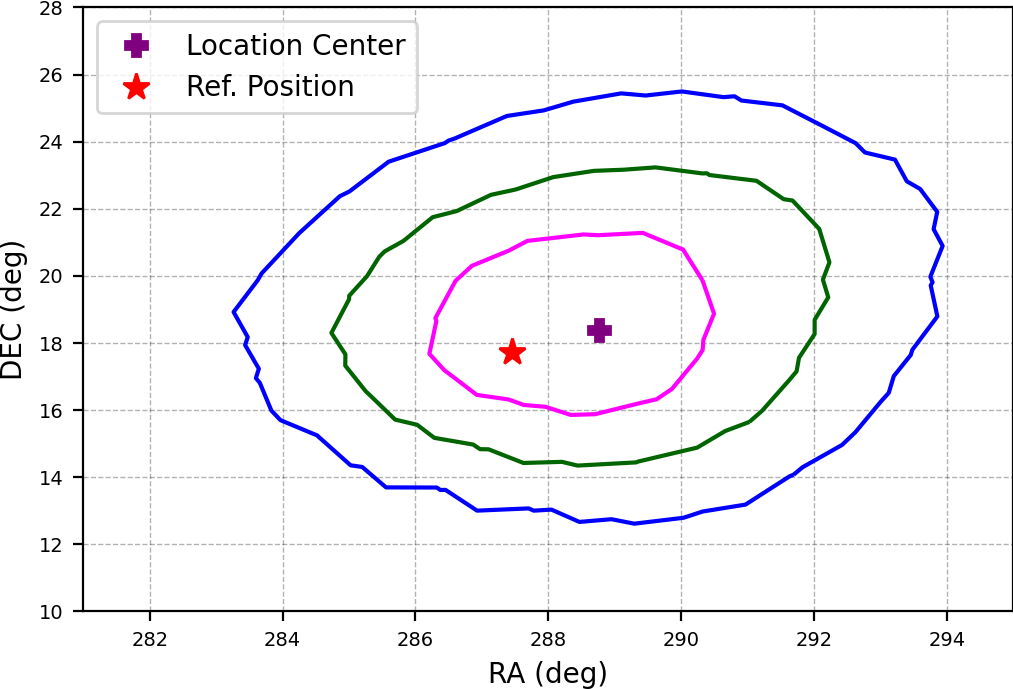}}
        \quad
        \\
        \subfigure[]{\includegraphics[height=3.0cm]{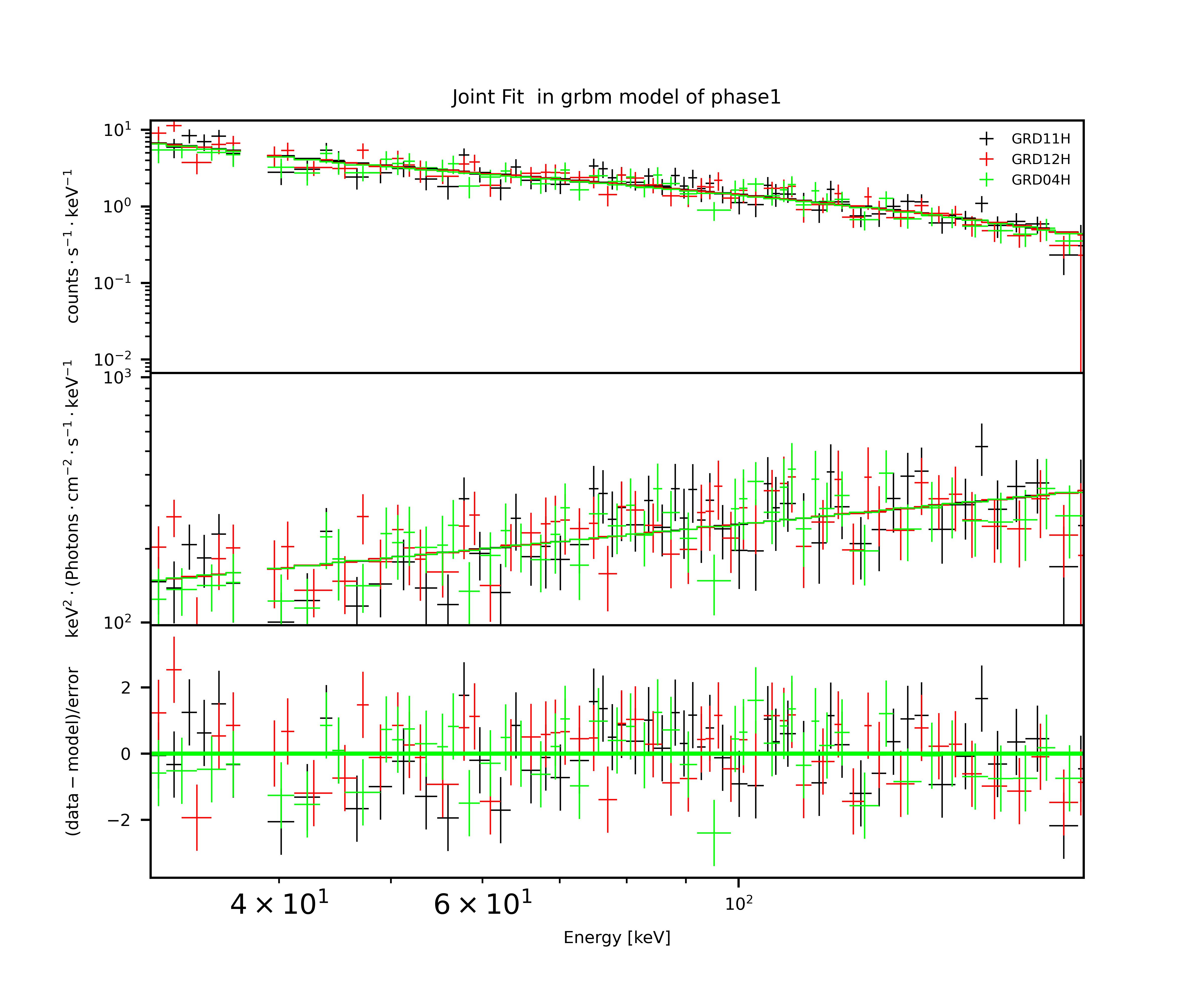}}
        \quad
        \subfigure[]{\includegraphics[height=3.0cm]{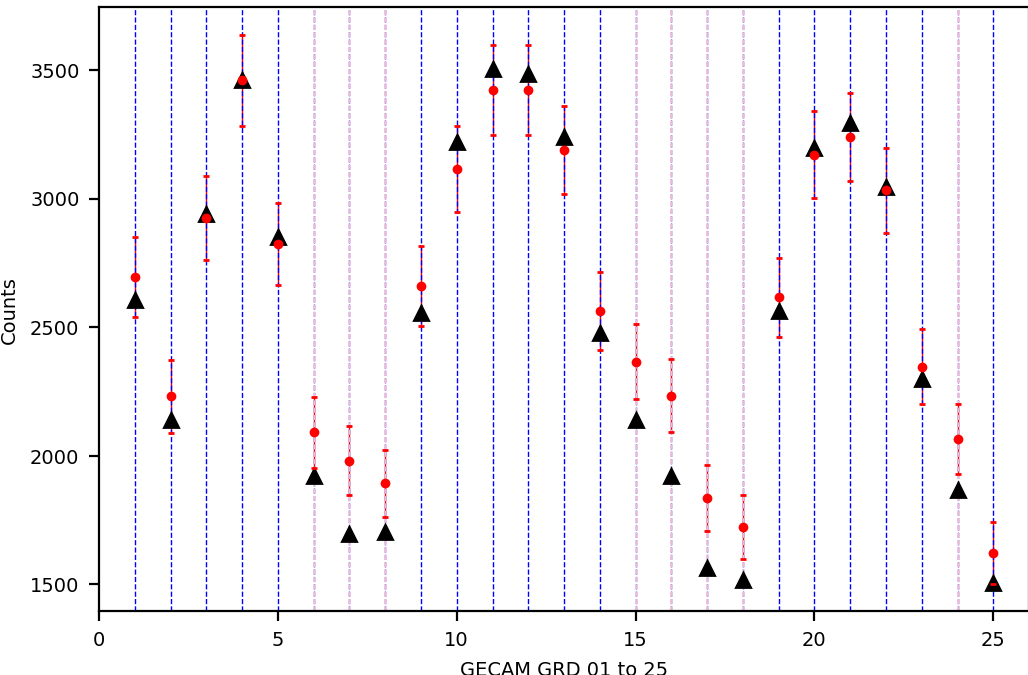}}
        \quad
        \subfigure[]{\includegraphics[height=3.0cm]{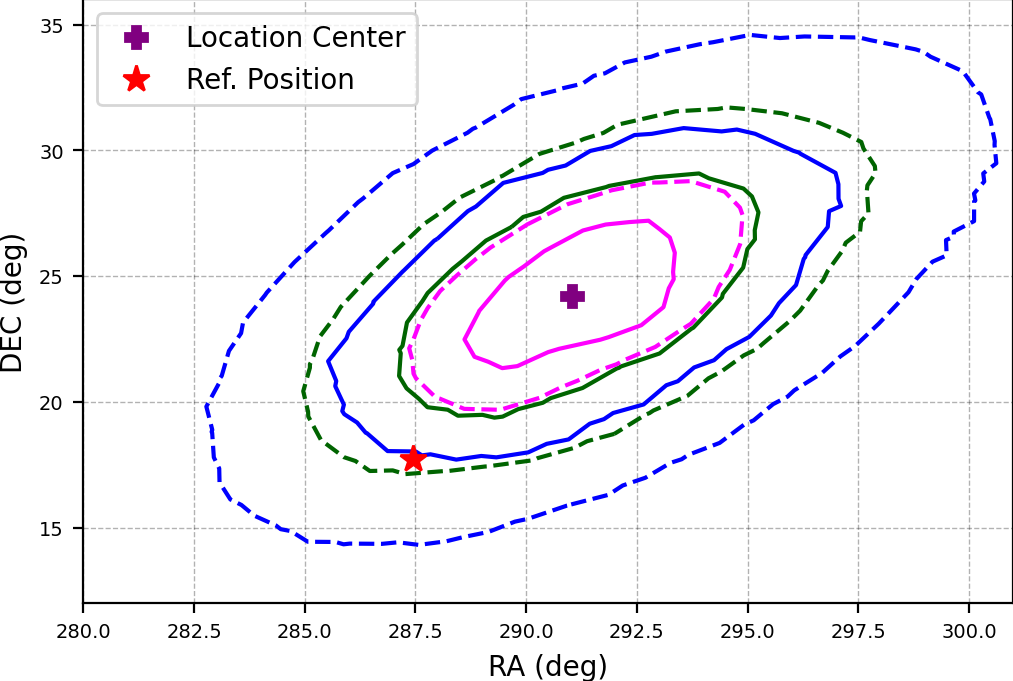}}
        \quad
        \caption{ GECAM localization for GRB 220511A (STEP 1 to 3 in Section 3). (a) The light curve of GRD \# 21 high gain which contains the majority of net (burst) counts. The vertical magenta dashed lines and blue shadow regions are the manually selected source interval and background interval for localization analysis. The horizontal red line represents the background and its uncertainty. (b) The localization credible region with fixed templates (see Table \ref{TABLE_FixedTemplate}) for all 25 GRDs. The solid magenta[green][blue] contour represent the 68.27\%[95.45\%][99.73\%] location HPD statistical credible region. The purple cross is the location center while the red star is the reference position. (c) The refined spectral fitting result for the location center of FIX localization with cutoffpl. The three GRDs (GRD \# 04, 11, and 12) within $60^{\circ}$ from the location center of FIX and with detection significance $>5\sigma$ are used for the spectral fitting. (d) The consistency check of the measured data and the expected counts (the sum of estimated background and expected source counts). The source contribution is derived from the refined spectrum. The detectors with derivation $<$ 3 $\sigma$ are used for localization with the refined spectral template. The black triangle markers are the expected counts for each detector. The red circles and error bars are the measured counts and 3 $\sigma$ data errors for each detector. The blue lines show the selected detectors. (e) The location HPD credible region with the refined template for selected detectors. The dashed magenta[green][blue] contour is the 68.27\%[95.45\%][99.73\%] total location error (see Section 5). }
        \label{fig2a}
    \end{figure*}

    \begin{figure*}[htb]
        \centering
        \subfigure[]{\includegraphics[height=3.0cm]{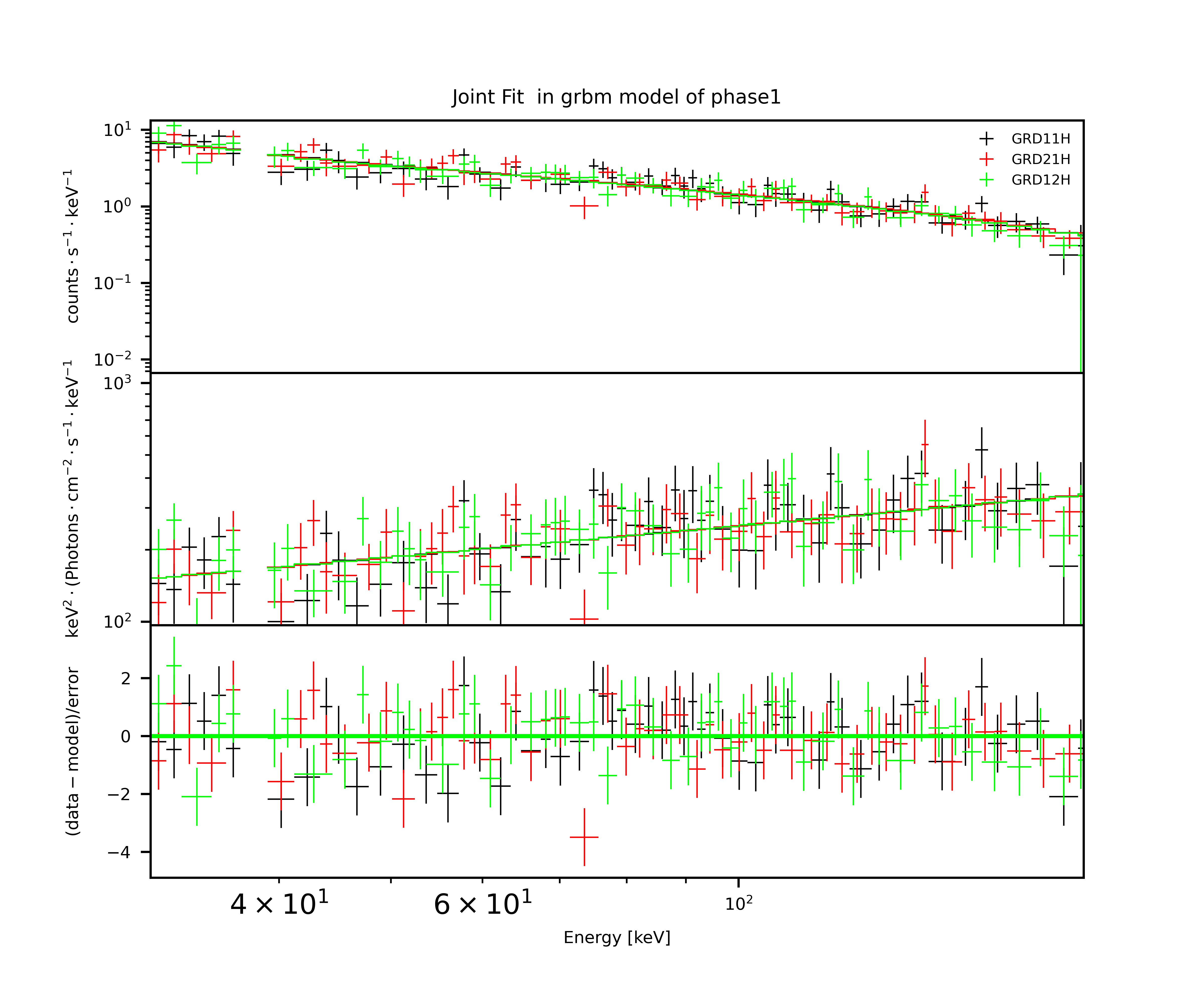}}
        \quad
        \subfigure[]{\includegraphics[height=3.0cm]{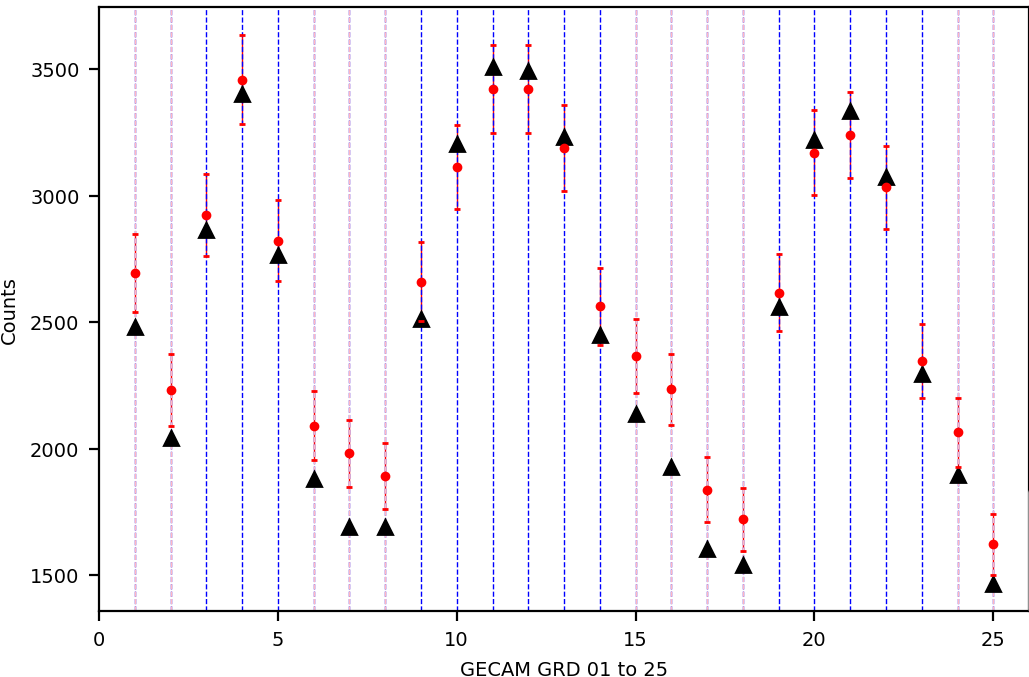}}
        \quad
        \subfigure[]{\includegraphics[height=3.0cm]{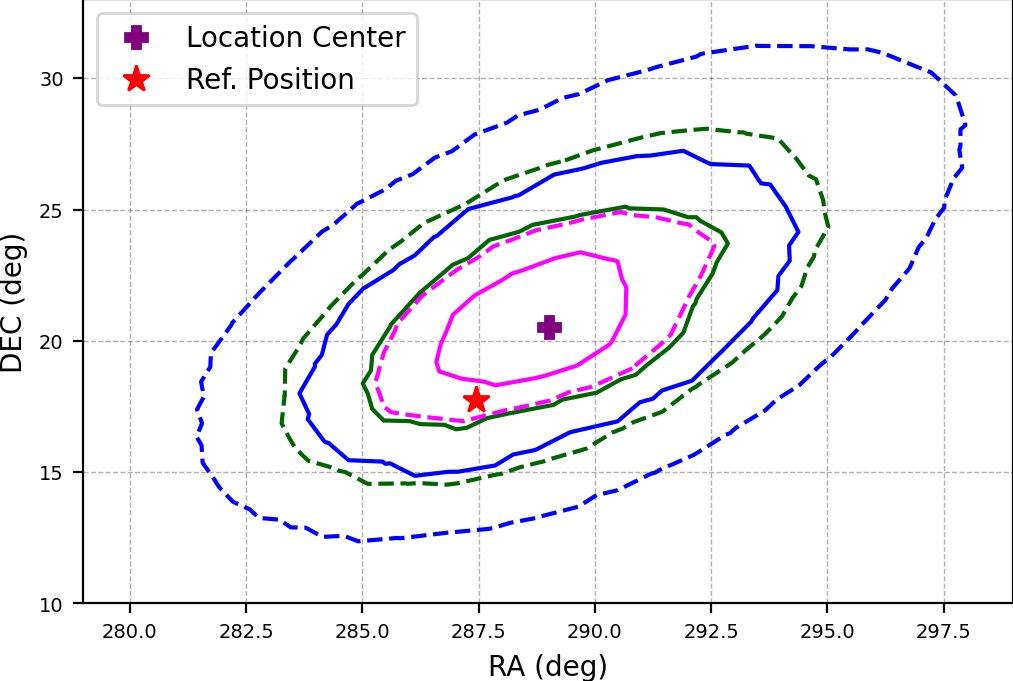}}
        \quad
        \\
        \subfigure[]{\includegraphics[height=2.5cm]{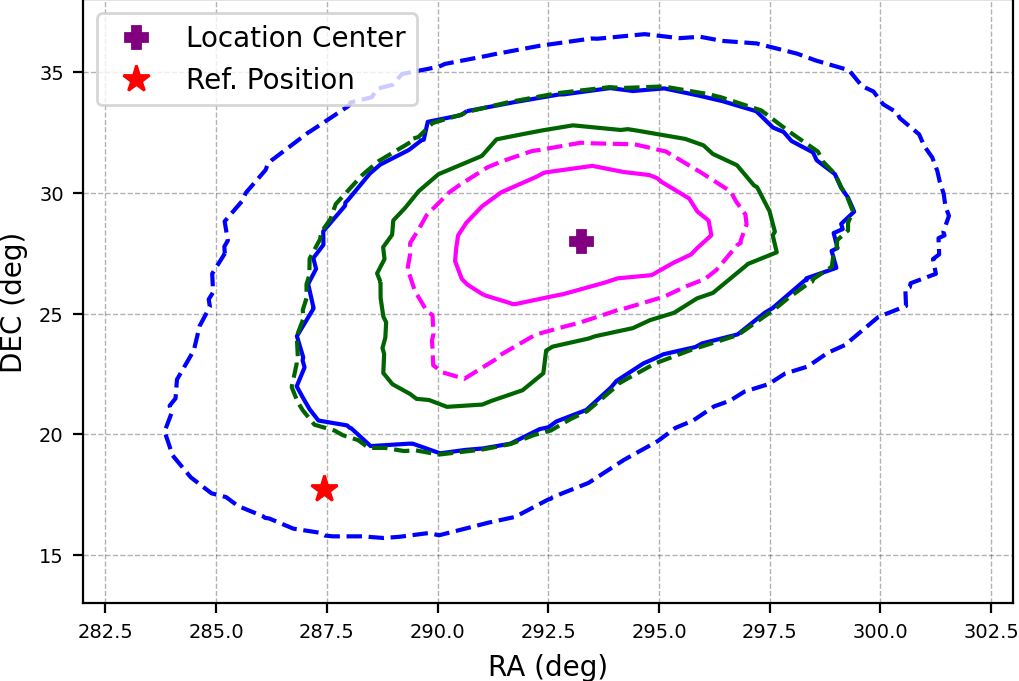}}
        \quad
        \caption{ GECAM localization for GRB 220511A (STEP 4 in Section 3). (a) The spectral fitting for the \textbf{accurate position based on the reference location}. (b) The consistency check of the measured data and the expected counts (the sum of estimated background and expected source counts). The source contribution is derived from this spectral fitting results. (c) The location HPD credible region with this spectral template for selected detectors. (d) To illustrate that using all detectors will introduce localization derivation, the location credible region with this spectral parameter for all 25 GRDs is shown. }
        \label{fig2b}
    \end{figure*}

    \begin{figure*}[htb]
        \centering
        \subfigure[]{\includegraphics[height=6.0cm]{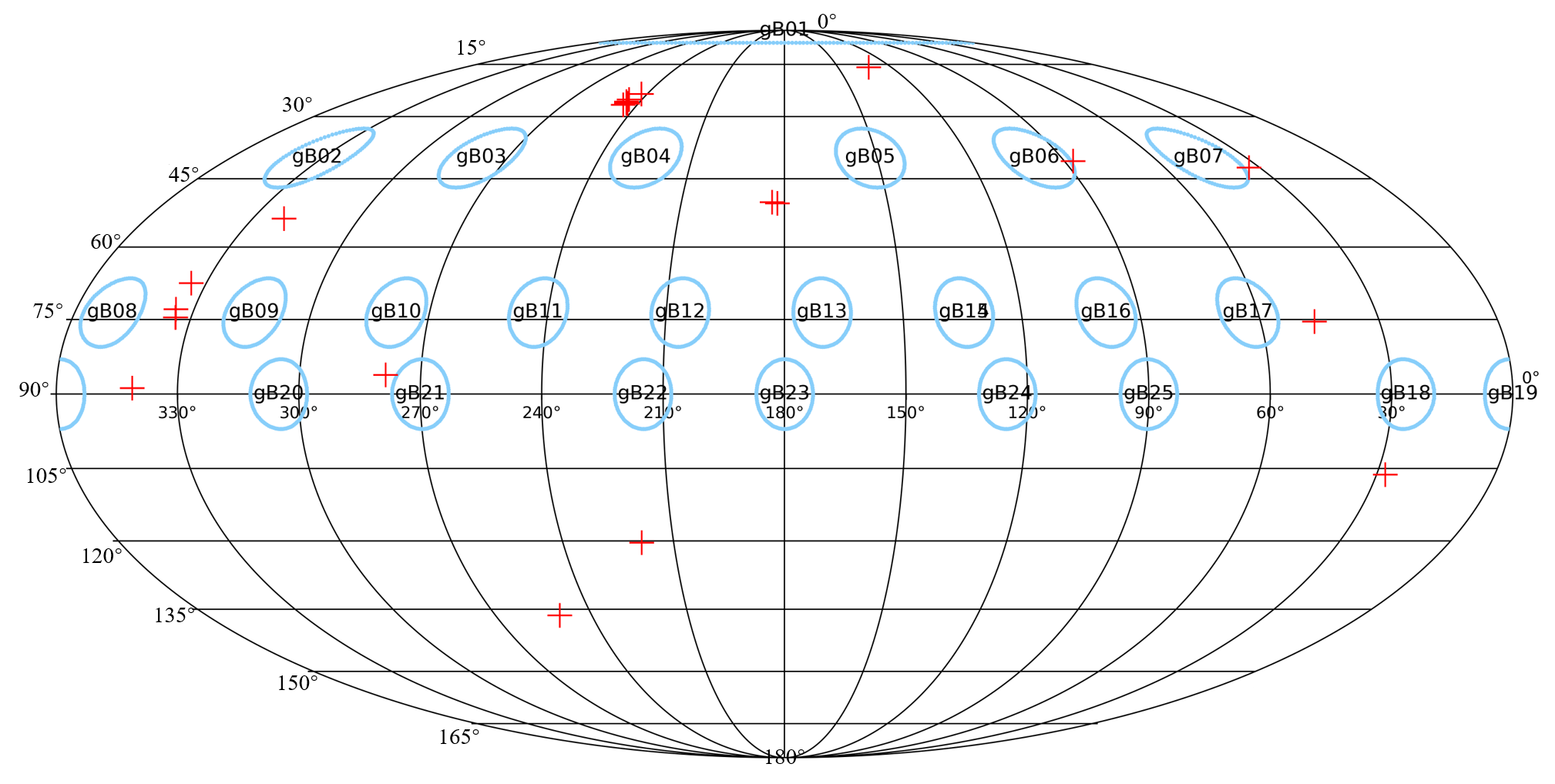}}
        \quad
        \caption{ The distribution of the incidence zenith/azimuth angle for the reference locations of 23 bright bursts (red crosses) in GECAM payload coordinates. }
        \label{fig3a}
    \end{figure*}

    \begin{figure*}[htb]
        \centering
        \subfigure[]{\includegraphics[height=6.0cm]{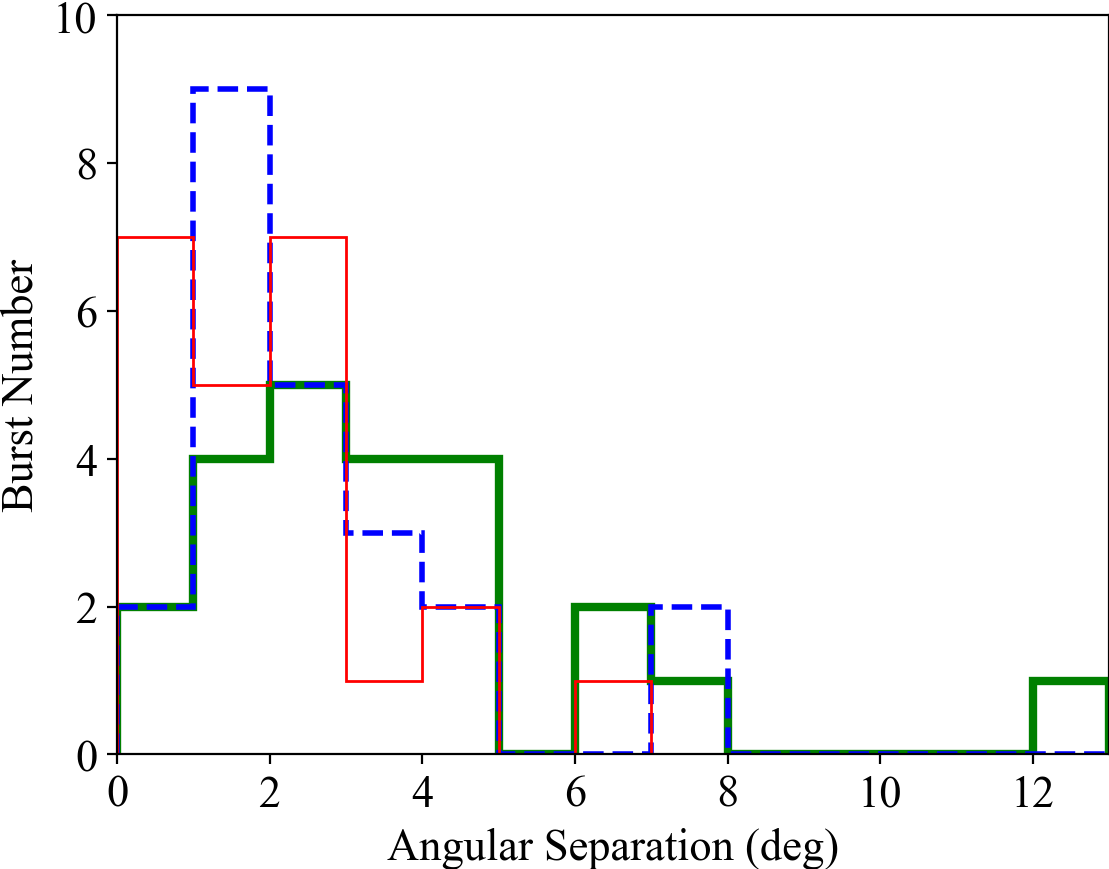}}
        \quad
        \caption{ The distribution of the angular separation between the reference positions and the location centers of FIX (green), RFD (blue), and APR (red) localization. }
        \label{fig3b}
    \end{figure*}

    \begin{figure*}[htb]
        \centering
        \includegraphics[width=\columnwidth]{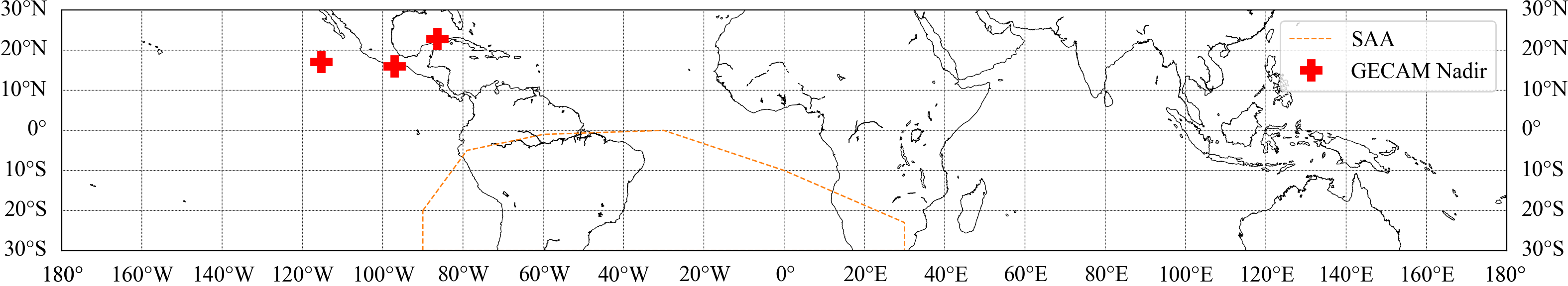}
        \caption{ Geographic distribution of the GECAM nadir points of three bright TGFs (red crosses) which are used for localization analysis. }
        \label{fig4a}
    \end{figure*}

    \begin{figure*}
        \centering
        \subfigure[]{\includegraphics[width=10cm]{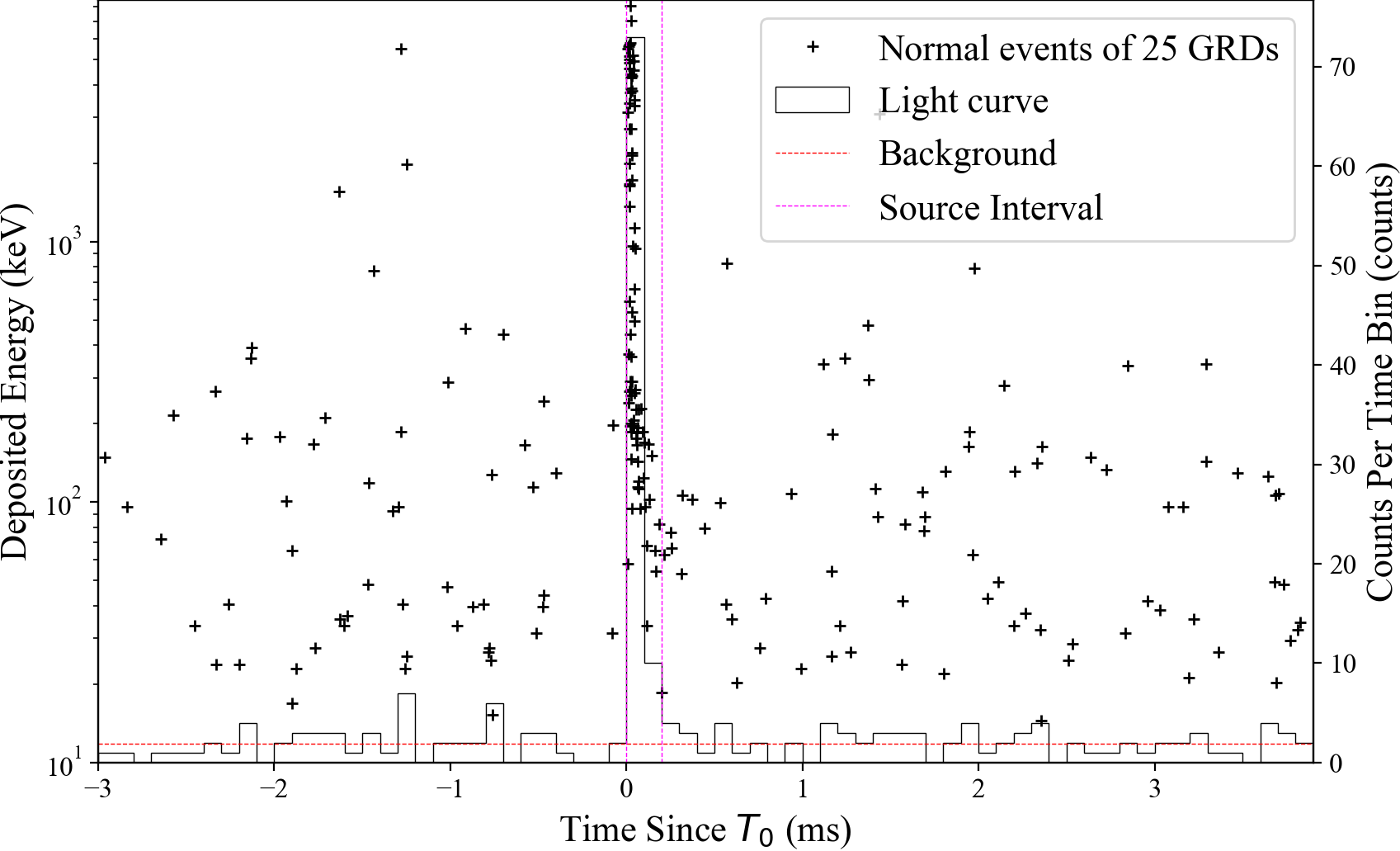}}
        \quad
        \subfigure[]{\includegraphics[width=10cm]{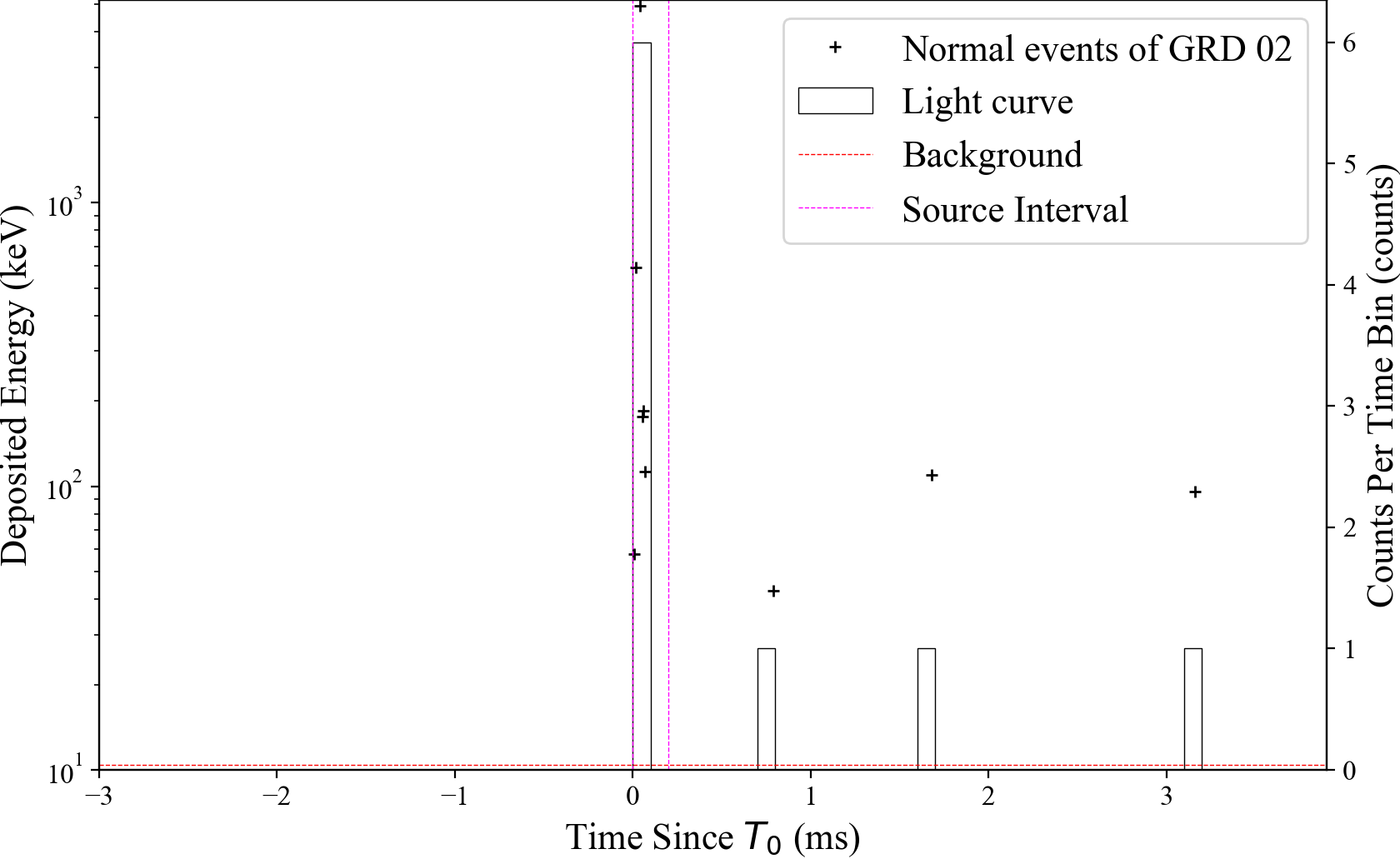}}
        \quad
        \subfigure[]{\includegraphics[width=10cm]{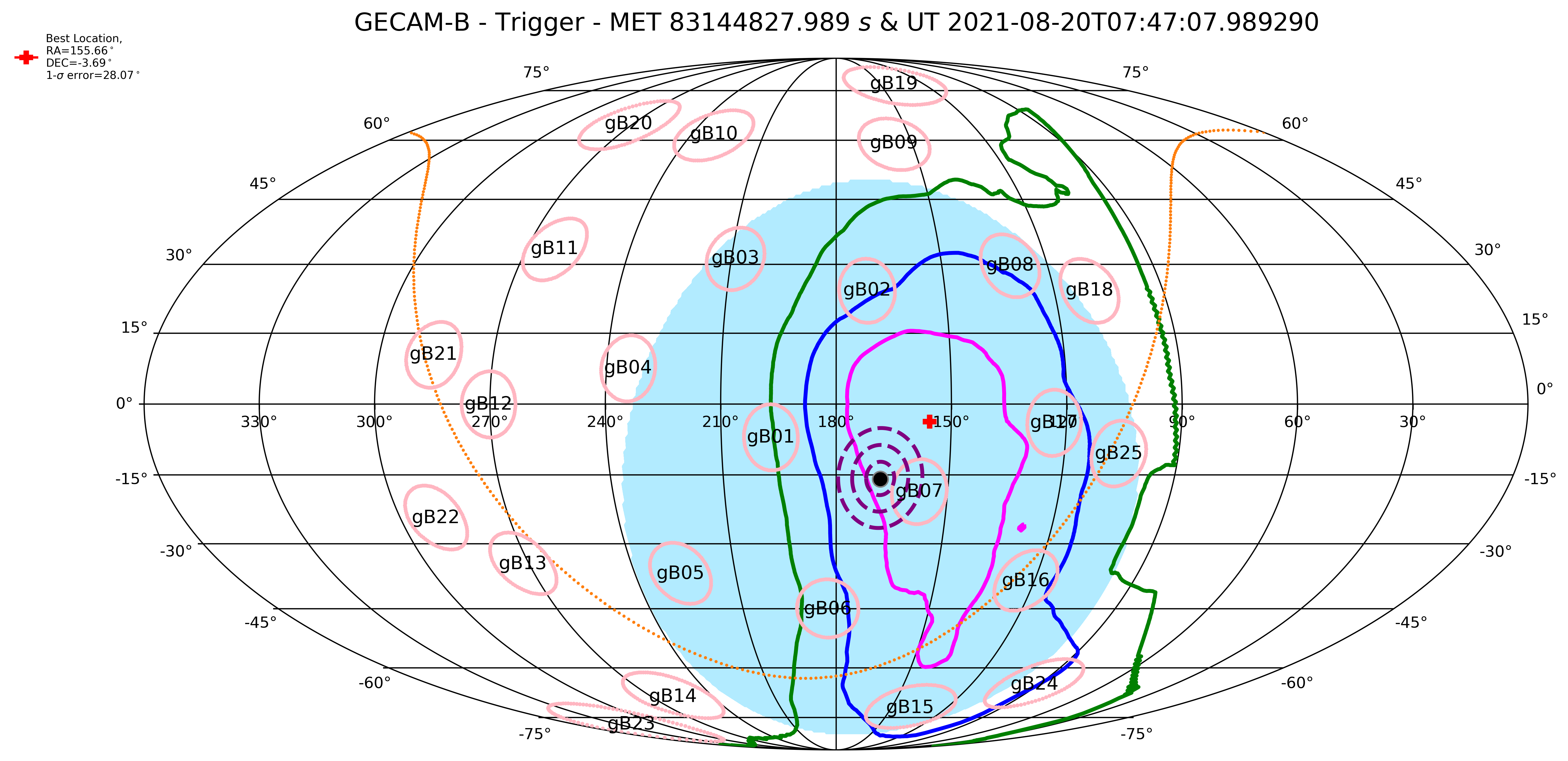}}
        \quad
        \caption{ GECAM localization for TGF 20210820. (a) The light curve and time-energy scatter plot of all 25 detectors. (b) The light curve and time-energy scatter plot of GRD \# 02 which contains the majority of net counts in the discovery bin. The black marker shows the normal events of GRDs. The dashed red line is the background and the dashed fuchsia line is the source interval. (c) The location sky map. The solid fuchsia[blue][green] contour is the 68.27\%[95.45\%][99.73\%] location HPD statistical credible region. The geocentric is inside the 95.45\% credible region which indicates its HPD cumulative probability $< 95.45\%$. The dashed purple curves are 400/800/1200 km equidistant lines from GECAM nadir. The red cross and black circle marker are the location center and geocentric, respectively. The blue shadow is Earth apparent radius. }
        \label{fig4b}
    \end{figure*}

    \begin{figure*}
        \centering
        \subfigure[]{\includegraphics[height=3.5cm]{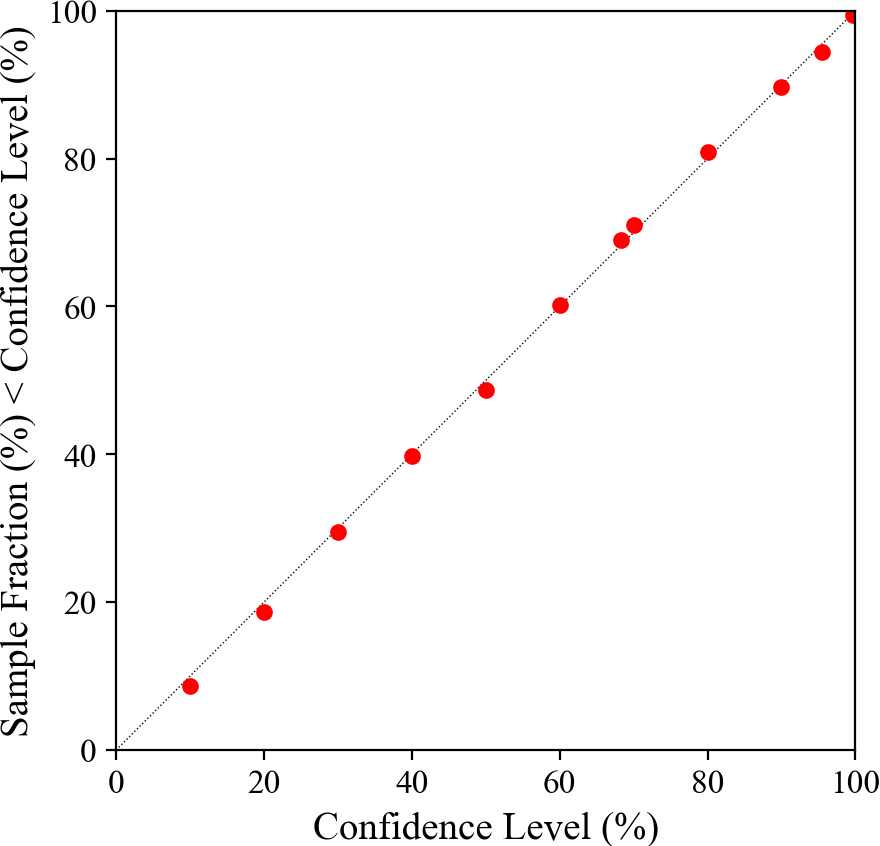}}
        \quad
        \subfigure[]{\includegraphics[height=3.5cm]{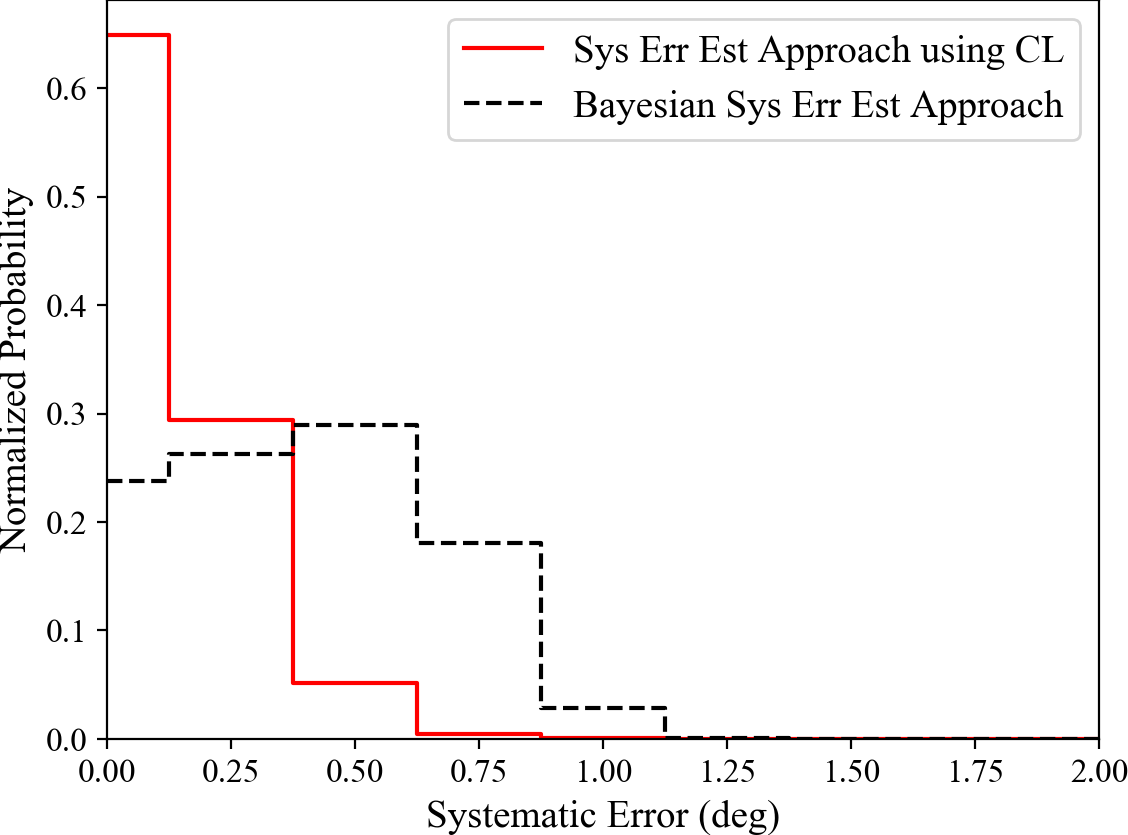}}
        \quad
        \subfigure[]{\includegraphics[height=3.5cm]{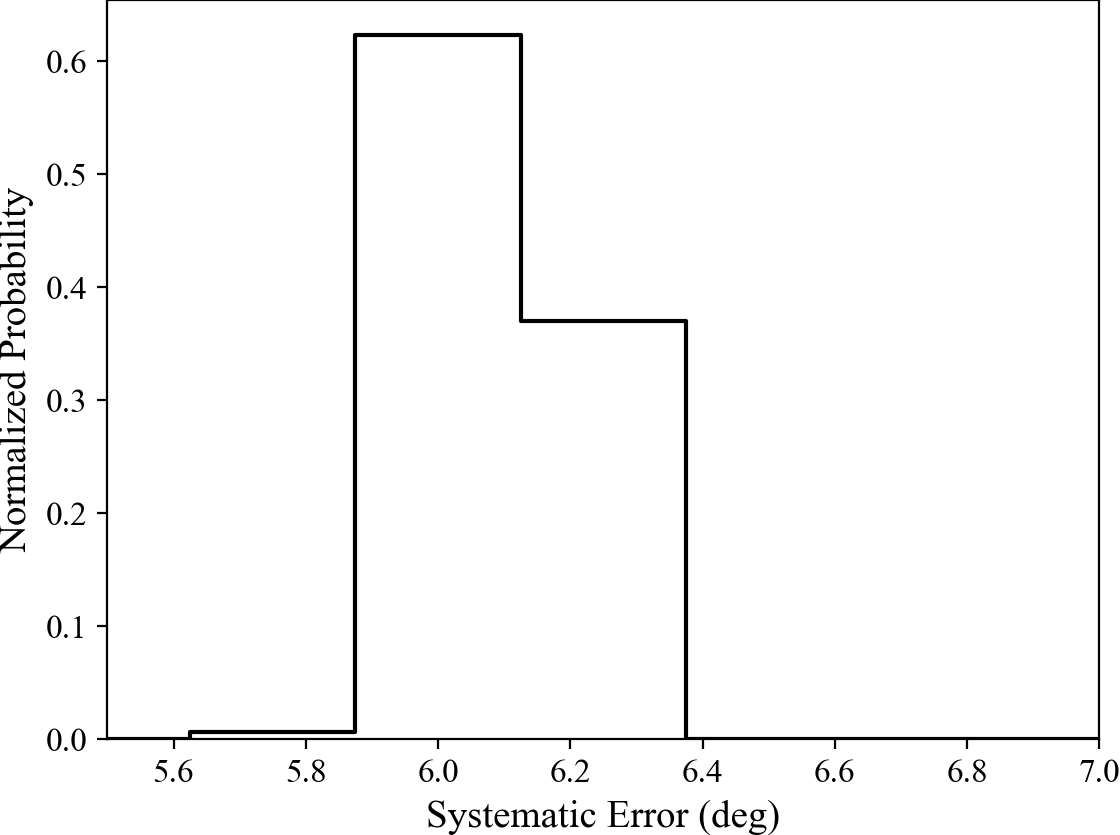}}
        \quad
        \subfigure[]{\includegraphics[height=3.5cm]{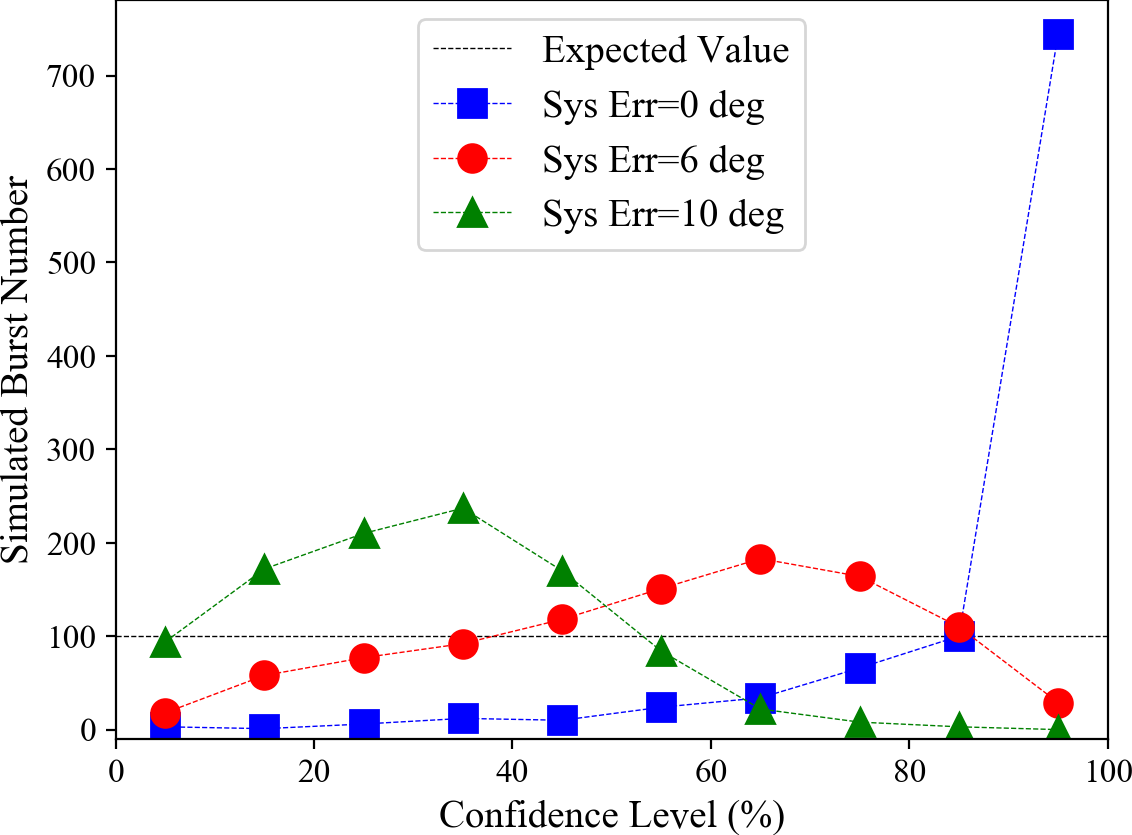}}
        \quad
        \subfigure[]{\includegraphics[height=3.5cm]{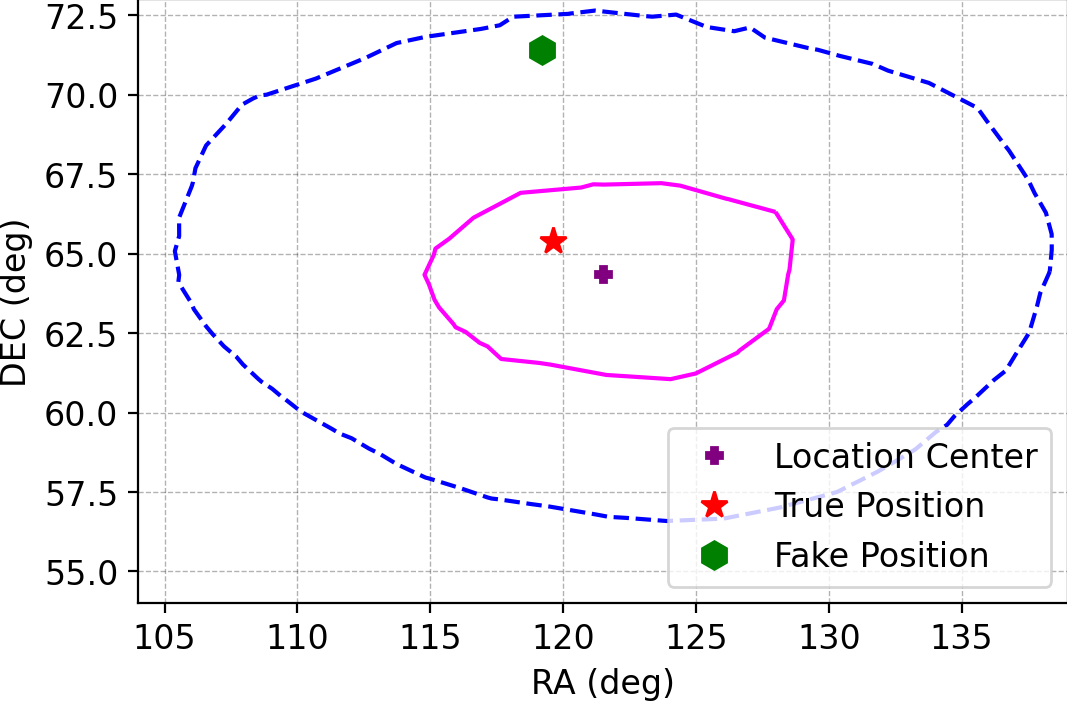}}
        \quad
        \caption{ (a) The HPD cumulative confidence level distribution (i.e. the simulated bursts' fraction within the cumulative probability $N\%$ for truth position $<$ the corresponding confidence level $N\%$). The dashed line represents the one-one line. The confidence level points are $10\%$ to $90\%$ step by $10\%$ as well as $68.27\%$, $95.45\%$ and $99.73\%$. (b) The probability distribution of systematic error estimation for the Bayesian approach and our confidence-based estimation approach using the simulated bursts only consist of statistical error. (c) The probability distribution of systematic uncertainties estimation for our confidence-based estimation approach using the simulated bursts only contains a statistical error. But the "true position" used for calculating cumulative probability adjusts to a "fake position" ($\rm RA=119.23^{\circ}$, $\rm DEC=+71.42^{\circ}$) which deviates by 6 deg away from the true position. (d) The distribution of simulated bursts number in each confidence level bin for different systematic error values. (e) Inspection of an individual simulated burst. The magenta and blue lines mark $68.27\%$ credible regions of statistical uncertainty and total (statistical+systematic) uncertainty, respectively. The purple cross, red star, and green hexagon represent the location center, true position, and fake position, respectively. }
        \label{fig5}
    \end{figure*}

    \begin{figure*}
        \centering
        \subfigure[]{\includegraphics[height=3.5cm]{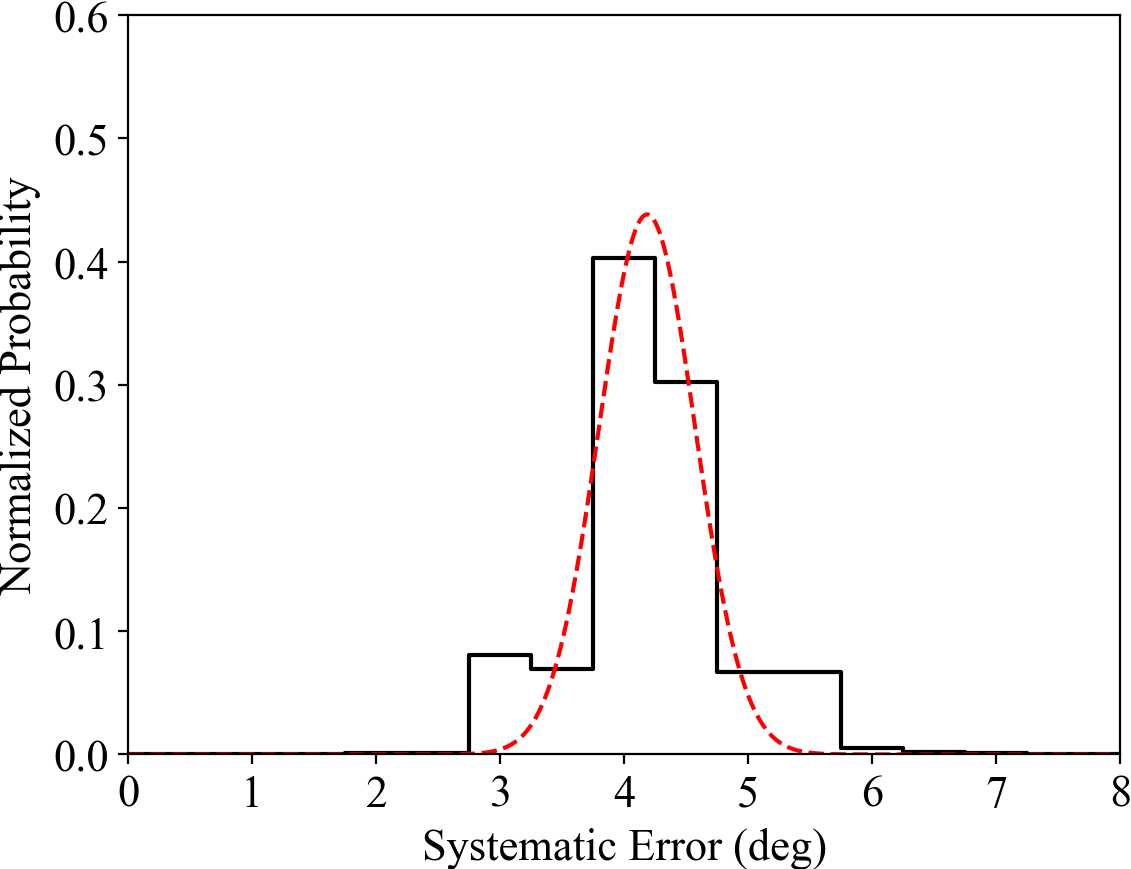}}
        \quad
        \subfigure[]{\includegraphics[height=3.5cm]{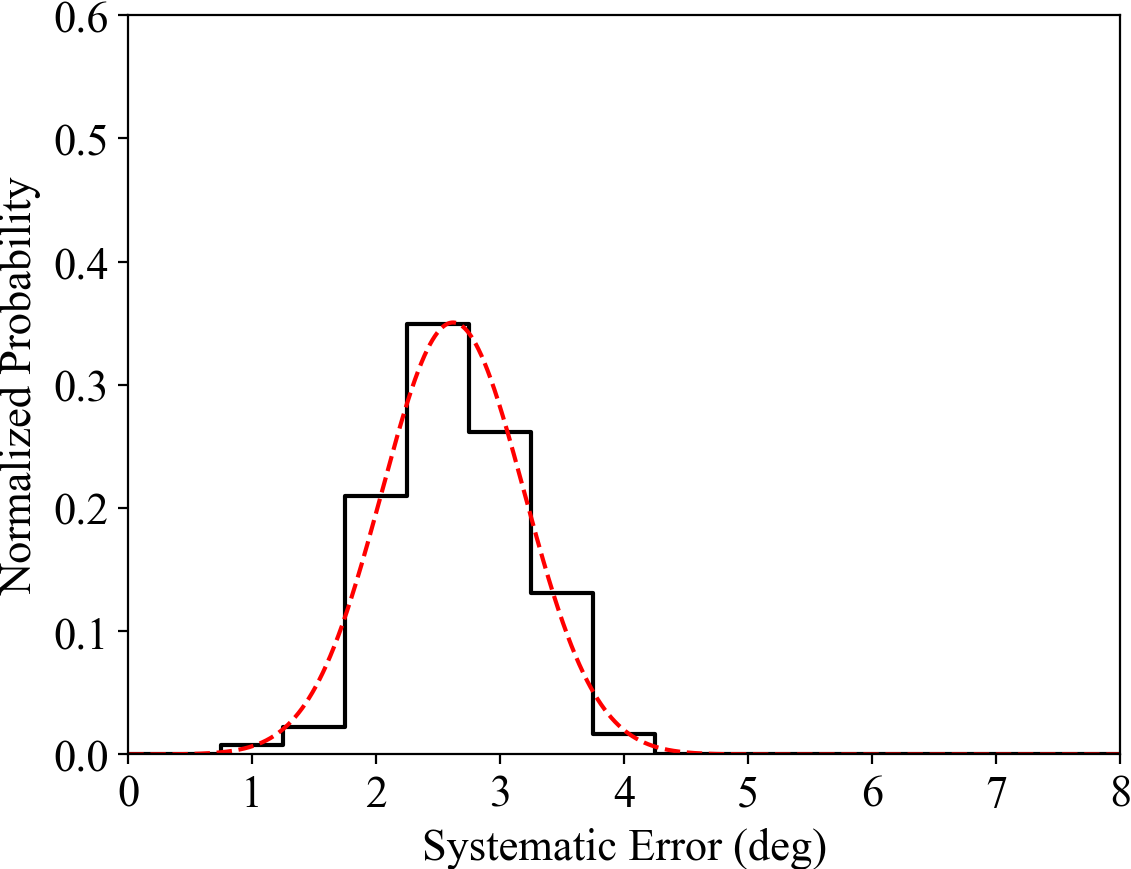}}
        \quad
        \subfigure[]{\includegraphics[height=3.5cm]{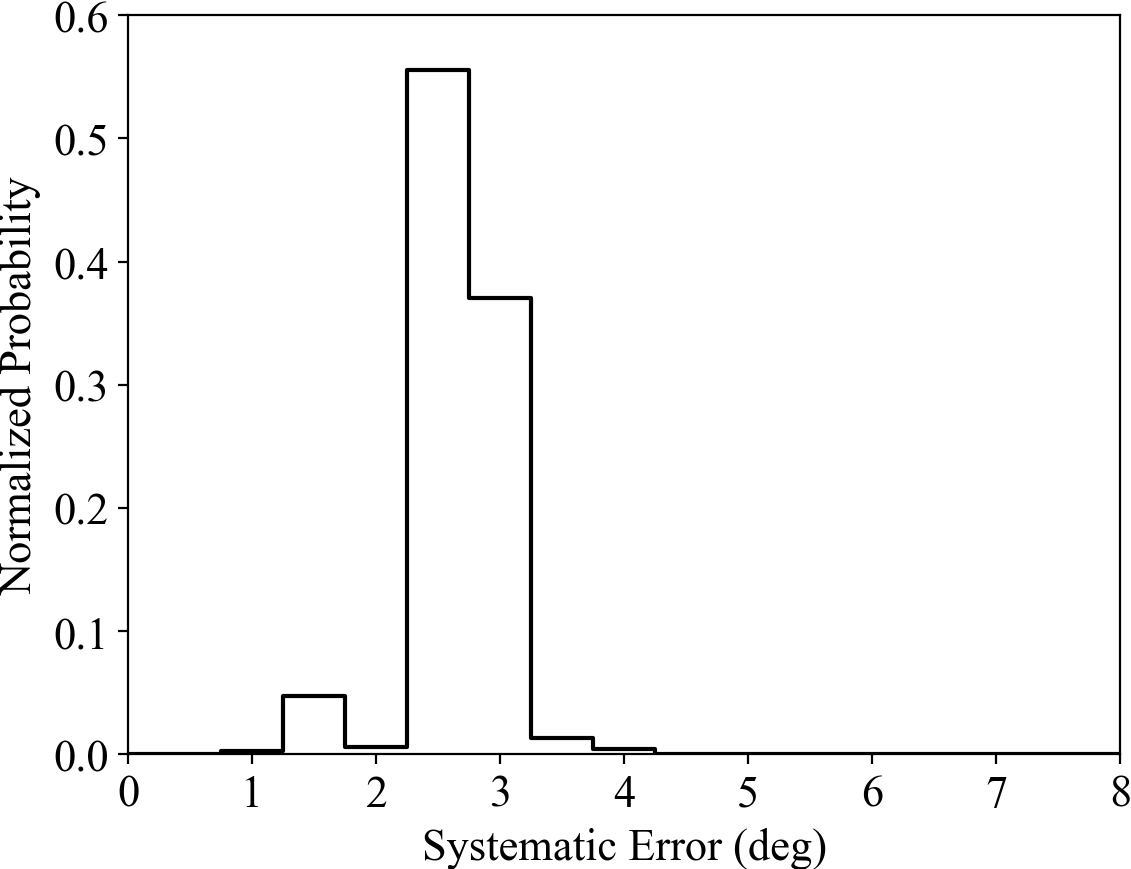}}
        \quad
        \\
        \subfigure[]{\includegraphics[height=3.5cm]{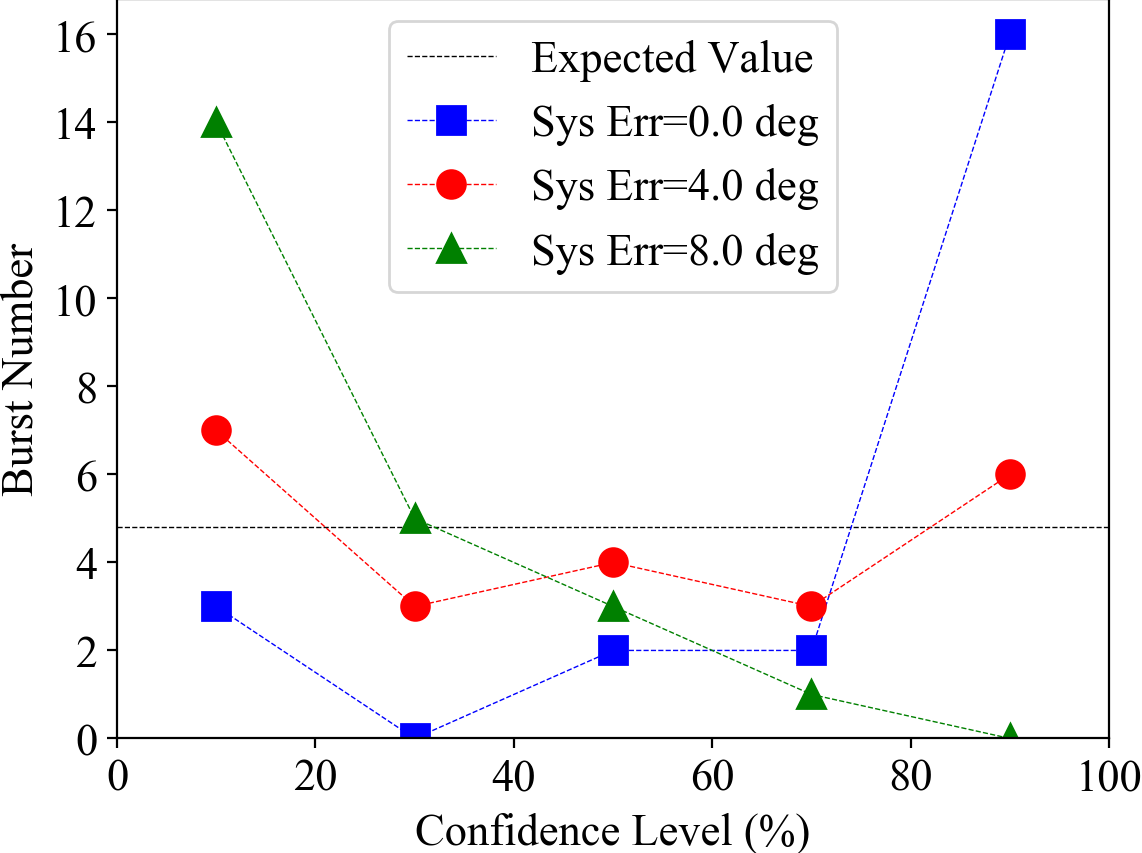}}
        \quad
        \subfigure[]{\includegraphics[height=3.5cm]{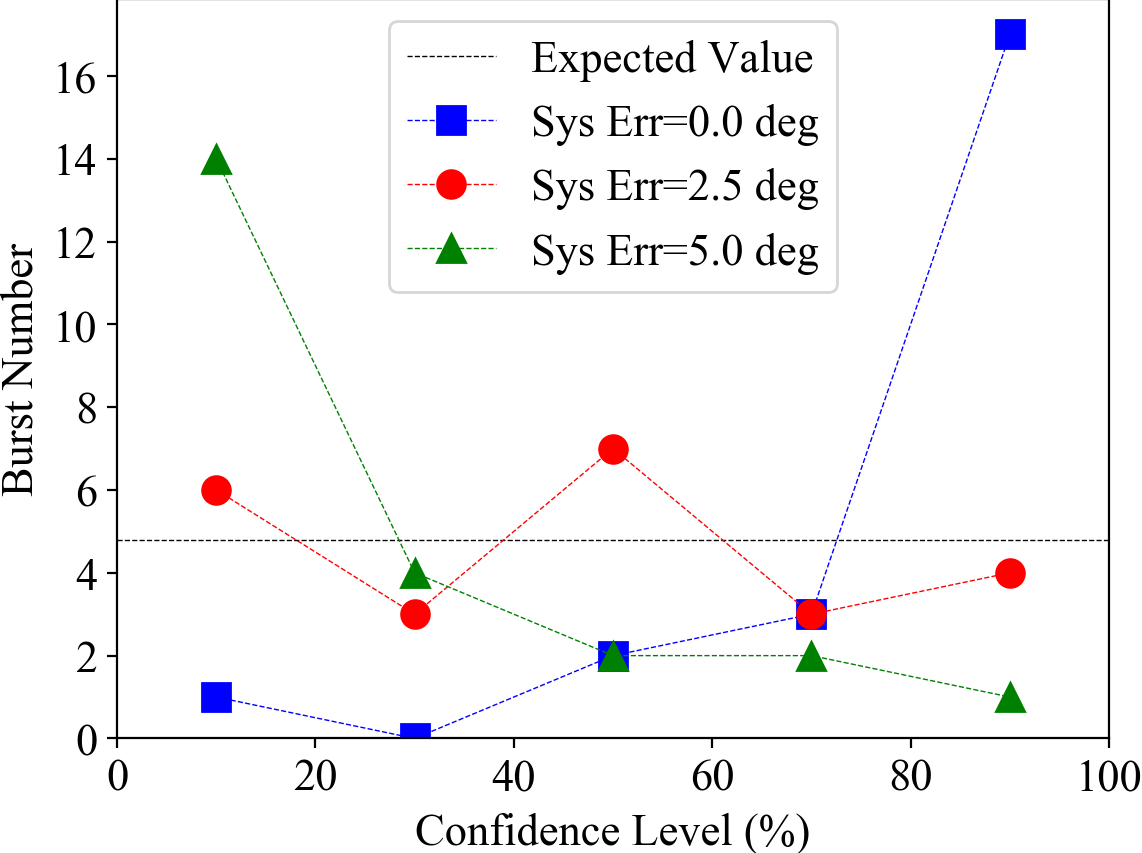}}
        \quad
        \subfigure[]{\includegraphics[height=3.5cm]{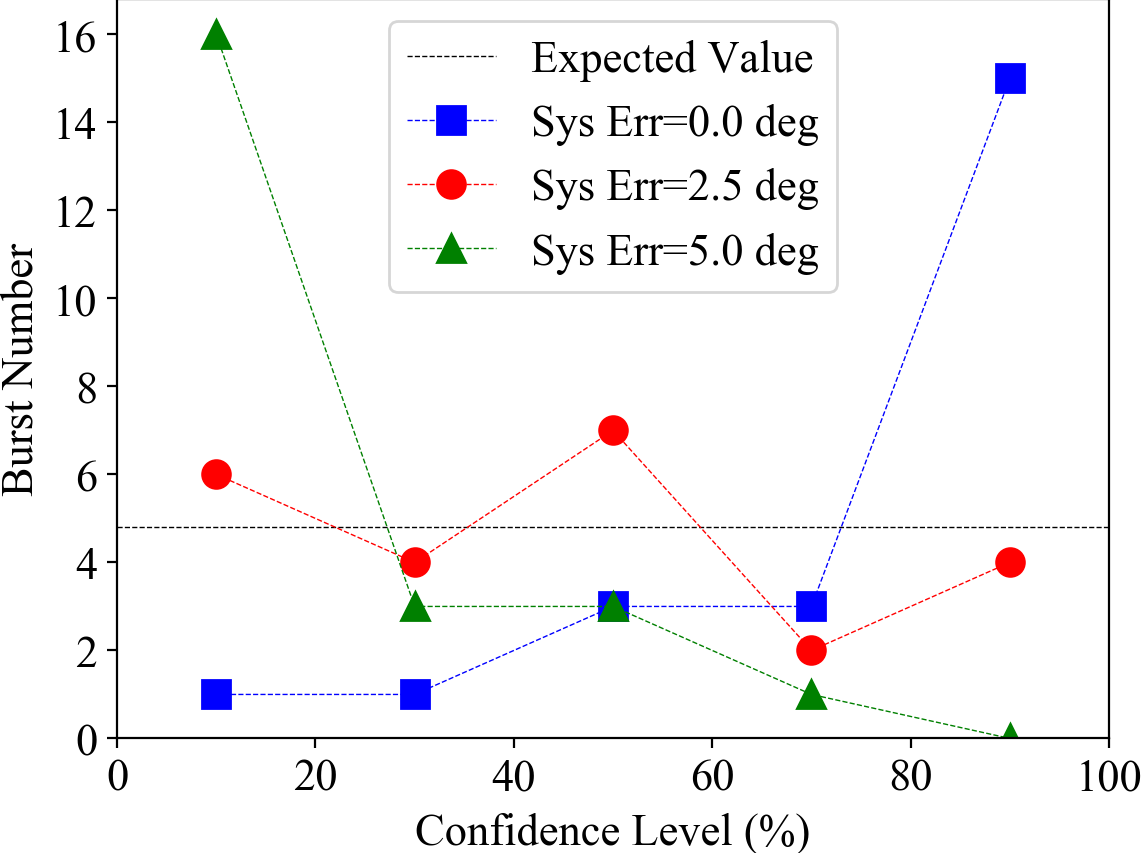}}
        \quad
        \caption{ The distribution of systematic uncertainties probability and bursts number in each confidence level bin for real observation of GECAM. (a) The probability distribution of systematic uncertainties for FIX localization. The dashed red curve shows a Gaussian fitting ($\mu$=4.18, $\sigma$=0.39). (b) The probability distribution of systematic uncertainties for RFD localization. The dashed red curve shows a Gaussian fitting ($\mu$=2.62, $\sigma$=0.58). (c) The probability distribution of systematic uncertainties for APR localization. We do not fit this distribution due to the few significant data points. The distribution of bursts number in each confidence level bin for different systematic error values with (d) FIX (e) RFD, and (f) APR localization is shown in the bottom panel. }
        \label{fig6}
    \end{figure*}

    \begin{table}
        \centering
        \caption{ Characteristics of the burst used in the localization simulation. The incident angle is $\rm Zenith=5.85^{\circ}$, $\rm Azimuth=22.50^{\circ}$ in GECAM payload coordinates which corresponds to $\rm RA=119.63^{\circ}$, $\rm DEC=+65.40^{\circ}$ (true position) at 2021-07-22T01:00:00 UTC. The background level is set to 1000 counts/s for each GRD, which is about the similar level as measured in orbit. Fluence is calculated in 10-1000 keV. }
        \label{TABLE_SimulationSetting}
        \begin{tabular}{cccccc}
            Source Intensity Type          & Medium Bright Burst \\
            \hline
            Spectral Model                 & Comptonized         \\
            Spectral Index                 & 1.50                \\
            $E_{\rm peak}$ (keV)           & 500                 \\
            Duration (s)                   & 10.0                \\
            Fluence $(\rm erg/\rm cm^{2})$ & $2.17\times10^{-5}$ \\
            \hline
        \end{tabular}
    \end{table}

    \begin{table}
        \centering
        \caption{ The fixed spectral parameters of the localization templates used in this research. }
        \label{TABLE_FixedTemplate}
        \begin{tabular}{cccccc}
            Templates & Model & Low-energy Index & $E_{peak}$ (keV) \\
            \hline
            Soft     & Comptonized function & -1.95 &   50 \\
            Moderate & Comptonized function & -1.15 &  350 \\
            Hard     & Comptonized function & -0.25 & 1000 \\
            \hline
        \end{tabular}
    \end{table}

    \begin{table*}[] 
        \caption{ List of GECAM localization results for flight location and Beidou ground location. Due to not triggered onboard or software failure, the flight location and Beidou ground location for some bursts are lacking. }
        \label{TABLE_LocRes_Part1}
        \centering
        \tiny
        \begin{tabular}{cccccccccccccccccccccccccccccc}
            \hline
            Burst & Trigger Time & \multicolumn{4}{c}{Flight Location} & \multicolumn{4}{c}{BD Ground Location} & \multicolumn{5}{c}{Reference Location} \\
            \cmidrule(r){3-6}  \cmidrule(r){7-10} \cmidrule(r){11-15}
            Name & (UT) & RA(${}^{\circ}$) & DEC(${}^{\circ}$) & ERR(${}^{\circ}$) & $\alpha$(${}^{\circ}$) & RA(${}^{\circ}$) & DEC(${}^{\circ}$) & ERR(${}^{\circ}$) & $\alpha$(${}^{\circ}$) & RA(${}^{\circ}$) & DEC(${}^{\circ}$) & $\theta$(${}^{\circ}$) & $\phi$(${}^{\circ}$) & Reference \\
            \hline

            GRB 210121A & 2021-01-21T18:41:48.800 &  22.1 & -49.5 &  1.3 &  4.6 &  23.3 & -47.7 & 2.0 & 4.5 &  17.0 & -46.4 &  40.9 & 267.2 & IPN$^1$ \\
            GRB 210511B & 2021-05-11T11:26:40.600 & 318.0 & +59.5 &  3.2 &  2.9 & 320.3 & +60.1 & 3.7 & 4.1 & 312.9 & +58.4 & 120.4 & 38.6  & IPN$^2$ \\
            GRB 210606B & 2021-06-06T22:41:08.100 &  85.5 & -16.5 &  1.0 &  2.4 &  87.8 & -18.3 & 1.4 & 1.6 &  88.0 & -16.7 &  15.9 & 312.7 & IPN$^3$ \\
            GRB 210619B & 2021-06-20T00:00:00.950 & 334.7 & +28.6 & 31.0 & 13.9 & 318.8 & +29.2 & 7.4 & 4.8 & 319.7 & +33.9 & 136.4 & 70.0  & \textit{Swift}-XRT$^4$ \\
            GRB 210822A & 2021-08-22T09:18:18.000 & 310.3 &  +4.5 &  1.0 &  5.9 & 298.3 &  +1.0 & 1.2 & 7.5 & 304.4 &  +5.3 & 106.2 & 207.8 & \textit{Swift}-XRT$^5$ \\
            GRB 210927B & 2021-09-27T23:54:45.600 & 240.3 & +69.5 &  9.9 &  8.3 & 249.6 & +70.4 & 3.3 & 5.3 & 263.0 & +73.8 &  42.5 & 213.4 & IPN$^6$ \\
            GRB 211120A & 2021-11-20T23:05:20.600 & 311.3 & +41.3 &  1.0 &  3.2 & 305.0 & +41.4 & 1.3 & 7.6 & 315.1 & +42.9 &  53.9 & 141.2 & IPN$^7$ \\
            GRB 220511A & 2022-05-11T13:41:56.800 & 286.8 & +15.3 &  1.3 &  2.5 & 290.1 & +22.2 & 2.1 & 5.2 & 287.4 & +17.7 &  75.4 & 226.4 & \textit{Swift}-XRT$^8$ \\
            GRB 220514A & 2022-05-14T12:24:32.950 & 139.8 & +19.1 &  3.5 &  9.7 & 143.4 & +12.6 & 4.8 & 4.2 & 147.7 & +13.1 &  86.2 & 98.7  & INTEGRAL-IBAS$^9$ \\

            SFL 210508       & 2021-05-08T18:30:39.950 &  31.1 & +45.9 &  7.3 &  9.7 &  24.0 & +55.6 & 7.4  & 10.9 &  17.2 & +45.6 & 50.6 & 2.0 & - \\
            SFL 211028       & 2021-10-28T17:37:24.000 & 213.6 & -28.8 &  3.4 & 15.5 & 218.7 & -16.3 & 3.5  & 6.4  & 212.9 & -13.3 & 50.3 & 3.6 & - \\

            SGR 1935+2154(a) & 2021-07-07T00:33:31.700 & 292.7 & +23.8 &  1.0 &  2.1 & 292.0 & +24.3 & 3.1  & 2.9  & 293.7 & +21.9 & 26.6 & 64.3 & \textit{Swift}-BAT$^{10}$ \\
            SGR 1935+2154(b) & 2021-09-11T17:01:10.301 & 292.5 & +22.7 &  1.0 &  1.4 & 289.8 & +23.9 & 1.5  & 4.1  & 293.7 & +21.9 & 26.6 & 64.3 & \textit{Swift}-BAT$^{10}$ \\
            SGR 1935+2154(c) & 2021-09-12T00:34:37.450 & none  &  none & none & none & none  & none  & none & none & 293.7 & +21.9 & 26.6 & 64.3 & \textit{Swift}-BAT$^{10}$ \\
            SGR 1935+2154(d) & 2021-09-14T11:10:36.250 & none  &  none & none & none & none  & none  & none & none & 293.7 & +21.9 & 26.6 & 64.3 & \textit{Swift}-BAT$^{11}$ \\
            SGR 1935+2154(e) & 2021-09-22T20:12:16.502 & 291.2 &  22.2 &  5.6 &  2.3 & none  & none  & none & none & 293.7 & +21.9 & 26.6 & 64.3 & \textit{Swift}-BAT$^{11}$ \\
            SGR 1935+2154(f) & 2022-01-12T08:39:25.450 & 293.6 &  20.0 &  1.0 &  1.9 & none  & none  & none & none & 293.7 & +21.9 & 26.6 & 64.3 & \textit{Swift}-BAT$^{11}$ \\
            SGR 1935+2154(g) & 2022-01-14T20:07:03.050 & none  &  none & none & none & none  & none  & none & none & 293.7 & +21.9 & 26.6 & 64.3 & \textit{Swift}-BAT$^{10}$ \\
            SGR 1935+2154(h) & 2022-01-14T20:15:54.401 & none  &  none & none & none & none  & none  & none & none & 293.7 & +21.9 & 26.6 & 64.3 & \textit{Swift}-BAT$^{10}$ \\
            SGR 1935+2154(i) & 2022-01-14T20:21:05.151 & 295.3 &  21.0 &  1.0 &  1.7 & none  & none  & none & none & 293.7 & +21.9 & 26.6 & 64.3 & \textit{Swift}-BAT$^{10}$ \\
            SGR 1935+2154(j) & 2022-01-14T20:29:07.250 & 297.3 &  16.6 &  1.2 &  6.3 & 296.8 & +19.7 & 2.0  & 3.6  & 293.7 & +21.9 & 26.6 & 64.3 & \textit{Swift}-BAT$^{10}$ \\
            SGR 1935+2154(k) & 2022-01-15T17:21:59.304 & 295.5 &  21.1 &  2.7 &  1.9 & none  & none  & none & none & 293.7 & +21.9 & 26.6 & 64.3 & \textit{Swift}-BAT$^{10}$ \\
            SGR 1935+2154(l) & 2022-01-23T20:06:38.751 & 300.4 &  19.0 &  1.0 &  6.9 & none  & none  & none & none & 293.7 & +21.9 & 26.6 & 64.3 & \textit{Swift}-BAT$^{10}$ \\
            \hline
            \hline
        \end{tabular}

        \tiny
        \footnotesize{$^1$ IPN, Hurley et al., GCN 29348}\\
        \footnotesize{$^2$ IPN, Hurley et al., GCN 30002}\\
        \footnotesize{$^3$ IPN, Hurley et al., GCN 30154}\\
        \footnotesize{$^4$ \textit{Swift}-XRT, D'Avanzo et al., GCN 30261}\\
        \footnotesize{$^5$ \textit{Swift}-XRT, Page et al., GCN 30677}\\
        \footnotesize{$^6$ IPN, Kozyrev et al., GCN 30956}\\
        \footnotesize{$^7$ IPN, Kozyrev et al., GCN 31129}\\
        \footnotesize{$^8$ \textit{Swift}-XRT, D'Elia et al., GCN 32029}\\
        \footnotesize{$^9$ INTEGRAL-IBAS, Mereghetti et al., GCN 32041}\\
        \footnotesize{$^{10}$ \textit{Swift}-BAT, Ambrosi et al., GCN 26153}\\ 

    \end{table*} 

    \begin{table}[] 
        \caption{ List of GECAM localization results for FIX, RFD, and APR localization in this research. We do not use the circle region to characterize error, instead by the area of error region. However, to compare with other methods, we list the equivalent radius in this table. }
        \label{TABLE_LocRes_Part2}
        \centering
        \tiny 
        \begin{tabular}{cccccccccccccccccccccccccccccc}
            \hline
            Burst & \multicolumn{5}{c}{FIX} & \multicolumn{5}{c}{RFD} & \multicolumn{5}{c}{APR} \\
            \cmidrule(r){2-6}  \cmidrule(r){7-11} \cmidrule(r){12-16}
            Name & RA(${}^{\circ}$) & DEC(${}^{\circ}$) & ERR(${}^{\circ}$) & $\alpha$(${}^{\circ}$) & CS$^1$ (\%) & RA(${}^{\circ}$) & DEC(${}^{\circ}$) & ERR(${}^{\circ}$) & $\alpha$(${}^{\circ}$) & CS$^1$ (\%) & RA(${}^{\circ}$) & DEC(${}^{\circ}$) & ERR(${}^{\circ}$) & $\alpha$(${}^{\circ}$) & CS$^1$ (\%) \\
            \hline
            GRB 210121A      &  27.0 & -47.8 &  2.0 &  6.9 & 100.0 &  21.0 & -45.6 & 2.0 & 2.9 &  86.8 &  19.6 & -45.0 & 0.6 & 2.3 &  99.2 \\
            GRB 210511B      & 308.4 & +59.3 &  3.2 &  2.5 &  55.0 & 307.8 & +62.6 & 3.2 & 4.9 &   9.0 & 308.4 & +59.3 & 0.6 & 2.5 &  88.6 \\
            GRB 210606B      &  85.6 & -20.1 &  1.0 &  4.1 & 100.0 &  86.8 & -17.9 & 0.8 & 1.6 &  96.9 &  86.8 & -17.9 & 0.6 & 1.6 &  99.6 \\
            GRB 210619B      & 321.1 & +34.0 &  0.6 &  1.2 & 100.0 & 321.1 & +34.0 & 0.6 & 1.2 & 100.0 & 320.0 & +33.2 & 0.6 & 0.7 & 100.0 \\
            GRB 210822A      & 300.6 &  -0.1 &  0.6 &  6.6 & 100.0 & 302.5 &  +6.2 & 1.3 & 2.2 &  89.6 & 303.4 &  +4.1 & 0.6 & 1.6 &  94.6 \\
            GRB 210927B      & 248.8 & +72.8 &  2.2 &  4.2 &  97.6 & 248.8 & +71.7 & 1.7 & 4.7 &  99.8 & 248.8 & +71.7 & 0.6 & 4.7 & 100.0 \\
            GRB 211120A      & 305.9 & +40.7 &  1.1 &  7.2 & 100.0 & 304.8 & +41.3 & 1.6 & 7.8 &  99.7 & 307.1 & +41.1 & 0.6 & 6.2 & 100.0 \\
            GRB 220511A      & 288.8 & +18.4 &  2.3 &  1.4 &  12.3 & 291.0 & +24.2 & 2.2 & 7.3 &  99.7 & 289.0 & +20.5 & 0.6 & 3.2 &  74.8 \\
            GRB 220514A      & 148.1 & +13.4 &  2.6 &  0.5 &   4.5 & 149.4 &  +9.7 & 1.4 & 3.8 &  97.6 & 149.0 & +10.9 & 0.6 & 2.6 &  68.5 \\

            SFL 210508       &  47.3 & +19.0 & 11.0 &  2.4 &   4.3 &  44.1 & +19.8 & 2.0 & 3.0 &  98.1 &  46.6 & +16.6 & 0.6 & 1.2 &  23.9 \\
            SFL 211028       & 224.6 &  -7.3 &  7.9 & 13.0 &  88.5 & 210.4 & -12.8 & 3.7 & 2.5 &  61.4 & 213.5 & -11.0 & 0.6 & 2.3 &  53.0 \\

            SGR 1935+2154(a) & 289.0 & +20.9 &  0.6 &  4.5 & 100.0 & 292.9 & +22.5 & 1.0 & 1.0 &  98.8 & 292.9 & +22.5 & 0.6 & 1.0 &  99.9 \\
            SGR 1935+2154(b) & 294.7 & +21.3 &  0.6 &  1.0 &  78.8 & 292.5 & +23.1 & 0.8 & 1.7 &  58.8 & 293.6 & +22.2 & 0.6 & 0.3 &  71.1 \\
            SGR 1935+2154(c) & 297.9 & +21.5 &  2.6 &  3.9 &  99.6 & 295.8 & +22.4 & 1.6 & 2.0 &  85.4 & 296.9 & +22.5 & 0.6 & 2.9 &  98.8 \\
            SGR 1935+2154(d) & 295.5 & +19.7 &  2.2 &  2.7 &  63.3 & 295.5 & +19.7 & 3.5 & 2.7 &  78.1 & 293.4 & +21.5 & 0.6 & 0.5 &  10.5 \\
            SGR 1935+2154(e) & 293.0 & +20.0 &  1.1 &  2.1 &  99.3 & 292.6 & +21.9 & 1.3 & 1.0 &  48.9 & 289.2 & +22.3 & 0.6 & 4.2 & 100.0 \\
            SGR 1935+2154(f) & 291.3 & +25.0 &  0.6 &  3.8 & 100.0 & 294.7 & +21.2 & 0.6 & 1.1 &  81.2 & 293.9 & +22.2 & 0.6 & 0.3 &  44.2 \\
            SGR 1935+2154(g) & 291.0 & +24.2 &  0.8 &  3.4 & 100.0 & 293.7 & +22.8 & 0.8 & 0.9 &  73.4 & 292.0 & +23.3 & 0.6 & 2.2 & 100.0 \\
            SGR 1935+2154(h) & 290.1 & +23.8 &  1.3 &  3.9 &  99.9 & 290.2 & +22.6 & 1.1 & 3.4 &  96.2 & 291.0 & +22.9 & 0.6 & 2.7 & 100.0 \\
            SGR 1935+2154(i) & 293.9 & +22.7 &  0.6 &  0.8 & 100.0 & 293.0 & +22.3 & 0.6 & 0.8 & 100.0 & 293.9 & +22.7 & 0.6 & 0.8 & 100.0 \\
            SGR 1935+2154(j) & 296.1 & +20.3 &  1.0 &  2.7 & 100.0 & 291.0 & +22.0 & 1.1 & 2.6 &  93.4 & 292.6 & +20.3 & 0.6 & 1.9 &  99.9 \\
            SGR 1935+2154(k) & 295.0 & +21.9 &  1.0 &  1.2 &  50.7 & 295.0 & +21.9 & 0.8 & 1.2 &  65.8 & 294.1 & +21.5 & 0.6 & 0.5 &  49.8 \\
            SGR 1935+2154(l) & 298.8 & +21.5 &  0.6 &  4.7 & 100.0 & 292.5 & +21.4 & 0.6 & 1.2 & 100.0 & 294.4 & +21.3 & 0.6 & 0.8 & 100.0 \\
            \hline
            \hline
        \end{tabular}

        \tiny
        \footnotesize{$^1$ CS: The cumulative sum probability of reference location.}\\

    \end{table} 

    \begin{table*}[]
        \caption{ List of the spectral parameters for RFD and APR spectrum. }
        \label{TABLE_SpeRes}
        \centering
        \tiny
        \begin{tabular}{cccccccccccccccccccccccccccccccccccccccc}
            \hline
            Burst & \multicolumn{4}{c}{Spectral Fitting Result of Optimized} & \multicolumn{4}{c}{Spectral Fitting Result of Ideal} \\
            \cmidrule(r){2-5} \cmidrule(r){6-9}
            Name & Model & Low-energy Index & High-energy Index & $E_{\rm cut}$ & Model & Low-energy Index & High-energy Index & $E_{\rm cut}$ \\
            \hline
            GRB 210121A      & cutoffpl &  0.47 & none   & 287.69 & cutoffpl &  0.43 & none   & 276.58 \\
            GRB 210619B      & band     & -0.45 & -10.00 & 165.60 & band     & -0.64 & -1.67  & 157.46 \\
            GRB 210822A      & cutoffpl &  0.57 & none   & 216.15 & cutoffpl &  0.58 & none   & 219.03 \\
            GRB 210927B      & cutoffpl &  0.47 & none   & 179.92 & cutoffpl &  0.35 & none   & 137.03 \\
            GRB 211120A      & cutoffpl &  0.55 & none   & 144.89 & cutoffpl &  0.67 & none   & 187.95 \\
            GRB 210511B      & band     & -1.49 & -2.01  &  42.92 & band     & -1.06 & -1.66  & 45.57  \\
            GRB 210606B      & band     & -1.22 & -1.51  & 459.39 & band     & -1.23 & -1.51  & 502.56 \\
            GRB 220511A      & band     & -1.22 & none   & 333.28 & cutoffpl & -1.35 & -10.00 & 535.58 \\
            GRB 220514A      & band     & -0.67 & -1.82  &  56.53 & band     & -0.57 & -1.82  & 50.44  \\

            SFL 210508L      & powerlaw &  7.69 & none   &   none & powerlaw &  7.72 & none   & none   \\
            SFL 211028L      & powerlaw &  6.11 & none   &   none & powerlaw &  6.45 & none   & none   \\

            SGR 1935+2154(a) & cutoffpl &  0.44 & none   &  17.82 & cutoffpl &  0.43 & none   & 17.82  \\
            SGR 1935+2154(b) & cutoffpl &  0.33 & none   &  14.86 & cutoffpl &  0.55 & none   & 16.35  \\
            SGR 1935+2154(c) & cutoffpl & -0.48 & none   &  12.58 & cutoffpl & -0.48 & none   & 12.59  \\
            SGR 1935+2154(d) & cutoffpl & -1.55 & none   &   9.40 & cutoffpl & -1.55 & none   & 9.40   \\
            SGR 1935+2154(e) & cutoffpl & -0.08 & none   &  14.27 & cutoffpl & -0.08 & none   & 14.27  \\
            SGR 1935+2154(f) & powerlaw &  3.62 & none   &   none & powerlaw &  3.62 & none   & none   \\
            SGR 1935+2154(g) & cutoffpl &  0.61 & none   &  21.32 & cutoffpl &  0.69 & none   & 22.02  \\
            SGR 1935+2154(h) & cutoffpl &  1.06 & none   &  25.16 & cutoffpl &  1.06 & none   & 25.17  \\
            SGR 1935+2154(i) & cutoffpl &  0.57 & none   &  25.78 & cutoffpl &  0.57 & none   & 25.78  \\
            SGR 1935+2154(j) & cutoffpl & -0.07 & none   &  16.59 & cutoffpl & -0.08 & none   & 16.58  \\
            SGR 1935+2154(k) & cutoffpl &  0.50 & none   &  21.73 & cutoffpl &  0.50 & none   & 21.75  \\
            SGR 1935+2154(l) & cutoffpl & -0.06 & none   &  14.03 & cutoffpl & -0.06 & none   & 14.04  \\
            \hline
            \hline
        \end{tabular}
    \end{table*}

\clearpage 
\appendix

\section{The Bayesian Localization Method with PGSTAT Profile Likelihood} \label{SECTION_LocMethod}

    The measured counts (i.e. counts in a given detector or a channel of a detector) follow a simple Poisson distribution if the background is known exactly:

        \begin{align}
            P_{\rm Poisson}(S|B,M) = \frac{ (B+M)^S \cdot \exp(-(B+M)) }{ S! } ,
            \label{EQ_Pois_R}
        \end{align}
    where $B$, $M$, and $S$ are the expected background level, the expected source counts, and the measured counts. The total of the background $B$ and source contribution $M$ are the expected measured counts. 

    Based on this simple Poisson distribution, a Poisson likelihood and its logarithmic form for the localization method of fixed spectral templates can be written as:

        \begin{align}
            \mathcal L_{\rm P}(i) = \prod \limits_{j=1}^{n} \frac{ \left( b_{j} + f_{i} \cdot m_{j,i} \right) ^ { s_{j} } \cdot \exp( -( b_{j} + f_{i} \cdot m_{j,i} ) ) }{ s_{j} ! } ,
            \label{EQ_Pois_P}
        \end{align}

        \begin{align}
            \ln{ \mathcal L_{\rm P}(i) } = \sum_{j=1}^{n} [ s_{j} \cdot \ln( b_{j} + f_{i} \cdot m_{j,i} ) - ( b_{j} + f_{i} \cdot m_{j,i} ) -\ln s_{j}! ] ,
            \label{EQ_Pois_L}
        \end{align}
    where $s_{j}$ is the entirely measured counts registered in detector $j$, $b_{j}$ is the expectation value of the background. Here we regard the expected source contribution as $f_{i} \cdot m_{j,i}$, where $m_{j,i}$ is the specific spectral template for localization, which is a matrix of counts of each detector $j$ for each incident direction $i$ (the whole sky is pixelized by HEALPix), and $f_{i}$ is the normalization factor to account for the fluence ratio between the real burst and the preset fixed burst spectrum shape used to construct the template $m_{j,i}$.

    For measured data, the expected value of the background is unknown. Generally, we can obtain the estimated background $\hat{B}$ and its uncertainty $\sigma$ from background analysis (e.g. the polynomial fitting to the background intervals), and these background uncertainties should be considered in the likelihood of Poisson data. In this case, the Poisson data with Gaussian background (PGSTAT) statistic can be utilized, e.g. XSPEC \citep{common_STAT_XSPEC2022} and the BALROG algorithm \citep[see also Equation 9 to 10 in][]{Loc_BALROG_Michael2017}.

    In practice, the estimated background $\hat{B}$ and associated uncertainty $\sigma$ come from the polynomial fitting to the background intervals. Therefore, the background distribution is taken to be a truncated normal form:

        \begin{align}
            P(B|\hat{B},\sigma) = \frac{ \Theta(B) }{ \frac{1}{2} + \frac{1}{2} \rm{erf} ( \frac{\hat{B}\sigma}{\sqrt{2}} ) } \cdot \frac{1}{\sqrt{2\pi} \sigma} \cdot \exp( -\frac{ B-\hat{B} }{ 2 \sigma^{2} } ) ,
            \label{EQ_TruncatedBackground}
        \end{align}
    where erf is the standard error function and $\Theta$ is the Heaviside step function which imposes the absolute prior constraint that the background cannot be negative. Thus the background's truncated normal distribution should be considered in the likelihood of Poisson data:

        \begin{align}
            P(S|M,\hat{B},\sigma) & = \int_{0}^{\infty } dB ( \frac{ (B+M)^{S} \exp(-(B+M)) }{ S! } ) \cdot ( \frac{1}{\sqrt{2\pi}\sigma} \exp(-\frac{ (B-\hat{B})^{2} }{ 2 \sigma^{2} }) ) \\
                                   & = \int_{0}^{\infty } dB \frac{ (B+M)^{S} }{ \sqrt{2\pi} \cdot \sigma \cdot S! } \cdot \exp( -( B+M+\frac{ (B-\hat{B})^{2} }{ 2 \sigma^{2} } ) ) .
            \label{EQ_PGSTAT_1}
        \end{align}

    Due to the time-consuming fact of numerical estimating for the background marginalization (Equation \ref{EQ_PGSTAT_1}), here it is instead by a profile likelihood. For a given observed counts, the integral is taken to be the value of the integrand evaluated at its peak (i.e. for the value of the background, $\tilde{B}$, that maximizes the integrand). This is the approach implemented by \citep{common_STAT_XSPEC2022} to obtain the Poisson data with Gaussian background (PGSTAT) statistic that is an option in XSPEC. This is equivalent to assuming that:

        \begin{align}
            P(S|M,\hat{B},\sigma) \propto
            \begin{cases}
                \exp( -( \hat{B}+M - \frac{\sigma^{2}}{2} ) ) , S=0 \\
                (\tilde{B}+M)^{S} \cdot \exp( -( \tilde{B} + M + \frac{ (\tilde{B}-\hat{B})^{2} }{ 2 \sigma^{2} } ) ) , S>0 ,
            \end{cases}
            \label{EQ_PGSTAT_2}
        \end{align}

    Its logarithmic form can be written as:

        \begin{align}
            \ln P(S|M,\hat{B},\sigma) \propto
            \begin{cases}
                -( \hat{B} + M - \frac{\sigma^{2}}{2} ) , S=0 \\
                S \cdot \ln(\tilde{B}+M) -( \tilde{B} + M + \frac{ (\tilde{B}-\hat{B})^{2} }{ 2 \sigma^{2} } ) , S>0 .
            \end{cases}
            \label{EQ_PGSTAT_3}
        \end{align}
    where in the S \textgreater 0 case,

        \begin{align}
            \tilde{B} = \frac{1}{2} ( \hat{B} - M - \sigma^{2} + \sqrt{ ( M + \sigma^{2} - \hat{B} )^{2} - 4 ( M \sigma^{2} - S \sigma^{2} - \hat{B} M ) } ) ,
            \label{EQ_PGSTAT_4}
        \end{align}
    where $\tilde{B}$ maximizes integrand evaluated at its peak for given observational data $S$.

    Then, a logarithmic Poisson data with Gaussian background (PGSTAT) profile likelihood for spectral template localization can be constructed by Equations \ref{EQ_PGSTAT_2} to \ref{EQ_PGSTAT_4}:

        \begin{align}
            \ln \mathcal L_{\rm PG} = \sum_{j=1}^{n}
            \begin{cases}
                -( \hat{b}_{j} + f_{i} \cdot m_{j,i} - \frac{\sigma_{j,i}^{2}}{2} ) , S=0 \\
                s_{j} \cdot \ln(\tilde{b}_{j}+f_{i} \cdot m_{j,i}) -( \tilde{b}_{j} + f_{i} \cdot m_{j,i} + \frac{ (\tilde{b}_{j}-\hat{b}_{j})^{2} }{ 2 \sigma_{j}^{2} } ) , S>0 ,
            \end{cases}
            \label{EQ_PGSTAT_5}
        \end{align}
    and,

        \begin{align}
            & \tilde{b}_{j} = \frac{1}{2} ( \hat{b}_{j} - f_{i} \cdot m_{j,i} - \sigma_{j}^{2} ) + \nonumber \\
            & \frac{1}{2} \sqrt{ ( f_{i} \cdot m_{j,i} + \sigma_{j}^{2} - \hat{b}_{j} )^{2} - 4 ( f_{i} \cdot m_{j,i} \cdot \sigma_{j}^{2} - s_{j} \cdot \sigma_{j}^{2} - \hat{b}_{j} \cdot f_{i} \cdot m_{j,i} ) } ,
            \label{EQ_PGSTAT_6}
        \end{align}
    where $s_{j}$ is the total observed counts in detector $j$, $n$ is the total number of detectors and $b_{j}$ is the expectation value of the background. Here we regard $f_{i} \cdot m_{j,i}$ as the expected source contribution, where $m_{j,i}$ is the localization template of a specific spectrum, which is a matrix of counts of each detector $j$ for each incident direction $i$ (the whole sky is pixelized with HEALPix), and $f_{i}$ is the normalization factor to account for the fluence ratio between the real burst and the preset fixed burst spectrum used to generate the template $m_{j,i}$.

    During the localization process with the fixed templates, the $f_{i}$ could be derived from the maximization for each direction ($i$), thus the burst position (i.e. direction $i$) is the only parameter of interest, whose prior could be assumed to be uniform all over the celestial sphere: $P_{\rm prior}(i)=\frac{1}{N}$, where $N$ is the total number of the HEALPix pixels of all sky. With the parameter prior and likelihood as shown in Equation \ref{EQ_Pois_L}, we derive the location results (location center, probability map, and credible region) through the Bayesian inference.

    We summarize this Bayesian localization method based on PGSTAT profile likelihood as follows:

    \begin{itemize}

        \item[\textbf{Step 1:}] For each incident direction $i$, maximize the likelihood ($\ln \mathcal L(i) $) by adjusting the normalization factor $f_{i}$. The mathematical form of likelihood and logarithmic likelihood is equivalent to the maximization process.

        \item[\textbf{Step 2:}] Calculate the posterior probability through Bayesian inference. Thus the posterior distribution, $P(i|s)$, could be derived from the prior probability $P_{\rm prior}(i)$, conditional probability for a given direction $i$ to obtain the observed counts $s$ and evidence $P(s)$:

            \begin{equation}
            \begin{aligned}
                P(i|s) &= \frac{ P_{\rm prior}(i) \cdot P(s|i) }{ P(s) }                                     \\
                       &= \frac{ \frac{1}{N} \cdot P(s|i) }{ \sum_{i^{'}} { \frac{1}{N} \cdot P(s|i^{'}) } } \\
                       &= \frac{ P(s|i) }{ \sum_{i^{'}} { P(s|i^{'}) } } .
                \label{EQ_BYS_1}
            \end{aligned}
            \end{equation}

        By substituting the conditional probability ($P(s|i)$) with the likelihood ($ \mathcal L(i) $), one can get the posterior probability for each direction ($i$), which is also the localization probability map:

            \begin{align}
                P(i) = \frac{ \mathcal L(i) }{ \sum_{i^{'}}{ \mathcal L(i^{'}) } } .
                \label{EQ_BYS_2}
            \end{align}

        \item[\textbf{Step 3:}] For simplicity, we take the direction with the maximum $P(i)$ as the location center and the Bayesian credible region with $N\%$ Highest Posterior Density (HPD) as the $N\%$ confidence interval of the burst position.

    \end{itemize}

\section{The Bayesian Systematic Uncertainties Estimation Approach} \label{SECTION_GBMsysERR}

    A Bayesian approach is usually adopted to estimate the localization systematic uncertainties \citep{Loc_BATSE_Briggs1999, Loc_GBM_Connaughton2015, Loc_POLAR_Wang2021}. The models are based on the Fisher probability density function, which has been called the Gaussian distribution on the sphere \citep{common_STAT_Fisher1993}:

        \begin{equation}
            P(\alpha) \mathrm{d} \Omega = \frac{ k \cdot \exp( k \cdot \cos \alpha ) }{ 2 \pi \cdot ( \exp(k) - \exp(-k) ) } \mathrm{d} \Omega ,
            \label{EQ_SysErr_GBM_1}
        \end{equation}
    where $\alpha$ is the angular separation between the measured and true position, $k$ is termed as the concentration parameter, and $d \Omega$ is the solid angle. Considering $\sigma_{\rm TOTAL}$ to be the radius of the circle containing $68.27\%$ of the total probability, integrating Equation \ref{EQ_SysErr_GBM_1} relates $k$ and $\sigma_{\rm TOTAL}$ in radians \citep{Loc_BATSE_Briggs1999}:

        \begin{equation}
            k = \frac{ 1 }{ ( 0.66 \cdot \sigma_{\rm TOTAL} )^{2} } .
            \label{EQ_SysErr_GBM_2}
        \end{equation}

    The localization uncertainties ($\sigma_{\rm TOTAL}$) are consist of statistical uncertainties ($\sigma _{\rm STAT}$) and systematic uncertainties ($\sigma_{\rm SYS}$):

        \begin{equation}
            \sigma_{\rm TOTAL}^{2} = \sigma_{\rm STAT}^{2} + \sigma_{\rm SYS}^{2} .
            \label{EQ_SysErr_GBM_3}
        \end{equation}

    During the calculation, the solid angle $\mathrm{d} \Omega$ is replaced with angular separation $\mathrm{d} \alpha$ ($\mathrm{d} \Omega = 2 \pi \sin \alpha \mathrm{d} \alpha$):

        \begin{equation}
            P(\alpha) \mathrm{d} \alpha = \frac{ k \cdot \exp( k \cdot \cos \alpha ) \cdot 2 \pi \sin \alpha }{ 2 \pi \cdot ( \exp(k) - \exp(-k) ) } \mathrm{d} \alpha .
            \label{EQ_SysErr_GBM_4}
        \end{equation}

    Then integrating $\mathrm{d} \alpha$:

        \begin{equation}
            \int{ P(\alpha) \mathrm{d} \alpha } = \int{ \frac{ k \cdot \exp( k \cdot \cos \alpha ) \cdot 2 \pi \sin \alpha }{ 2 \pi \cdot ( \exp(k) - \exp(-k) ) } \mathrm{d} \alpha } ,
            \label{EQ_SysErr_GBM_5}
        \end{equation}
    and,

        \begin{equation}
            P(\alpha) = \frac{ k \cdot \exp( -1 + \cos \alpha ) \cdot \sin \alpha }{ 1 - \exp(-2k) } .
            \label{EQ_SysErr_GBM_6}
        \end{equation}

    For each burst, the probability $P_{\rm q,r}$ could be calculated by Equations \ref{EQ_SysErr_GBM_2}, \ref{EQ_SysErr_GBM_3} and \ref{EQ_SysErr_GBM_6} for a given assumed systematic error $r$, where $q=1,2,...,Q$ ($Q$ is the burst number). Then successively multiplying $P_{\rm q,r}$ along a systematic error $r$ and normalizing:

        \begin{align}
            P_{\rm B}^{'} = \prod_{q} P_{\rm q,r} (\alpha) ,
            \label{EQ_SysErr_GBM_7}
        \end{align}
    and,

        \begin{align}
            P_{\rm B} = \frac{ P_{\rm B}^{'} }{ \sum_{r} P_{\rm B}^{'} } .
            \label{EQ_SysErr_GBM_8}
        \end{align}

    The normalized probability $P_{\rm B}$ represents the systematic error's probability distribution.

\clearpage
\section{GECAM Localization Results for GRBs, SGRs, SFLs} \label{SECTION_LocRes_GRB}

    The GECAM localization result of 22 bright GRBs, SGRs, and SFLs (One is shown in Figures \ref{fig2a} and \ref{fig2b}), as shown in Table \ref{TABLE_LocRes_Part2}.

    \begin{figure*}
        \centering
        \subfigure[]{\includegraphics[height=3.5cm]{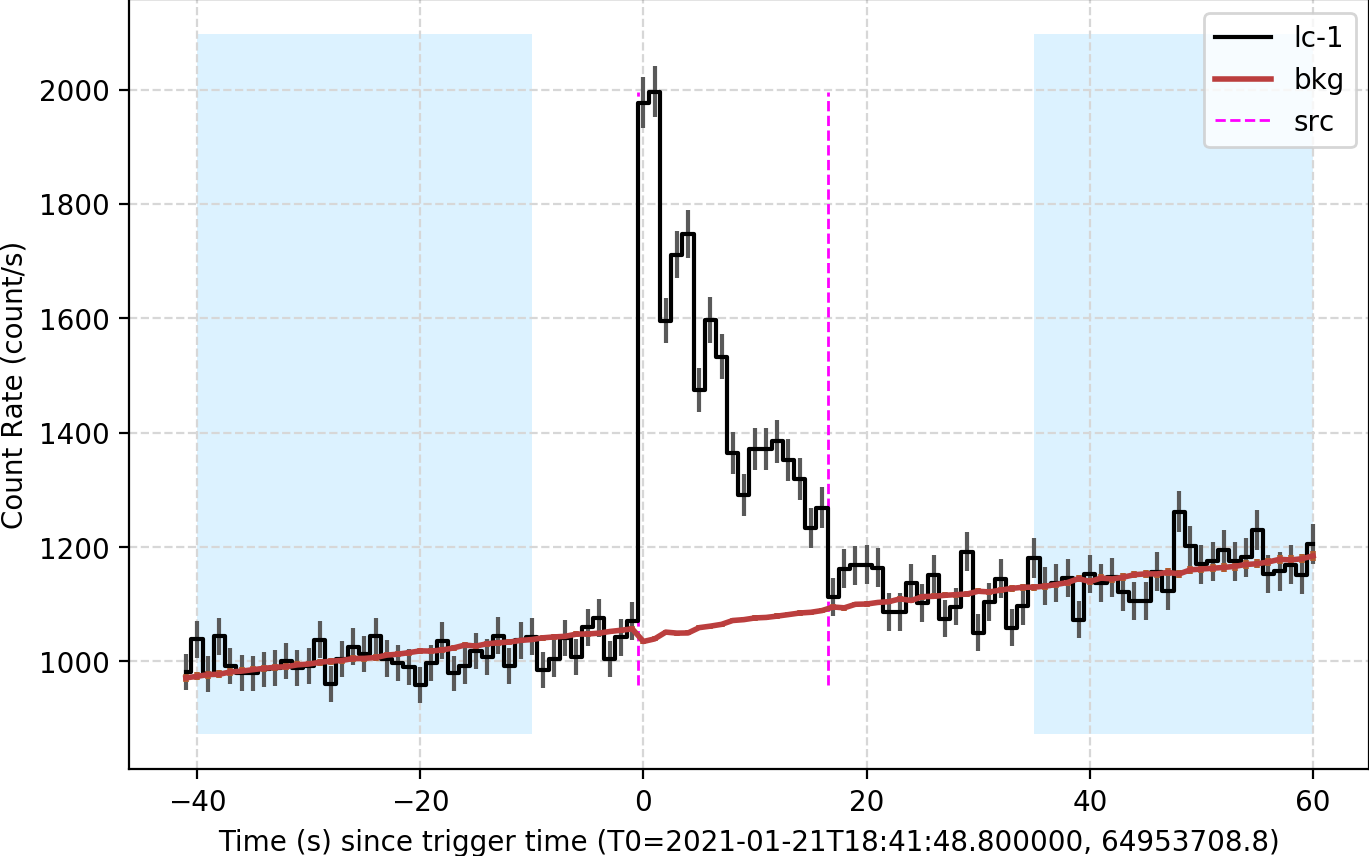}}
        \quad
        \subfigure[]{\includegraphics[height=3.5cm]{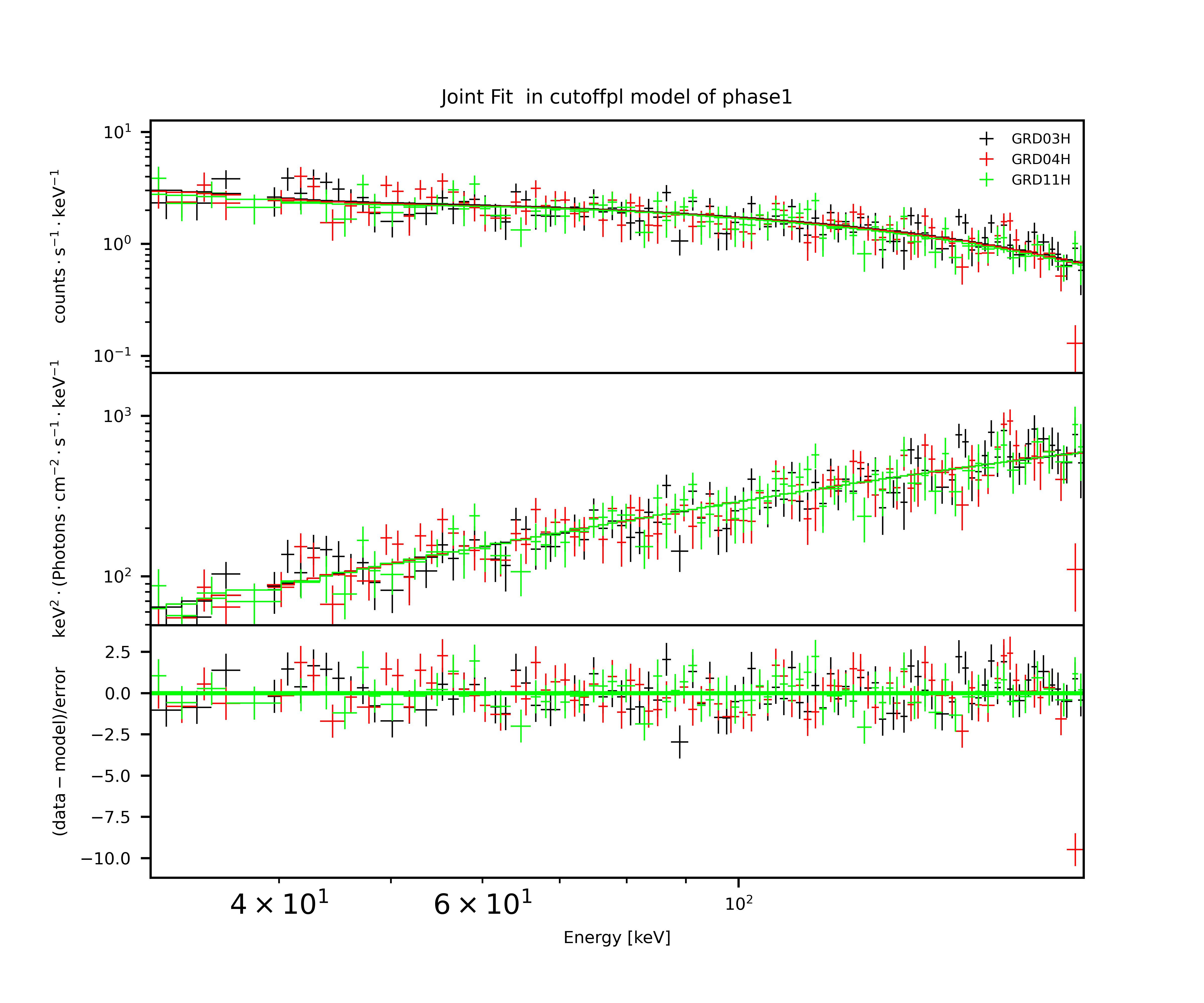}}
        \quad
        \subfigure[]{\includegraphics[height=3.5cm]{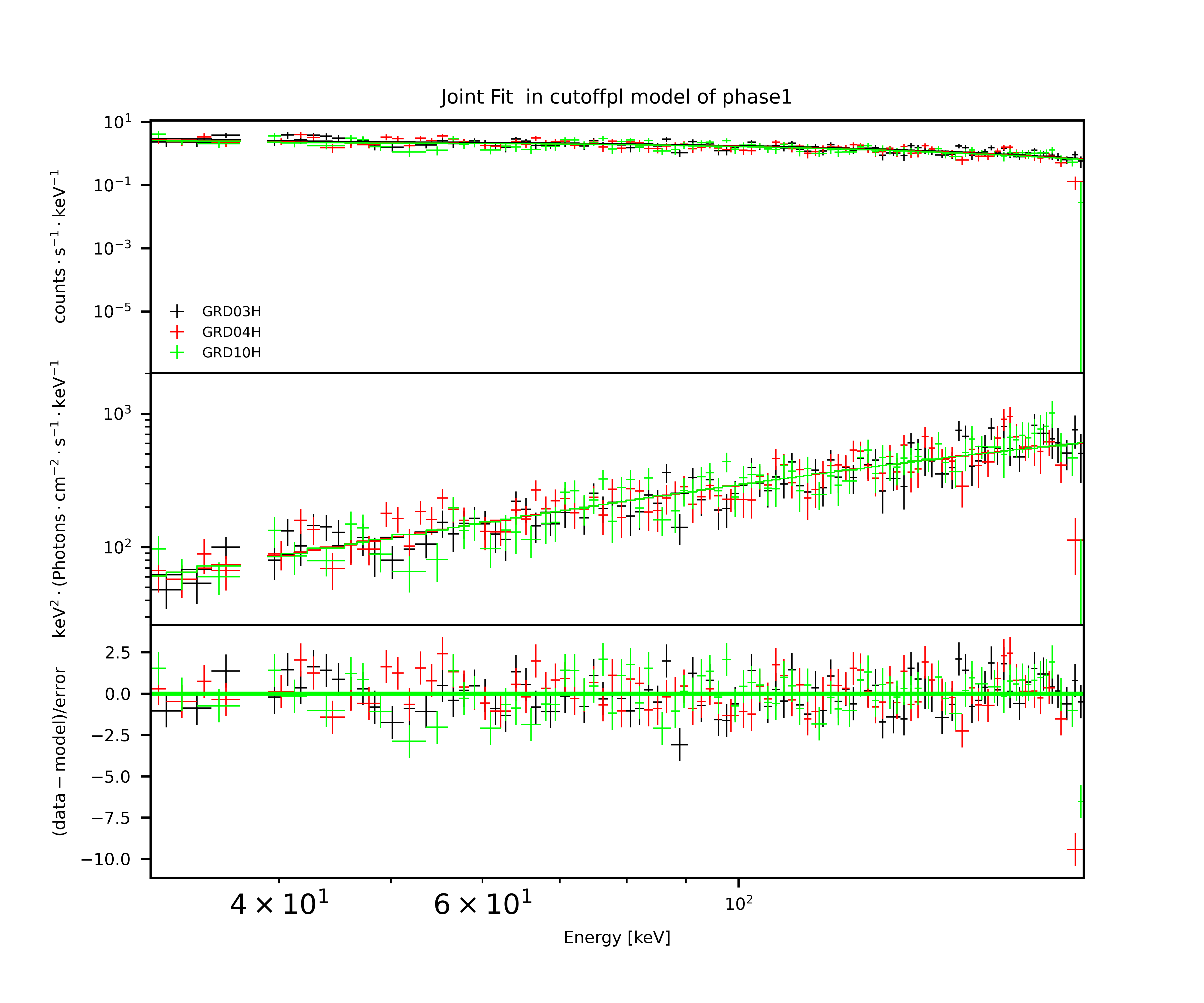}}
        \quad
        \\
        \subfigure[]{\includegraphics[width=4.5cm]{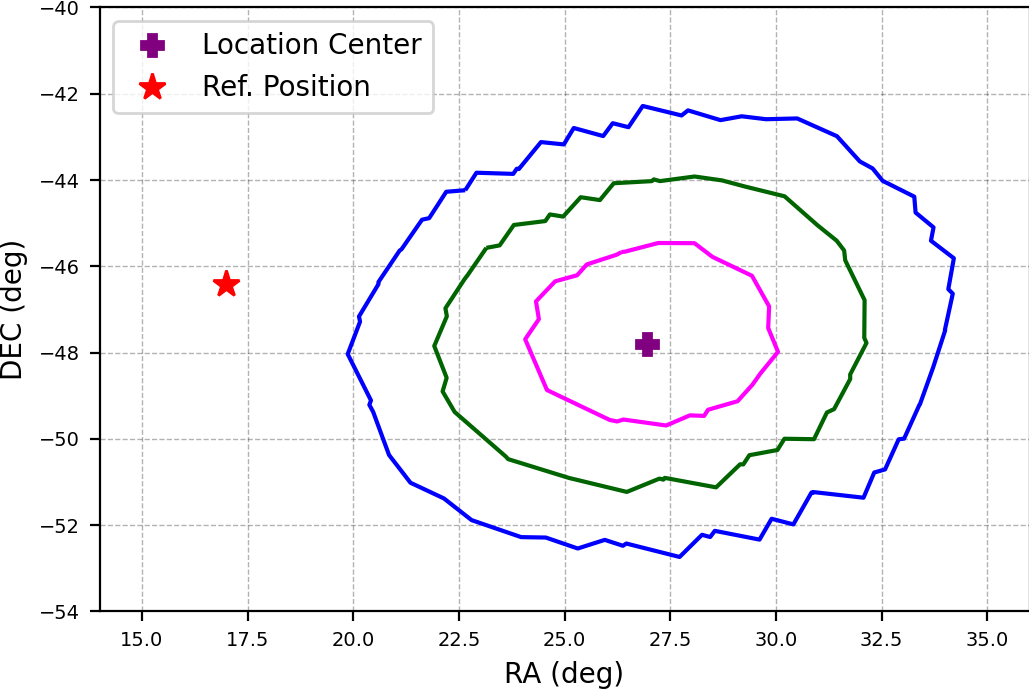}}
        \quad
        \subfigure[]{\includegraphics[width=4.5cm]{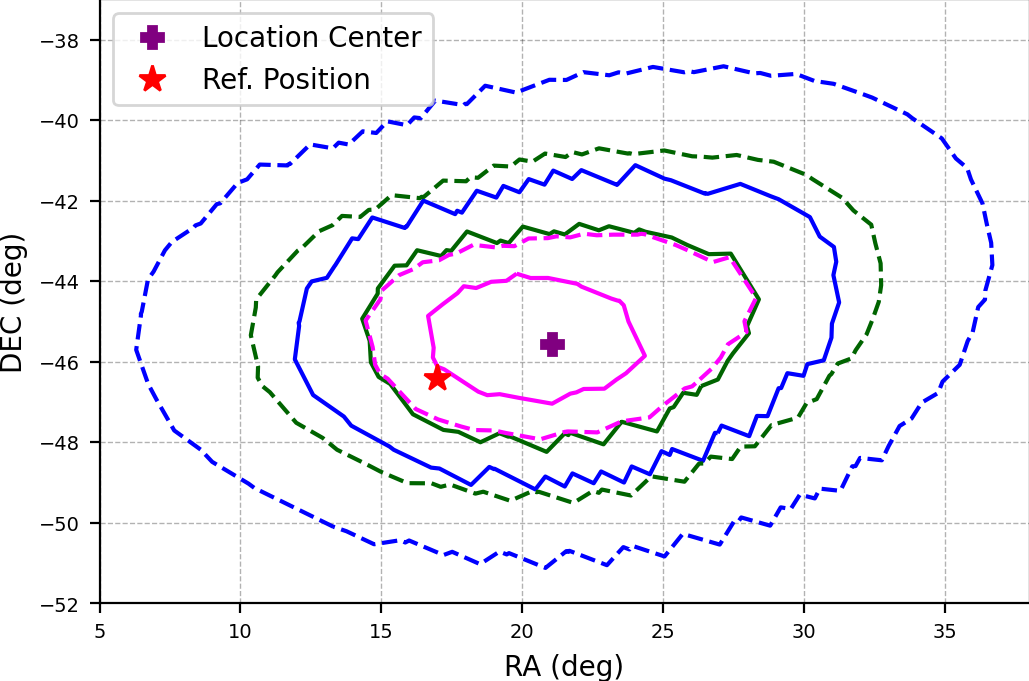}}
        \quad
        \subfigure[]{\includegraphics[width=4.5cm]{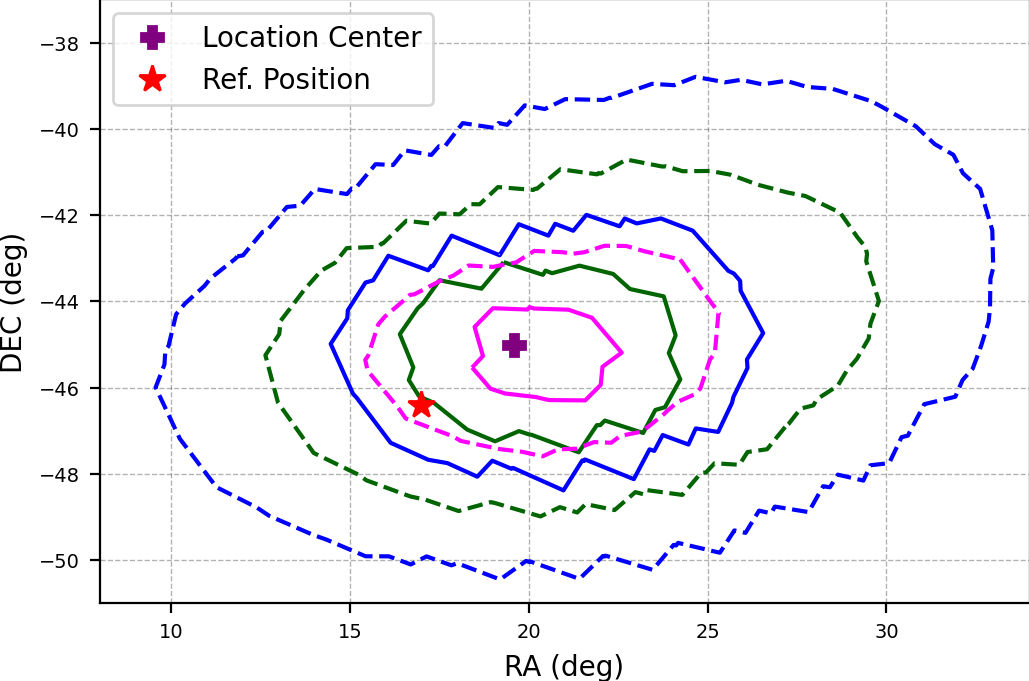}}
        \caption{ GECAM localization results of GRB 210121A. (a) The light curve of GRD \# 03 high gain which contains the majority of net (burst) counts. (b) The RFD spectral fitting result. (c) The RFD spectral fitting result. The location credible region of (d) FIX, (e) RFD, and (f) APR localization. The captions are the same as Figure \ref{fig2a}. }
        \label{fig2_GRB02}
    \end{figure*}


    \begin{figure*}
        \centering
        \subfigure[]{\includegraphics[height=3.5cm]{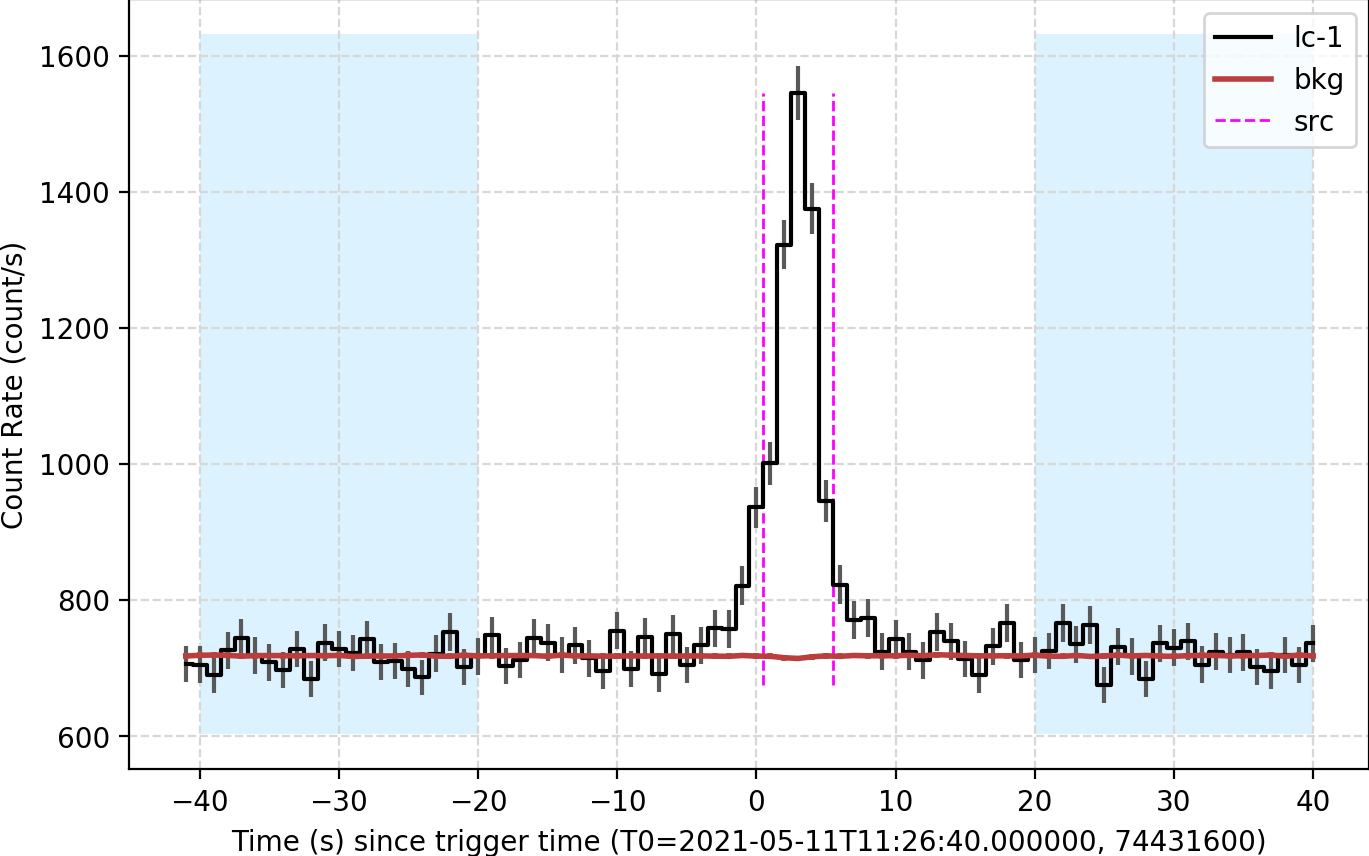}}
        \quad
        \subfigure[]{\includegraphics[height=3.5cm]{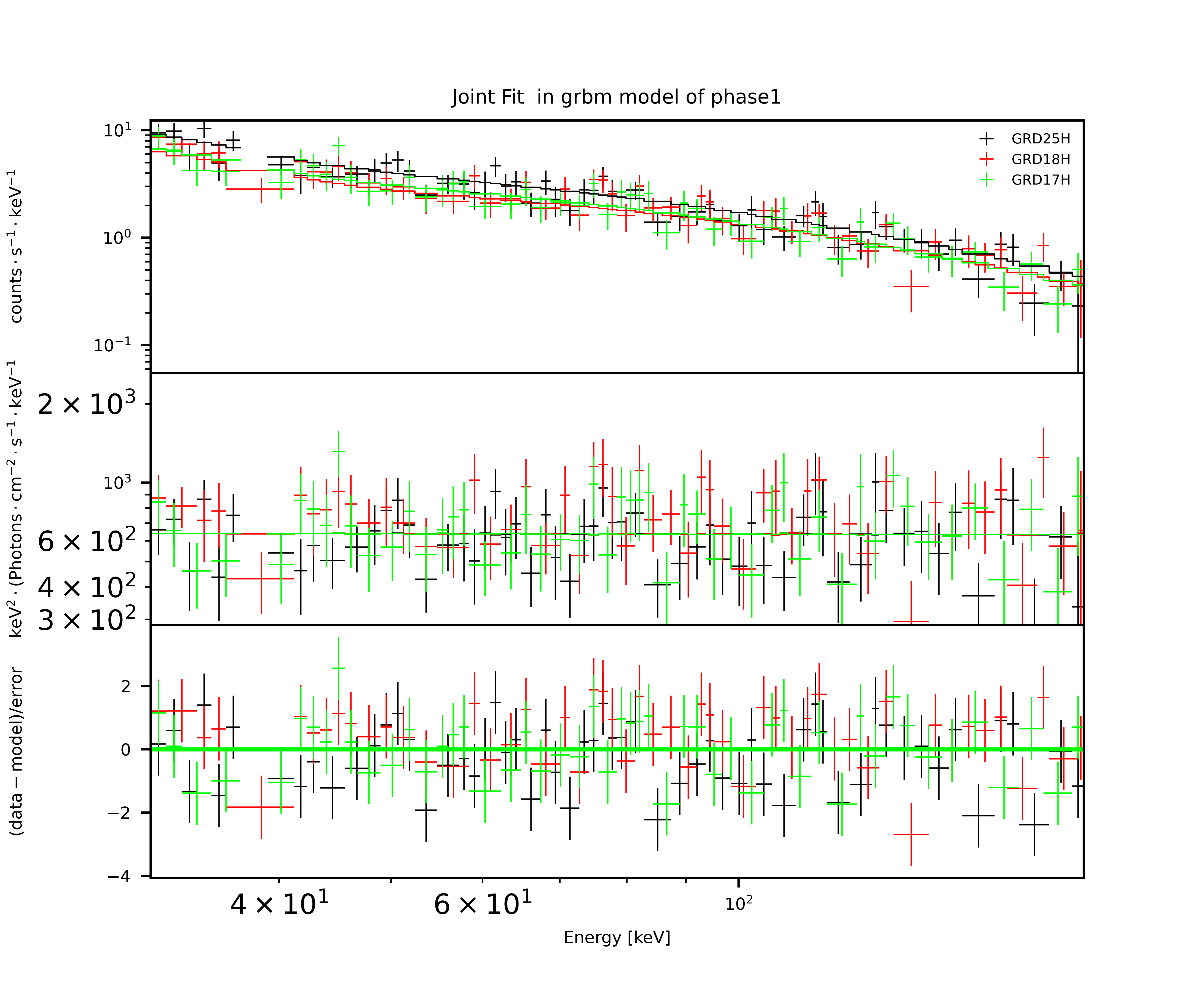}}
        \quad
        \subfigure[]{\includegraphics[height=3.5cm]{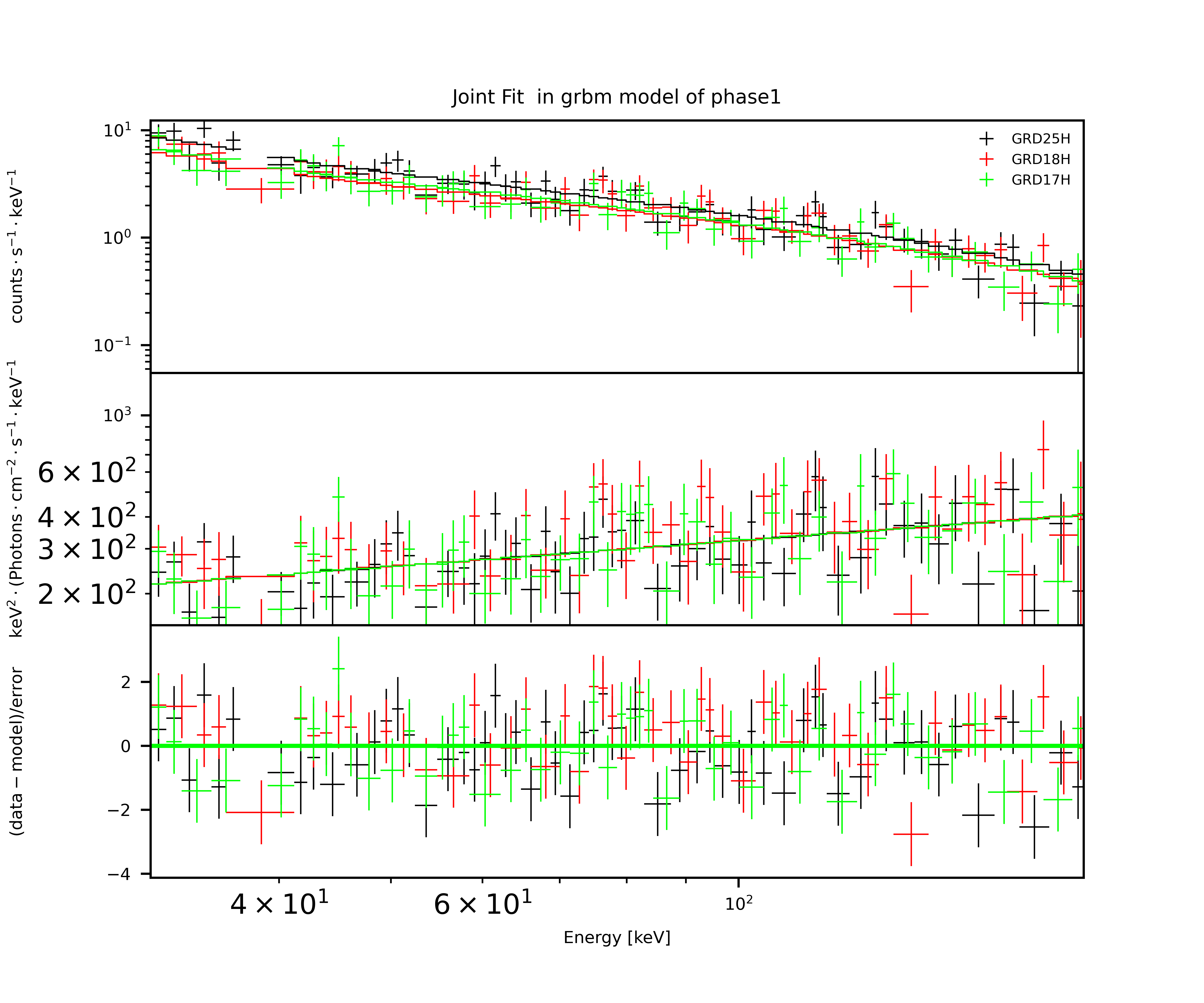}}
        \quad
        \\
        \subfigure[]{\includegraphics[width=4.5cm]{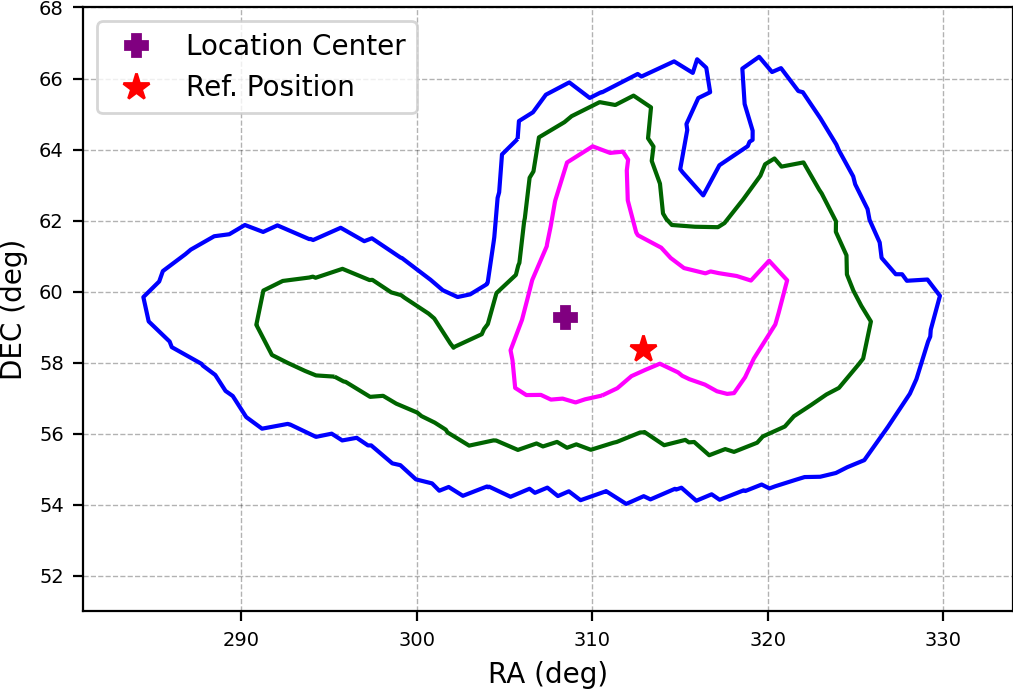}}
        \quad
        \subfigure[]{\includegraphics[width=4.5cm]{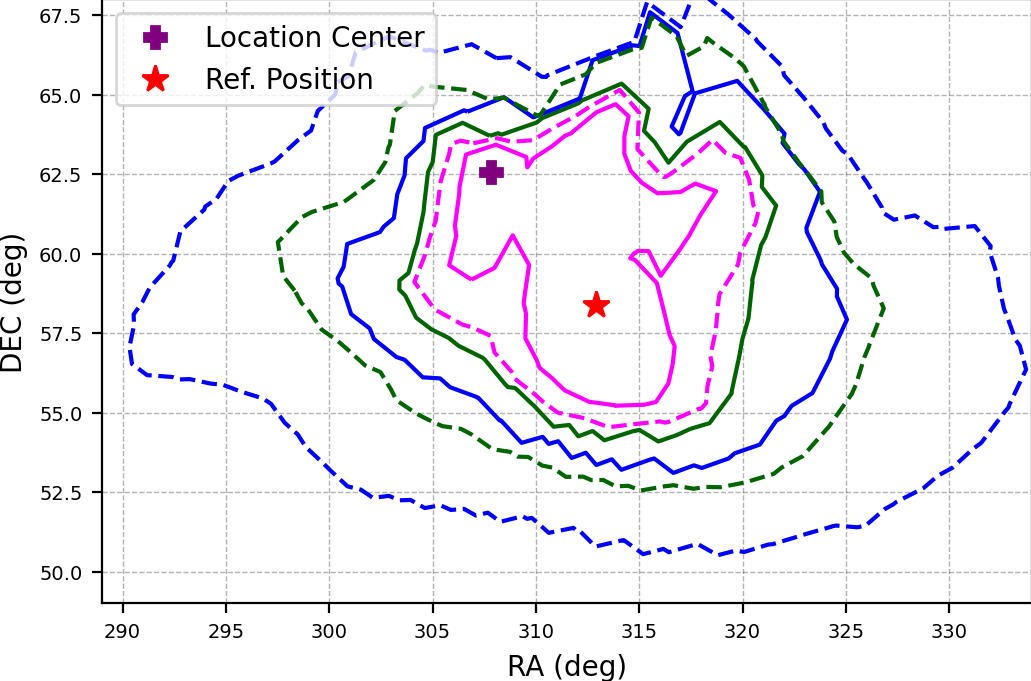}}
        \quad
        \subfigure[]{\includegraphics[width=4.5cm]{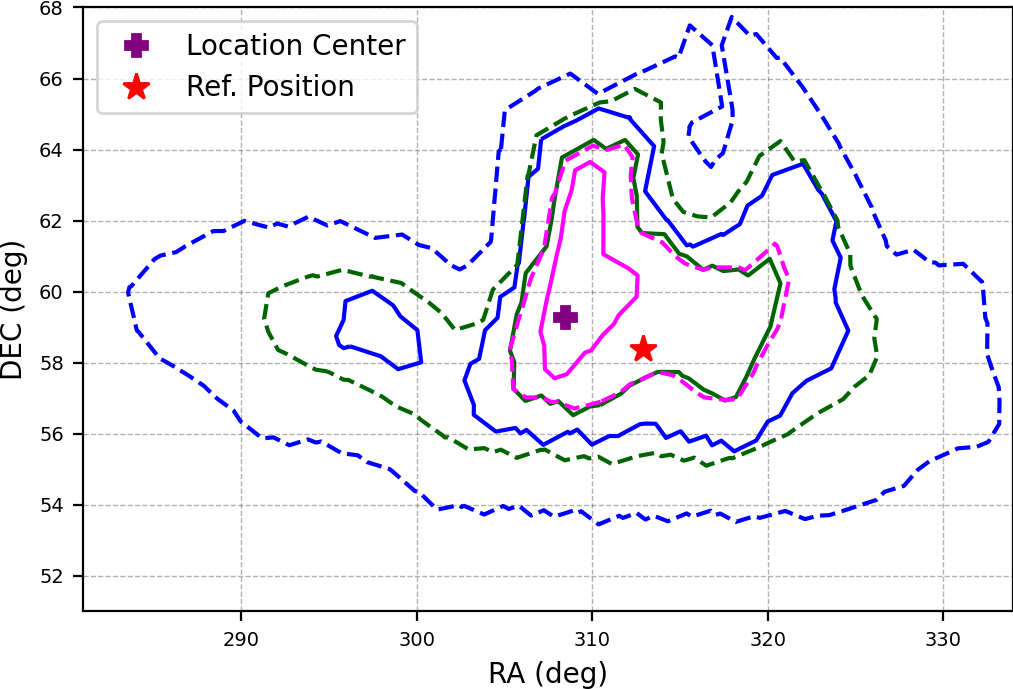}}
        \caption{ GECAM localization results of GRB 210511B. (a) The light curve of GRD \# 25 high gain which contains the majority of net (burst) counts. (b) The RFD spectral fitting result. (c) The RFD spectral fitting result. The location credible region of (d) FIX, (e) RFD, and (f) APR localization. The captions are the same as Figure \ref{fig2a}. }
        \label{fig2_03}
    \end{figure*}

    \begin{figure*}
        \centering
        \subfigure[]{\includegraphics[height=3.5cm]{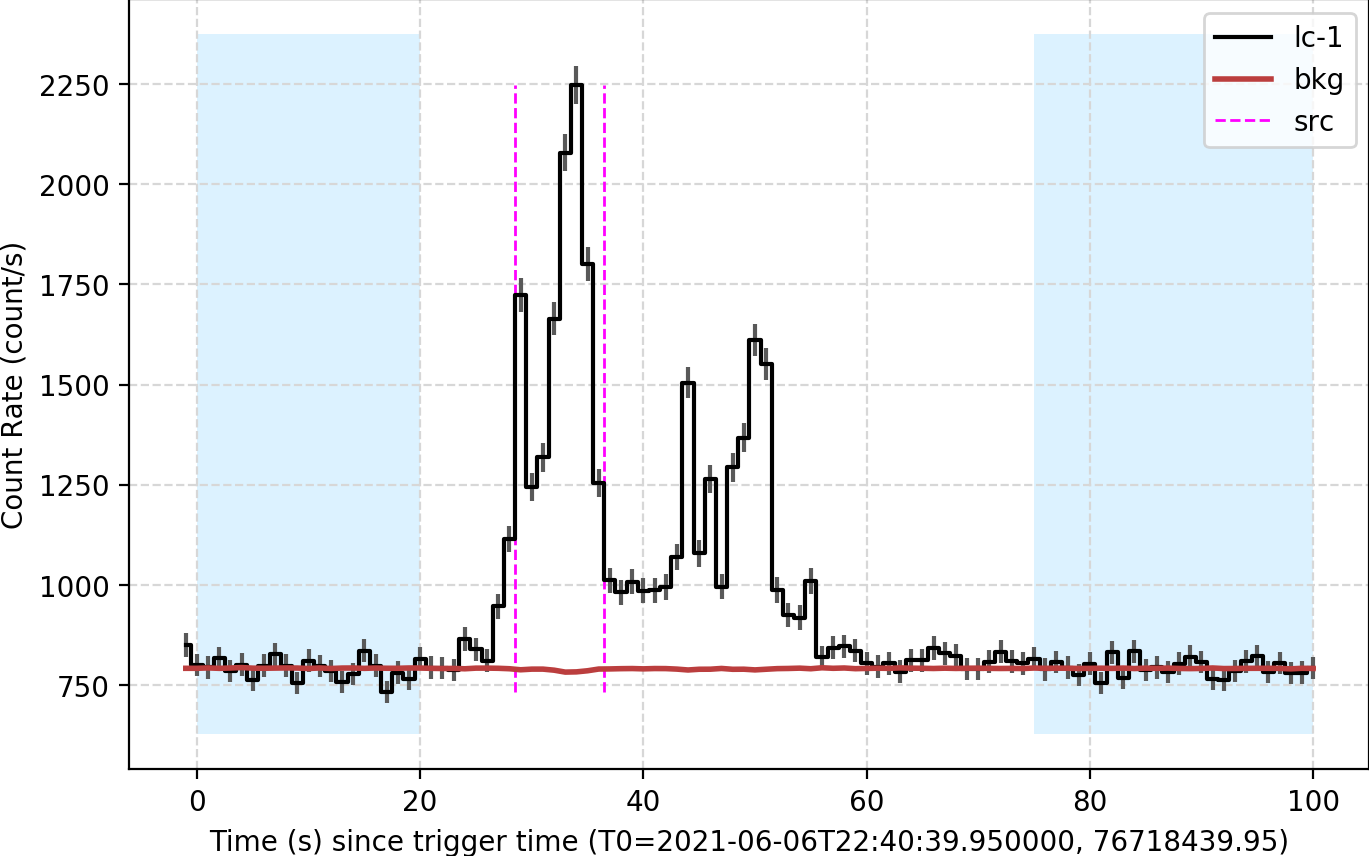}}
        \quad
        \subfigure[]{\includegraphics[height=3.5cm]{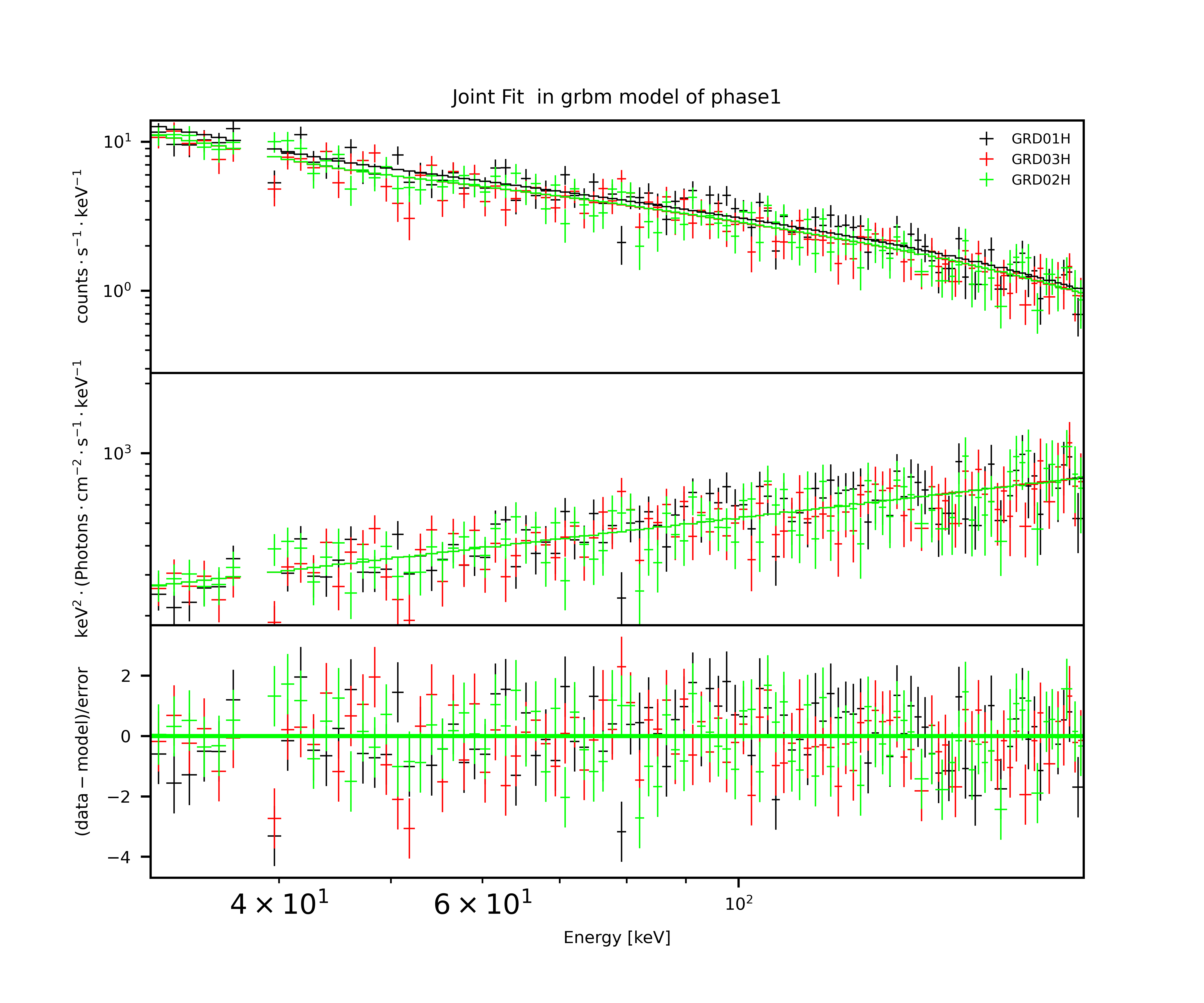}}
        \quad
        \subfigure[]{\includegraphics[height=3.5cm]{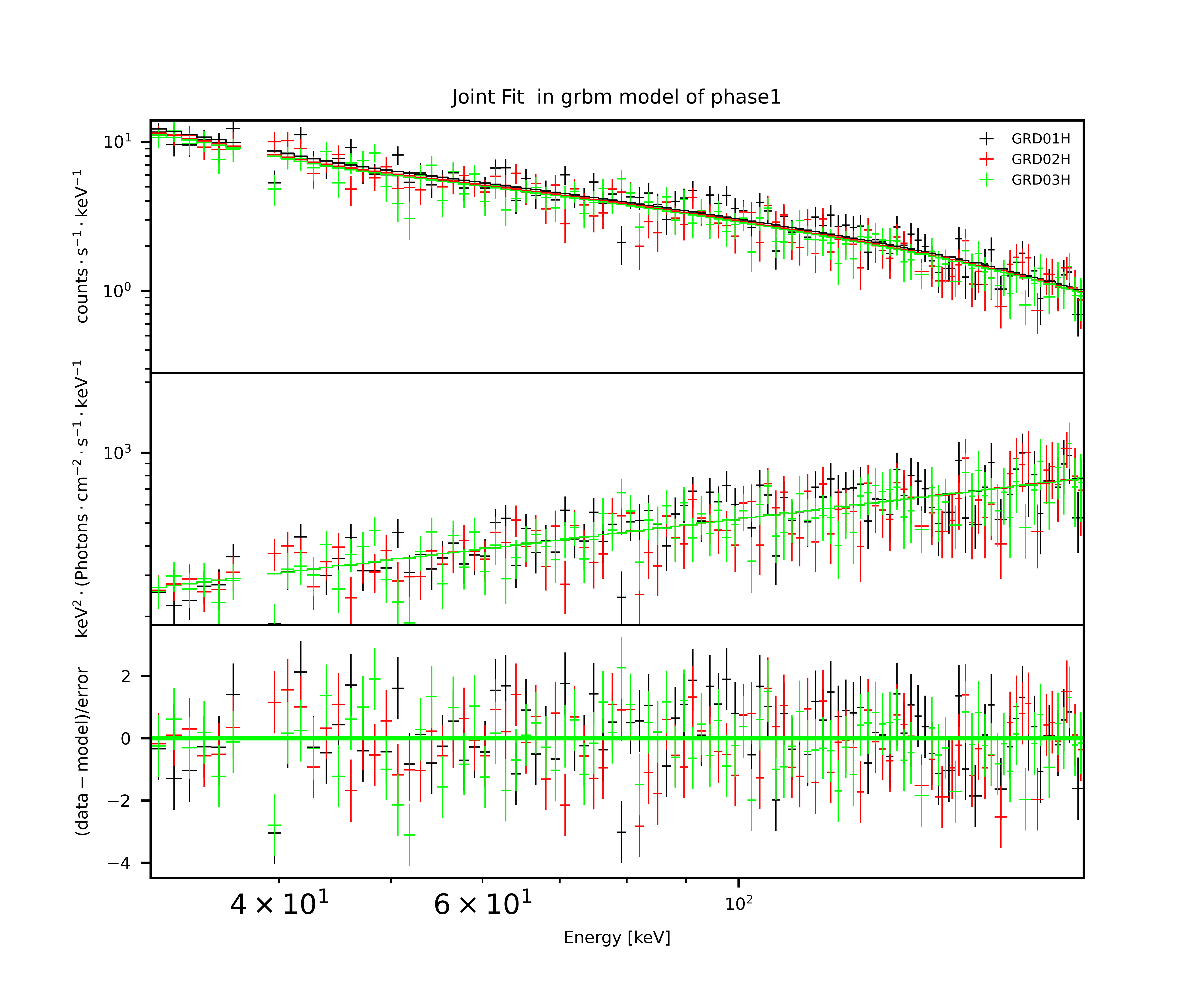}}
        \quad
        \\
        \subfigure[]{\includegraphics[width=4.5cm]{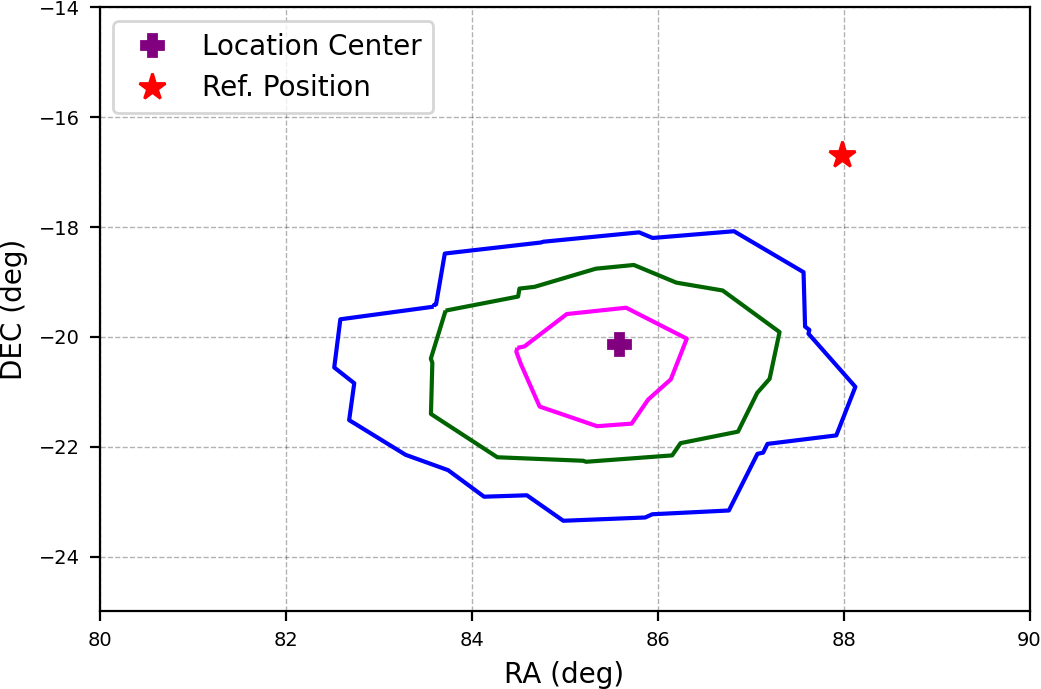}}
        \quad
        \subfigure[]{\includegraphics[width=4.5cm]{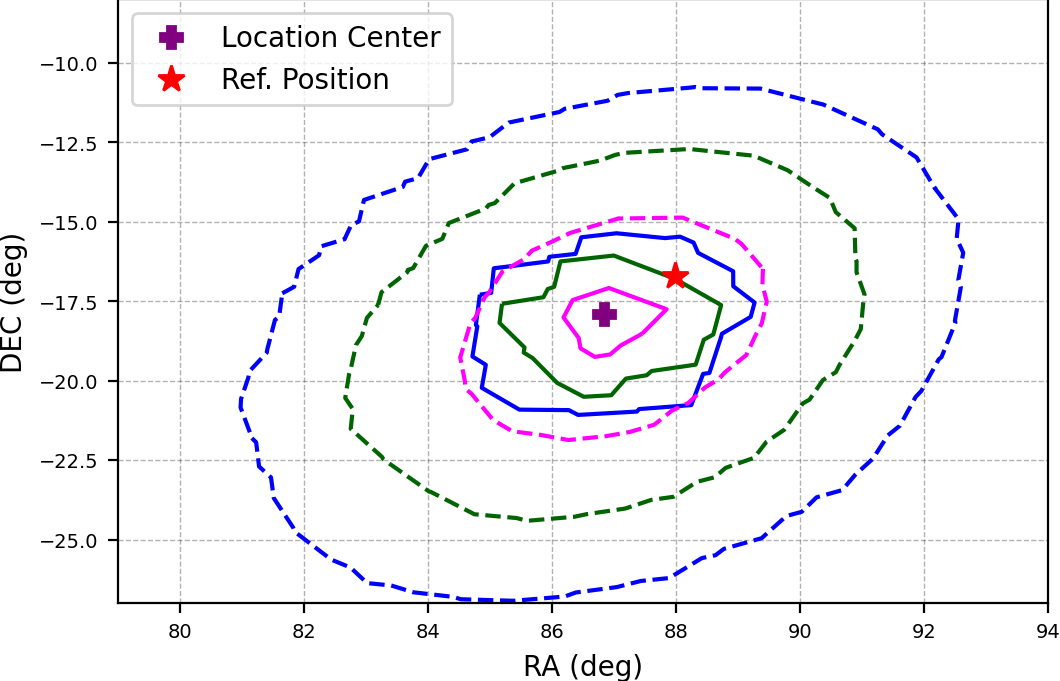}}
        \quad
        \subfigure[]{\includegraphics[width=4.5cm]{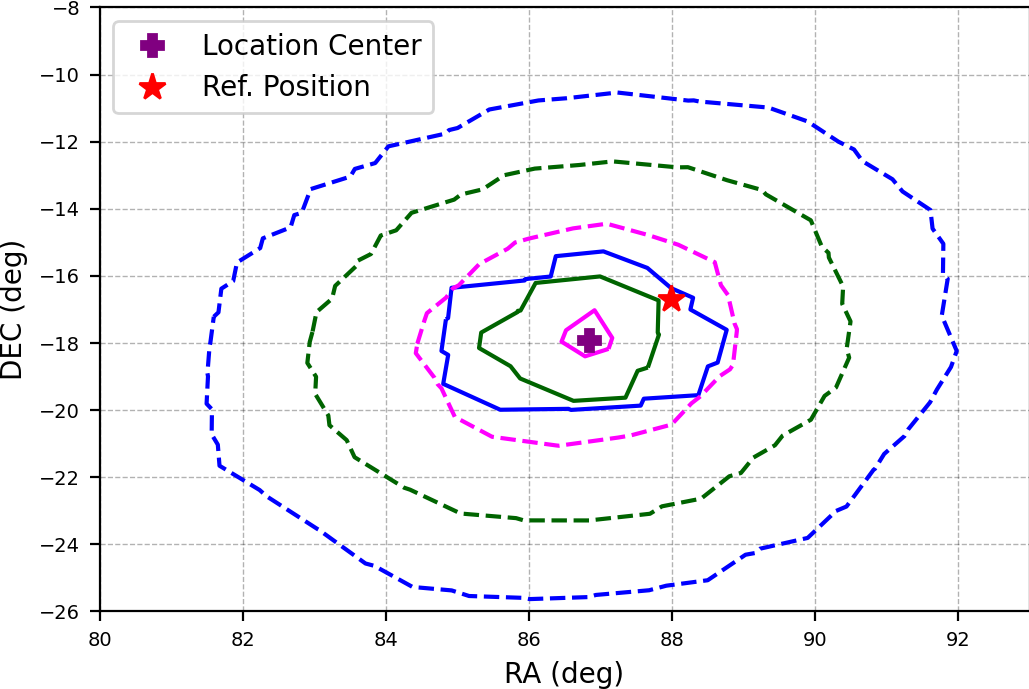}}
        \caption{ GECAM localization results of GRB 210606B. (a) The light curve of GRD \# 01 high gain which contains the majority of net (burst) counts. (b) The RFD spectral fitting result. (c) The RFD spectral fitting result. The location credible region of (d) FIX, (e) RFD, and (f) APR localization. The captions are the same as Figure \ref{fig2a}. }
        \label{fig2_04}
    \end{figure*}

    \begin{figure*}
        \centering
        \subfigure[]{\includegraphics[height=3.5cm]{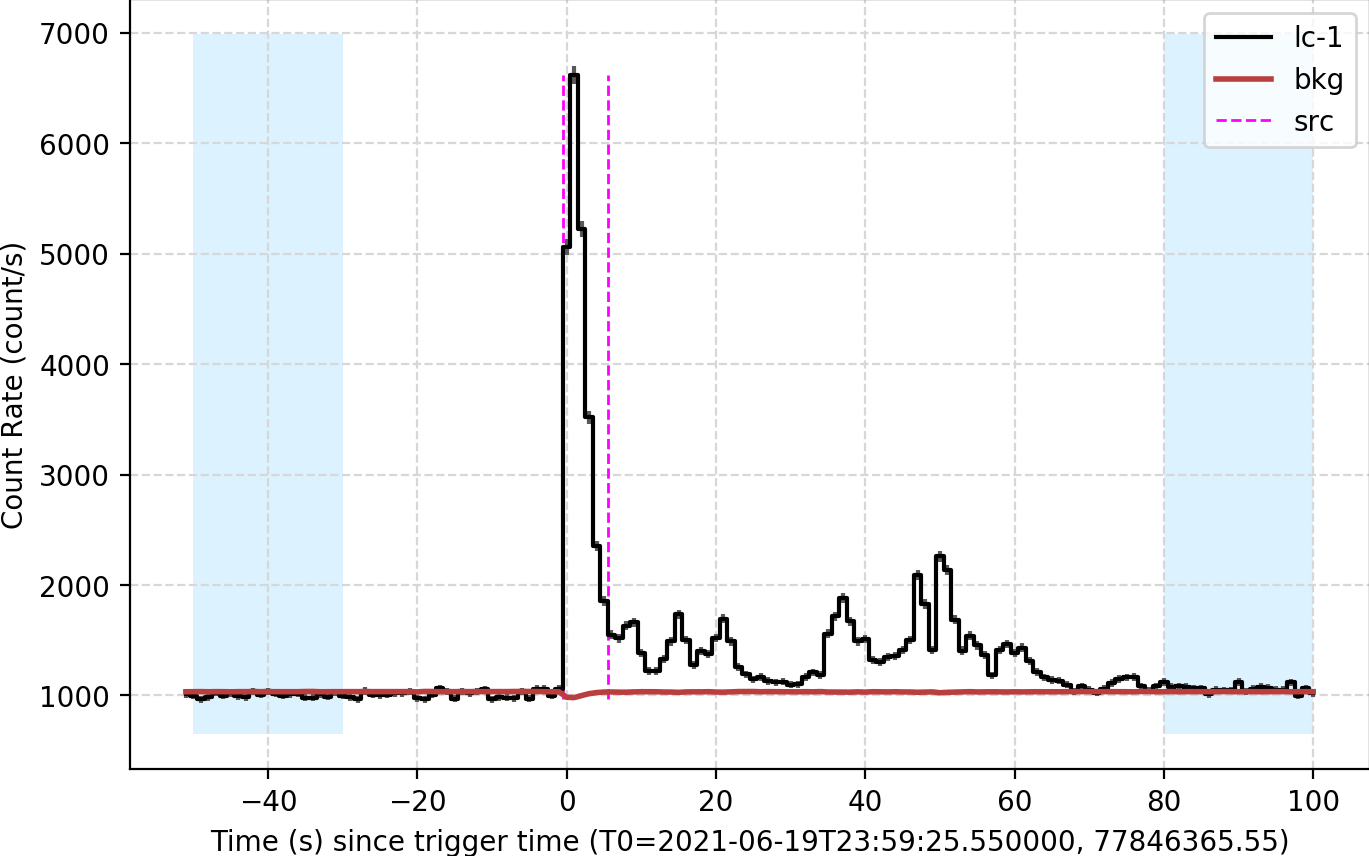}}
        \quad
        \subfigure[]{\includegraphics[height=3.5cm]{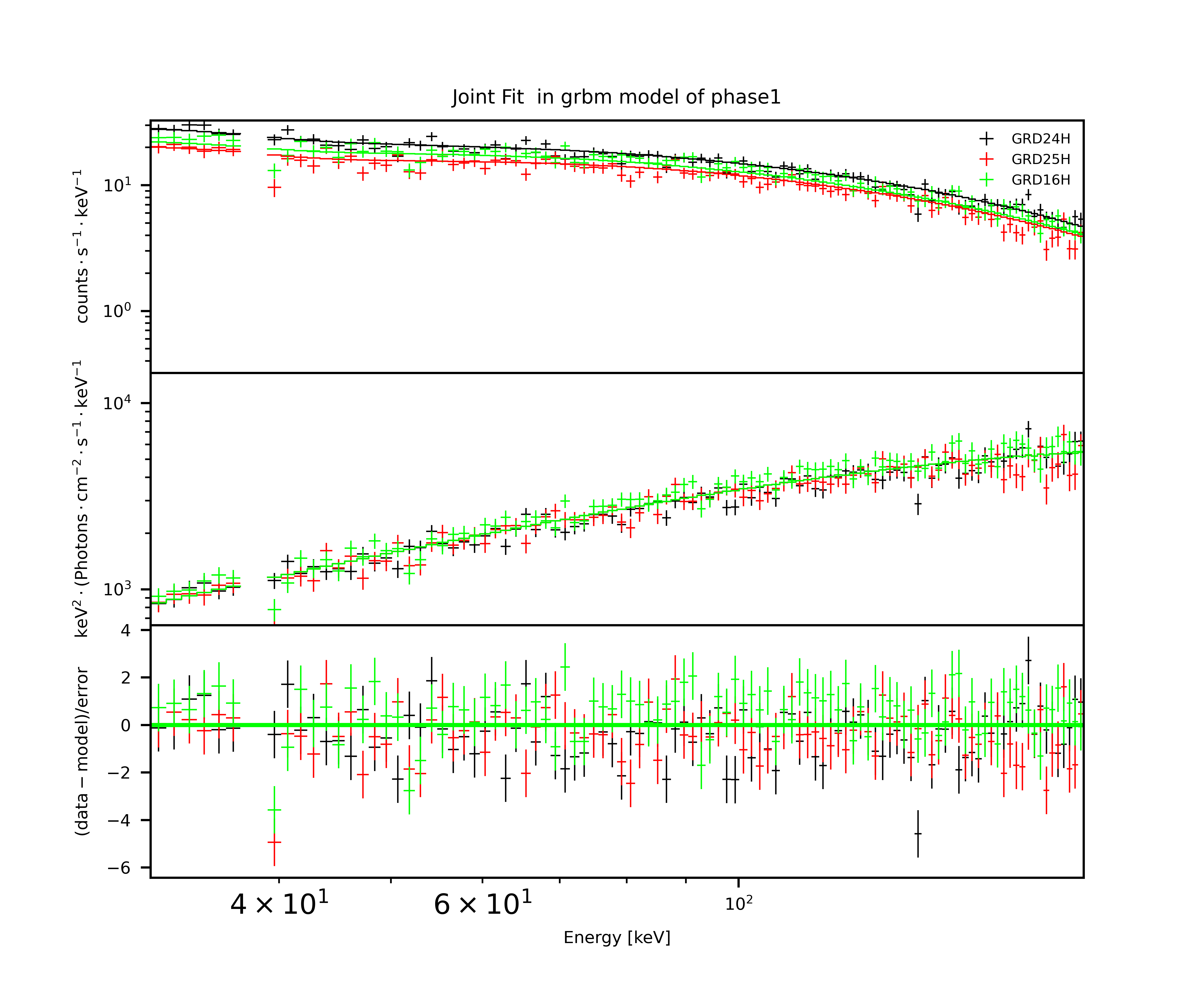}}
        \quad
        \subfigure[]{\includegraphics[height=3.5cm]{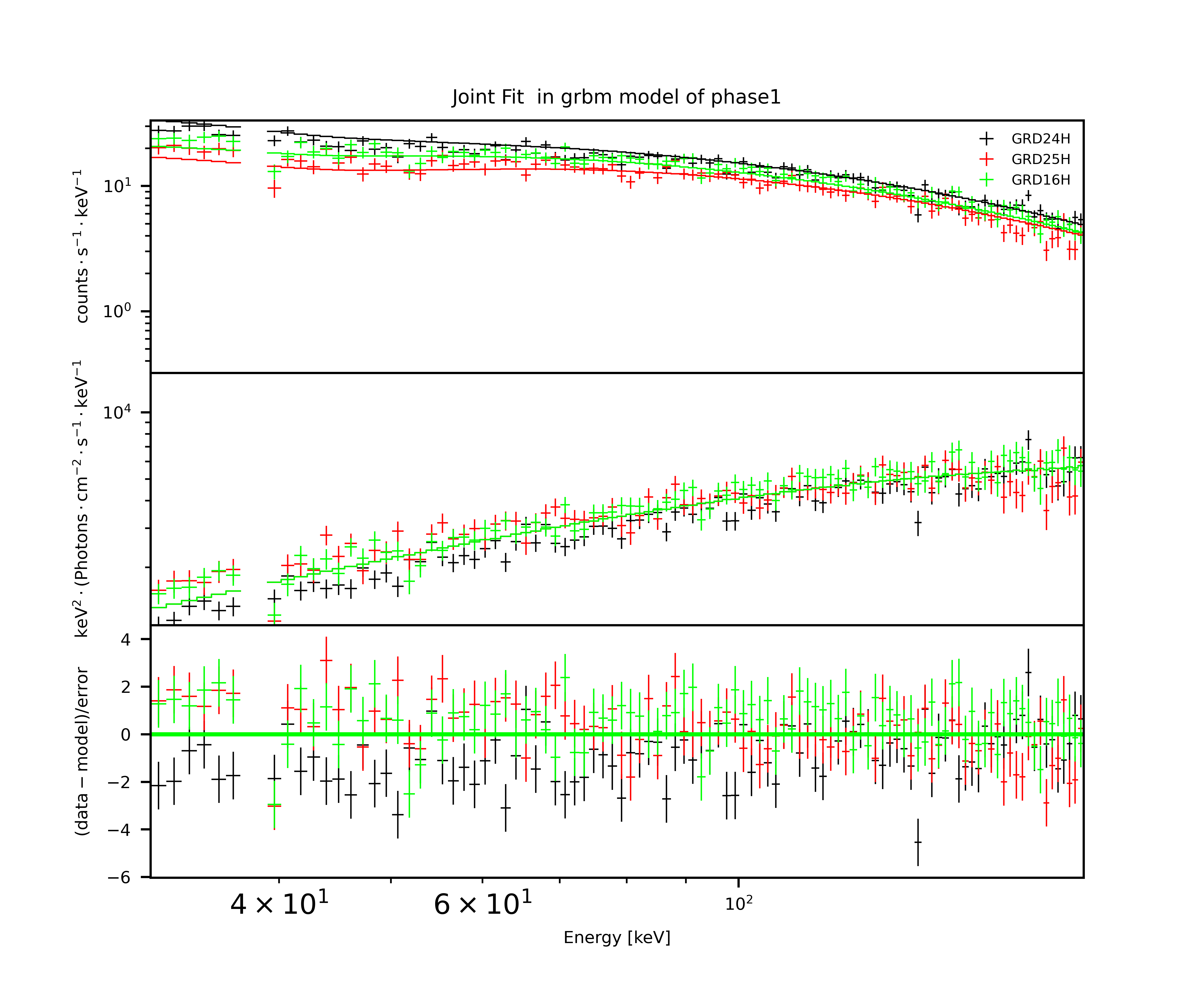}}
        \quad
        \\
        \subfigure[]{\includegraphics[width=4.5cm]{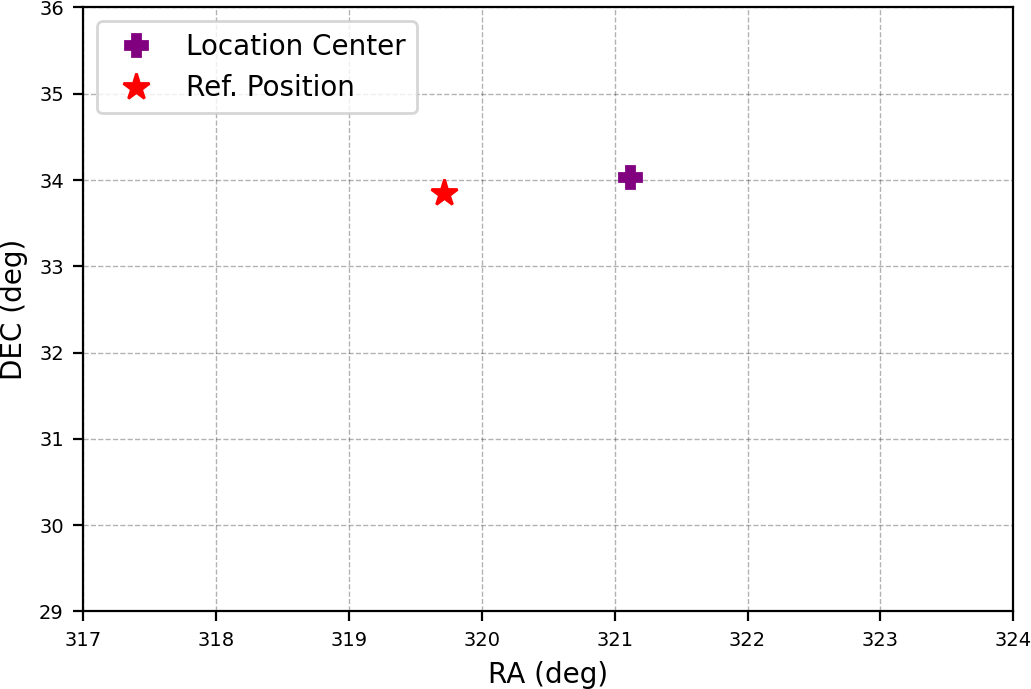}}
        \quad
        \subfigure[]{\includegraphics[width=4.5cm]{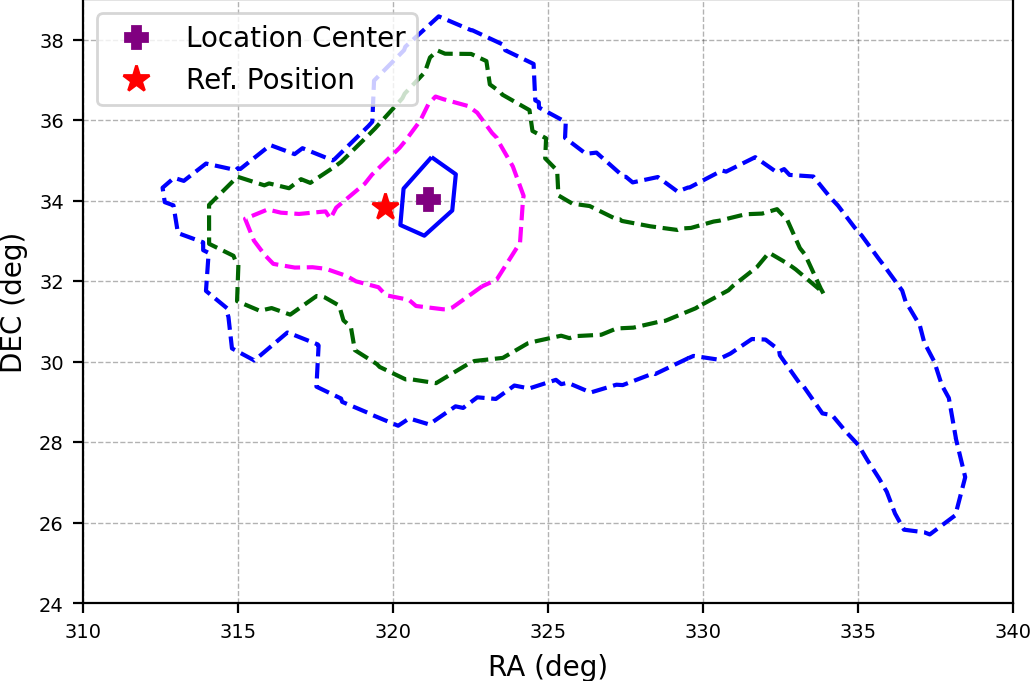}}
        \quad
        \subfigure[]{\includegraphics[width=4.5cm]{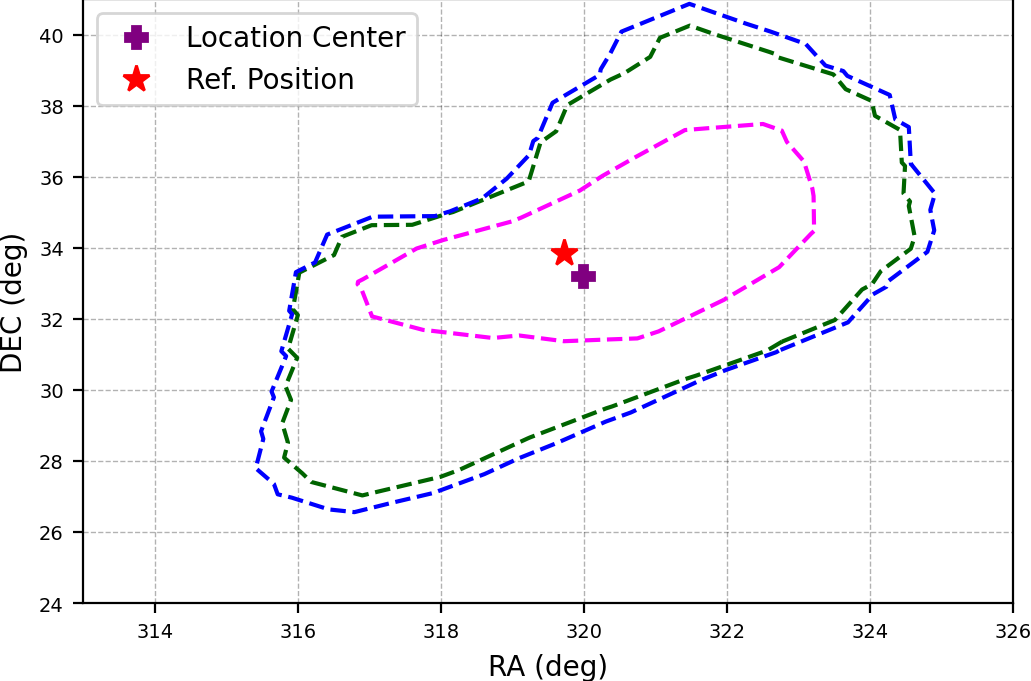}}
        \caption{ GECAM localization results of GRB 210619B. (a) The light curve of GRD \# 24 high gain which contains the majority of net (burst) counts. (b) The RFD spectral fitting result. (c) The RFD spectral fitting result. The location credible region of (d) FIX, (e) RFD, and (f) APR localization. No contour shown in (f) is due to the location probability of the location center $> 99.73\%$. The captions are the same as Figure \ref{fig2a}. }
        \label{fig2_05}
    \end{figure*}

    \begin{figure*}
        \centering
        \subfigure[]{\includegraphics[height=3.5cm]{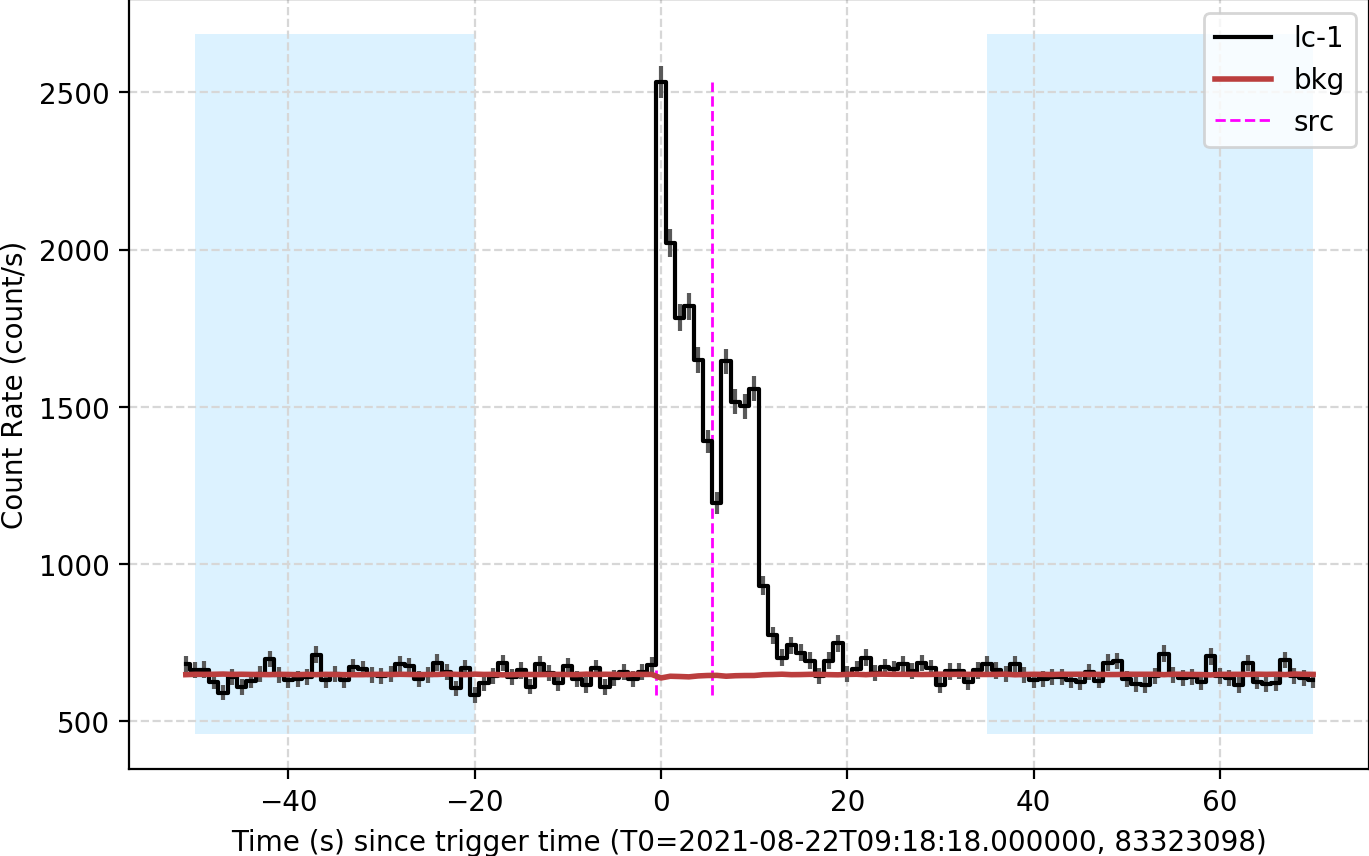}}
        \quad
        \subfigure[]{\includegraphics[height=3.5cm]{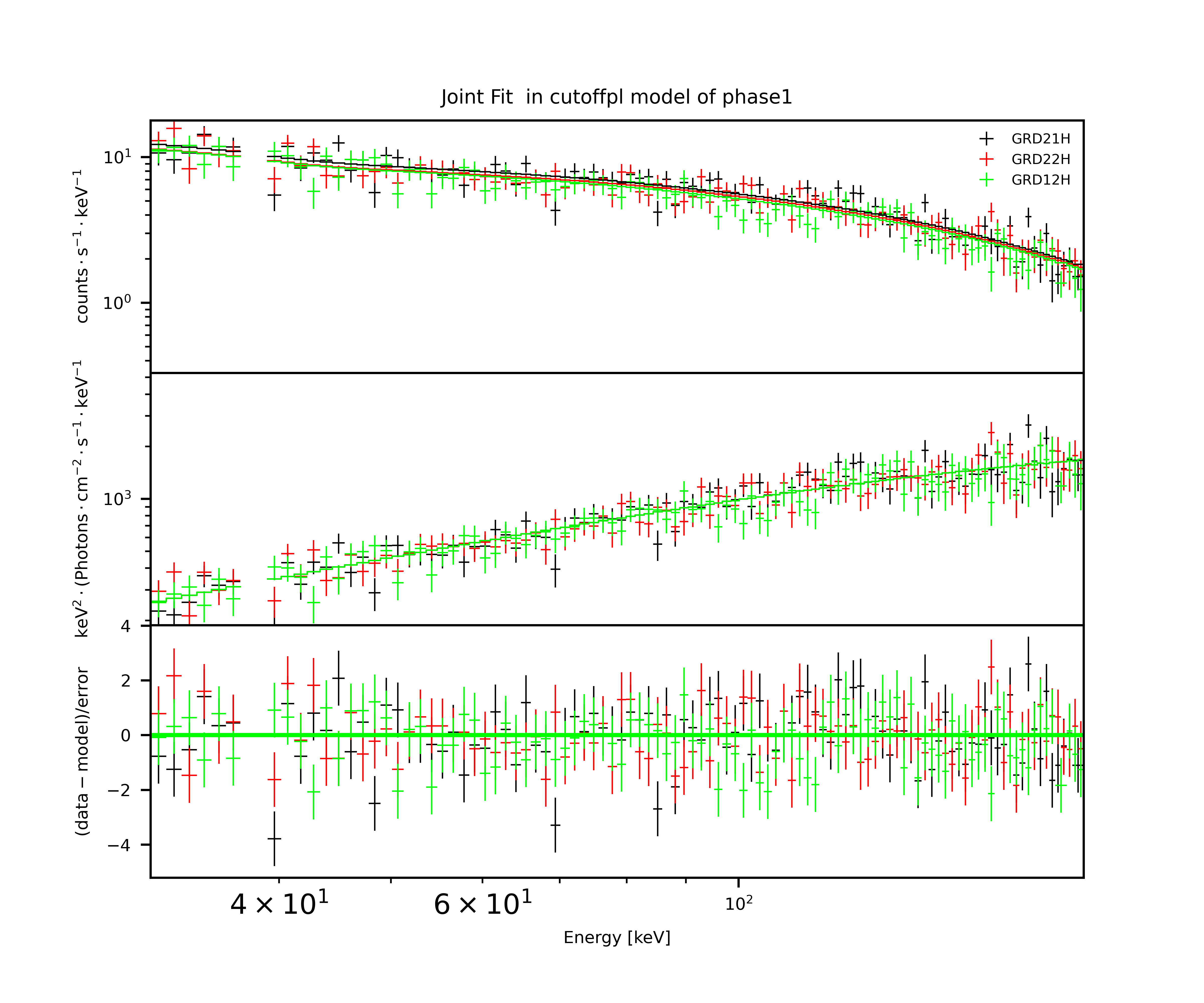}}
        \quad
        \subfigure[]{\includegraphics[height=3.5cm]{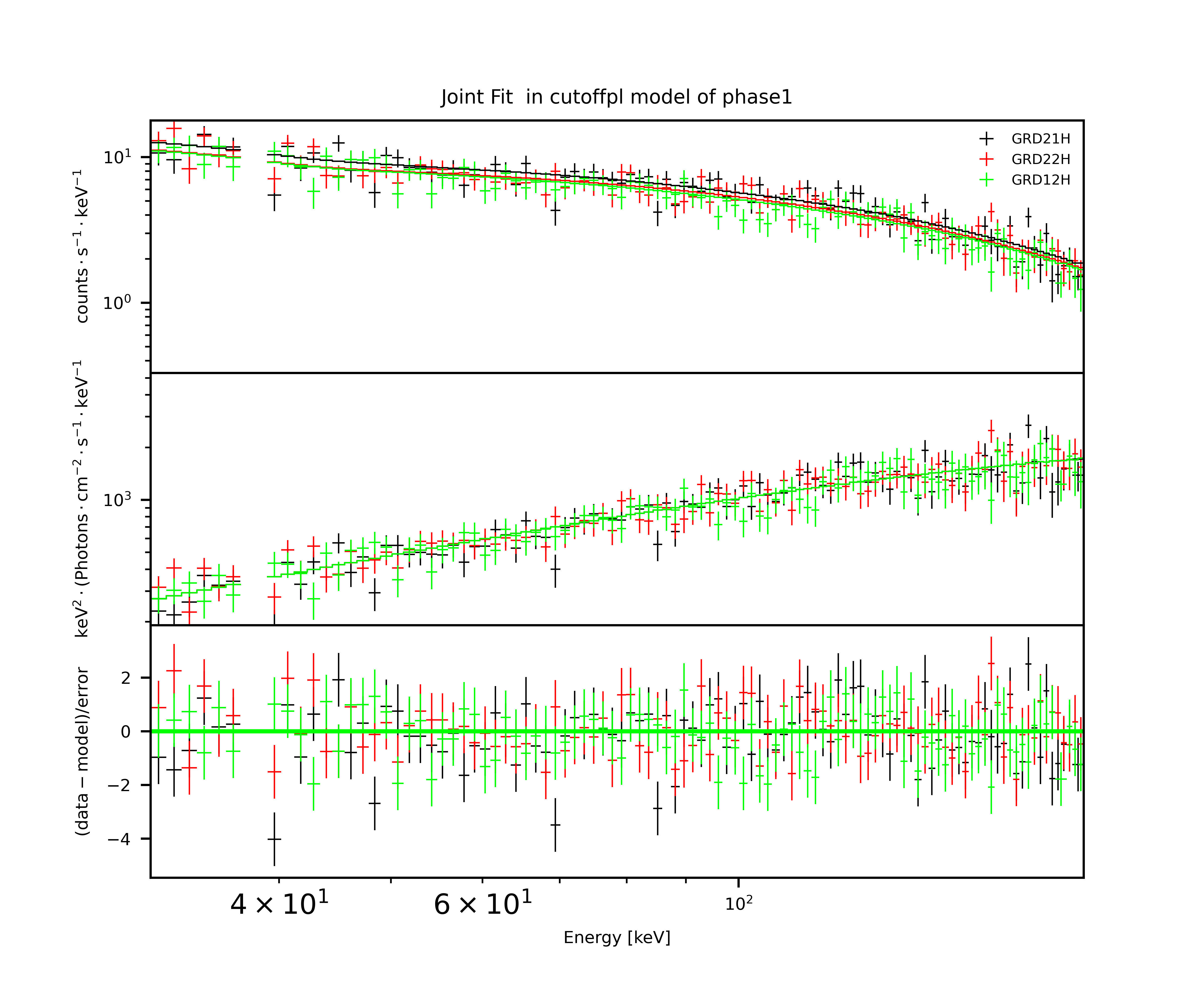}}
        \quad
        \\
        \subfigure[]{\includegraphics[width=4.5cm]{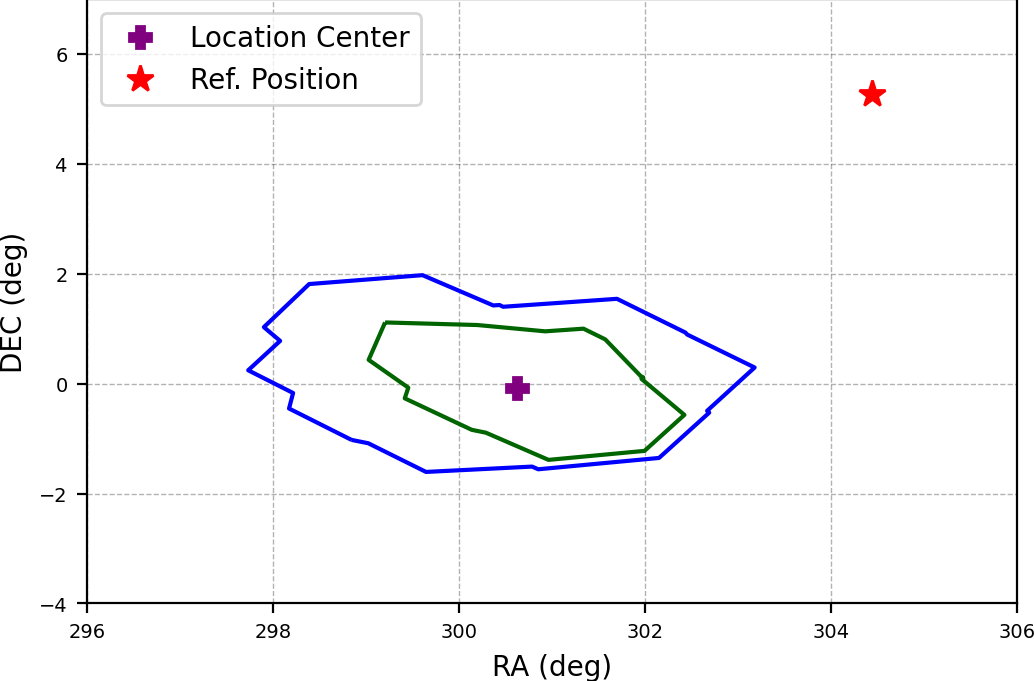}}
        \quad
        \subfigure[]{\includegraphics[width=4.5cm]{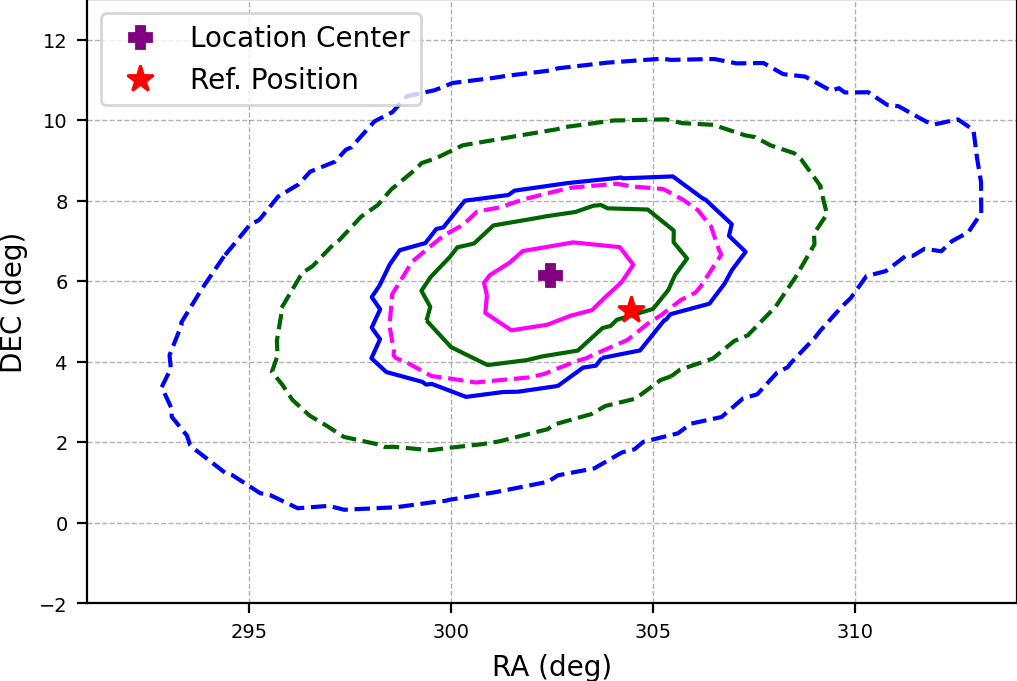}}
        \quad
        \subfigure[]{\includegraphics[width=4.5cm]{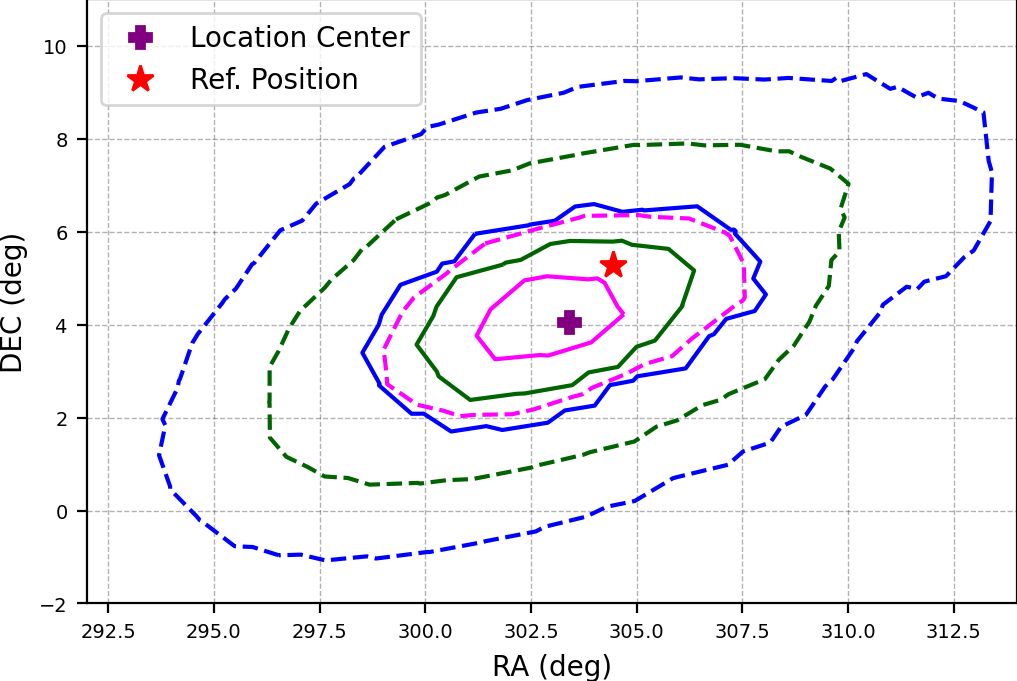}}
        \caption{ GECAM localization results of GRB 210822A (see also Table \ref{TABLE_LocRes_Part2}, \ref{TABLE_LocRes_Part2}, and \ref{TABLE_SpeRes}). (a) The light curve of GRD \# 21 high gain which contains the majority of net (burst) counts. (b) The RFD spectral fitting result. (c) The RFD spectral fitting result. The location credible region of (d) FIX, (e) RFD, and (f) APR localization. The captions are the same as Figure \ref{fig2a}. }
        \label{fig2_06}
    \end{figure*}

    \begin{figure*}
        \centering
        \subfigure[]{\includegraphics[height=3.5cm]{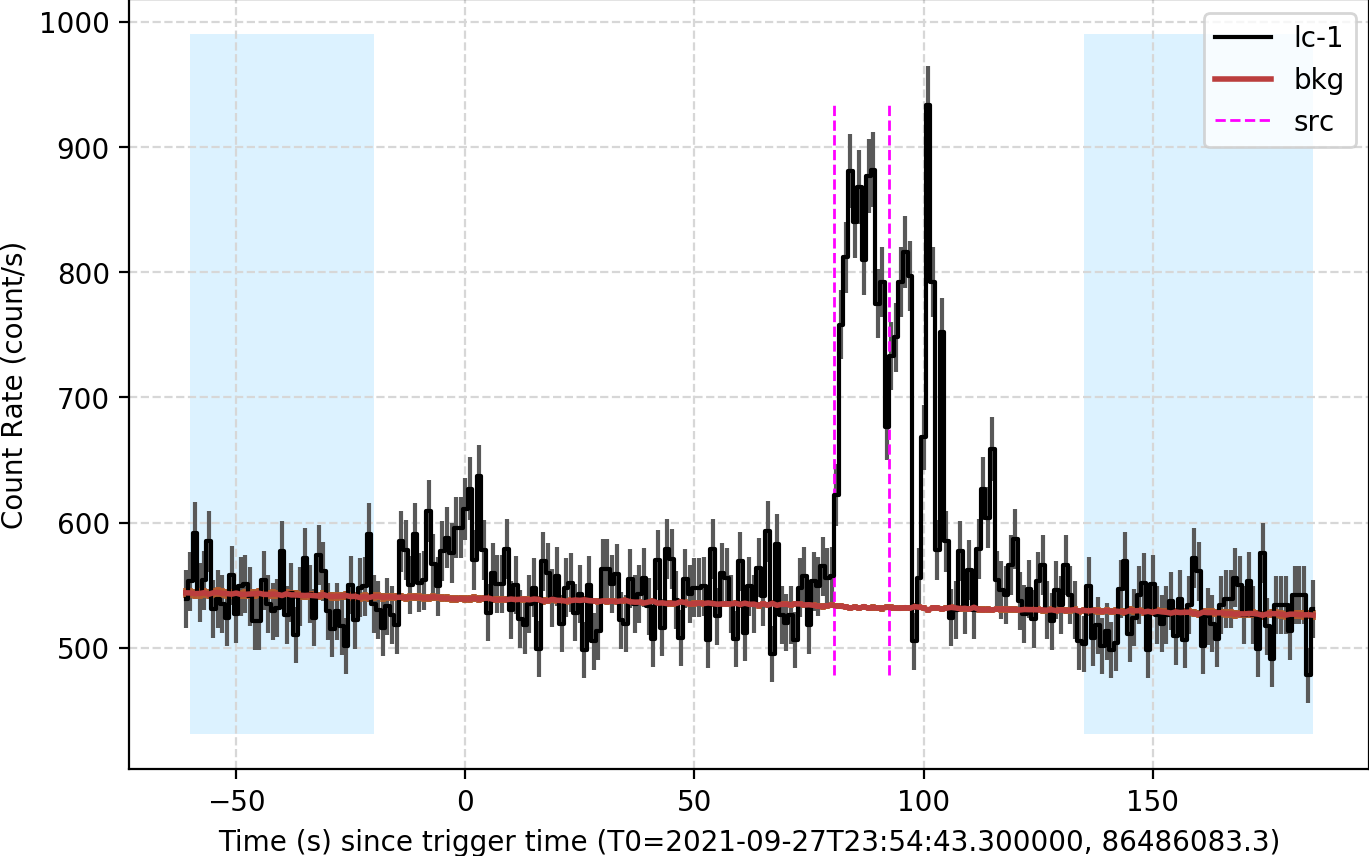}}
        \quad
        \subfigure[]{\includegraphics[height=3.5cm]{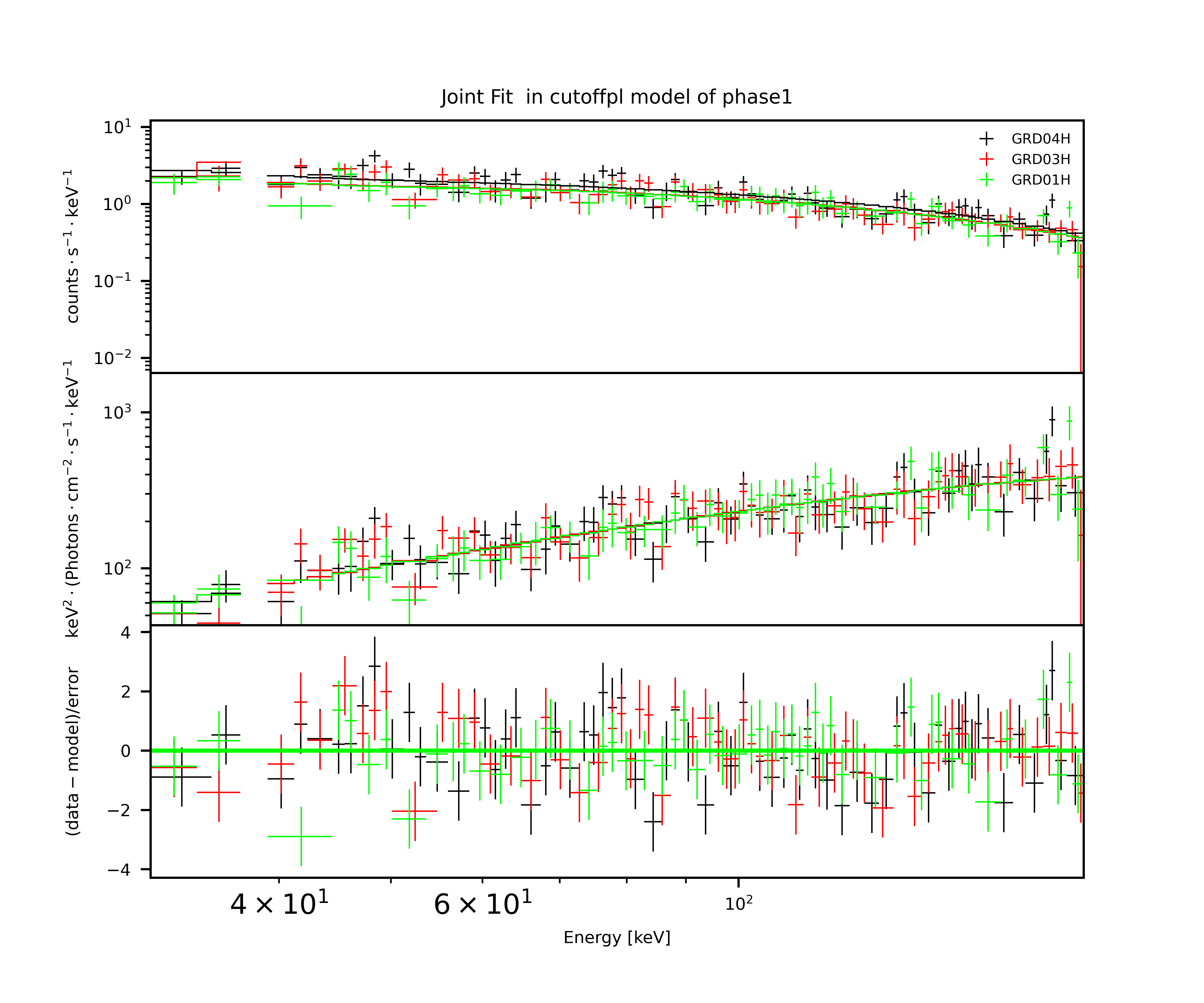}}
        \quad
        \subfigure[]{\includegraphics[height=3.5cm]{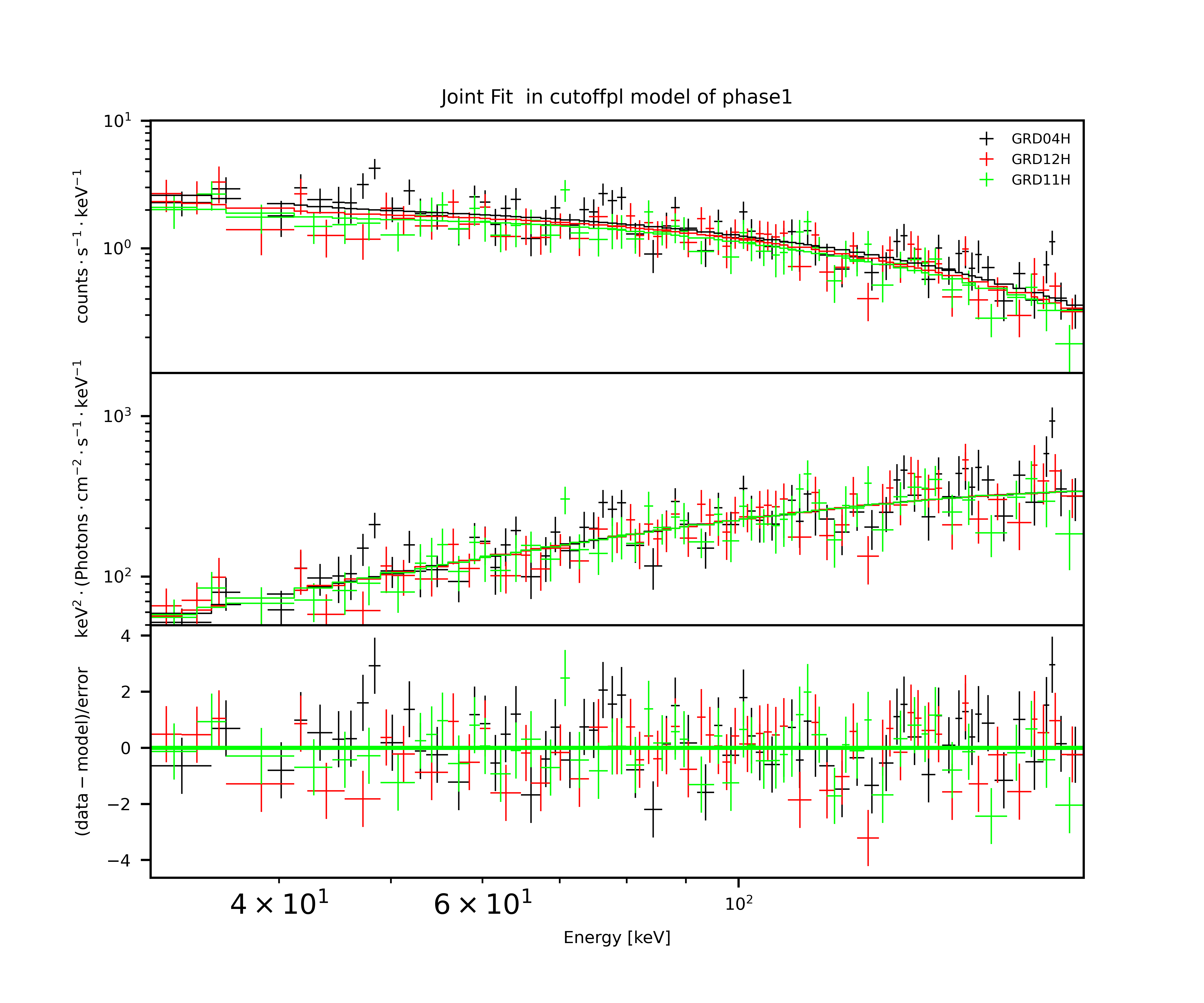}}
        \quad
        \\
        \subfigure[]{\includegraphics[width=4.5cm]{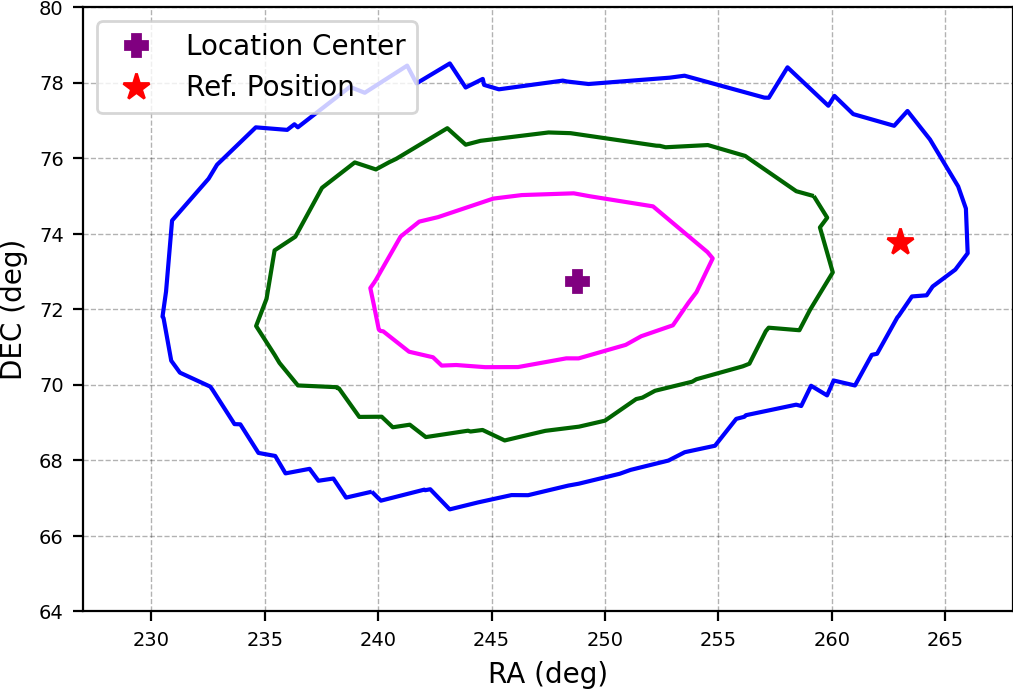}}
        \quad
        \subfigure[]{\includegraphics[width=4.5cm]{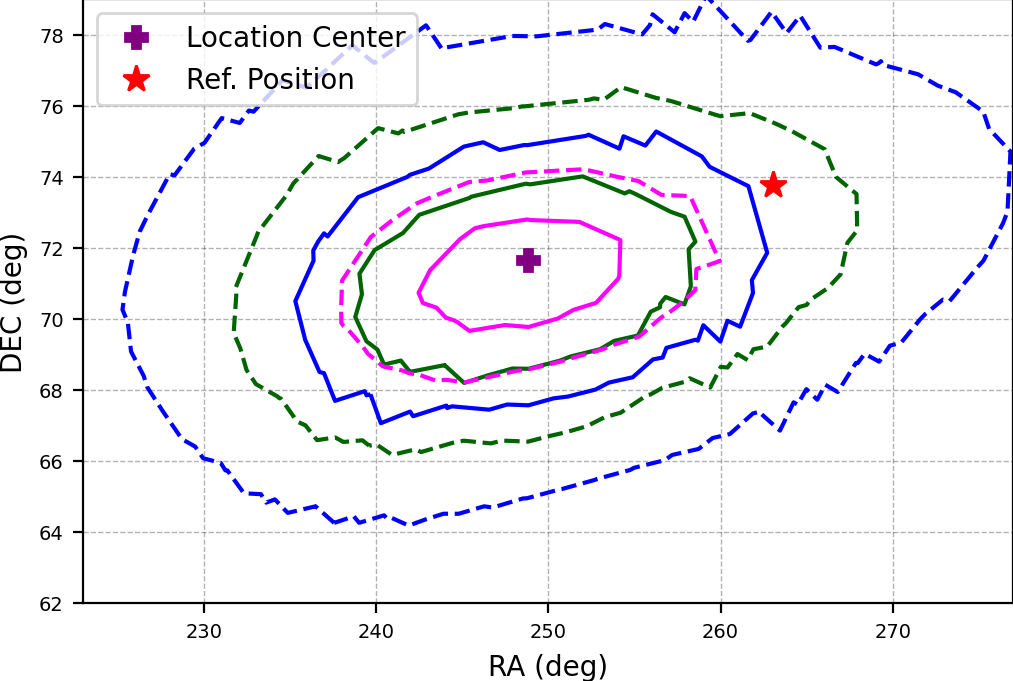}}
        \quad
        \subfigure[]{\includegraphics[width=4.5cm]{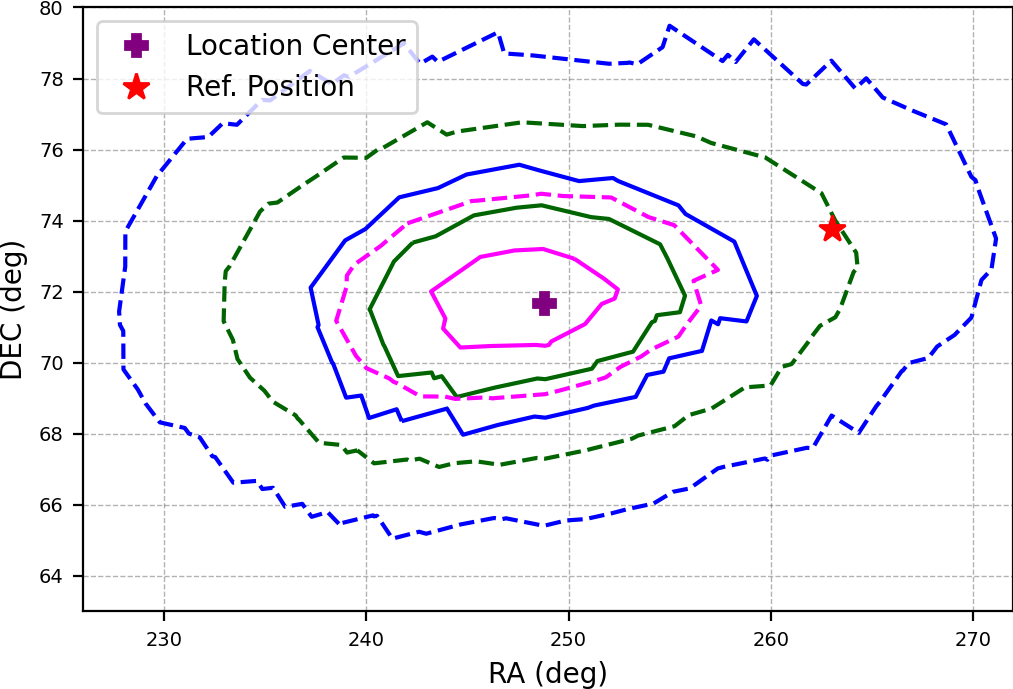}}
        \caption{ GECAM localization results of GRB 210927B. (a) The light curve of GRD \# 04 high gain which contains the majority of net (burst) counts. (b) The RFD spectral fitting result. (c) The RFD spectral fitting result. The location credible region of (d) FIX, (e) RFD, and (f) APR localization. The captions are the same as Figure \ref{fig2a}. }
        \label{fig2_07}
    \end{figure*}

    \begin{figure*}
        \centering
        \subfigure[]{\includegraphics[height=3.5cm]{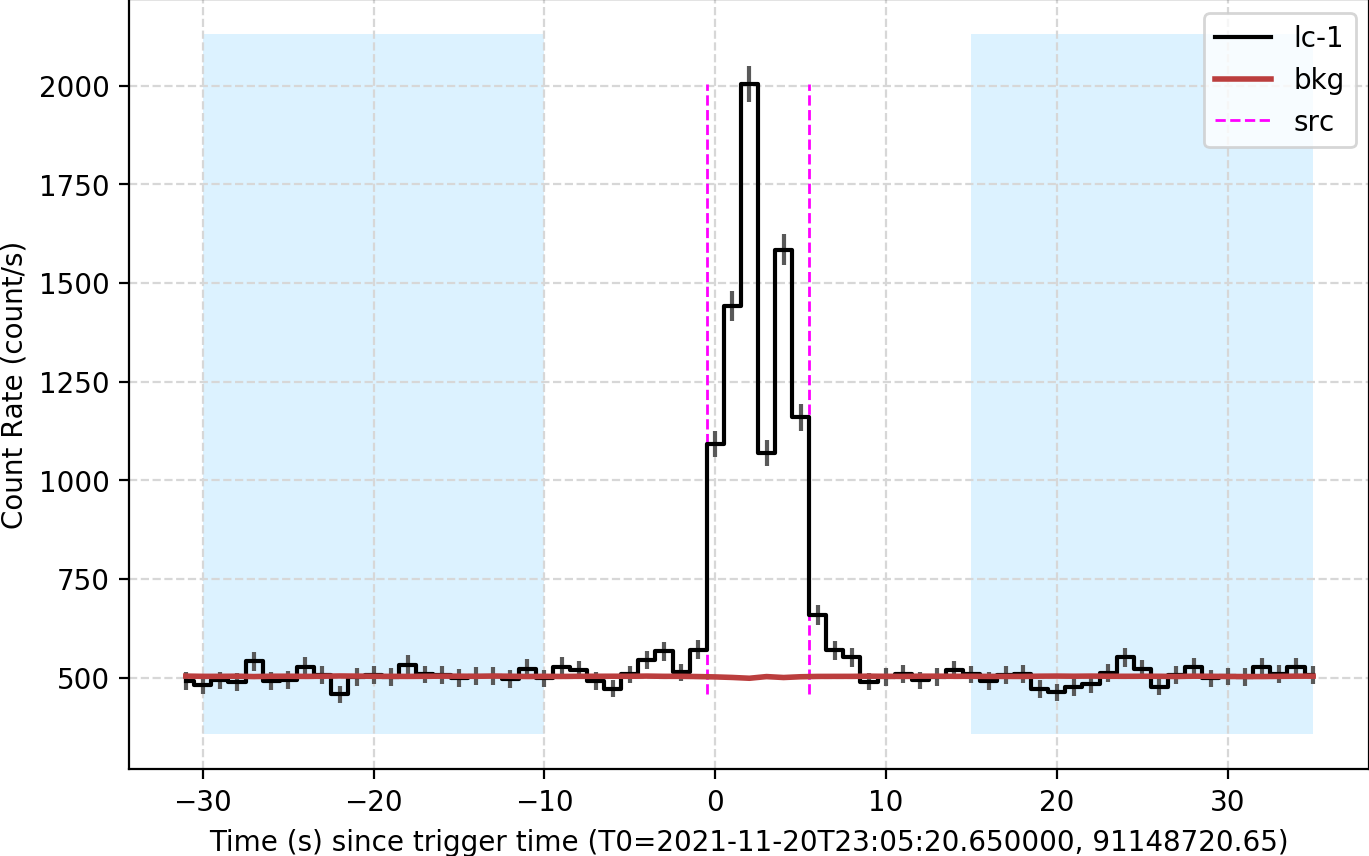}}
        \quad
        \subfigure[]{\includegraphics[height=3.5cm]{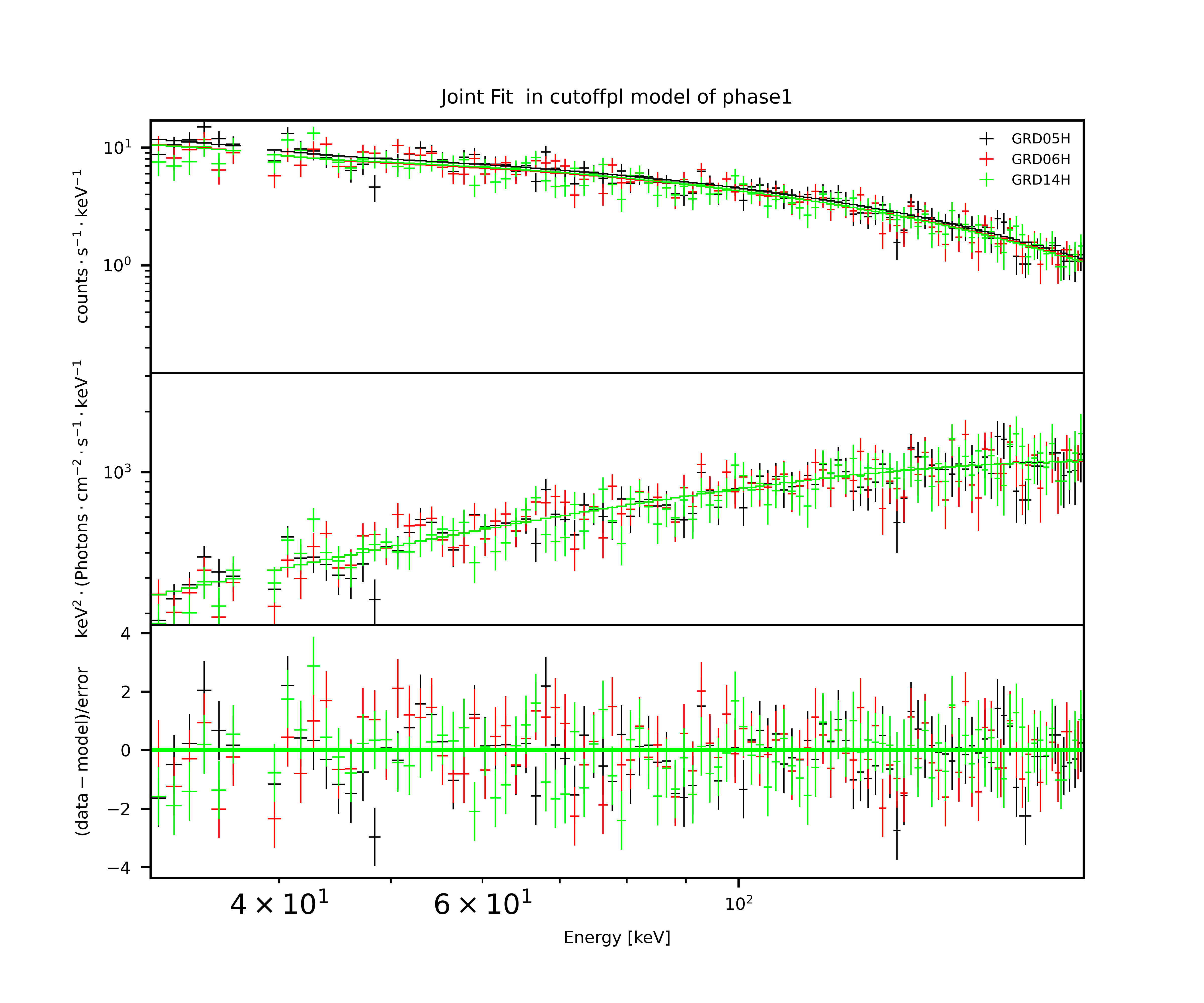}}
        \quad
        \subfigure[]{\includegraphics[height=3.5cm]{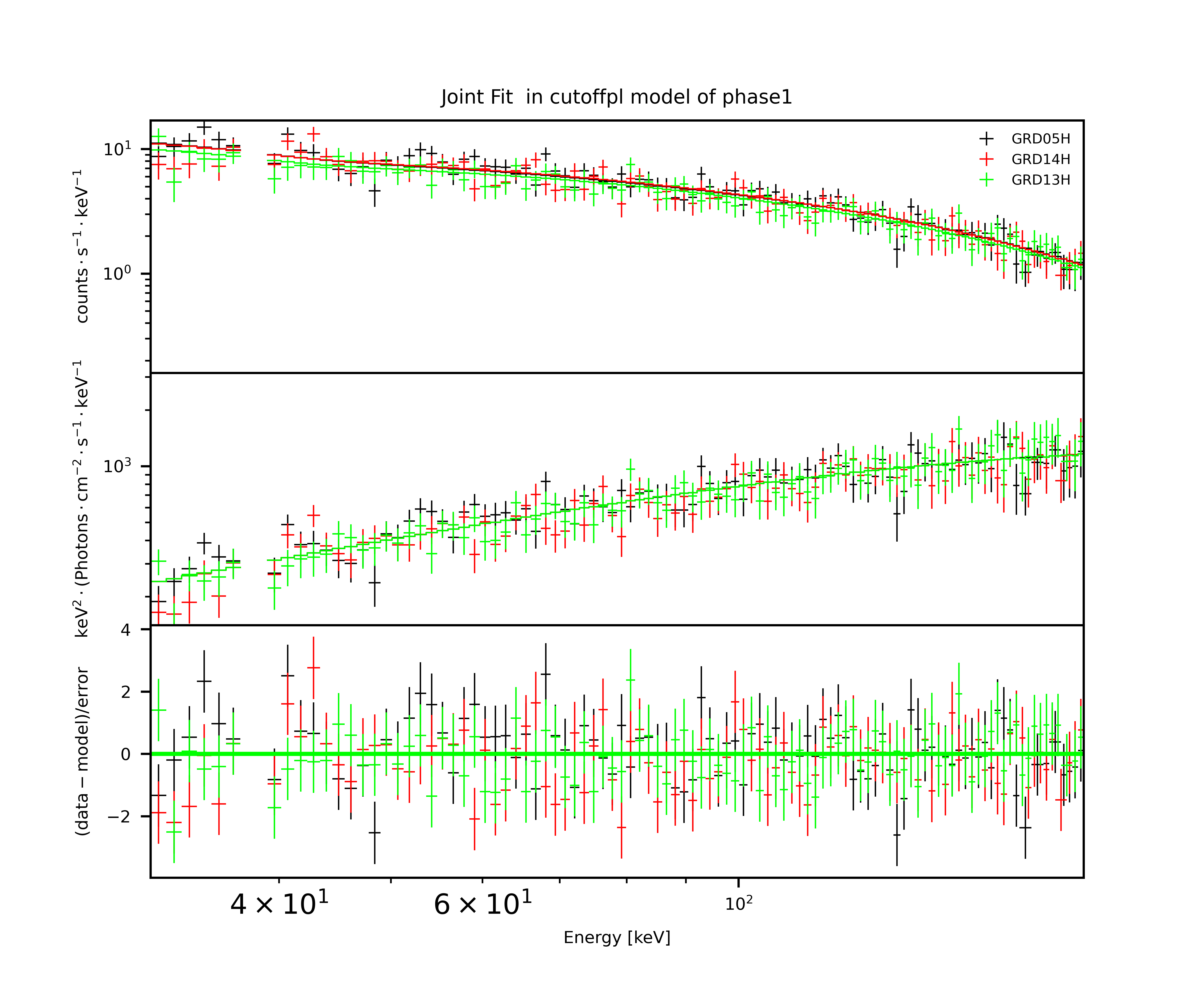}}
        \quad
        \\
        \subfigure[]{\includegraphics[width=4.5cm]{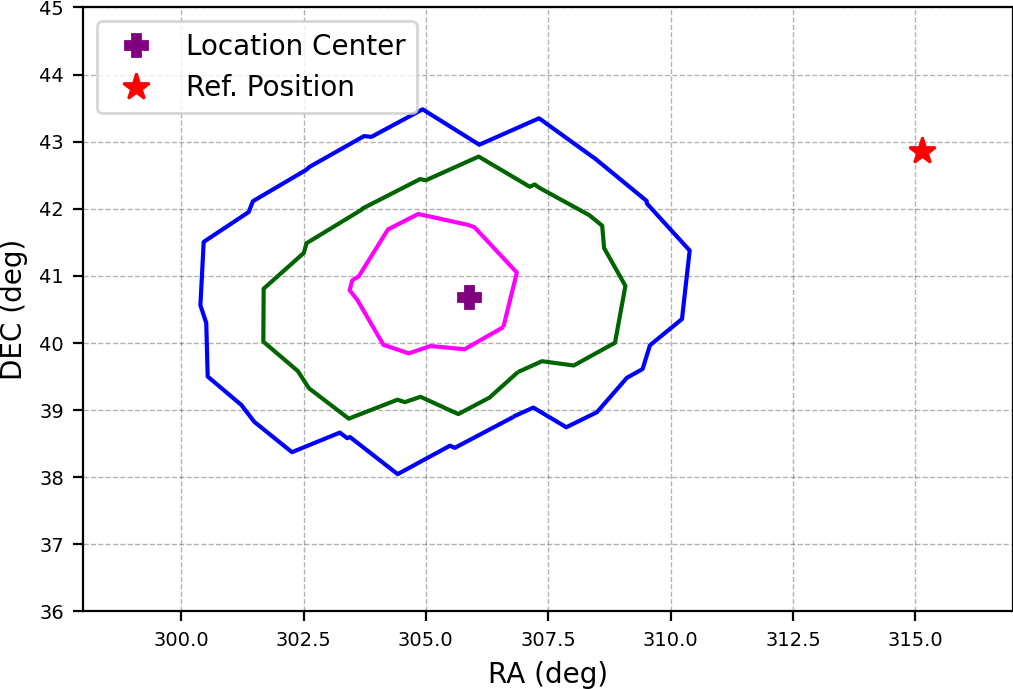}}
        \quad
        \subfigure[]{\includegraphics[width=4.5cm]{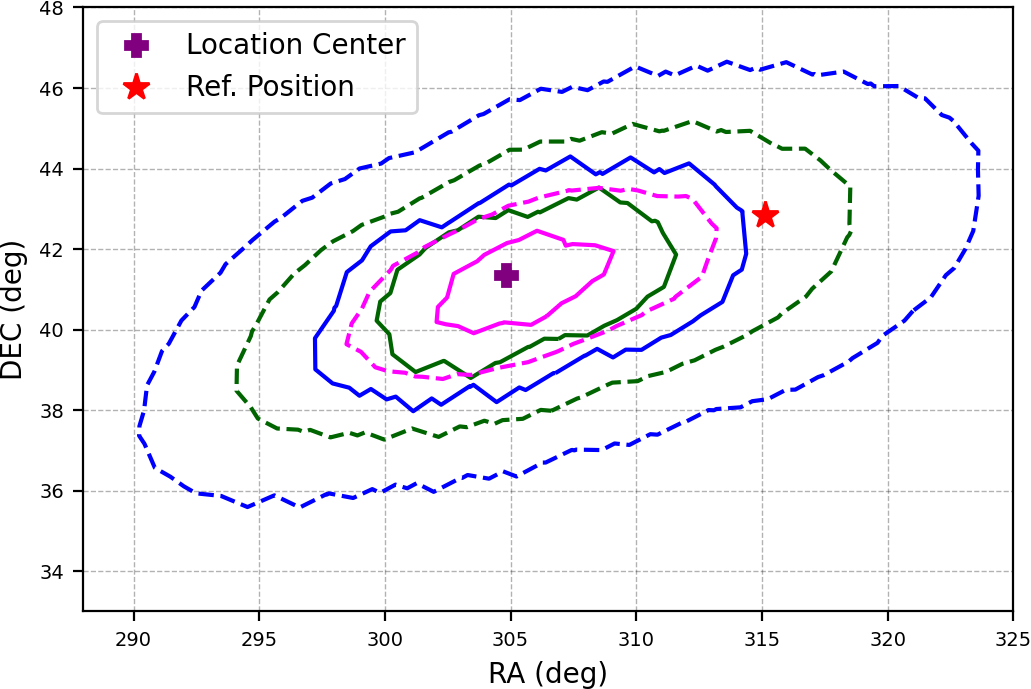}}
        \quad
        \subfigure[]{\includegraphics[width=4.5cm]{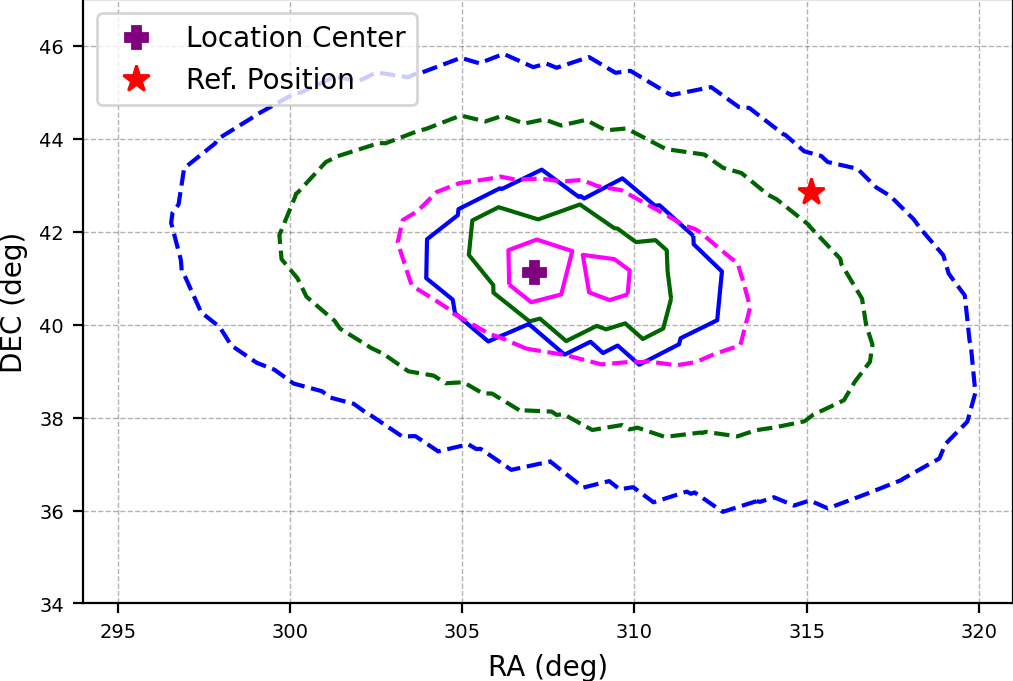}}
        \caption{ GECAM localization results of GRB 211120A. (a) The light curve of GRD \# 05 high gain which contains the majority of net (burst) counts. (b) The RFD spectral fitting result. (c) The RFD spectral fitting result. The location credible region of (d) FIX, (e) RFD, and (f) APR localization. The captions are the same as Figure \ref{fig2a}. }
        \label{fig2_08}
    \end{figure*}

    \begin{figure*}
        \centering
        \subfigure[]{\includegraphics[height=3.5cm]{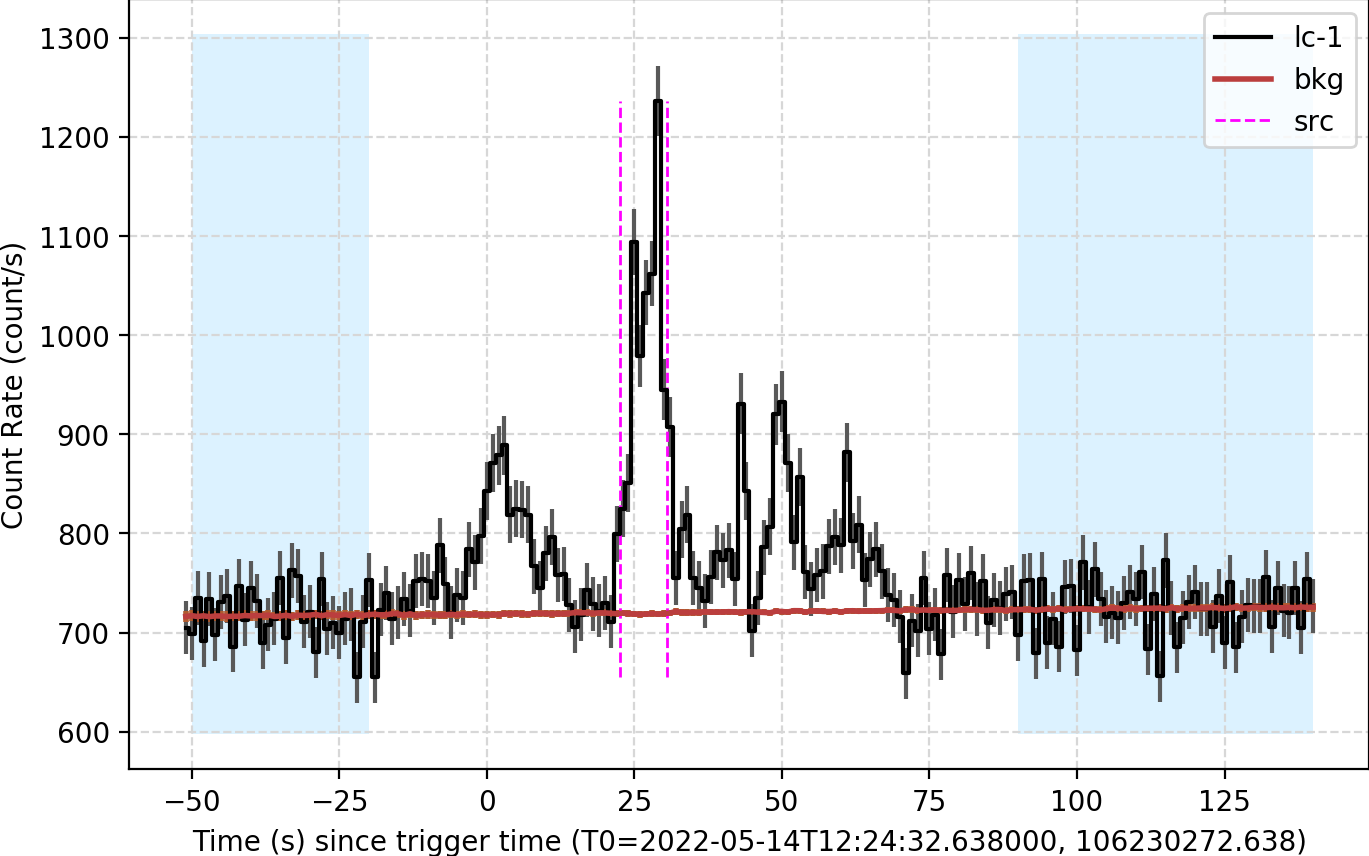}}
        \quad
        \subfigure[]{\includegraphics[height=3.5cm]{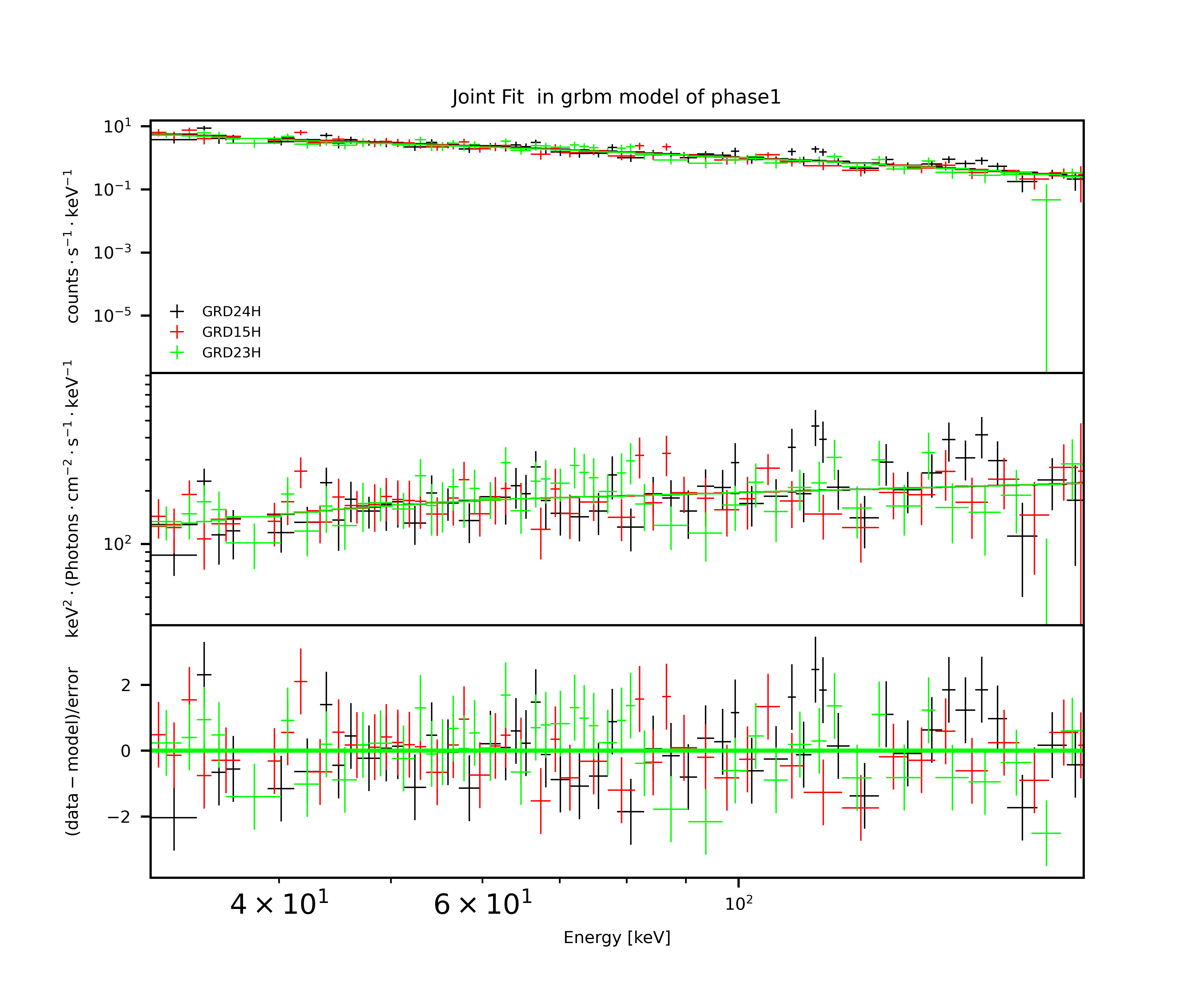}}
        \quad
        \subfigure[]{\includegraphics[height=3.5cm]{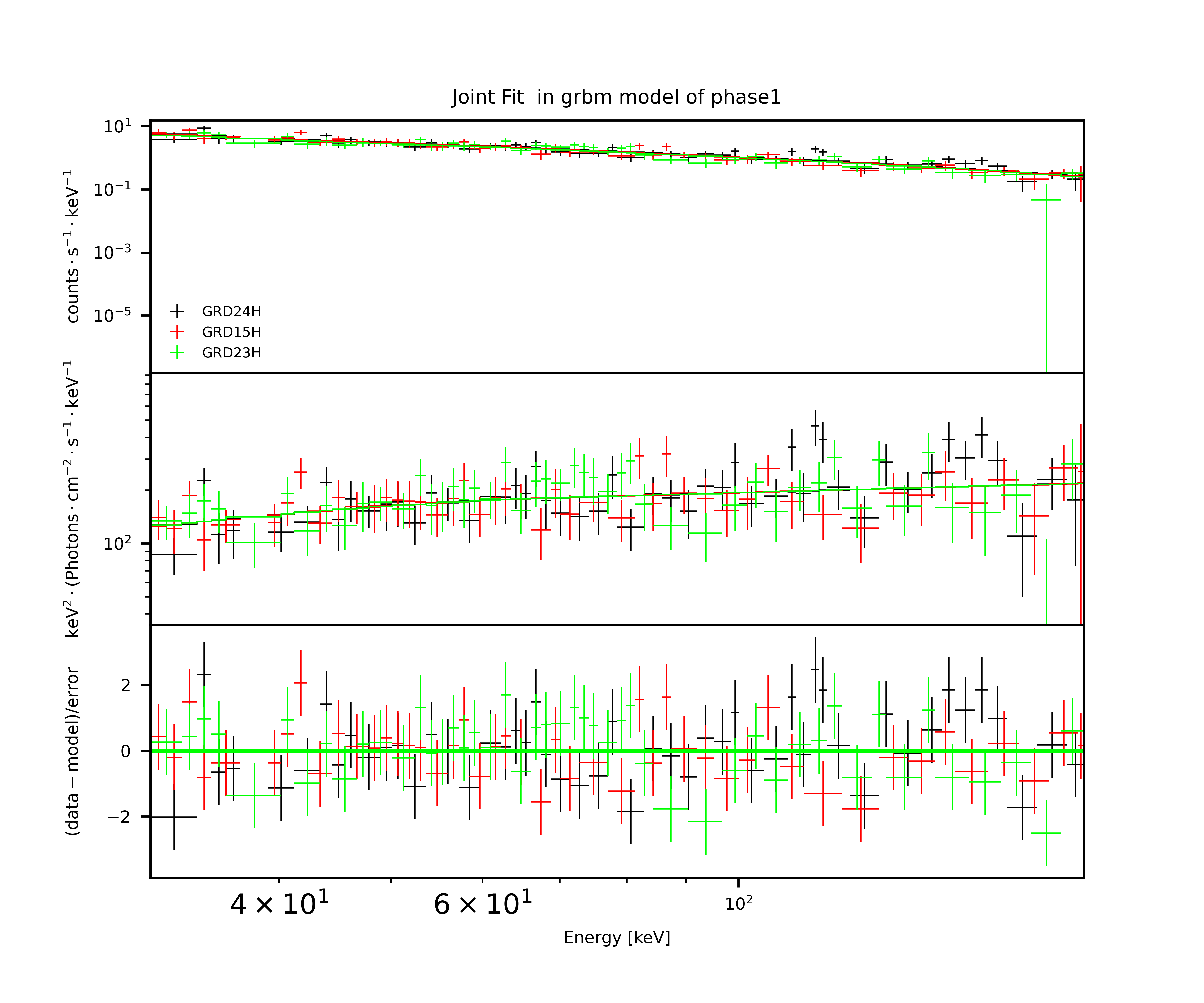}}
        \quad
        \\
        \subfigure[]{\includegraphics[width=4.5cm]{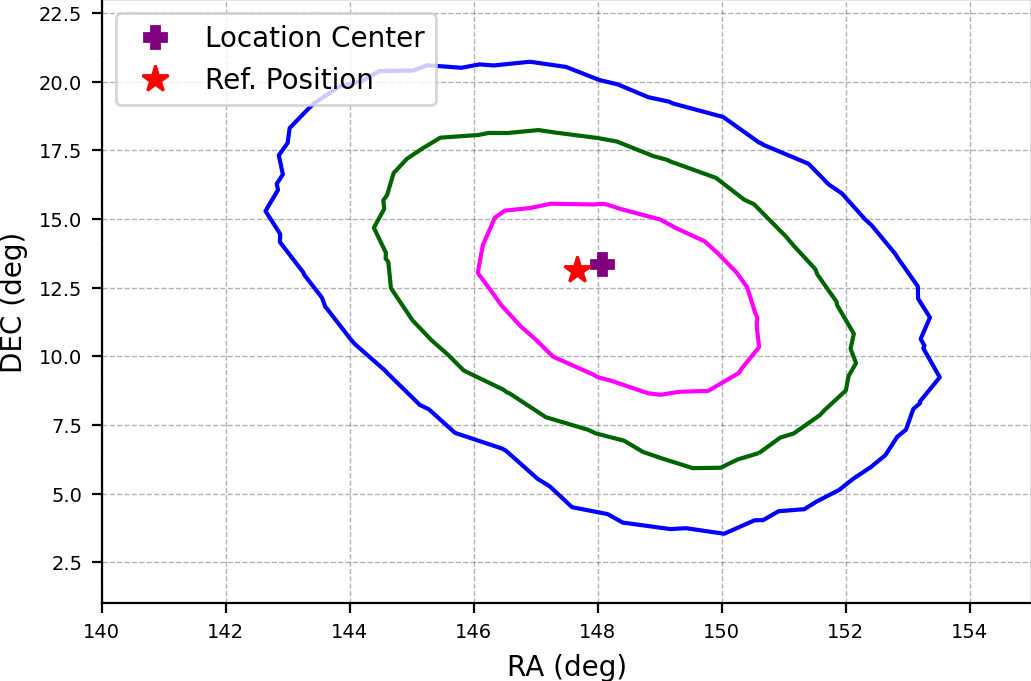}}
        \quad
        \subfigure[]{\includegraphics[width=4.5cm]{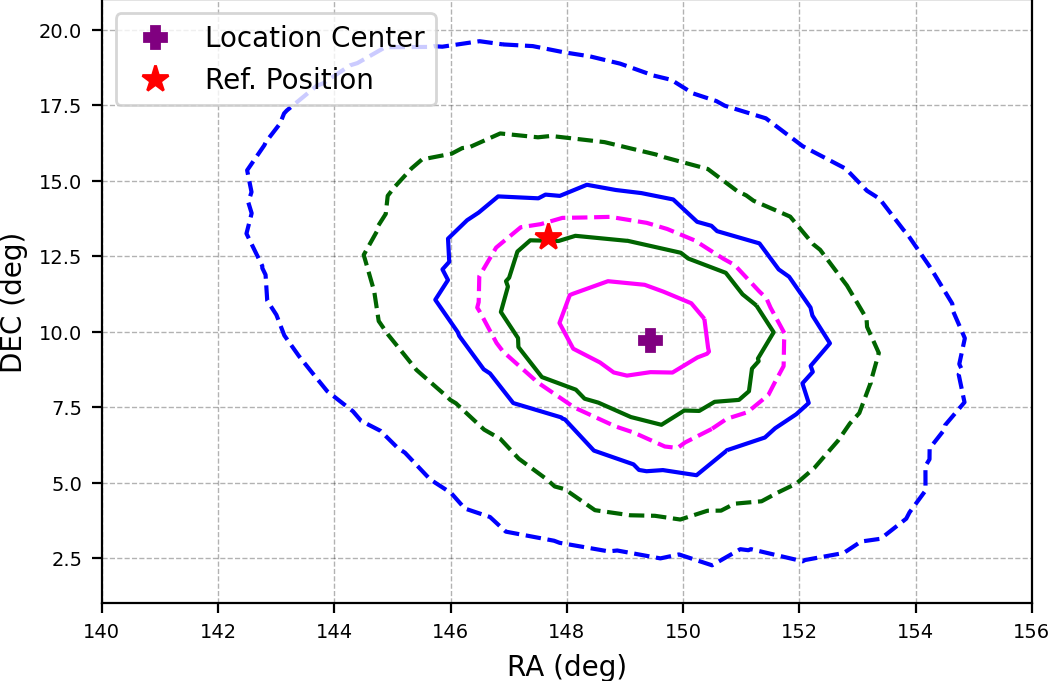}}
        \quad
        \subfigure[]{\includegraphics[width=4.5cm]{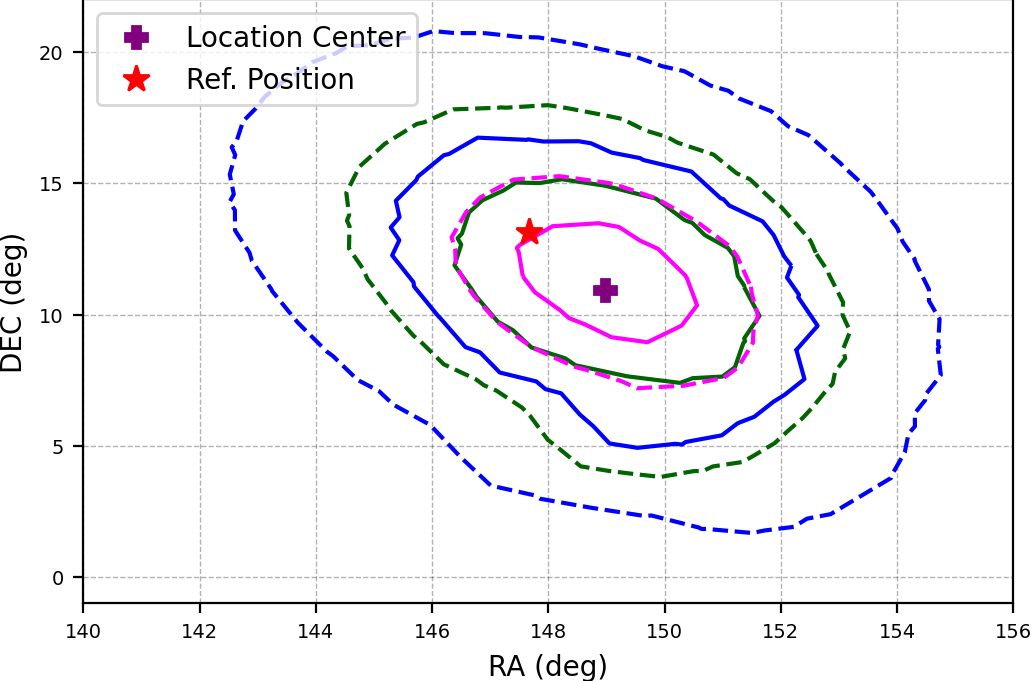}}
        \caption{ GECAM localization results of GRB 220514A. (a) The light curve of GRD \# 24 high gain which contains the majority of net (burst) counts. (b) The RFD spectral fitting result. (c) The RFD spectral fitting result. The location credible region of (d) FIX, (e) RFD, and (f) APR localization. The captions are the same as Figure \ref{fig2a}. }
        \label{fig2_GRB09}
    \end{figure*}

    \begin{figure*}
        \centering
        \subfigure[]{\includegraphics[height=3.5cm]{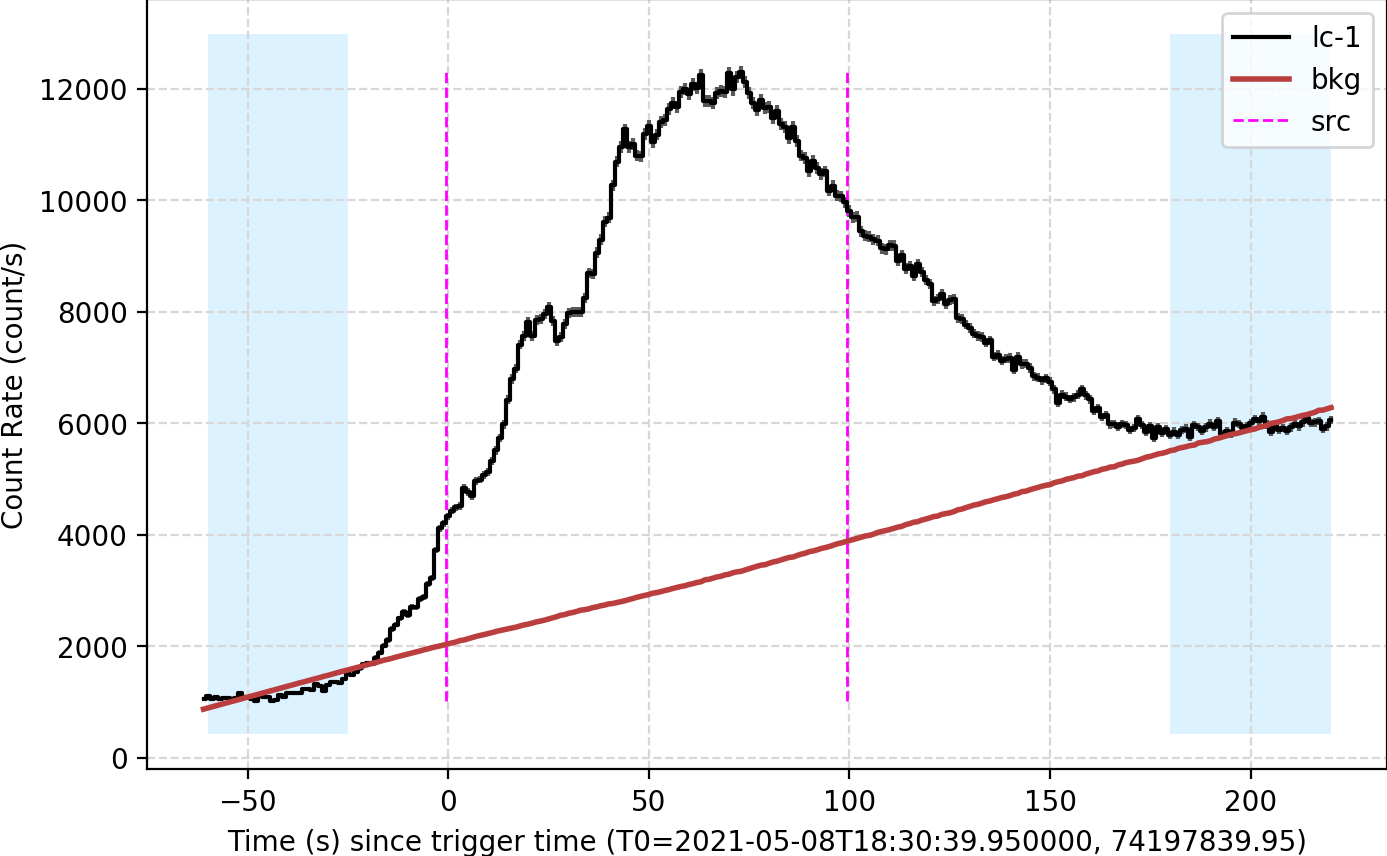}}
        \quad
        \subfigure[]{\includegraphics[height=3.5cm]{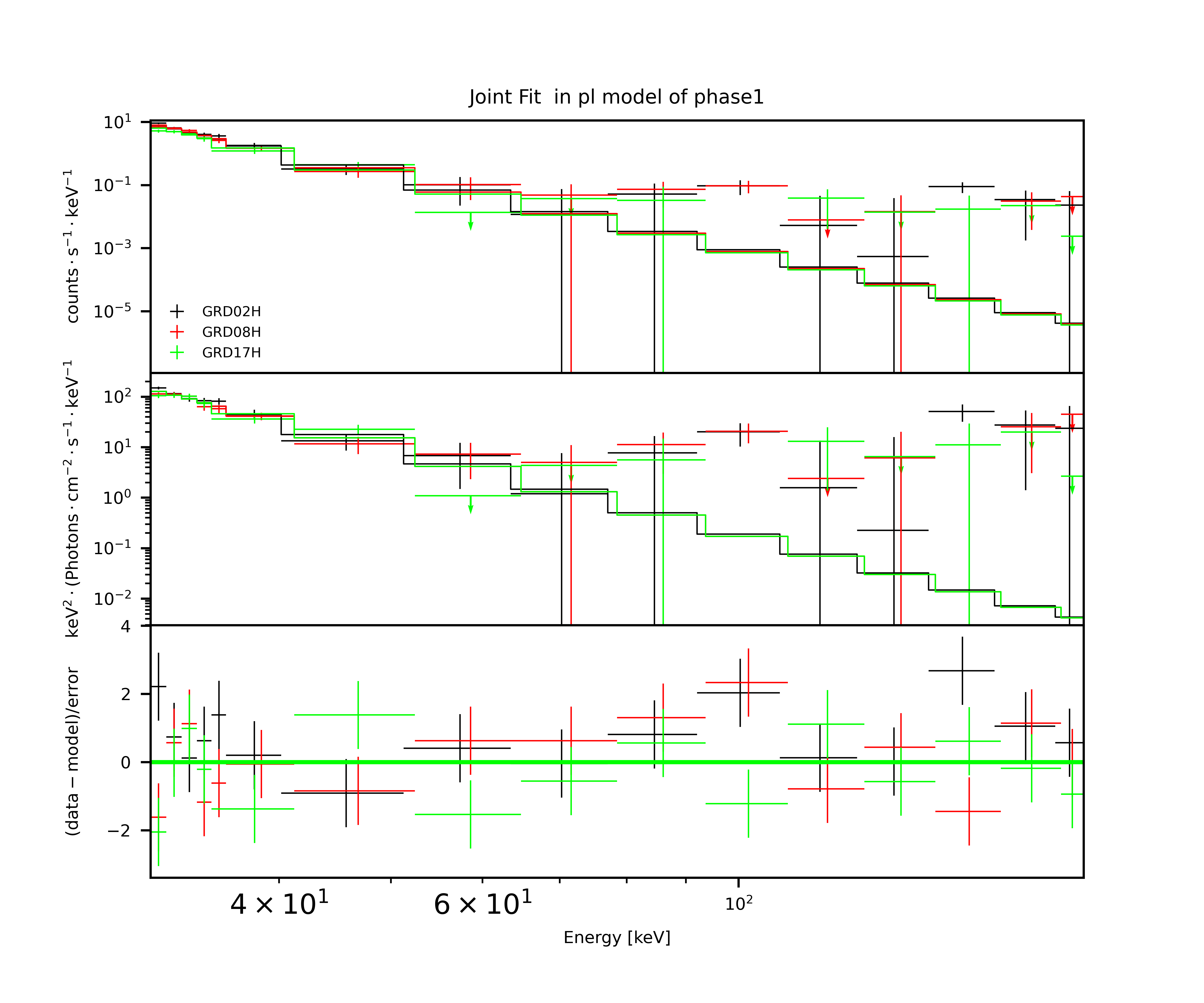}}
        \quad
        \subfigure[]{\includegraphics[height=3.5cm]{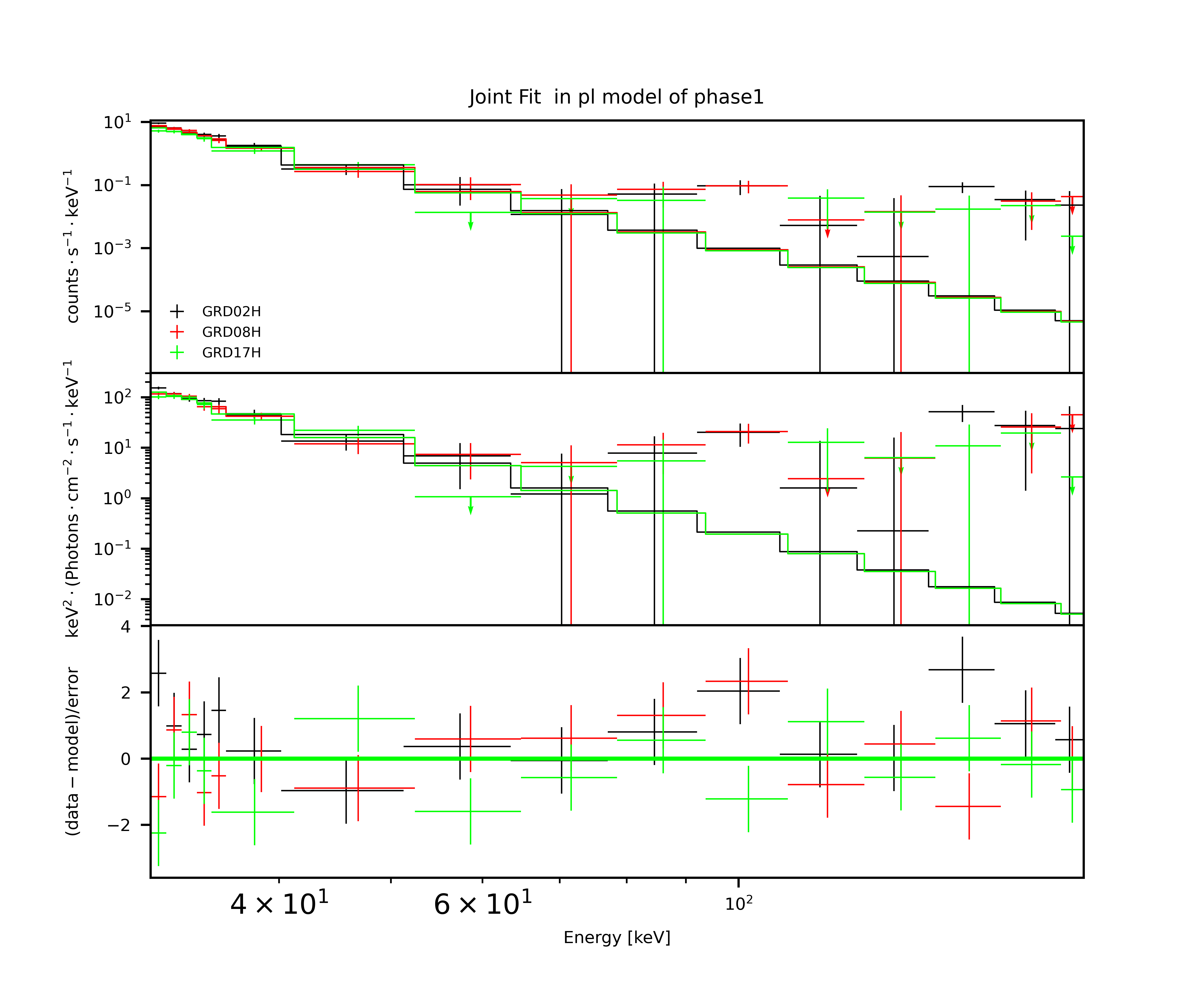}}
        \quad
        \\
        \subfigure[]{\includegraphics[width=4.5cm]{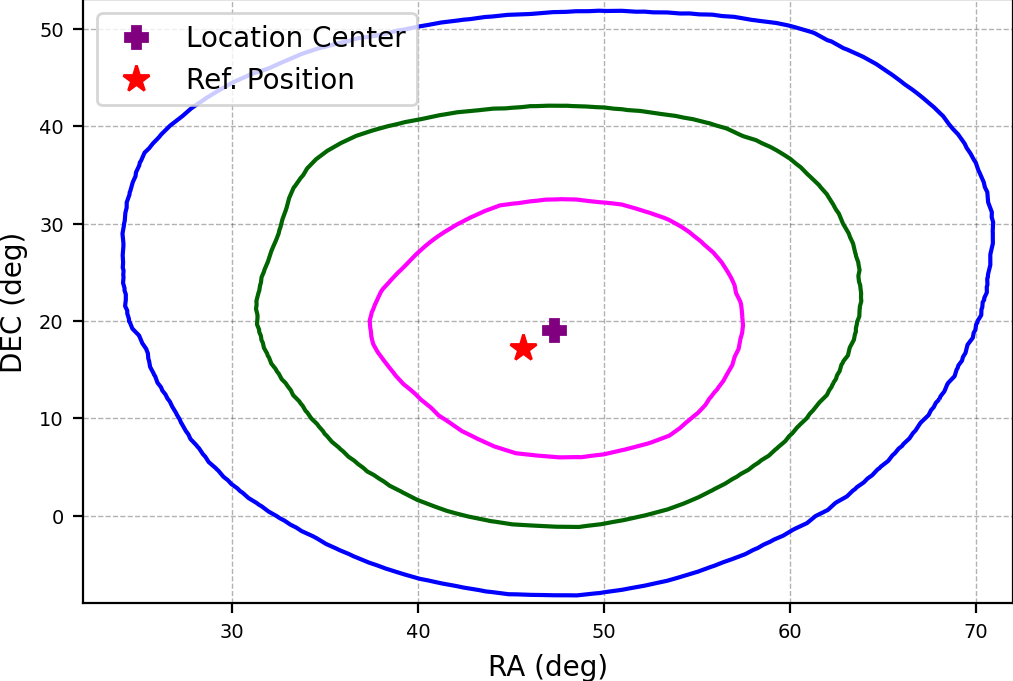}}
        \quad
        \subfigure[]{\includegraphics[width=4.5cm]{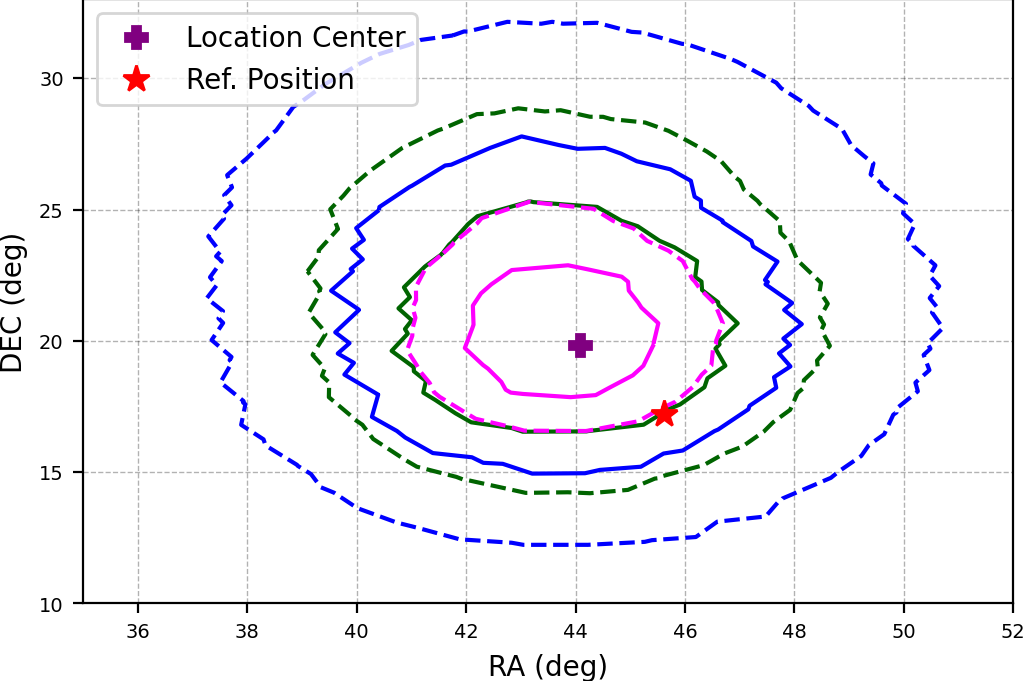}}
        \quad
        \subfigure[]{\includegraphics[width=4.5cm]{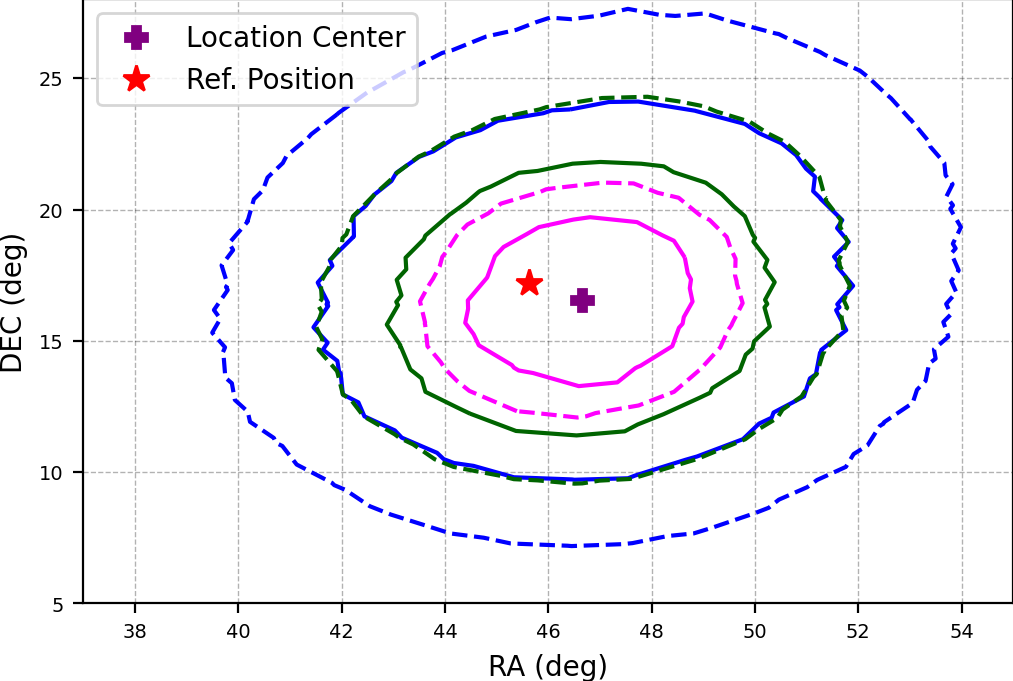}}
        \caption{ GECAM localization results of SFL 210508. (a) The light curve of GRD \# 02 high gain which contains the majority of net (burst) counts. (b) The RFD spectral fitting result. (c) The RFD spectral fitting result. The location credible region of (d) FIX, (e) RFD, and (f) APR localization. The captions are the same as Figure \ref{fig2a}. }
        \label{fig2_SFL210508}
    \end{figure*}

    \begin{figure*}
        \centering
        \subfigure[]{\includegraphics[height=3.5cm]{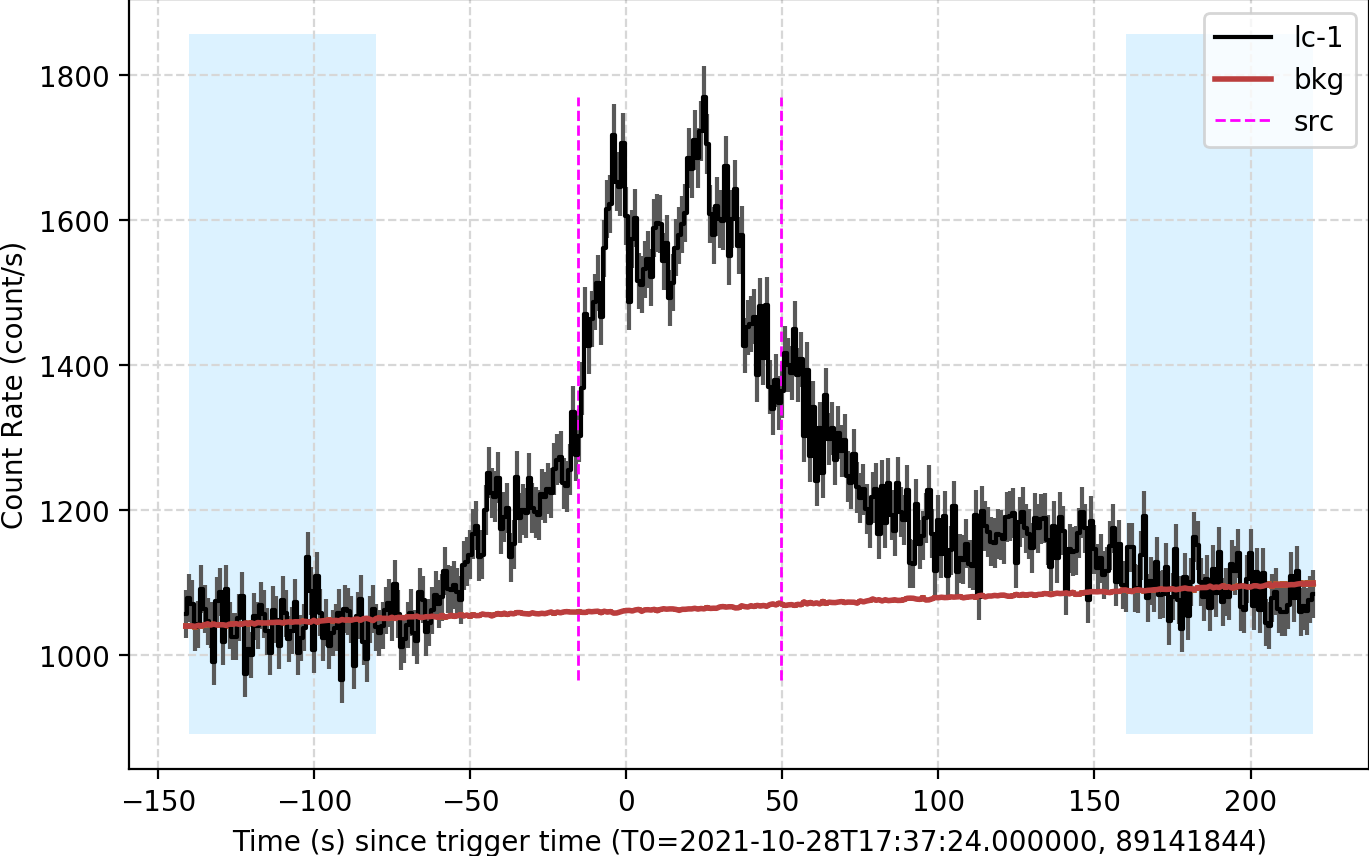}}
        \quad
        \subfigure[]{\includegraphics[height=3.5cm]{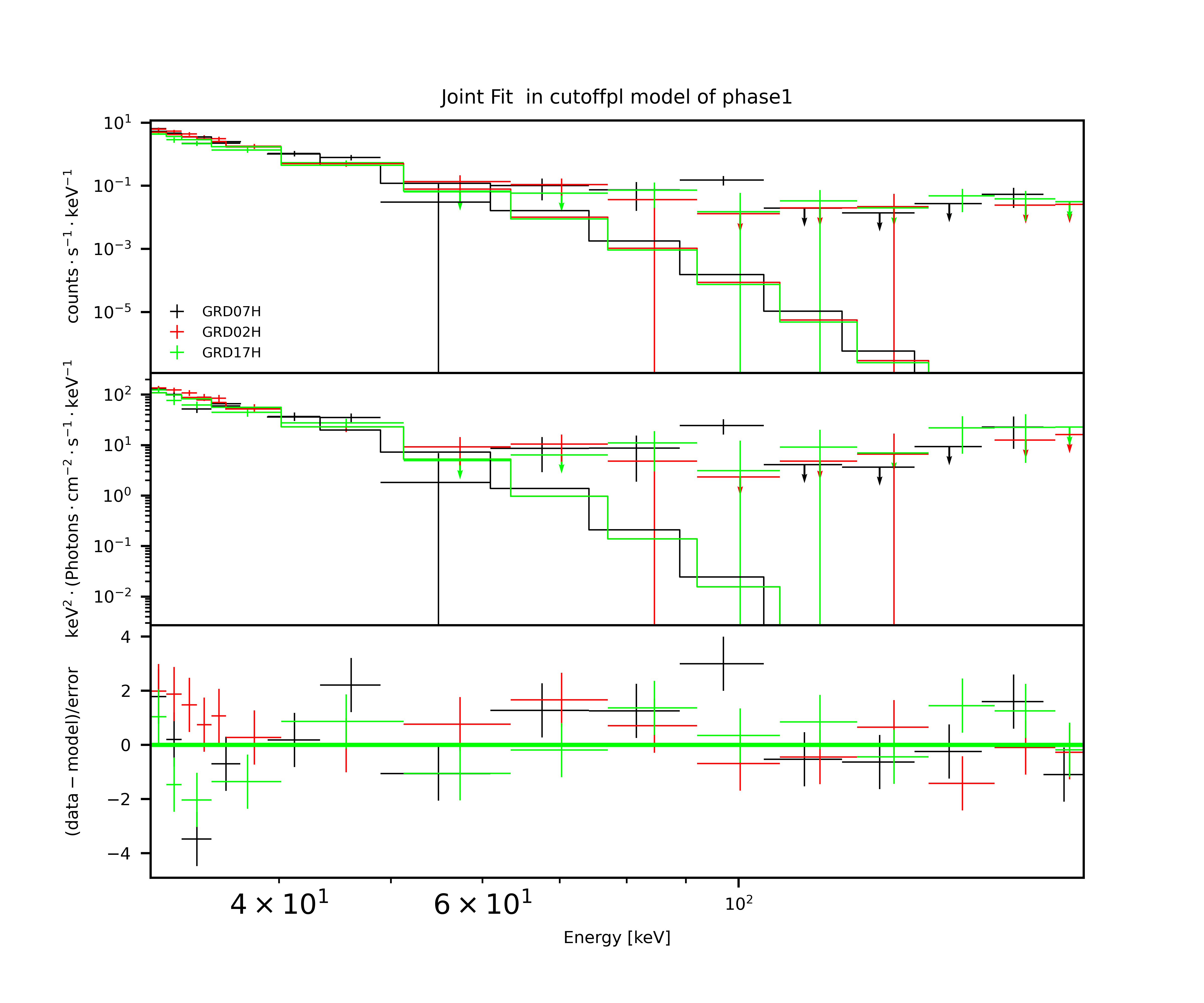}}
        \quad
        \subfigure[]{\includegraphics[height=3.5cm]{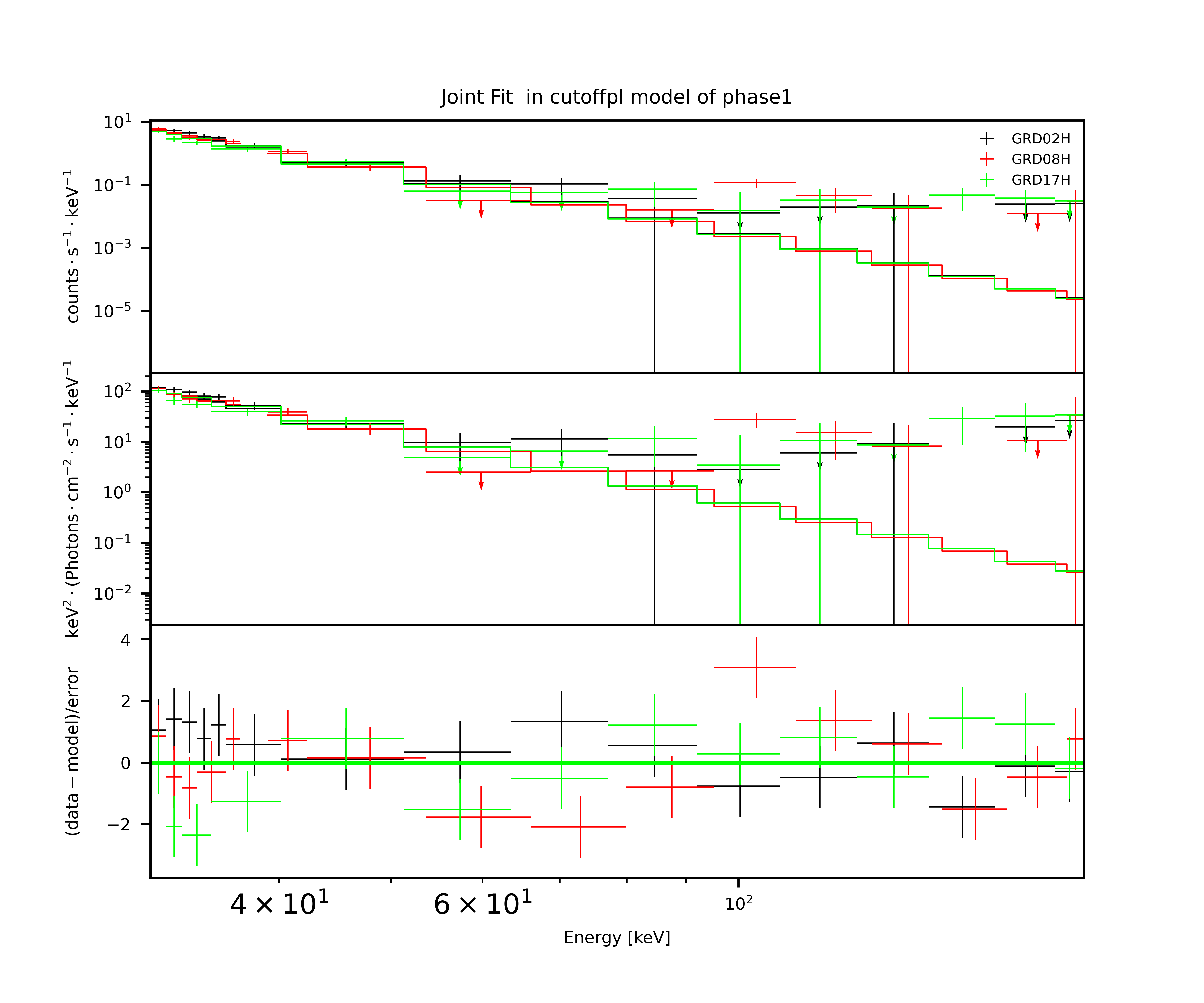}}
        \quad
        \\
        \subfigure[]{\includegraphics[width=4.5cm]{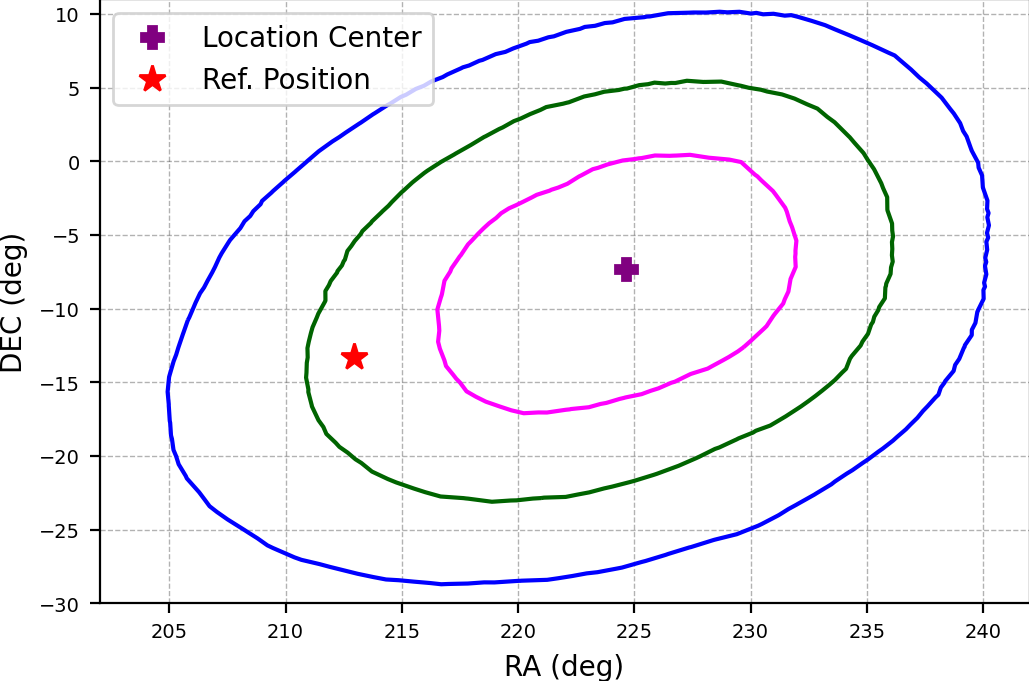}}
        \quad
        \subfigure[]{\includegraphics[width=4.5cm]{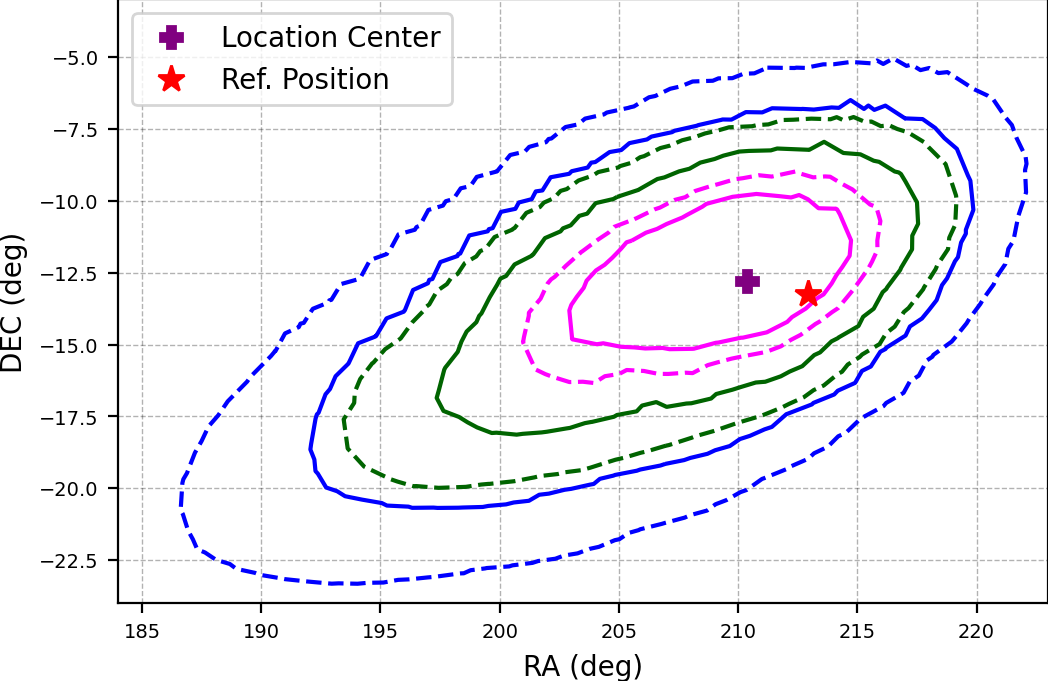}}
        \quad
        \subfigure[]{\includegraphics[width=4.5cm]{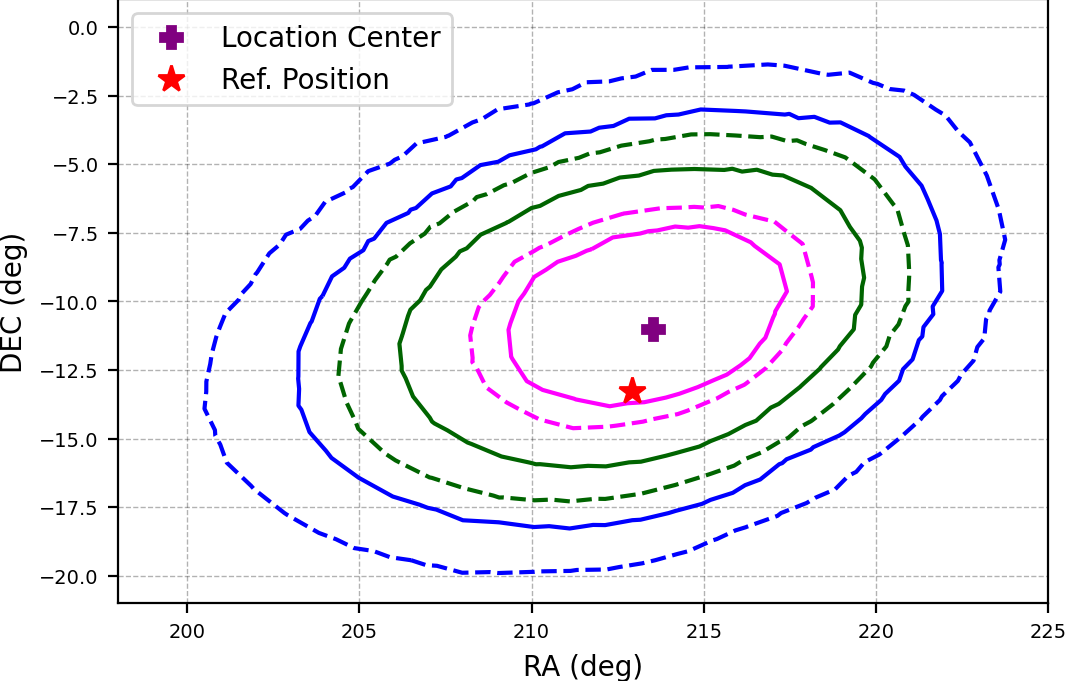}}
        \caption{ GECAM localization results of SFL 210528. (a) The light curve of GRD \# 07 high gain which contains the majority of net (burst) counts. (b) The RFD spectral fitting result. (c) The RFD spectral fitting result. The location credible region of (d) FIX, (e) RFD, and (f) APR localization. The captions are the same as Figure \ref{fig2a}. }
        \label{fig2_SFL210528}
    \end{figure*}

    \begin{figure*}
        \centering
        \subfigure[]{\includegraphics[height=3.5cm]{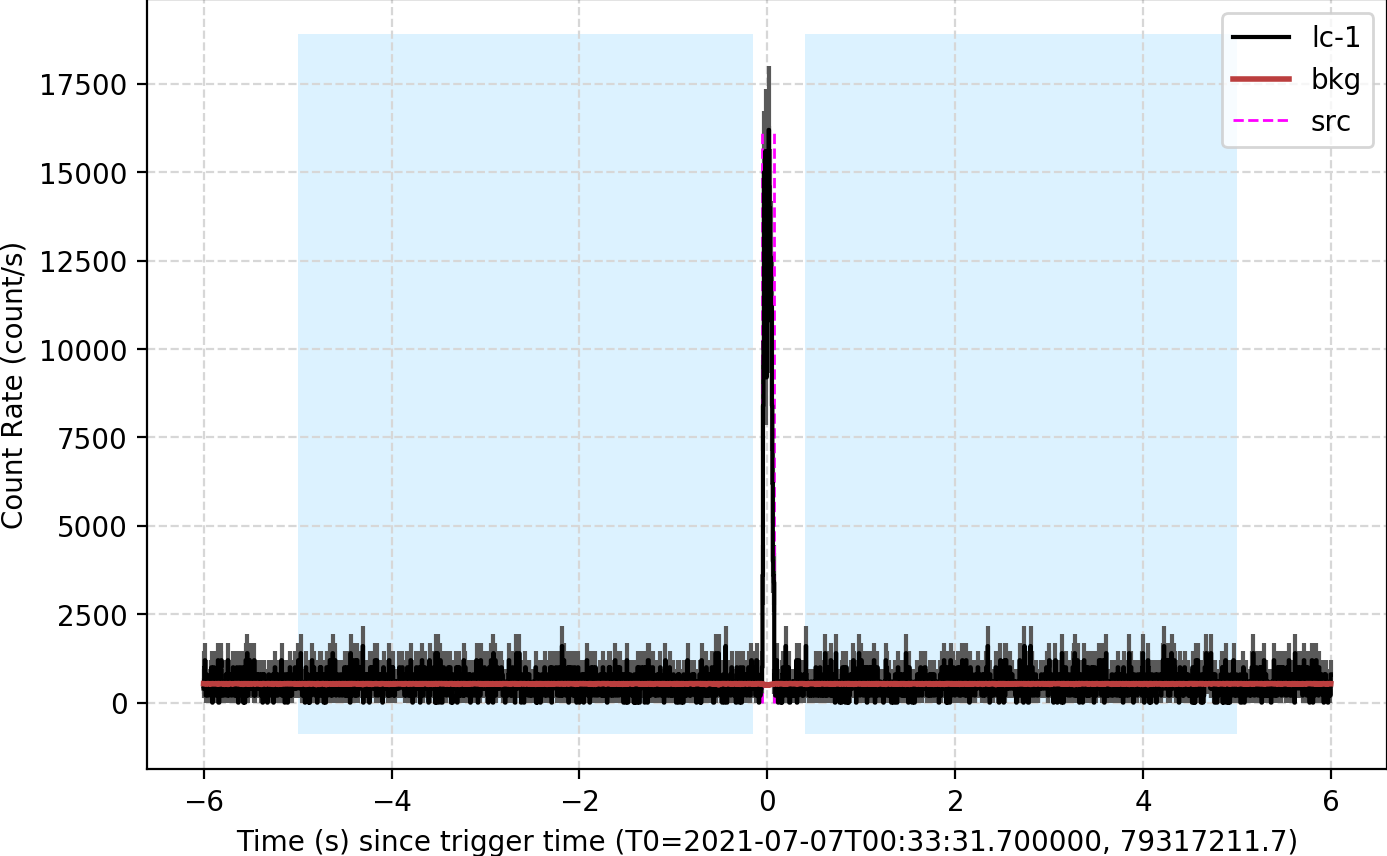}}
        \quad
        \subfigure[]{\includegraphics[height=3.5cm]{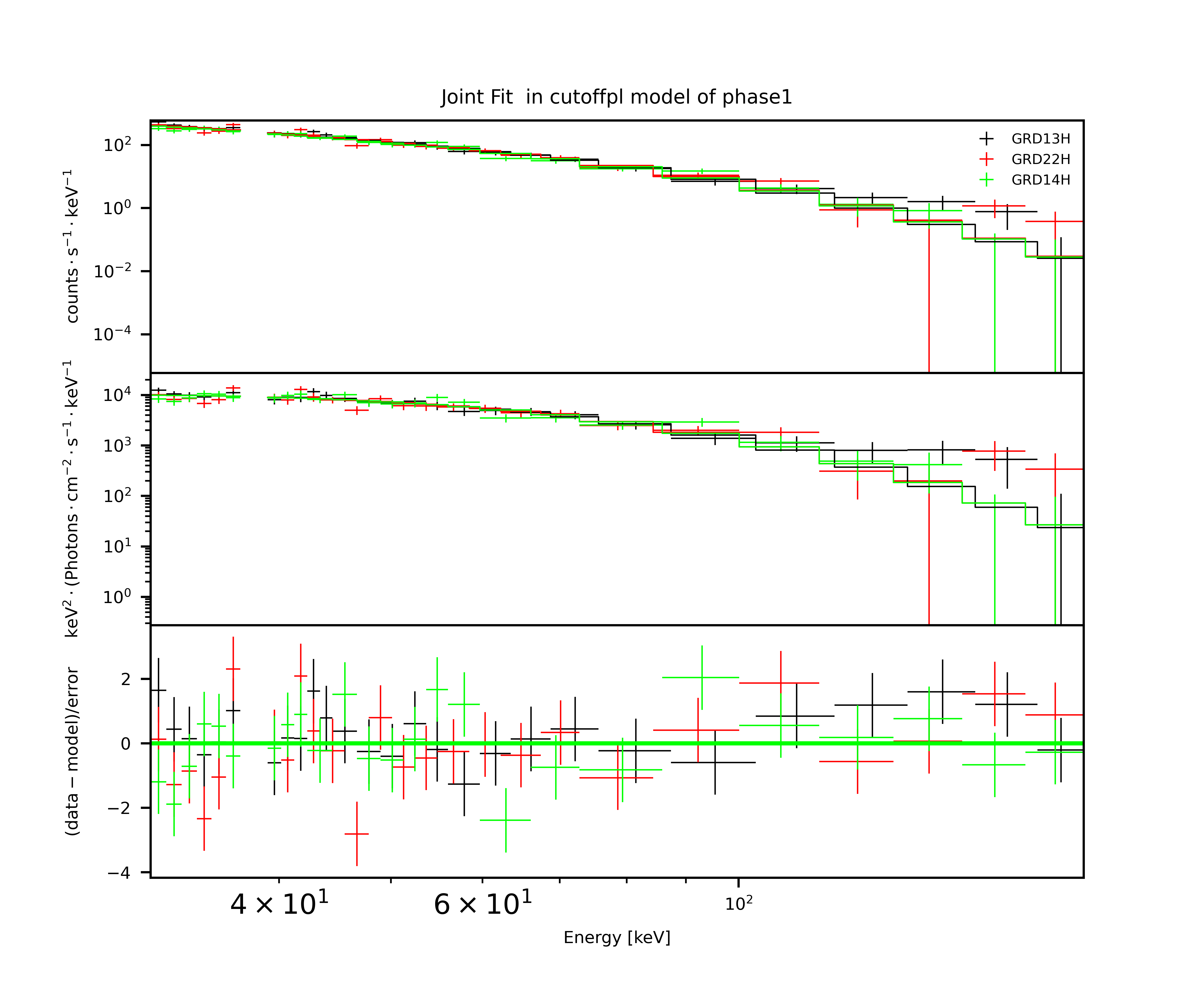}}
        \quad
        \subfigure[]{\includegraphics[height=3.5cm]{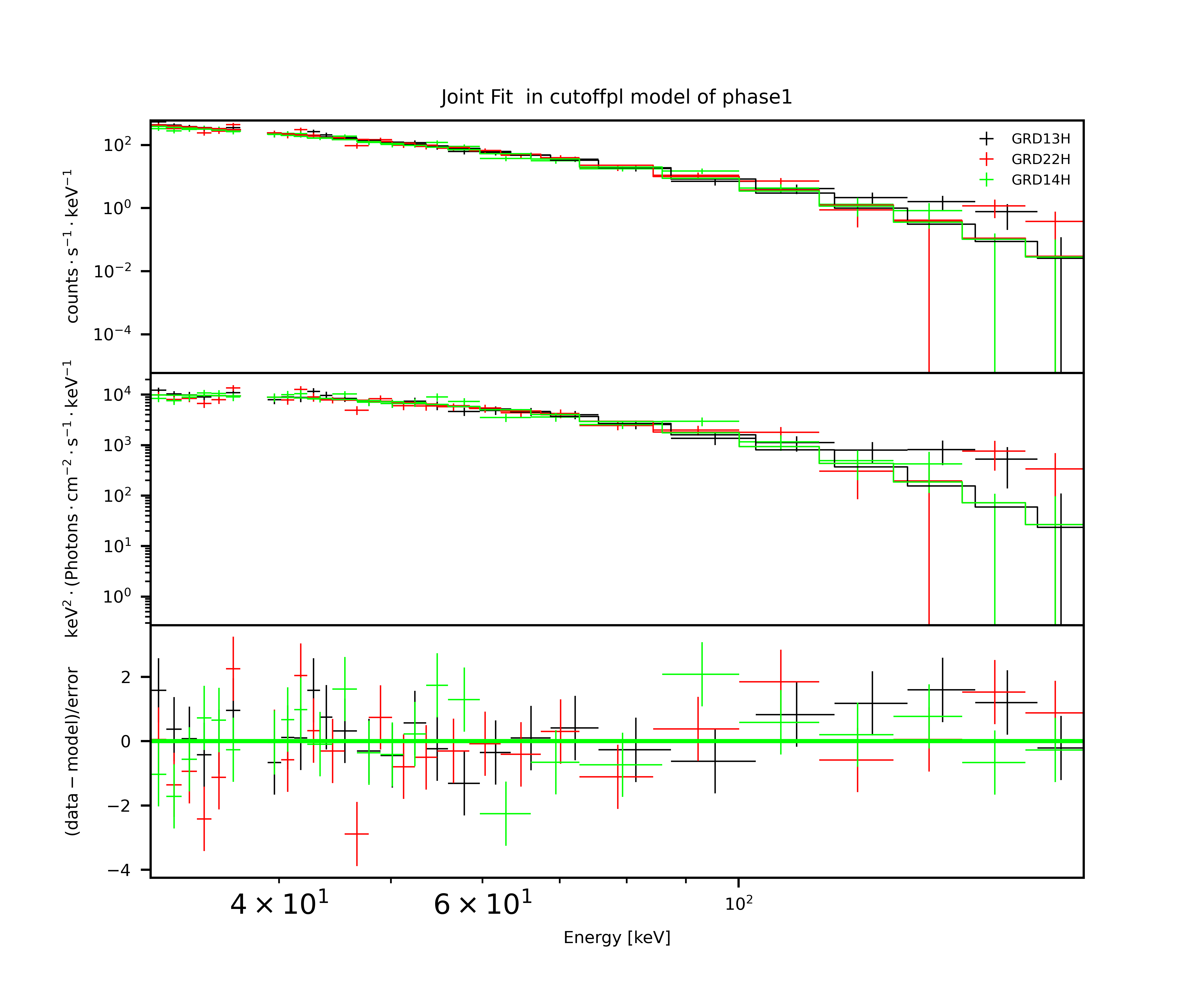}}
        \quad
        \\
        \subfigure[]{\includegraphics[width=4.5cm]{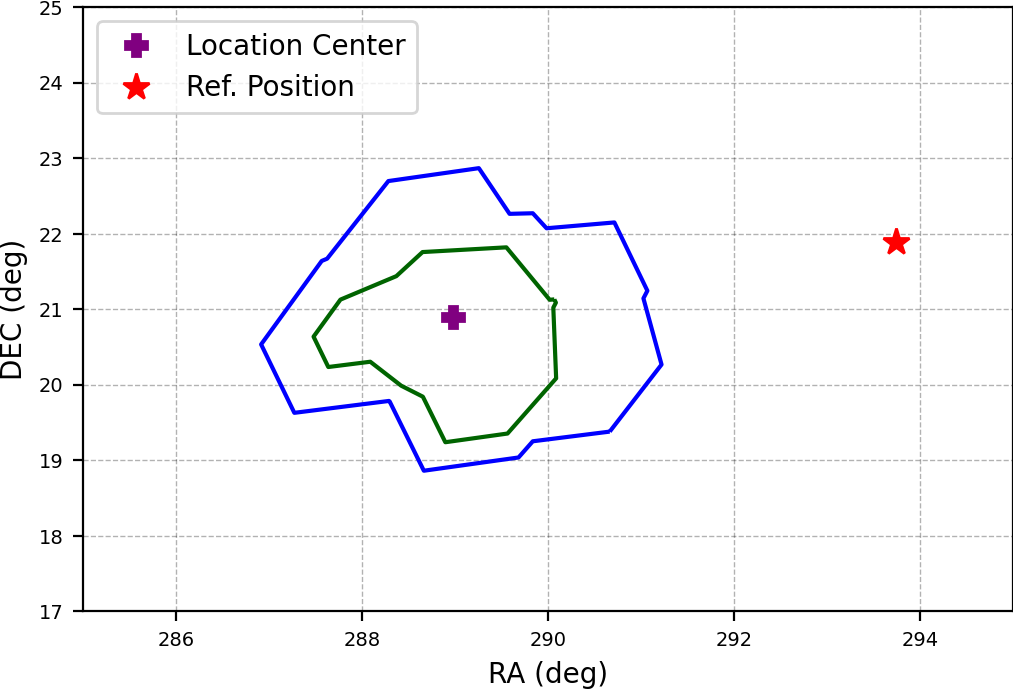}}
        \quad
        \subfigure[]{\includegraphics[width=4.5cm]{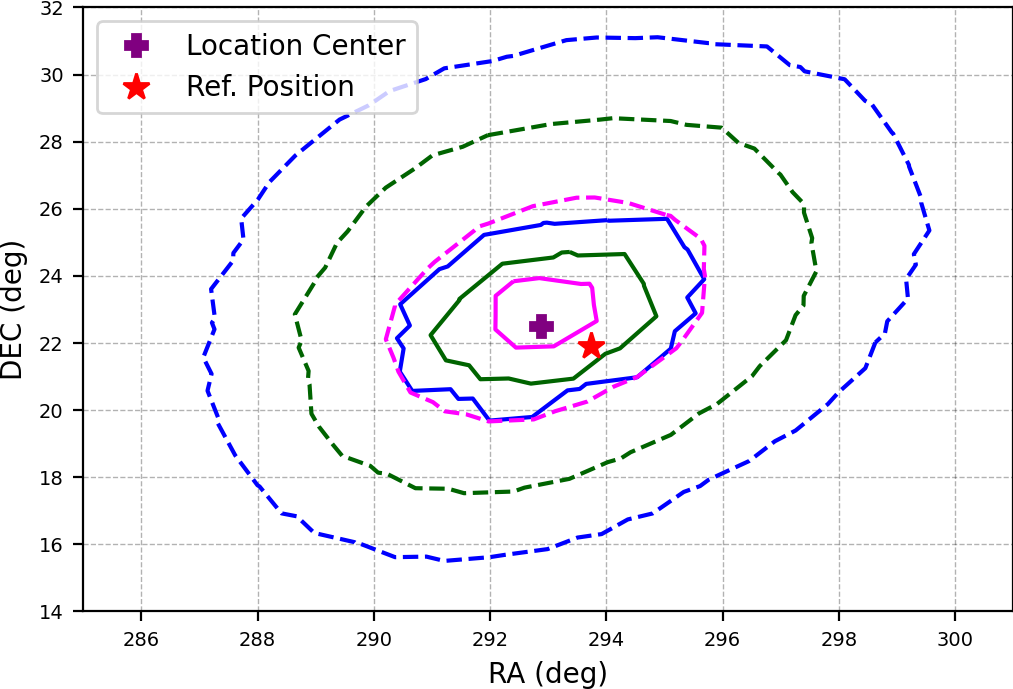}}
        \quad
        \subfigure[]{\includegraphics[width=4.5cm]{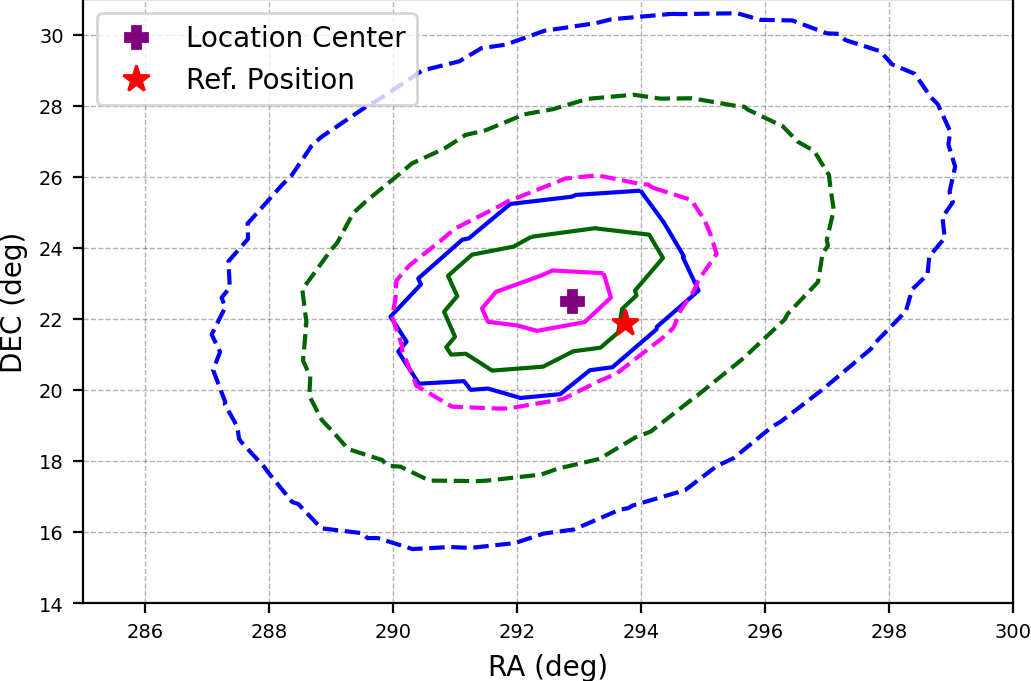}}
        \caption{ GECAM localization results of SGR 1935+2154 (UT 2021-07-07T00-33-31.700). (a) The light curve of GRD \# 13 high gain which contains the majority of net (burst) counts. (b) The RFD spectral fitting result. (c) The RFD spectral fitting result. The location credible region of (d) FIX, (e) RFD, and (f) APR localization. The captions are the same as Figure \ref{fig2a}. }
        \label{fig2_SGR1935a}
    \end{figure*}

    \begin{figure*}
        \centering
        \subfigure[]{\includegraphics[height=3.5cm]{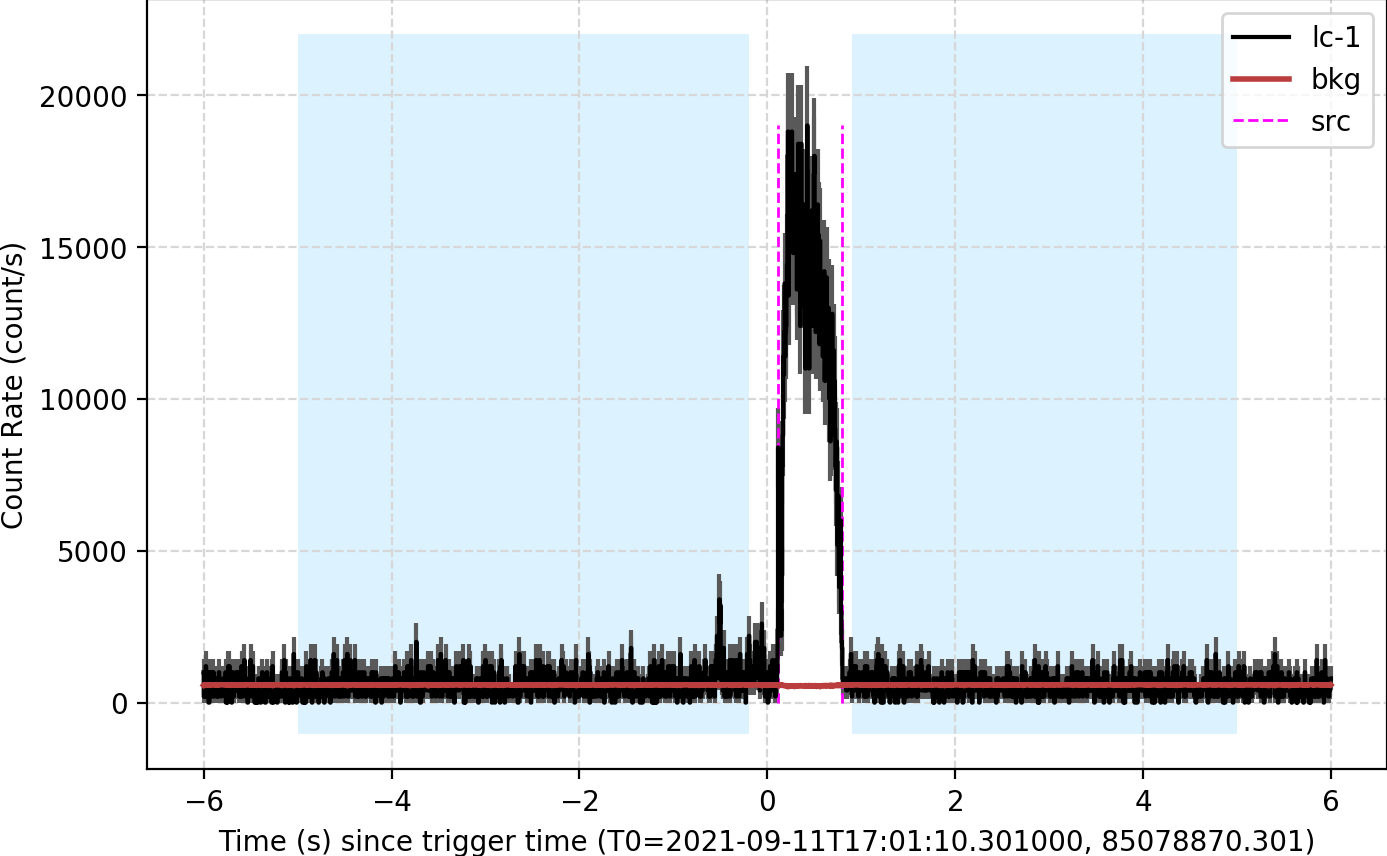}}
        \quad
        \subfigure[]{\includegraphics[height=3.5cm]{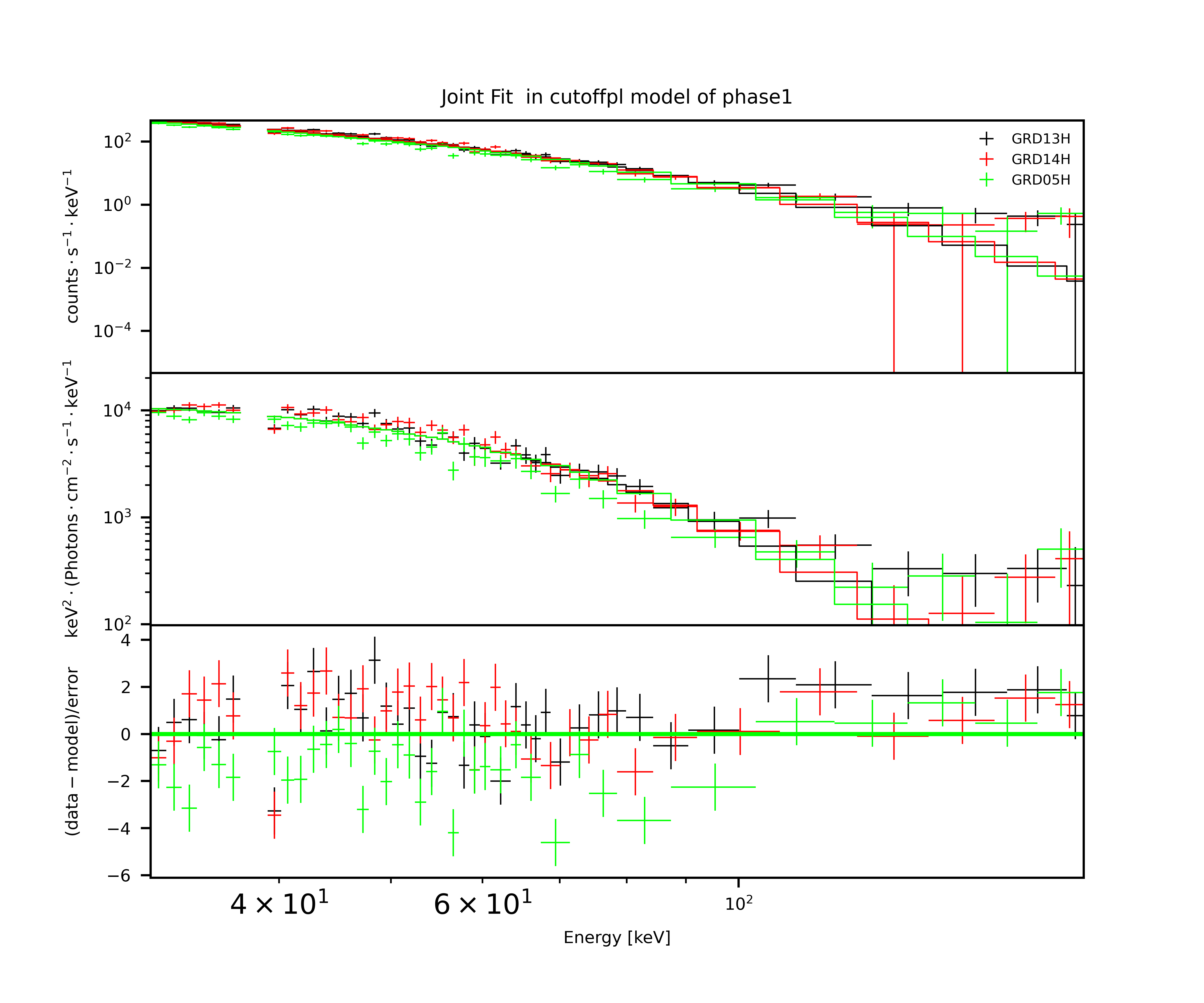}}
        \quad
        \subfigure[]{\includegraphics[height=3.5cm]{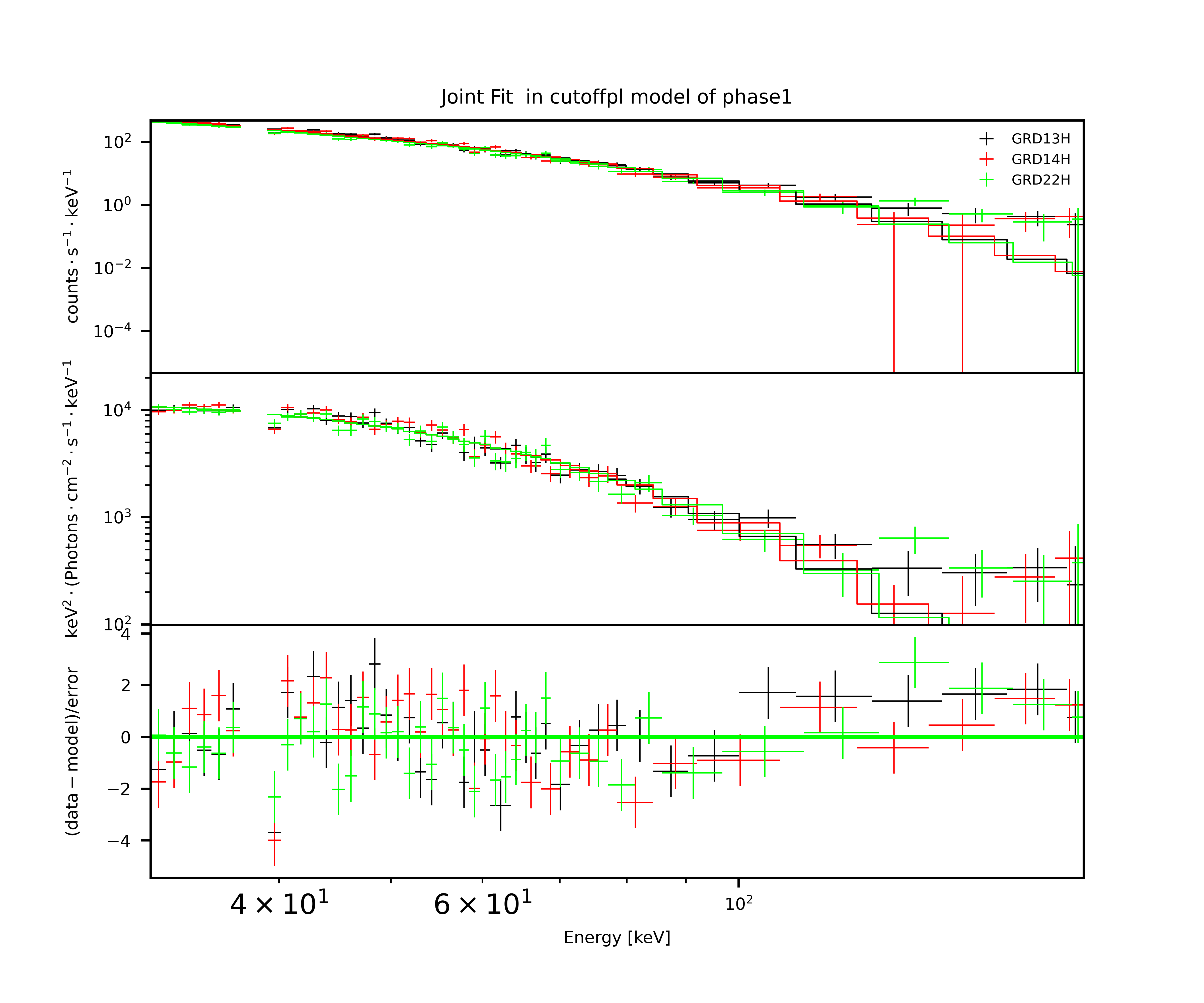}}
        \quad
        \\
        \subfigure[]{\includegraphics[width=4.5cm]{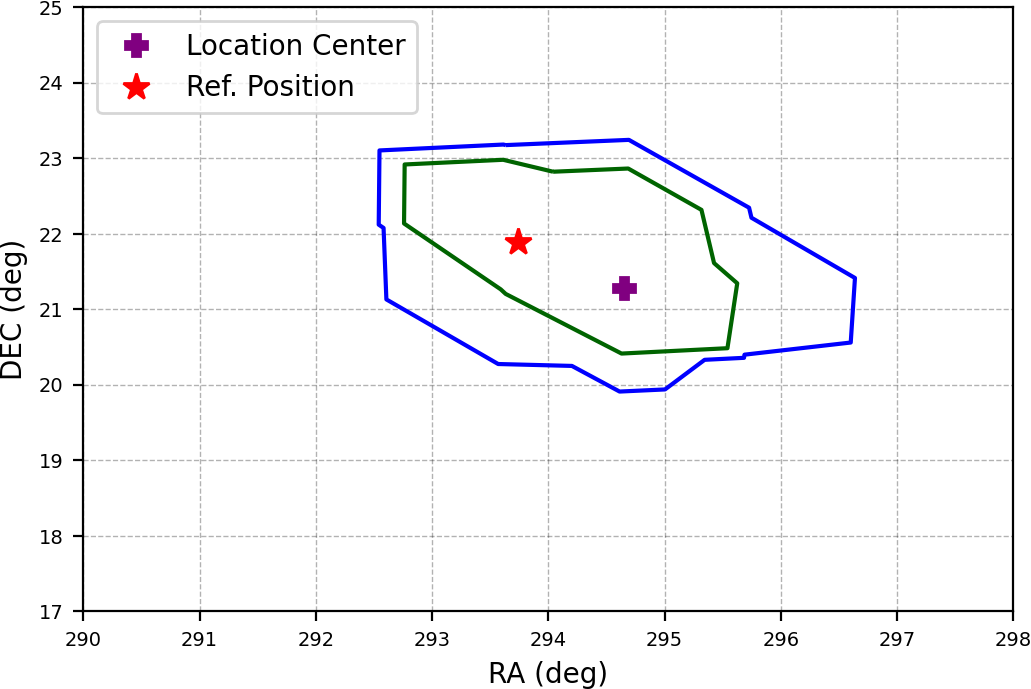}}
        \quad
        \subfigure[]{\includegraphics[width=4.5cm]{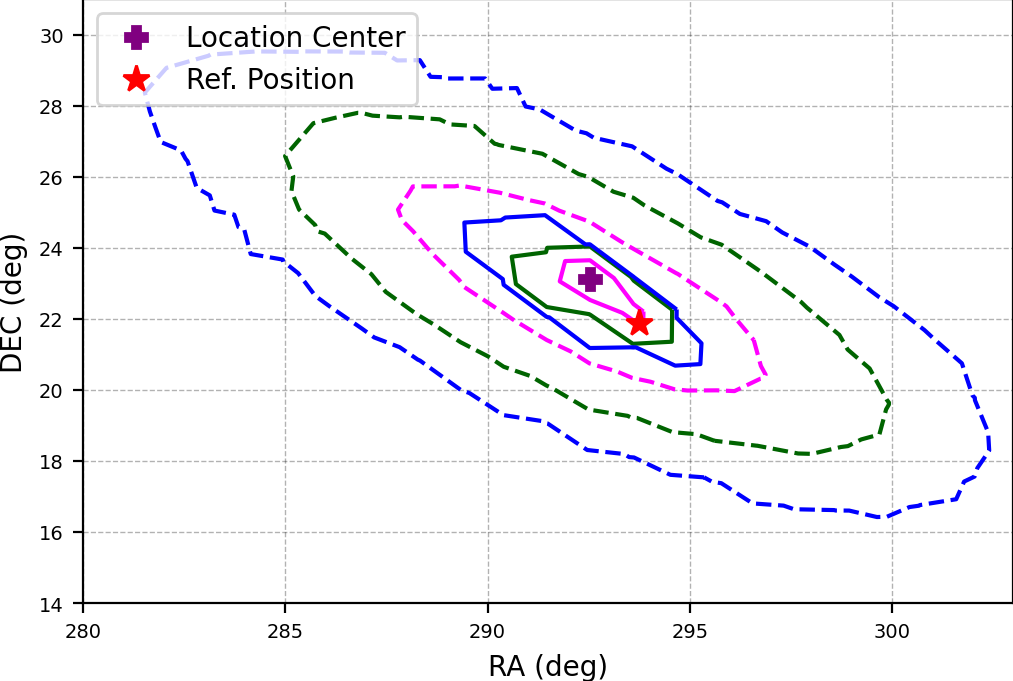}}
        \quad
        \subfigure[]{\includegraphics[width=4.5cm]{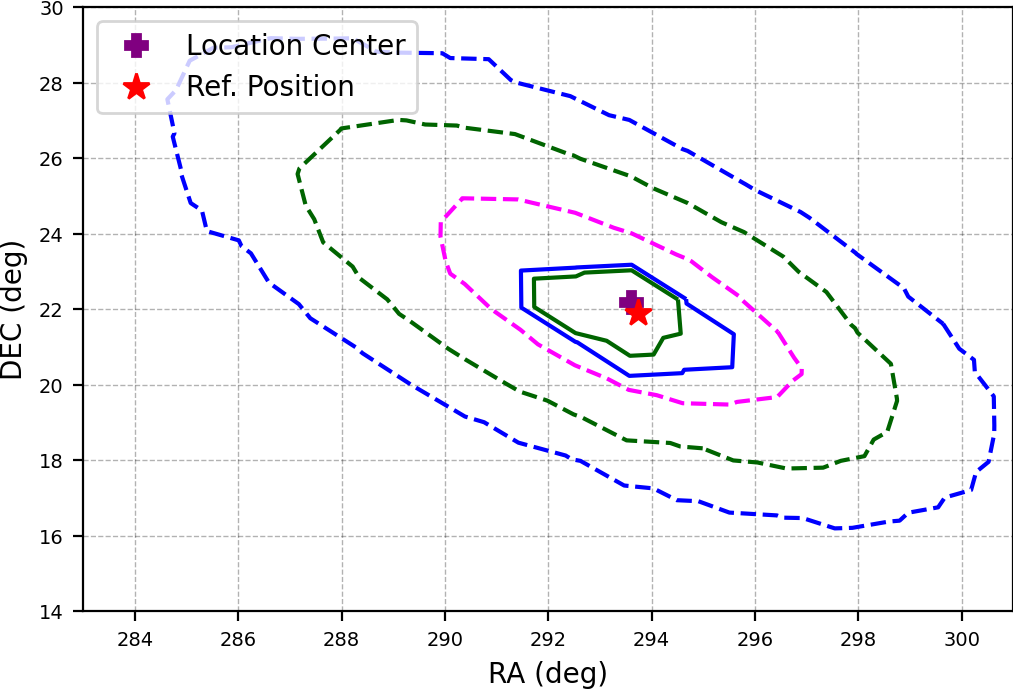}}
        \caption{ GECAM localization results of SGR 1935+2154 (UT 2021-09-11T17-01-10.301). (a) The light curve of GRD \# 13 high gain which contains the majority of net (burst) counts. (b) The RFD spectral fitting result. (c) The RFD spectral fitting result. The location credible region of (d) FIX, (e) RFD, and (f) APR localization. The captions are the same as Figure \ref{fig2a}. }
        \label{fig2_SGR1935b}
    \end{figure*}

    \begin{figure*}
        \centering
        \subfigure[]{\includegraphics[height=3.5cm]{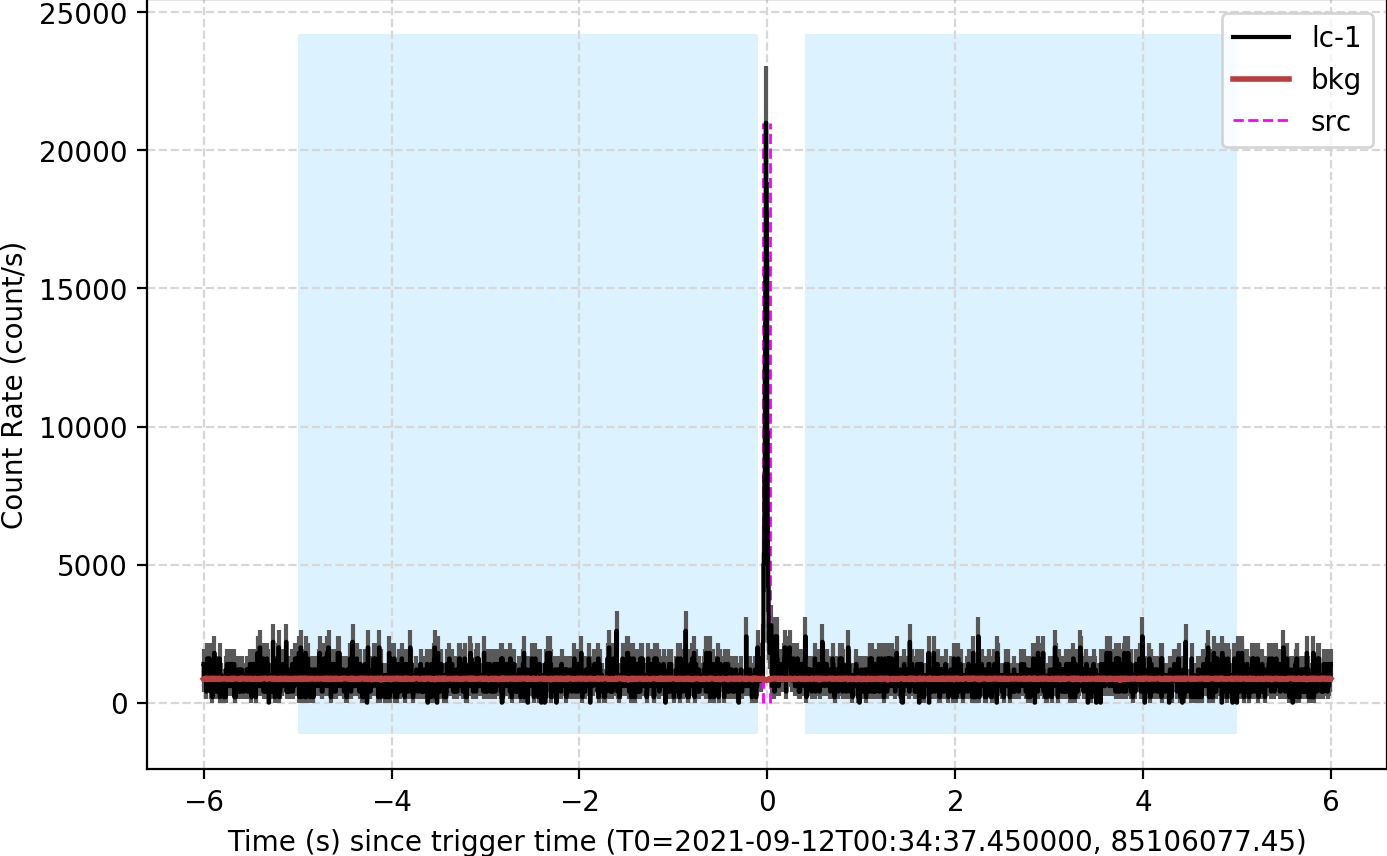}}
        \quad
        \subfigure[]{\includegraphics[height=3.5cm]{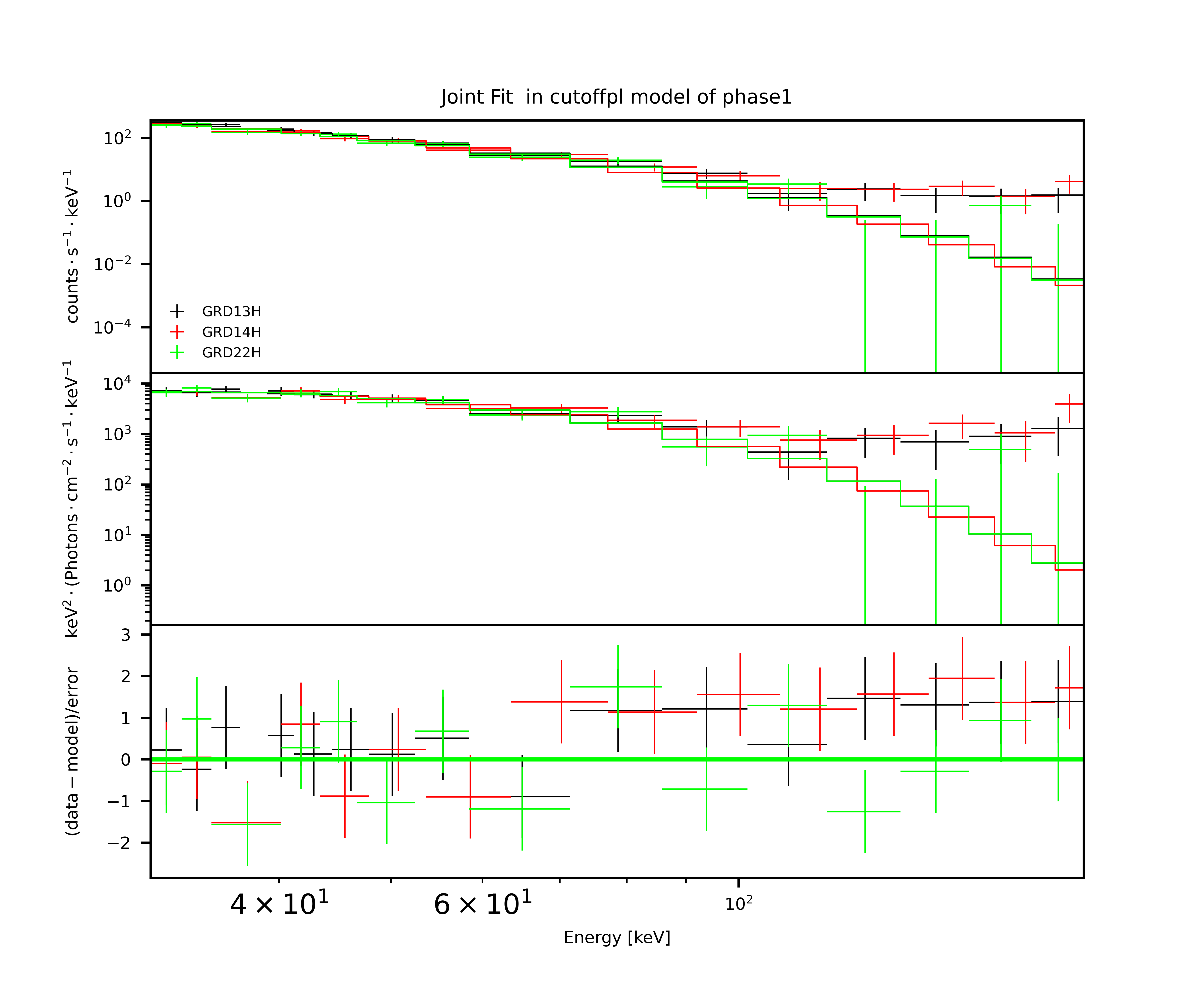}}
        \quad
        \subfigure[]{\includegraphics[height=3.5cm]{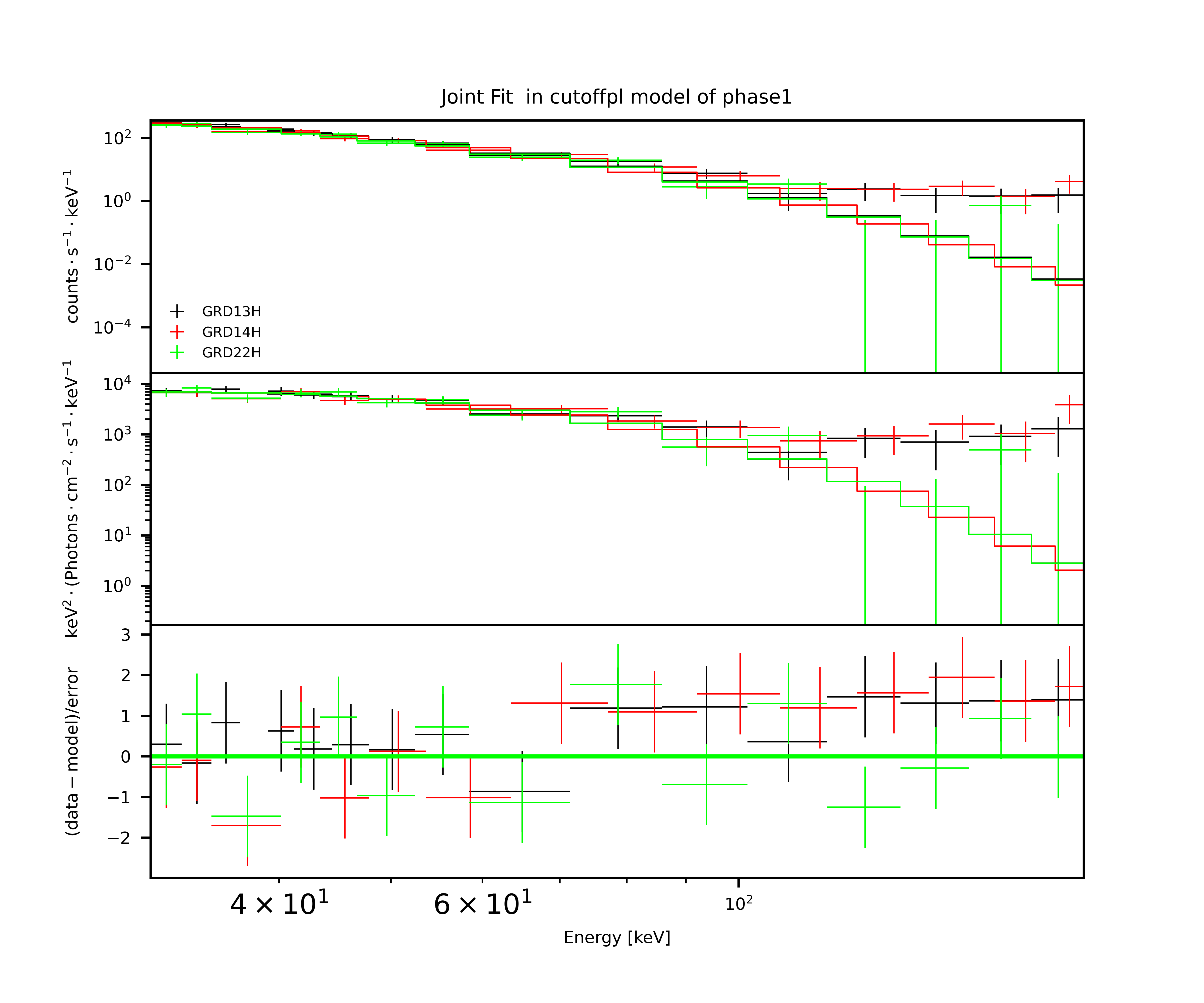}}
        \quad
        \\
        \subfigure[]{\includegraphics[width=4.5cm]{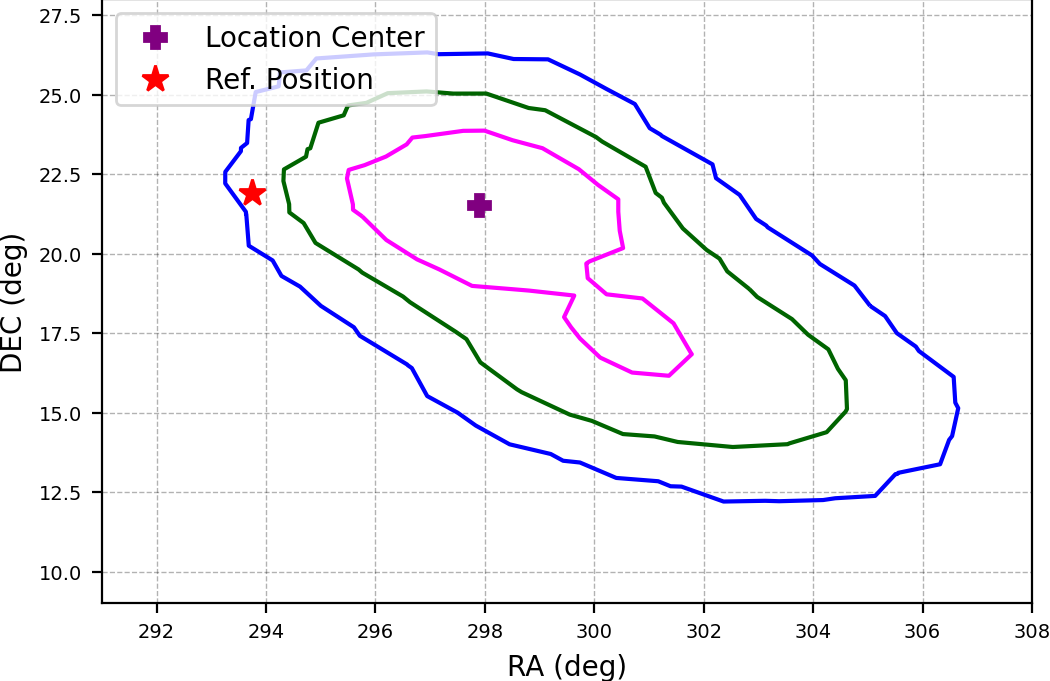}}
        \quad
        \subfigure[]{\includegraphics[width=4.5cm]{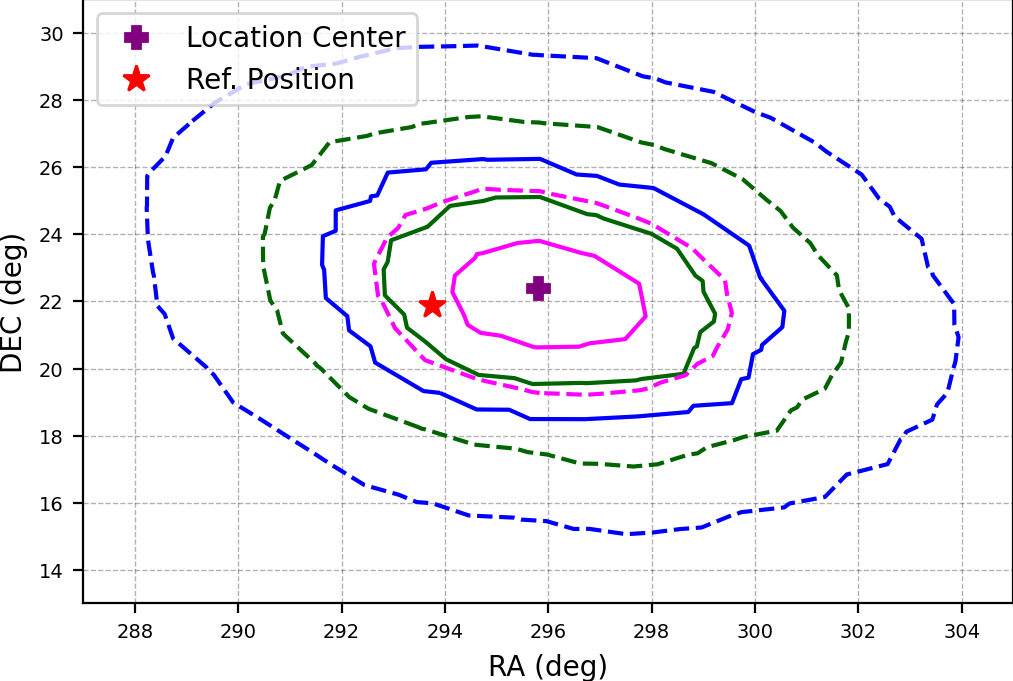}}
        \quad
        \subfigure[]{\includegraphics[width=4.5cm]{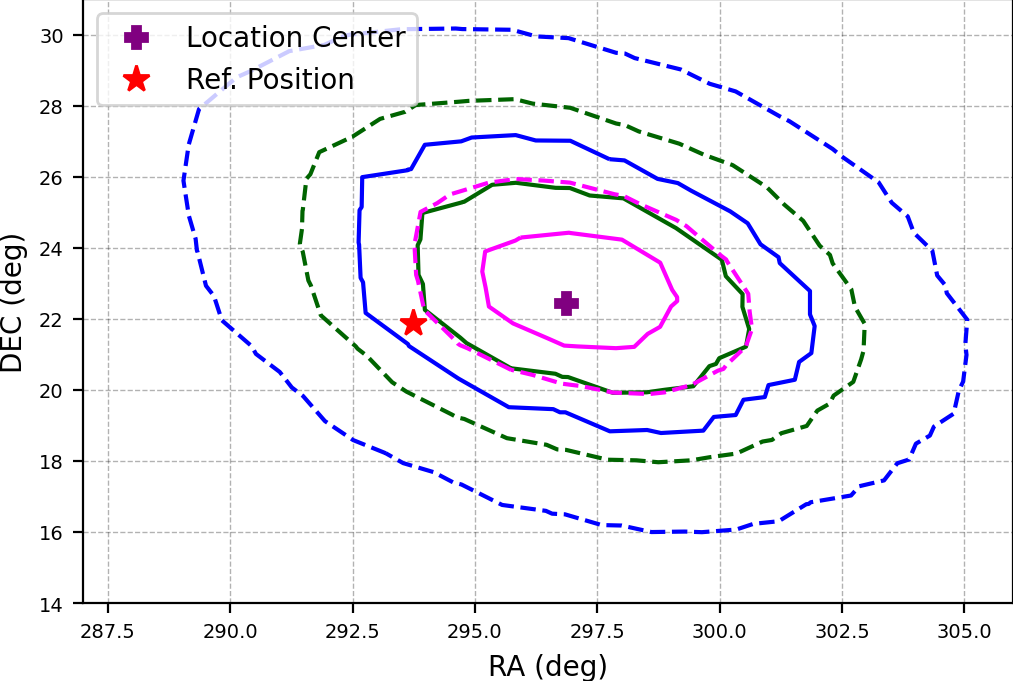}}
        \caption{ GECAM localization results of SGR 1935+2154 (UT 2021-09-12T00-34-37.450). (a) The light curve of GRD \# 13 high gain which contains the majority of net (burst) counts. (b) The RFD spectral fitting result. (c) The RFD spectral fitting result. The location credible region of (d) FIX, (e) RFD, and (f) APR localization. The captions are the same as Figure \ref{fig2a}. }
        \label{fig2_SGR1935c}
    \end{figure*}

    \begin{figure*}
        \centering
        \subfigure[]{\includegraphics[height=3.5cm]{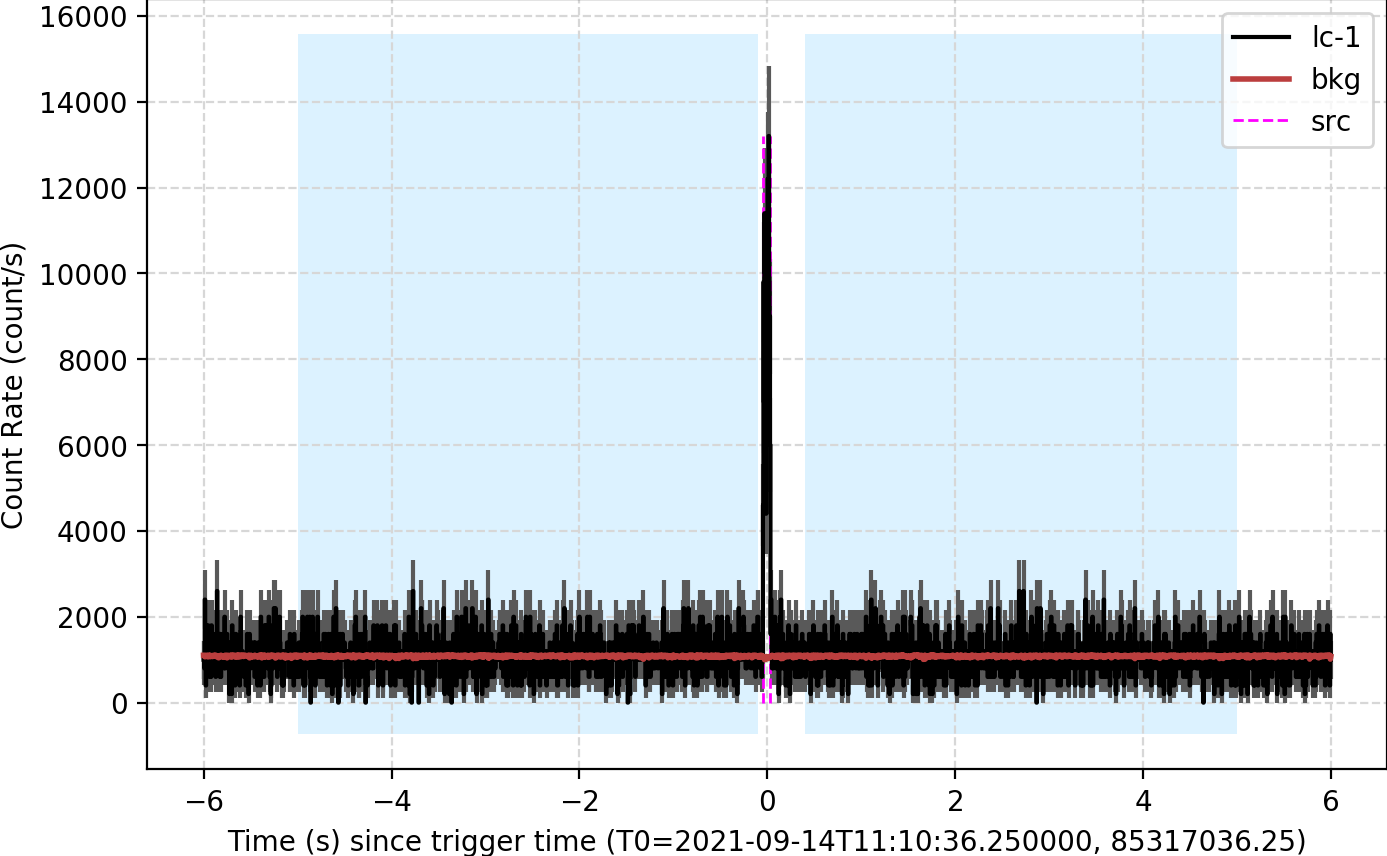}}
        \quad
        \subfigure[]{\includegraphics[height=3.5cm]{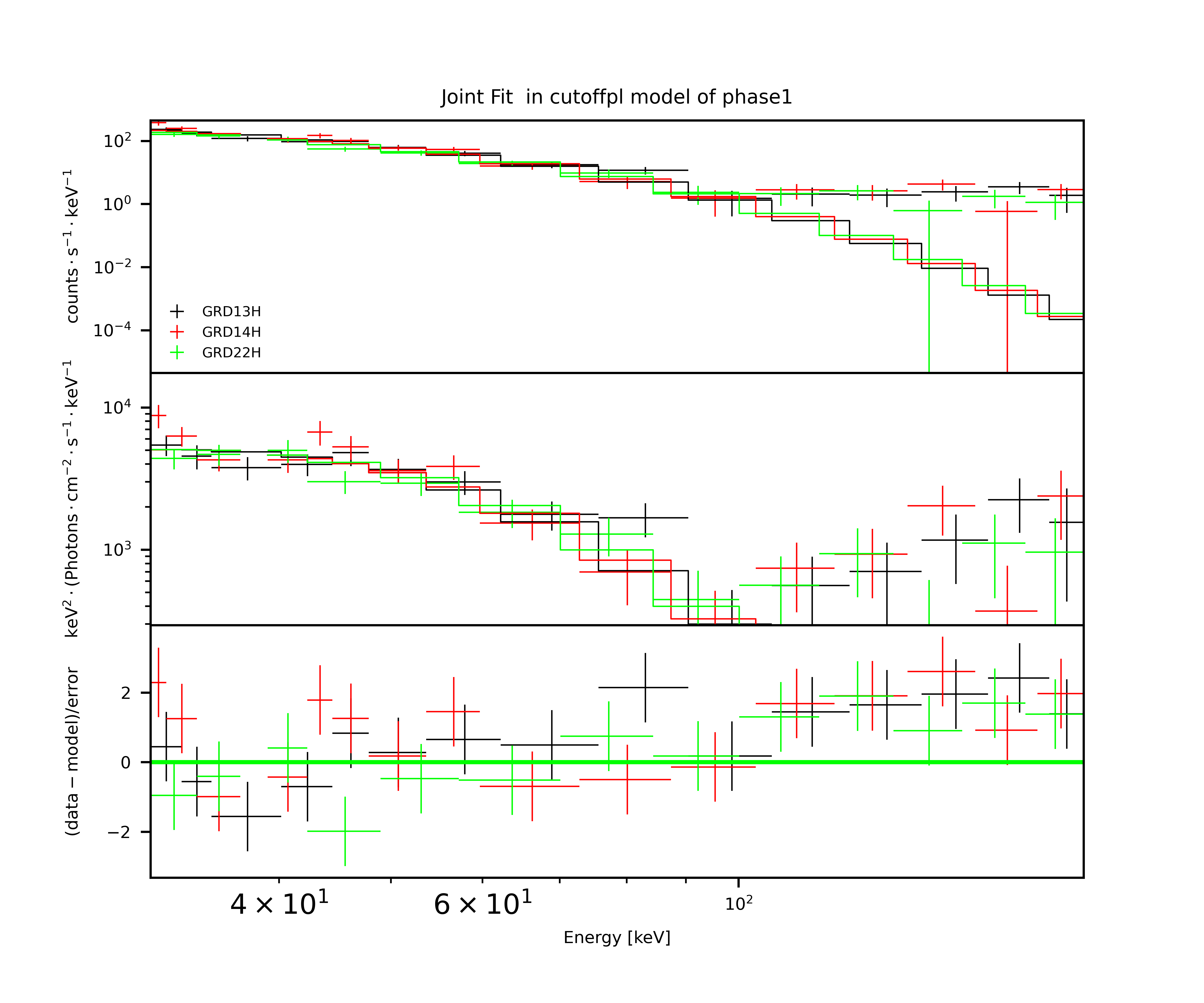}}
        \quad
        \subfigure[]{\includegraphics[height=3.5cm]{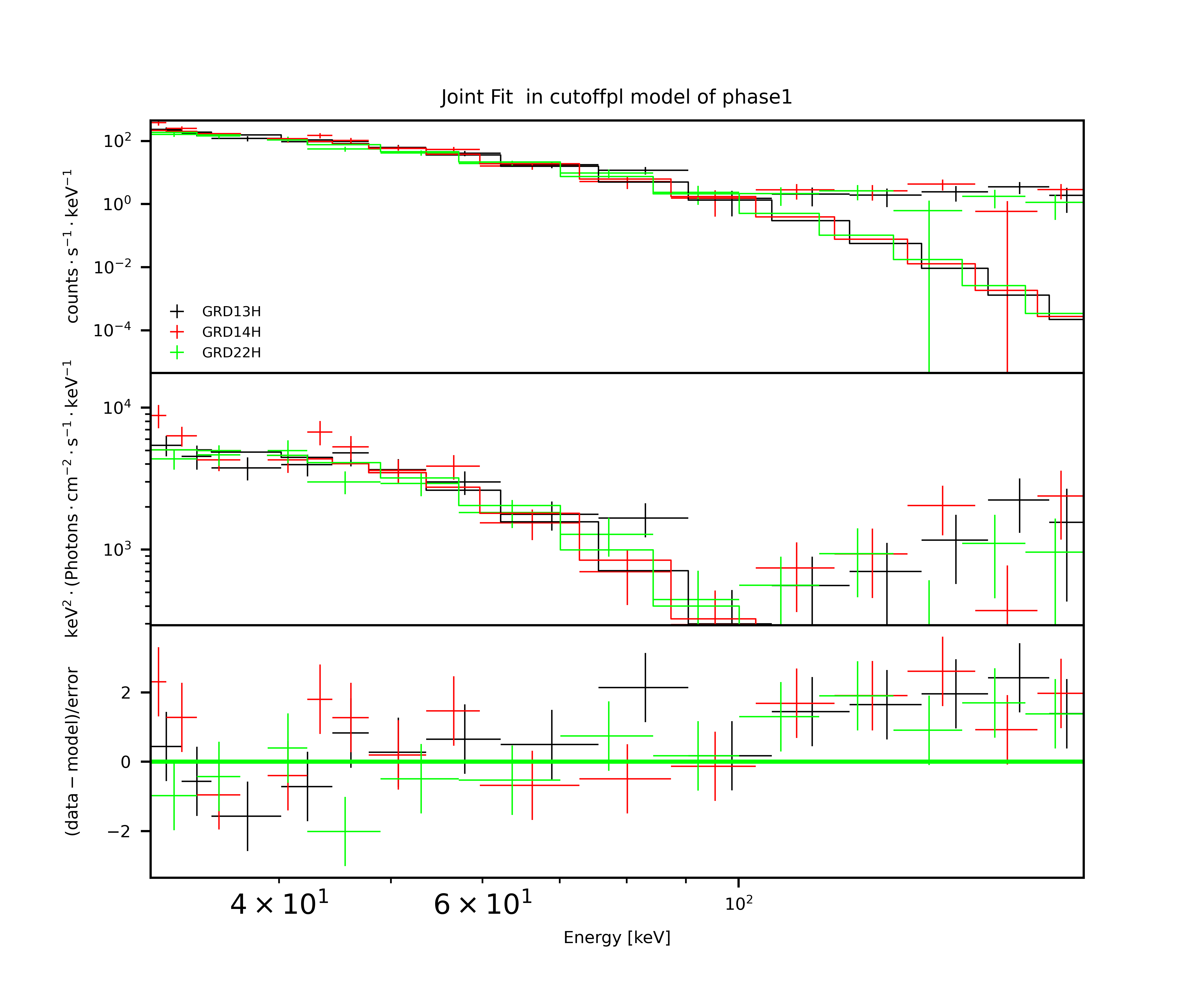}}
        \quad
        \\
        \subfigure[]{\includegraphics[width=4.5cm]{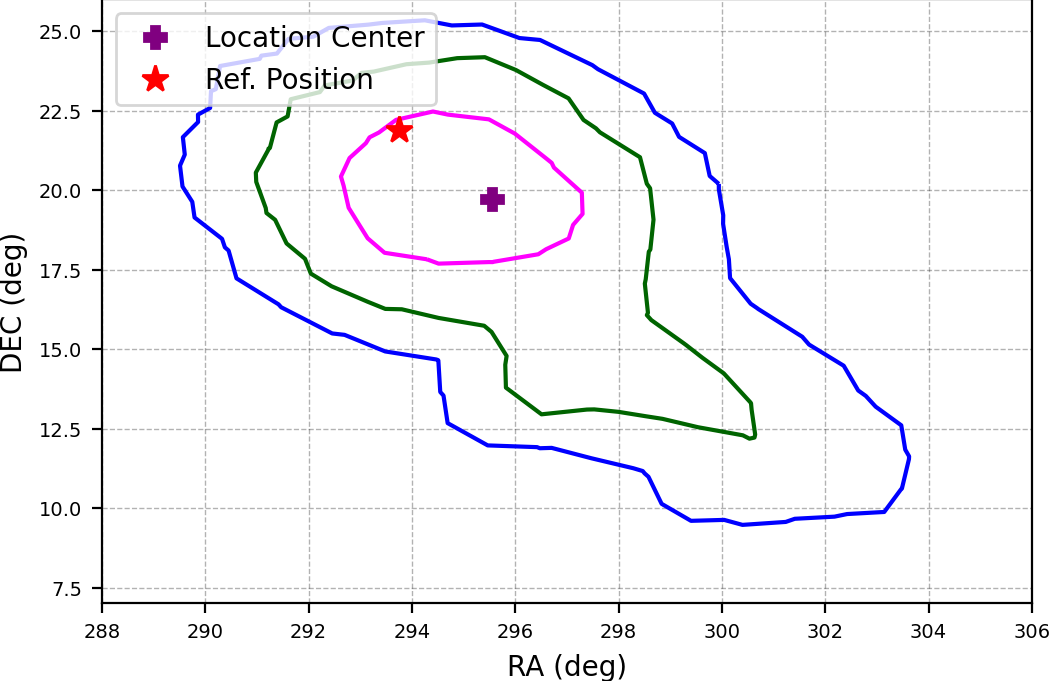}}
        \quad
        \subfigure[]{\includegraphics[width=4.5cm]{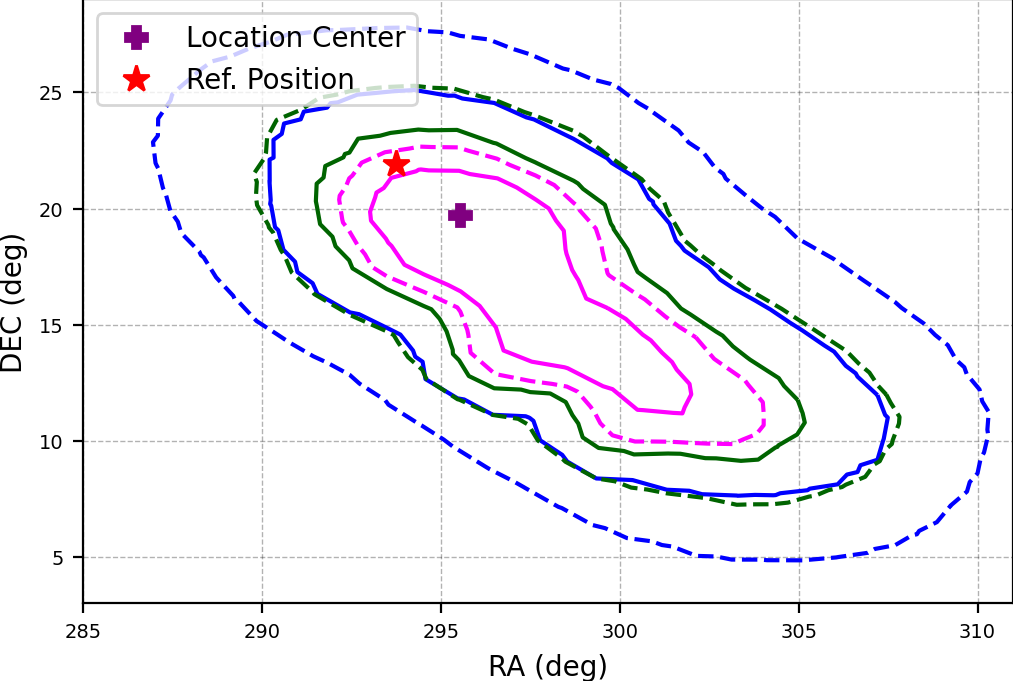}}
        \quad
        \subfigure[]{\includegraphics[width=4.5cm]{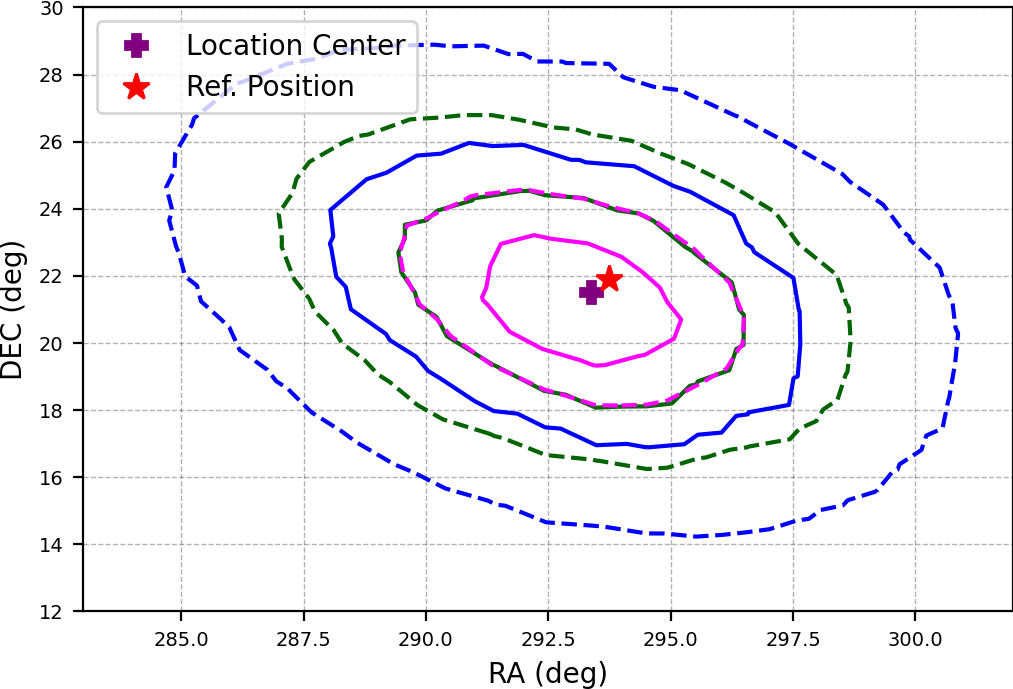}}
        \caption{ GECAM localization results of SGR 1935+2154 (UT 2021-09-14T11-10-36.250). (a) The light curve of GRD \# 14 high gain which contains the majority of net (burst) counts. (b) The RFD spectral fitting result. (c) The RFD spectral fitting result. The location credible region of (d) FIX, (e) RFD, and (f) APR localization. The captions are the same as Figure \ref{fig2a}. }
        \label{fig2_SGR1935d}
    \end{figure*}

    \begin{figure*}
        \centering
        \subfigure[]{\includegraphics[height=3.5cm]{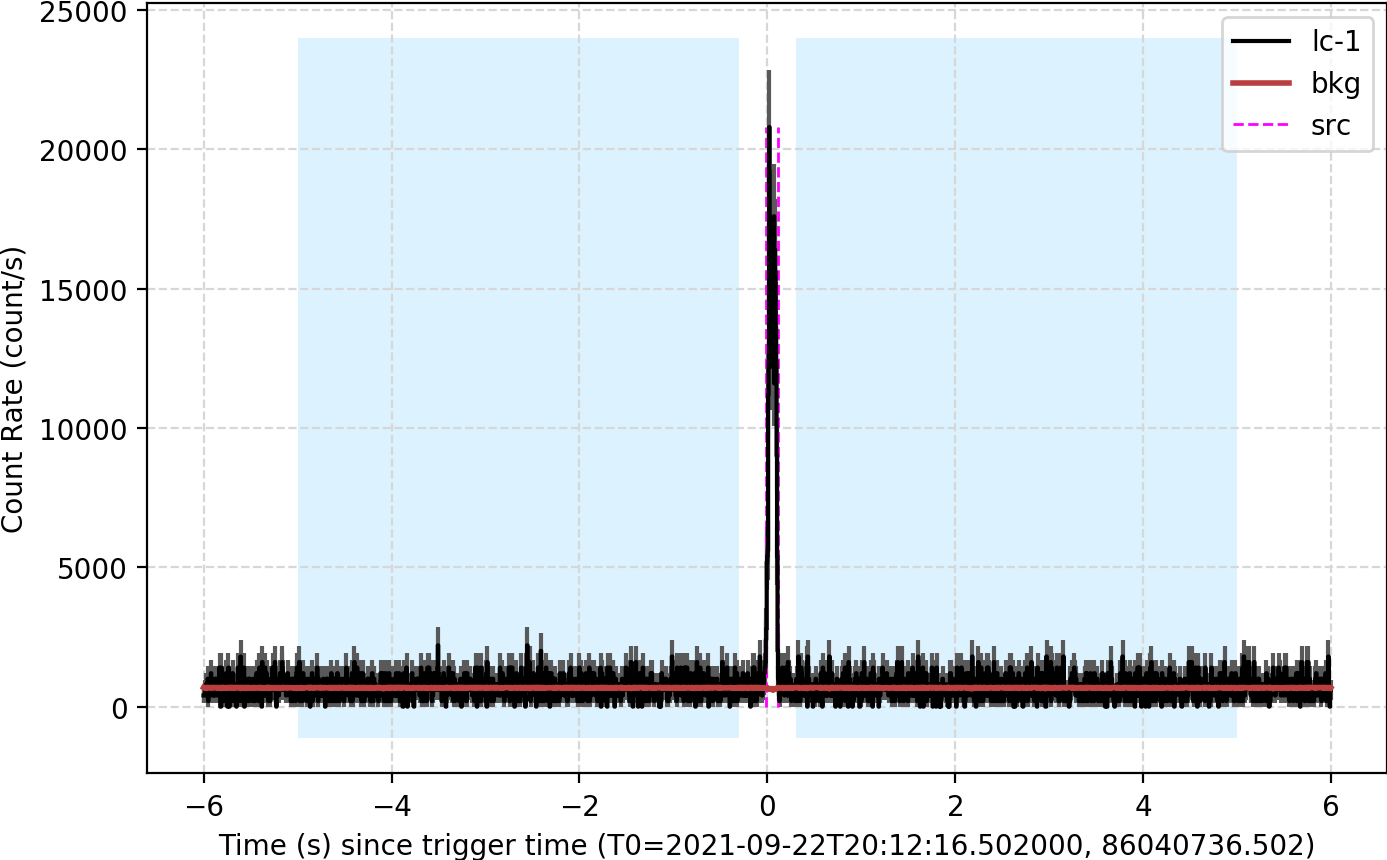}}
        \quad
        \subfigure[]{\includegraphics[height=3.5cm]{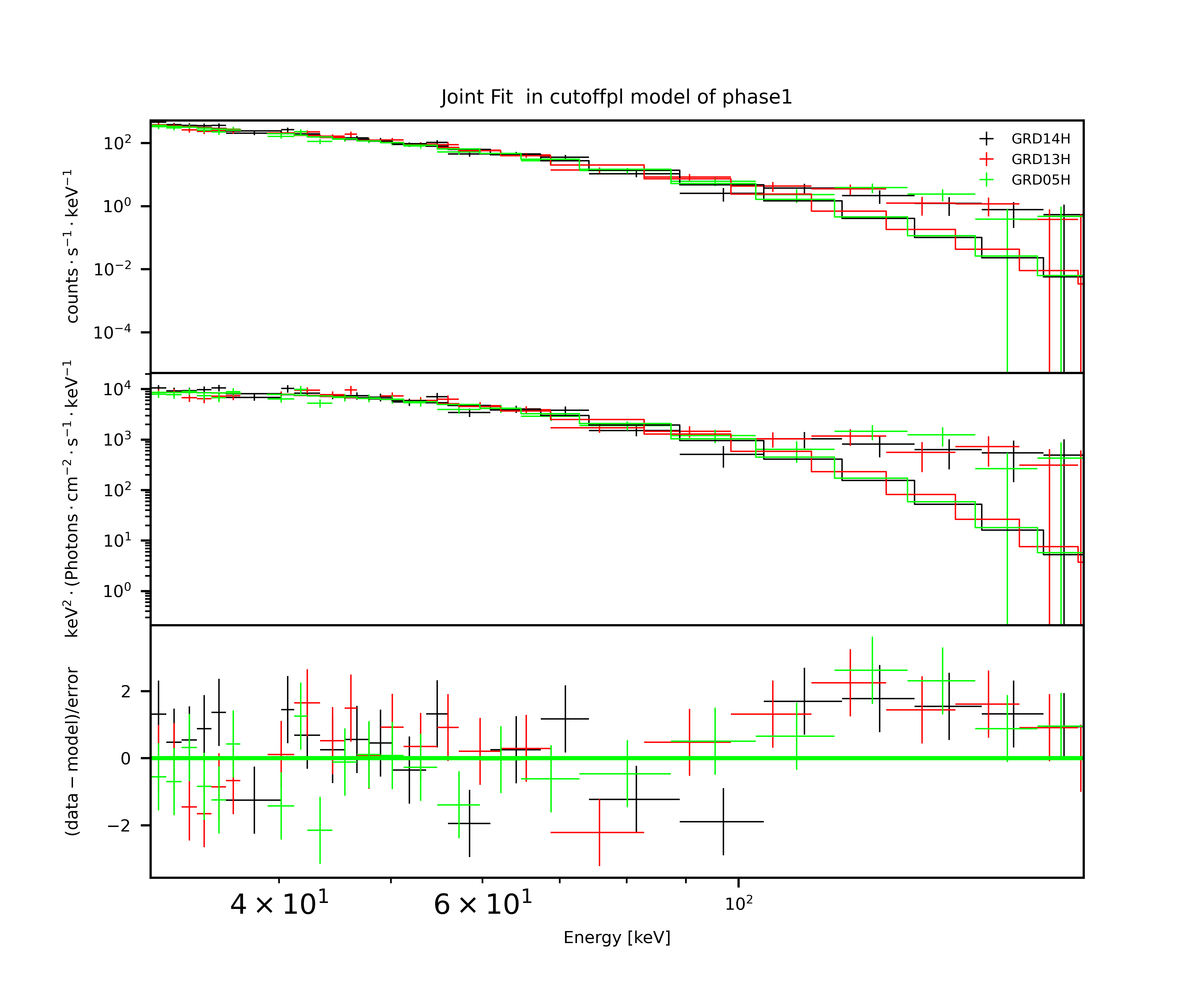}}
        \quad
        \subfigure[]{\includegraphics[height=3.5cm]{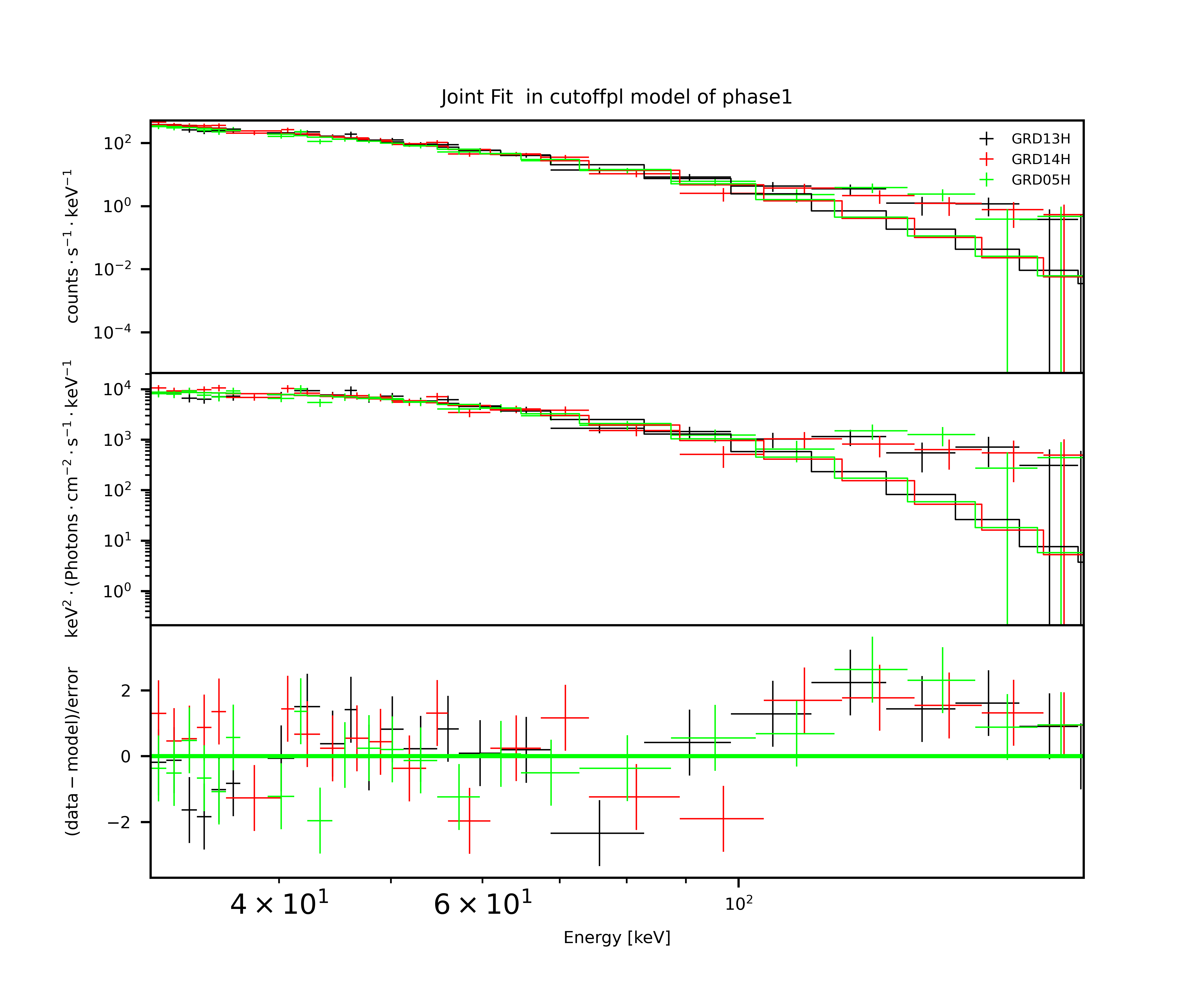}}
        \quad
        \\
        \subfigure[]{\includegraphics[width=4.5cm]{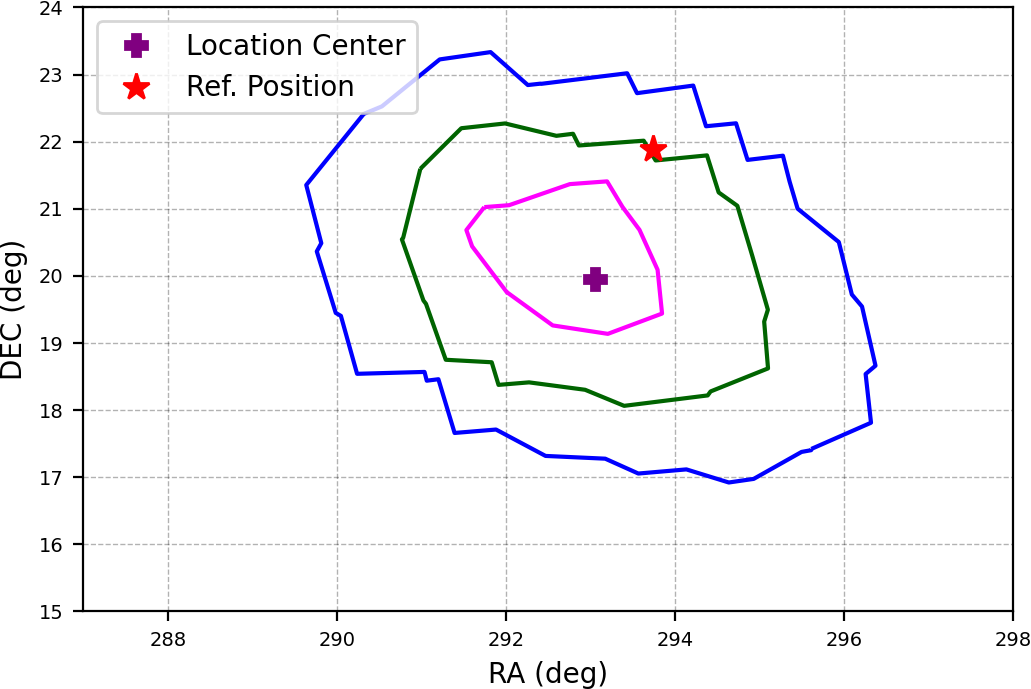}}
        \quad
        \subfigure[]{\includegraphics[width=4.5cm]{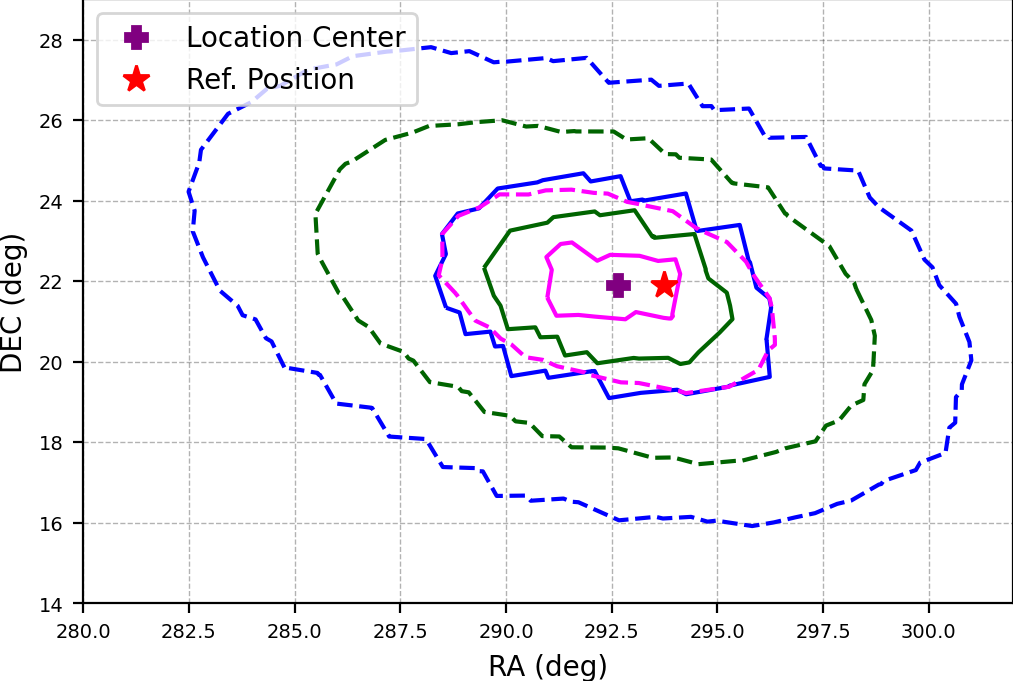}}
        \quad
        \subfigure[]{\includegraphics[width=4.5cm]{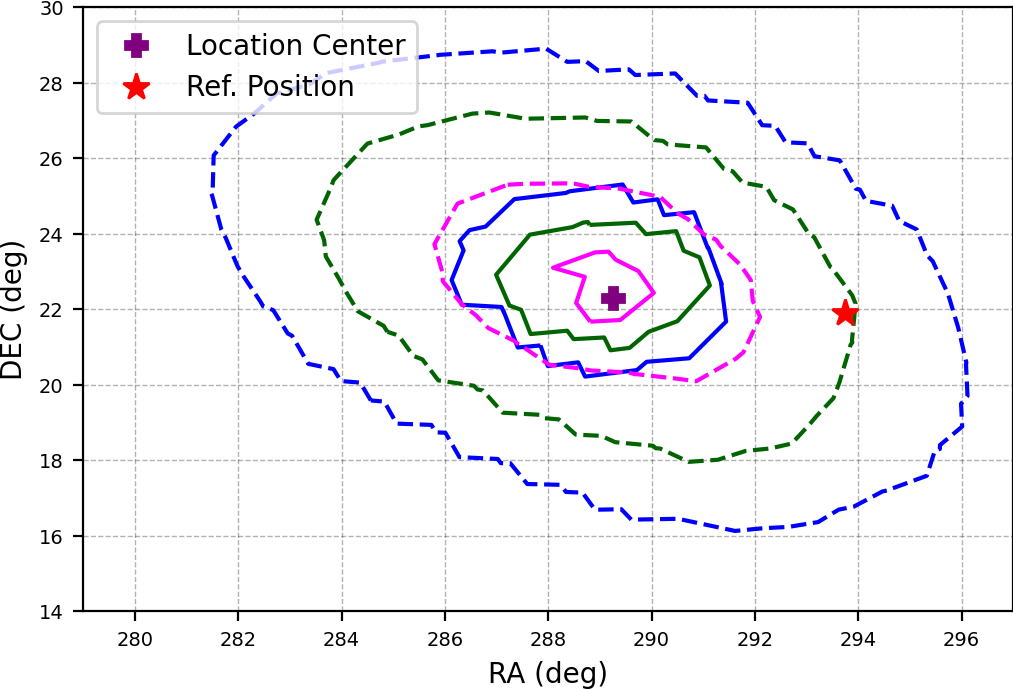}}
        \caption{ GECAM localization results of SGR 1935+2154 (UT 2021-09-22T20-12-16.502). (a) The light curve of GRD \# 14 high gain which contains the majority of net (burst) counts. (b) The RFD spectral fitting result. (c) The RFD spectral fitting result. The location credible region of (d) FIX, (e) RFD, and (f) APR localization. The captions are the same as Figure \ref{fig2a}. }
        \label{fig2_SGR1935e}
    \end{figure*}

    \begin{figure*}
        \centering
        \subfigure[]{\includegraphics[height=3.5cm]{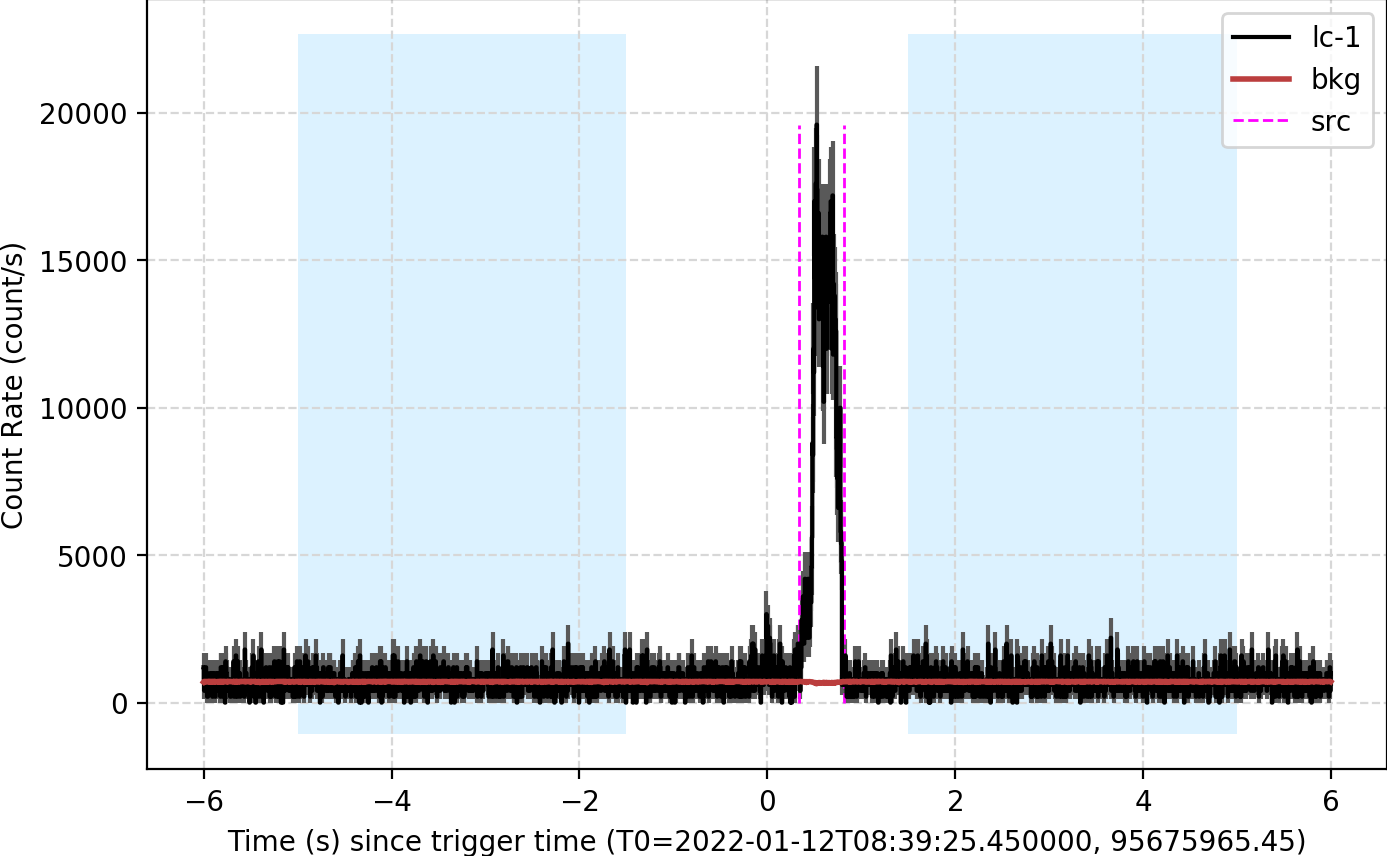}}
        \quad
        \subfigure[]{\includegraphics[height=3.5cm]{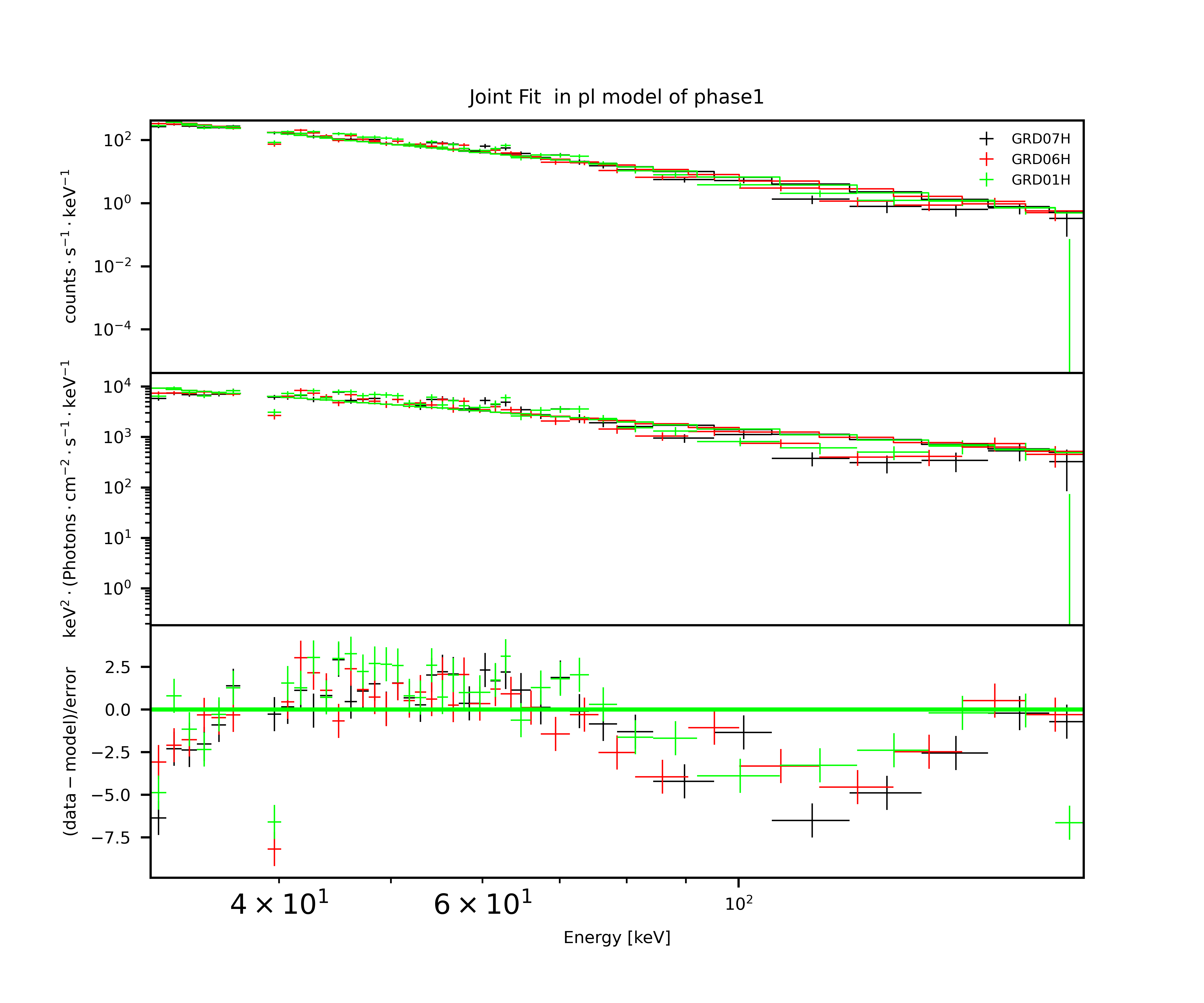}}
        \quad
        \subfigure[]{\includegraphics[height=3.5cm]{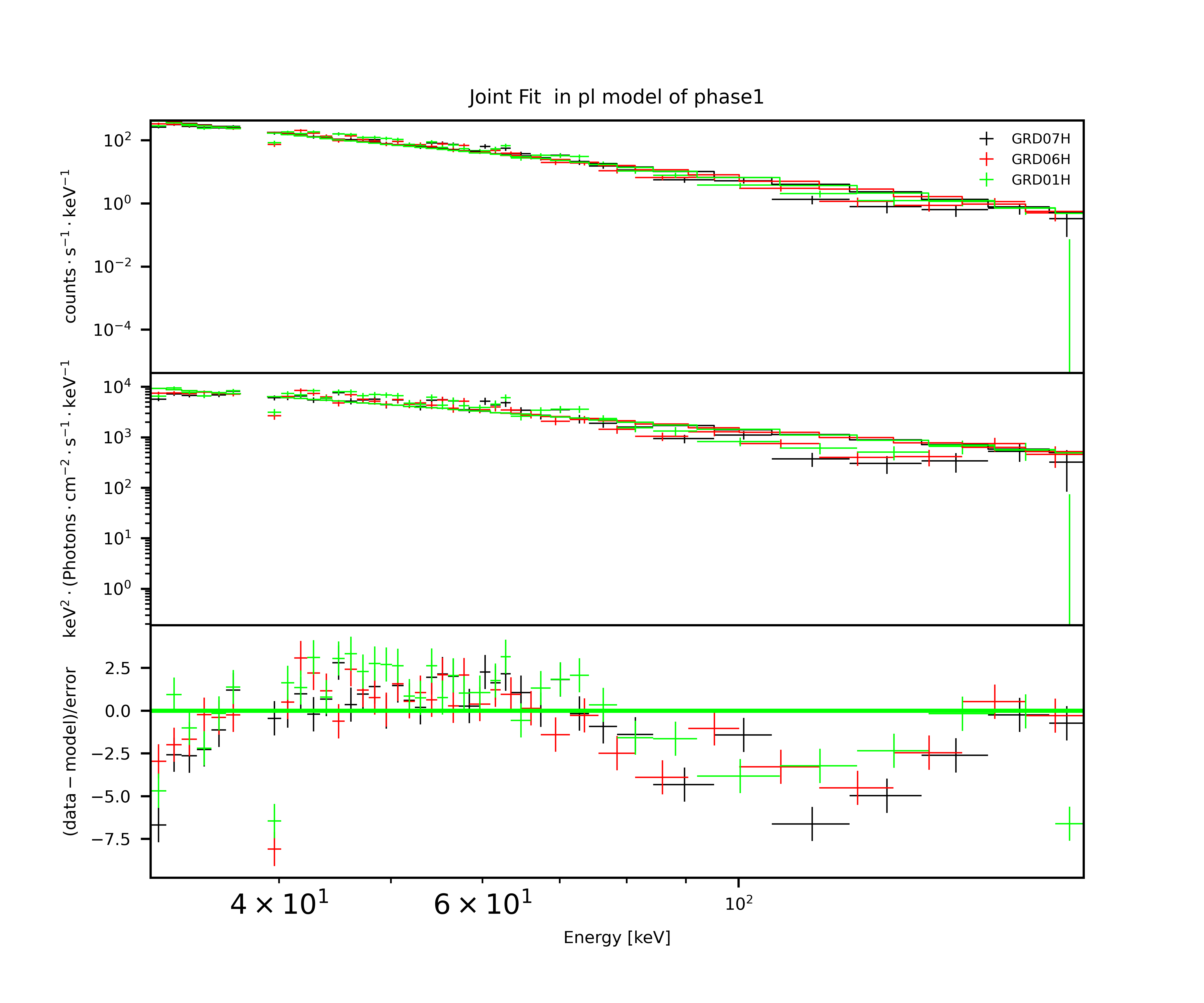}}
        \quad
        \\
        \subfigure[]{\includegraphics[width=4.5cm]{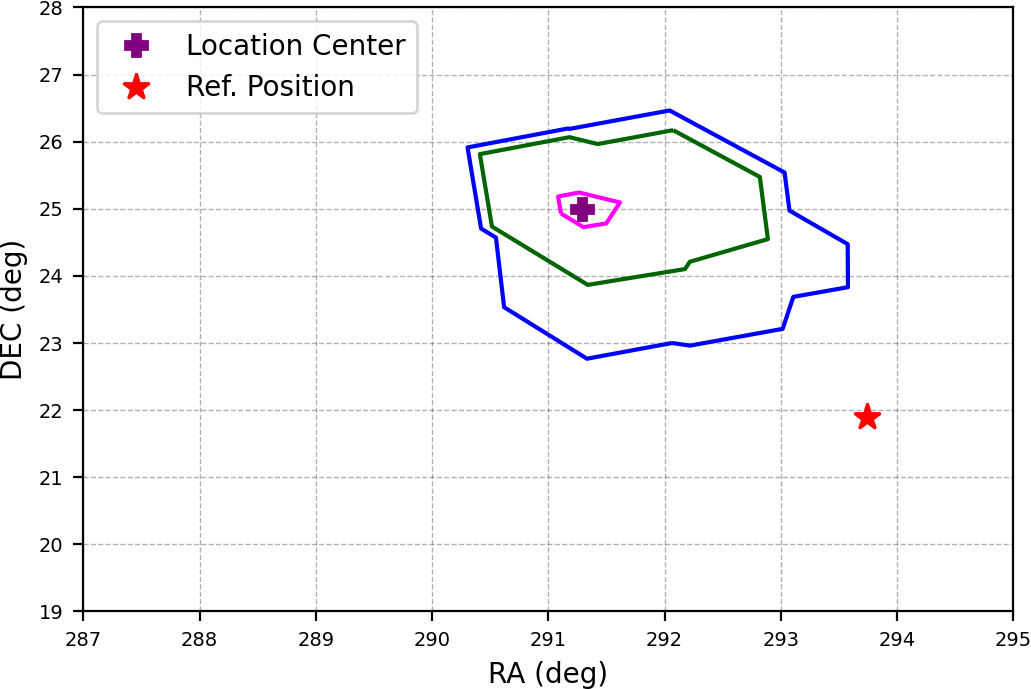}}
        \quad
        \subfigure[]{\includegraphics[width=4.5cm]{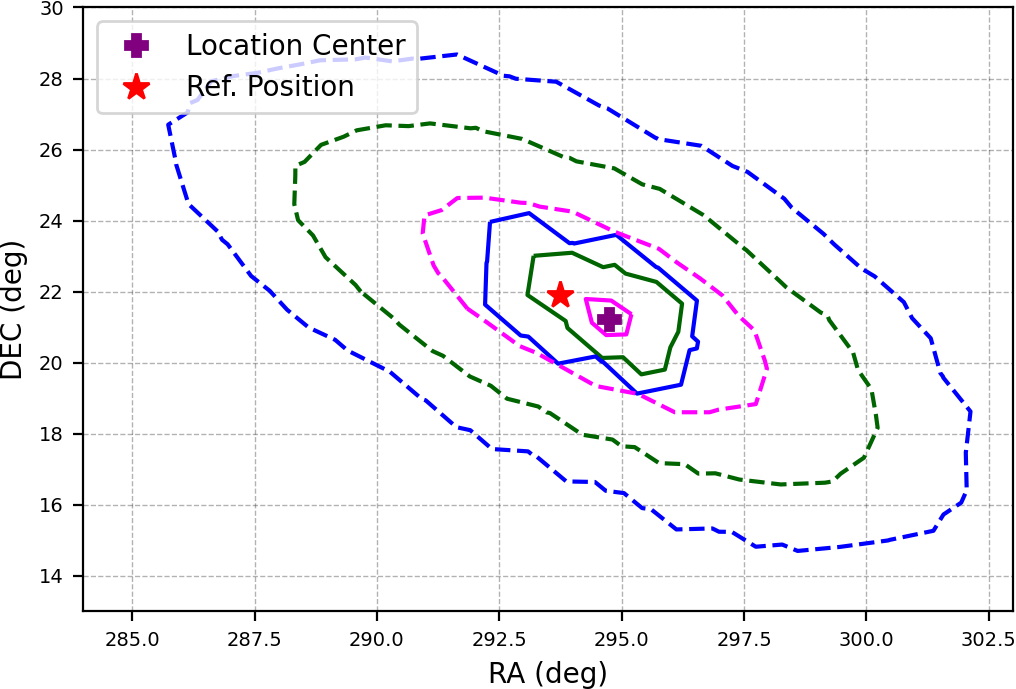}}
        \quad
        \subfigure[]{\includegraphics[width=4.5cm]{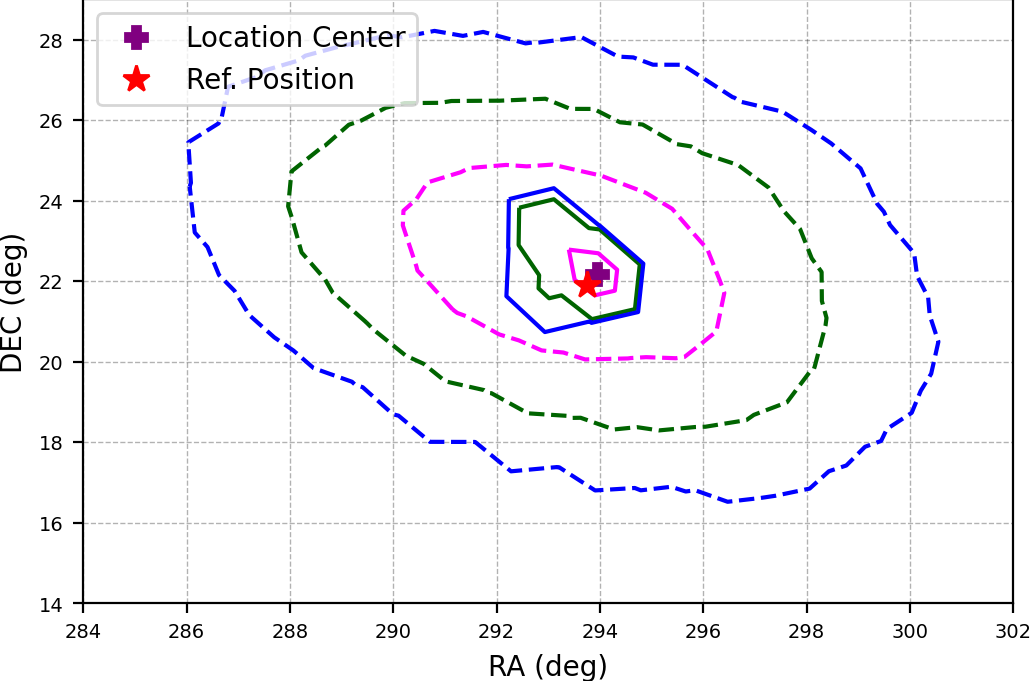}}
        \caption{ GECAM localization results of SGR 1935+2154 (UT 2022-01-12T08-39-25.450). (a) The light curve of GRD \# 01 high gain which contains the majority of net (burst) counts. (b) The RFD spectral fitting result. (c) The RFD spectral fitting result. The location credible region of (d) FIX, (e) RFD, and (f) APR localization. The captions are the same as Figure \ref{fig2a}. }
        \label{fig2_SGR1935f}
    \end{figure*}

    \begin{figure*}
        \centering
        \subfigure[]{\includegraphics[height=3.5cm]{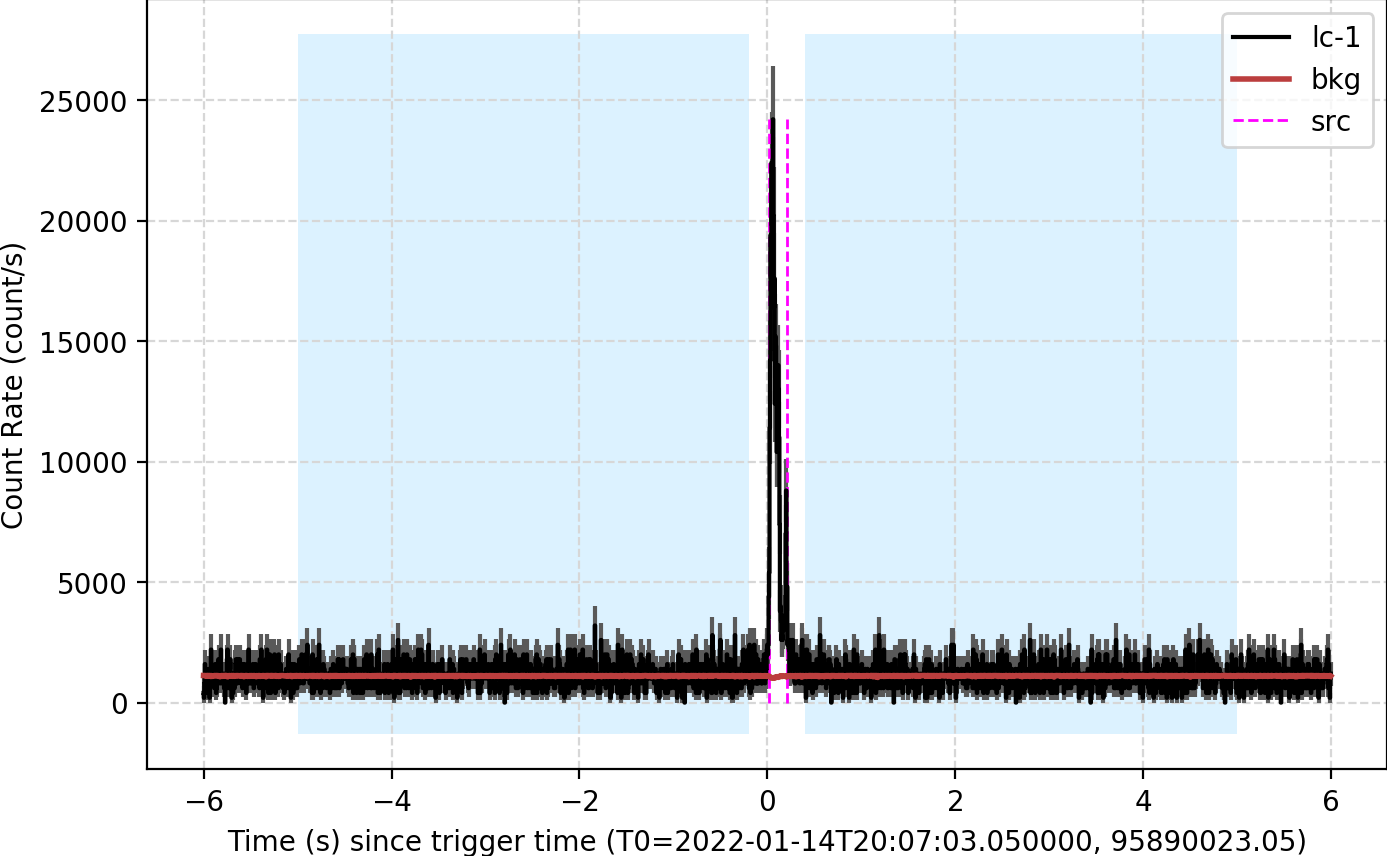}}
        \quad
        \subfigure[]{\includegraphics[height=3.5cm]{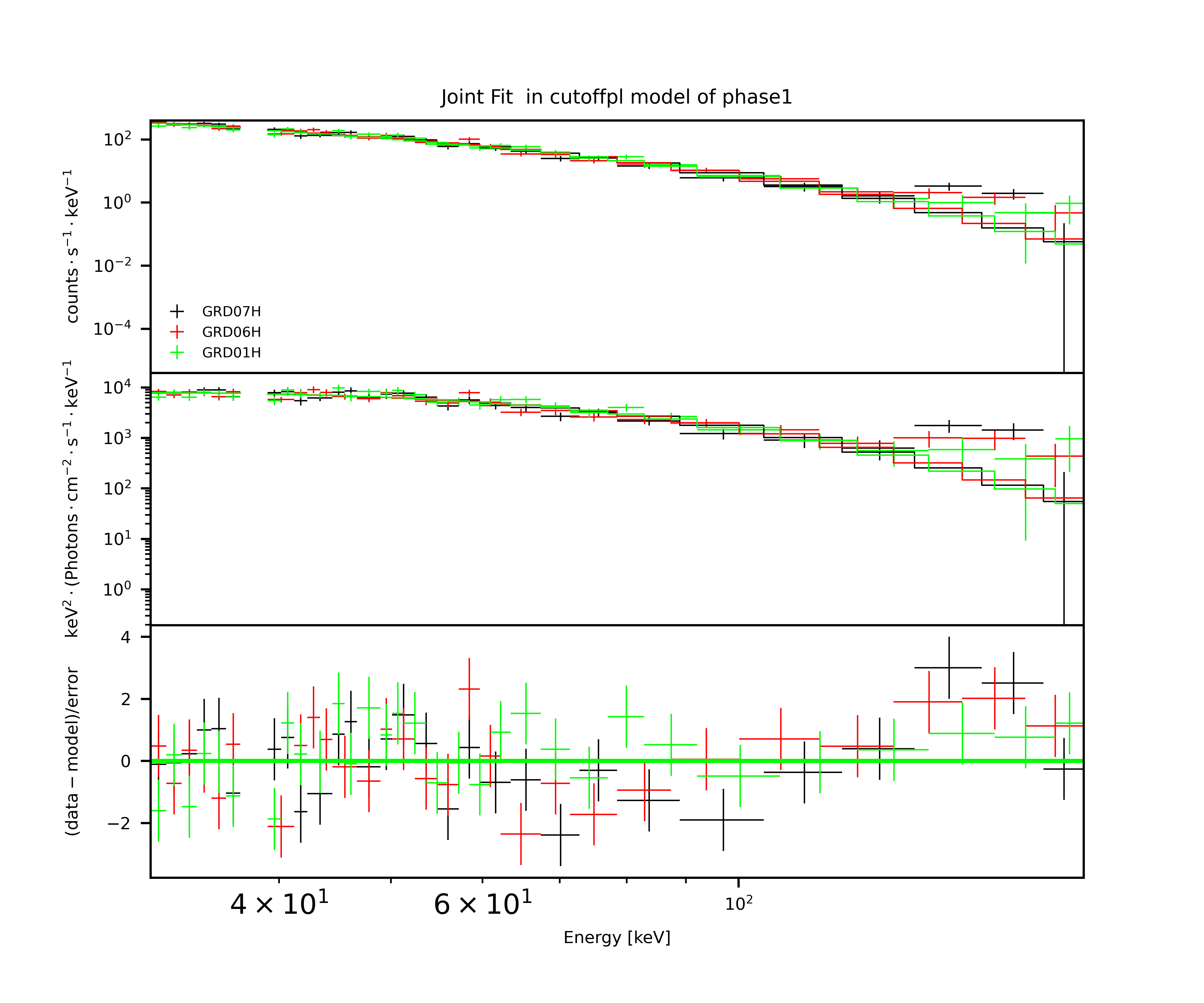}}
        \quad
        \subfigure[]{\includegraphics[height=3.5cm]{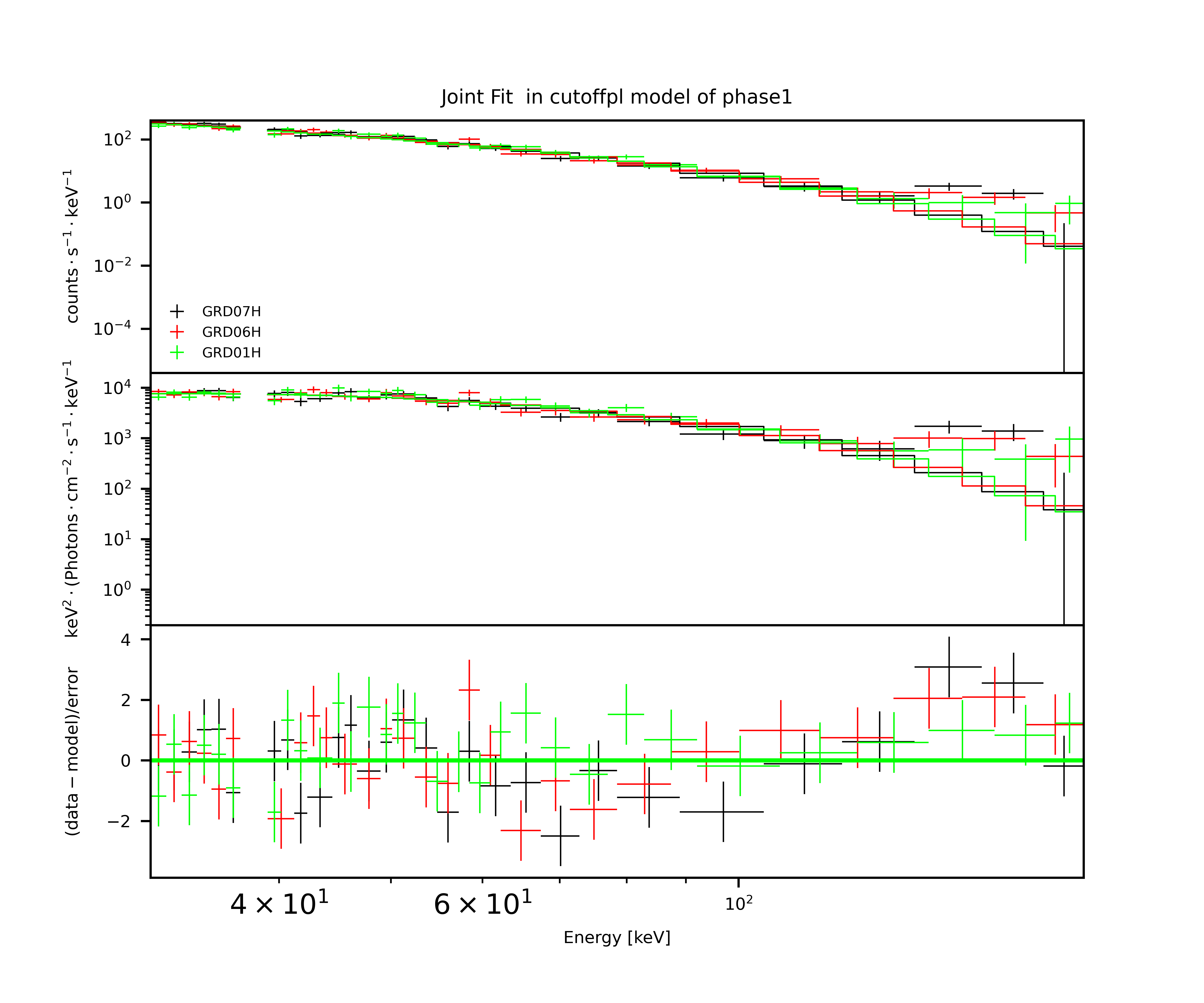}}
        \quad
        \\
        \subfigure[]{\includegraphics[width=4.5cm]{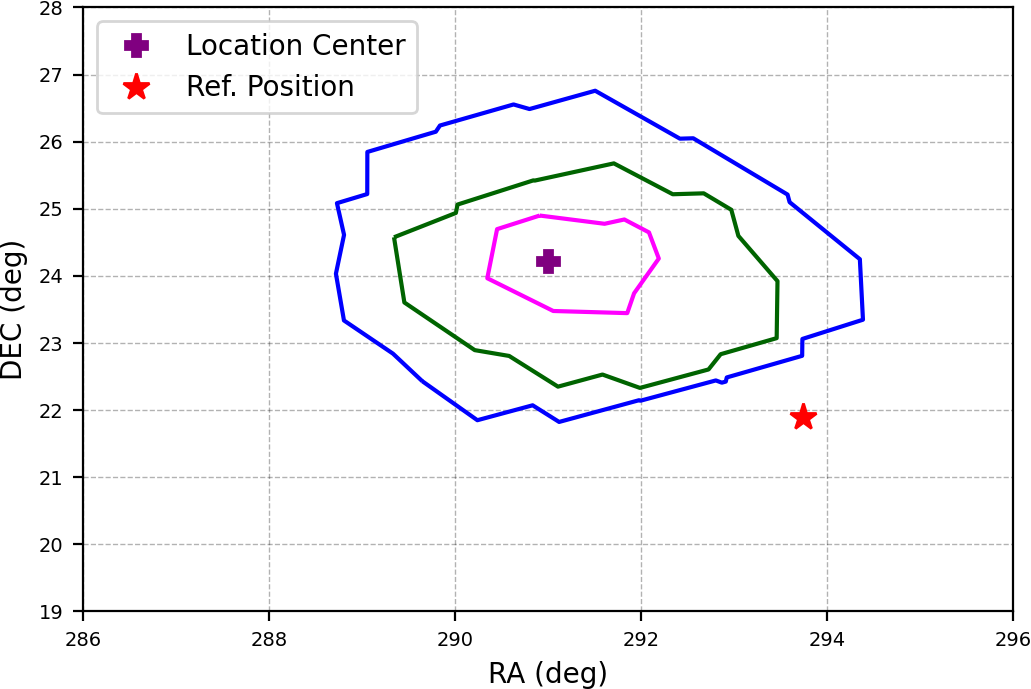}}
        \quad
        \subfigure[]{\includegraphics[width=4.5cm]{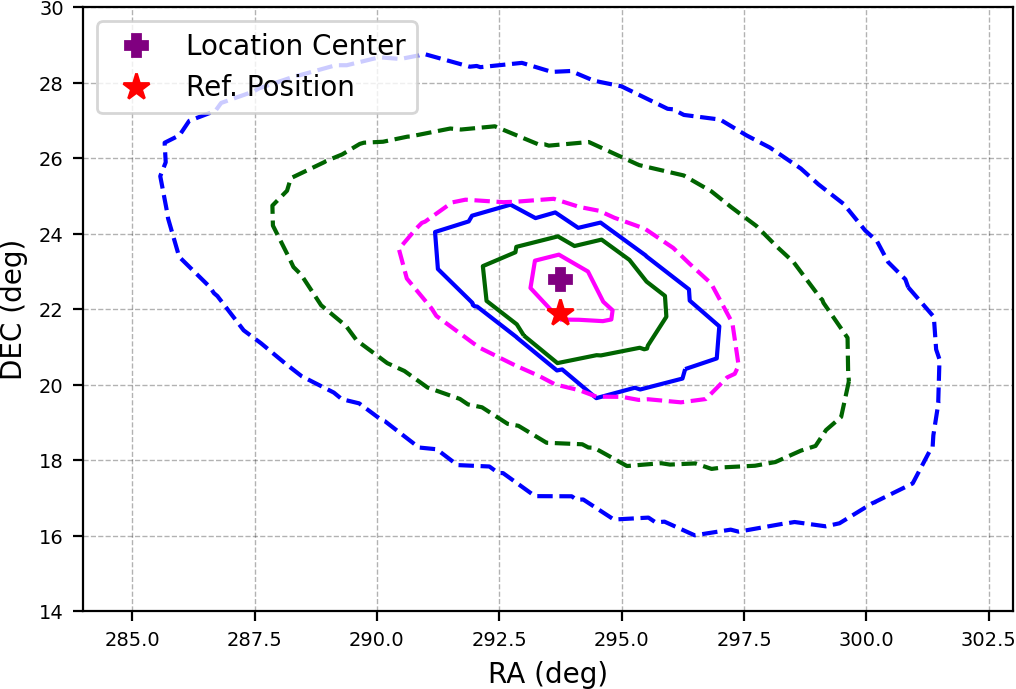}}
        \quad
        \subfigure[]{\includegraphics[width=4.5cm]{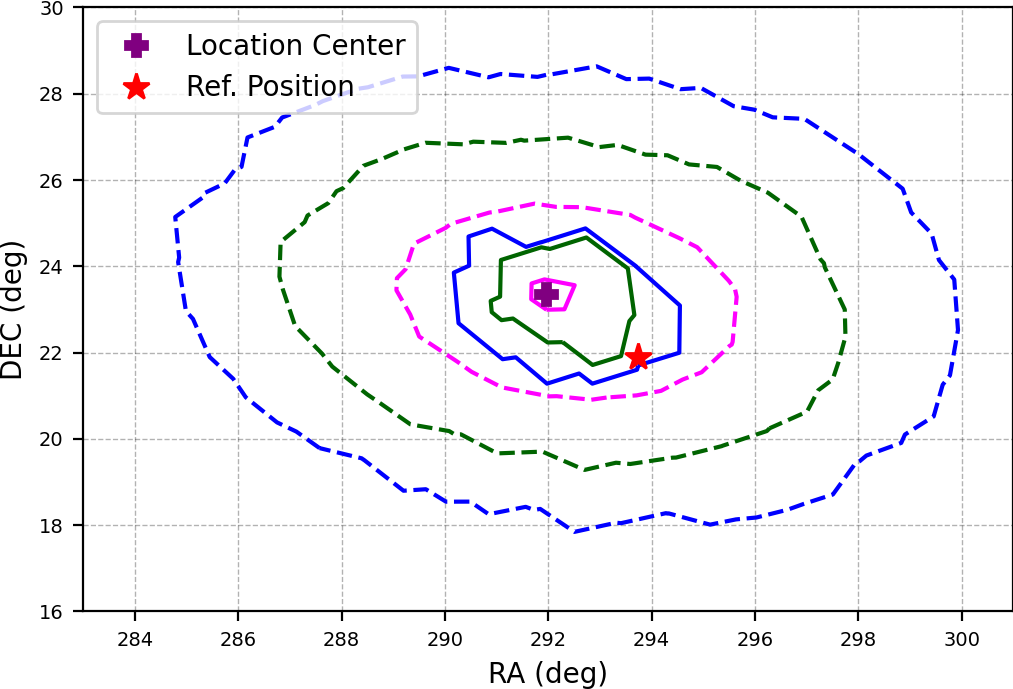}}
        \caption{ GECAM localization results of SGR 1935+2154 (UT 2022-01-14T20-07-03.050). (a) The light curve of GRD \# 07 high gain which contains the majority of net (burst) counts. (b) The RFD spectral fitting result. (c) The RFD spectral fitting result. The location credible region of (d) FIX, (e) RFD, and (f) APR localization. The captions are the same as Figure \ref{fig2a}. }
        \label{fig2_SGR1935g}
    \end{figure*}

    \begin{figure*}
        \centering
        \subfigure[]{\includegraphics[height=3.5cm]{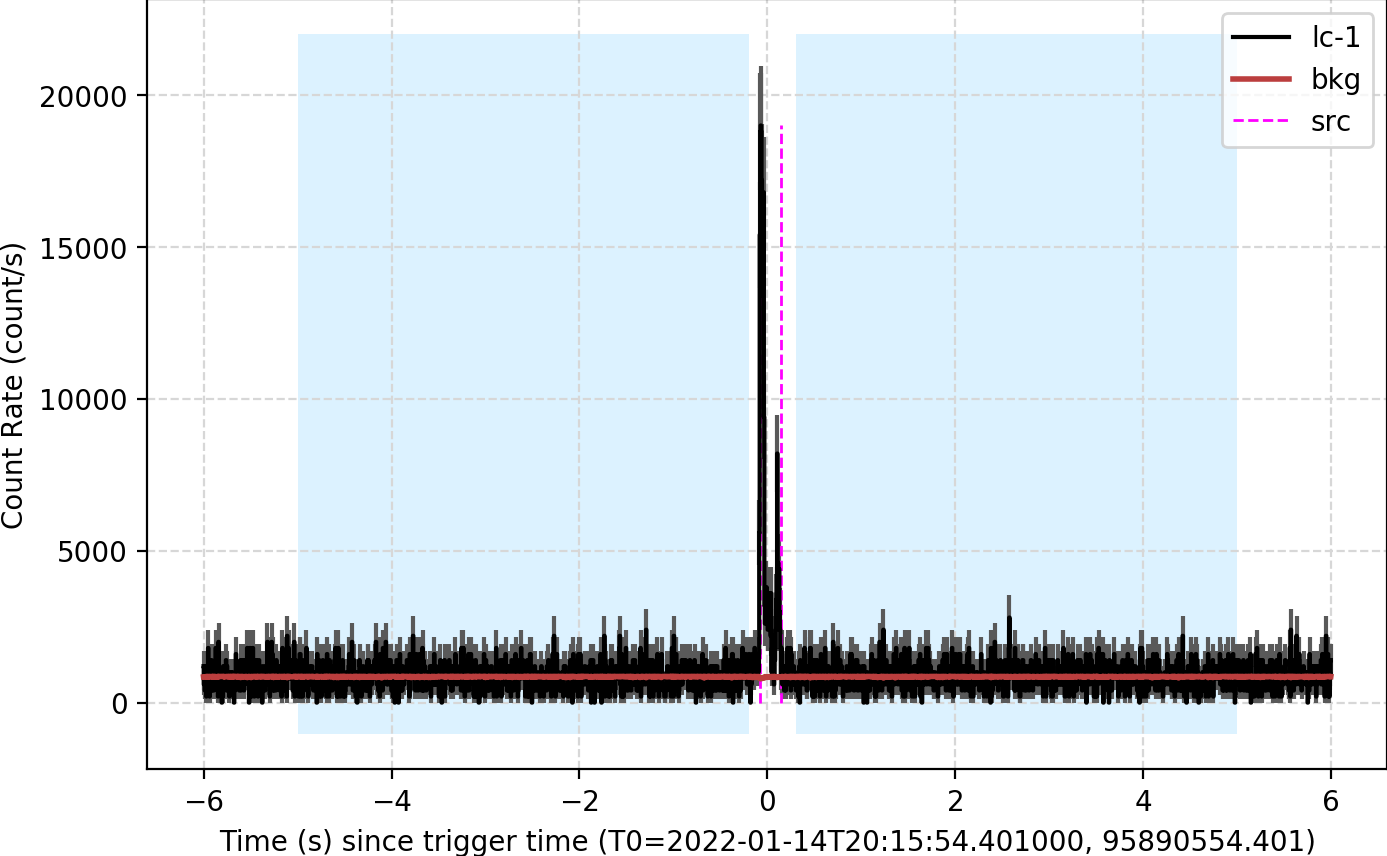}}
        \quad
        \subfigure[]{\includegraphics[height=3.5cm]{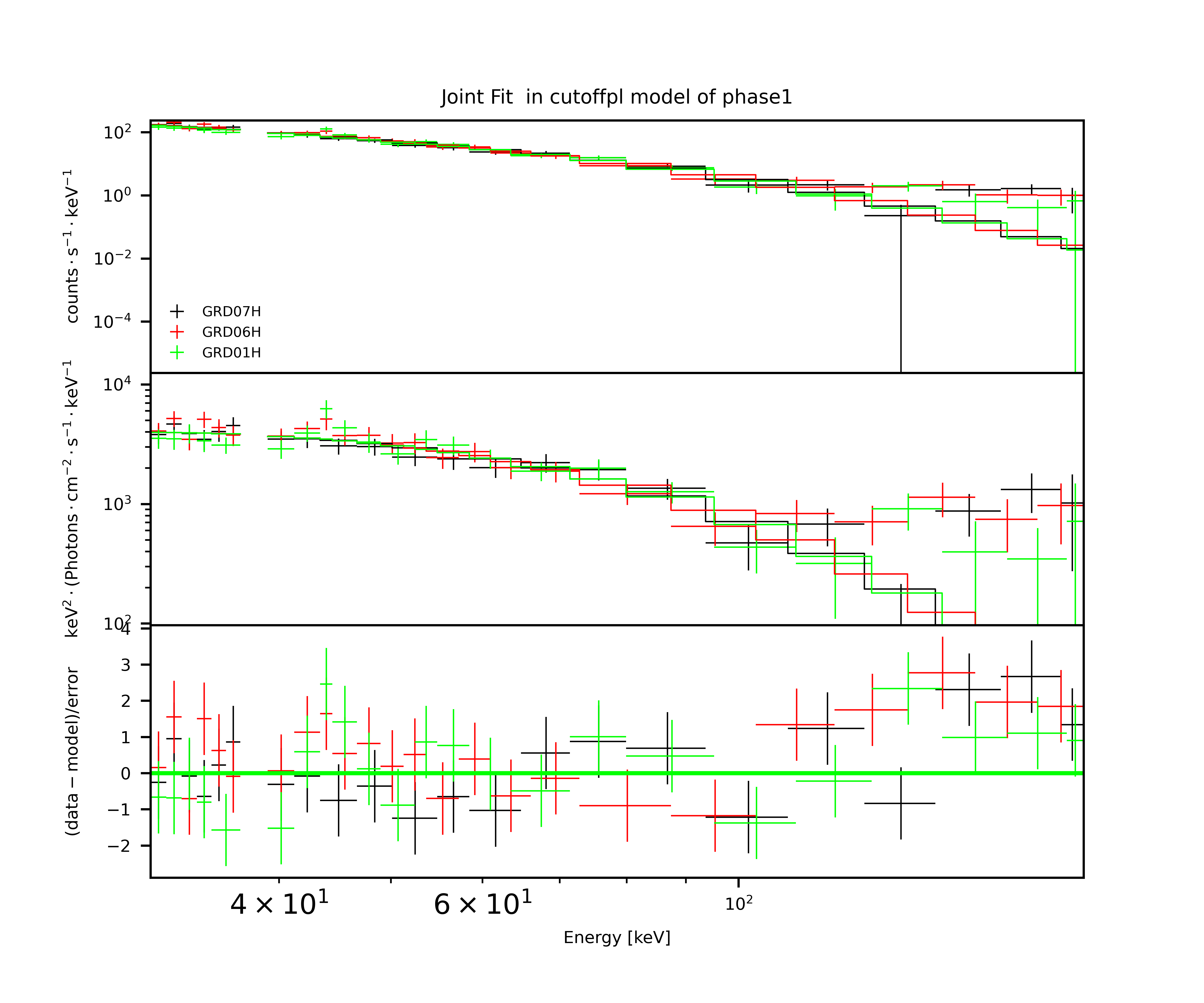}}
        \quad
        \subfigure[]{\includegraphics[height=3.5cm]{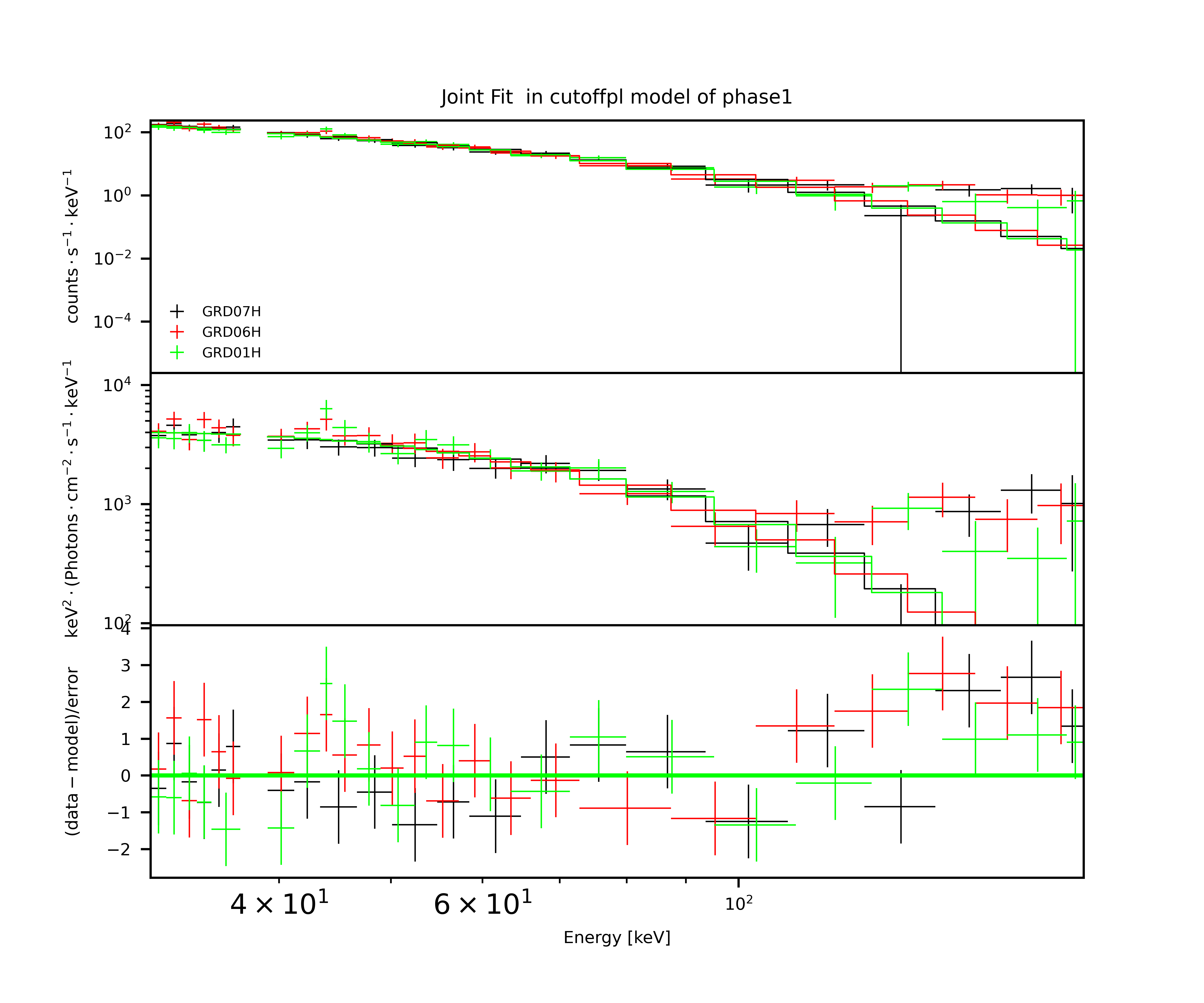}}
        \quad
        \\
        \subfigure[]{\includegraphics[width=4.5cm]{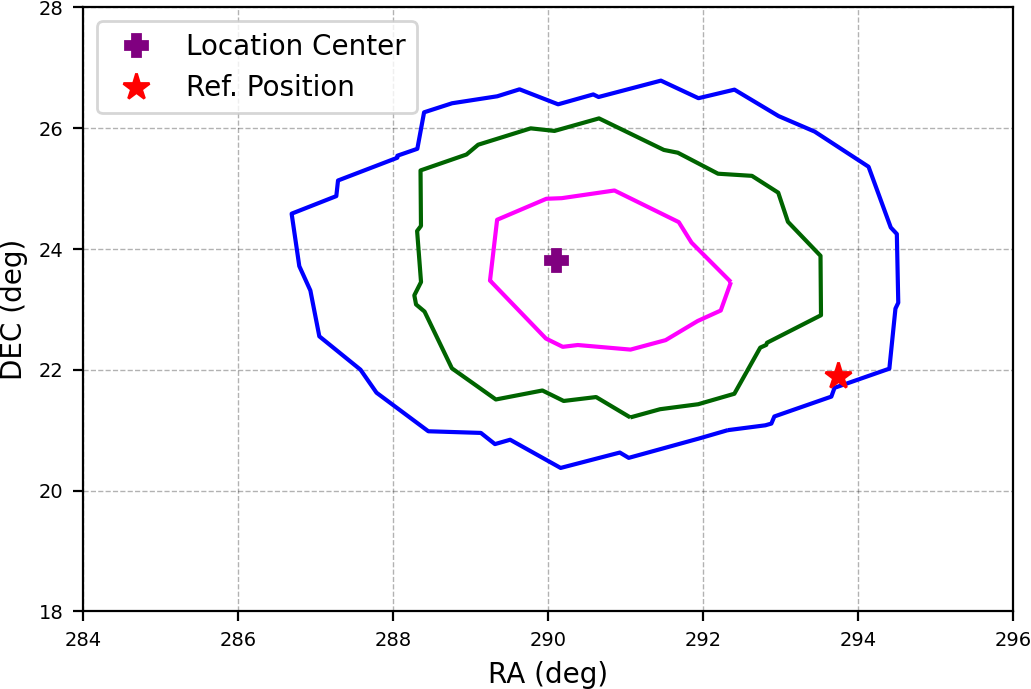}}
        \quad
        \subfigure[]{\includegraphics[width=4.5cm]{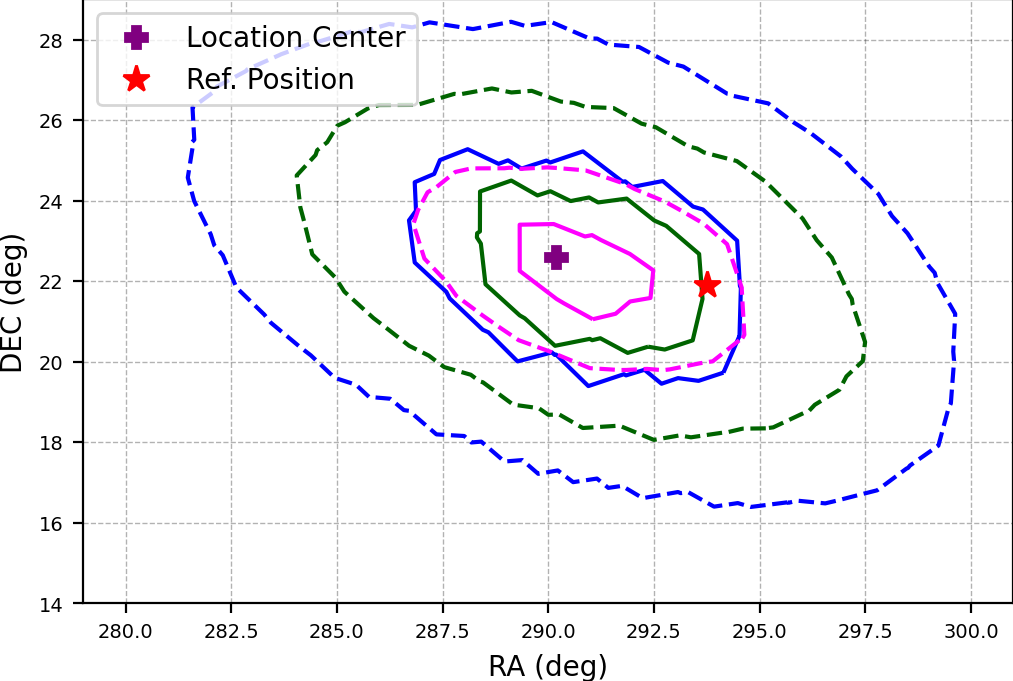}}
        \quad
        \subfigure[]{\includegraphics[width=4.5cm]{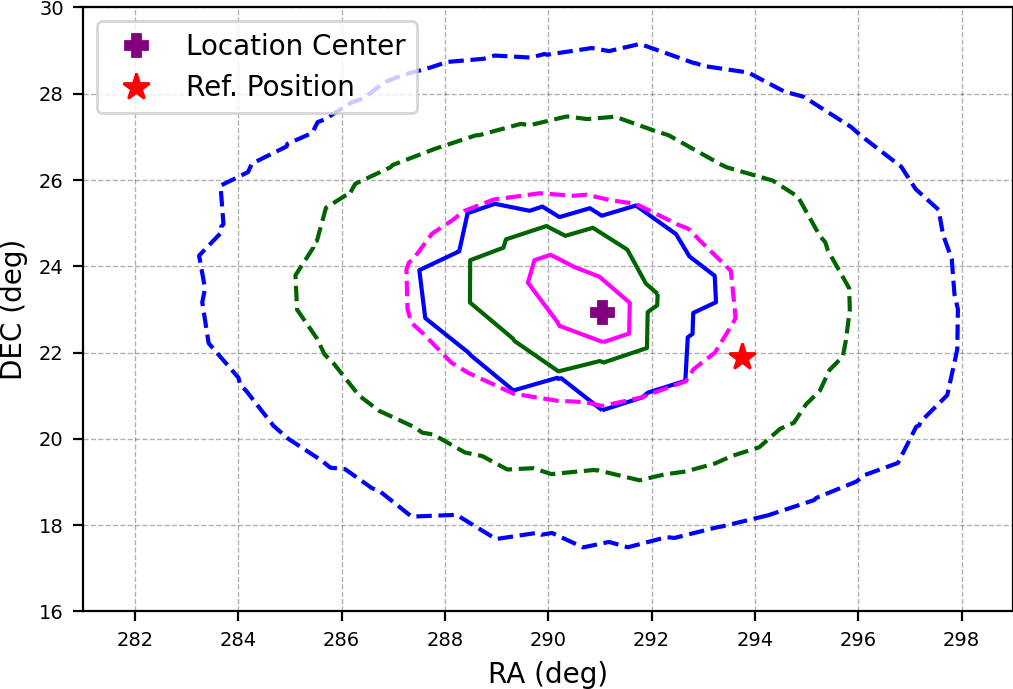}}
        \caption{ GECAM localization results of SGR 1935+2154 (UT 2022-01-14T20-15-54.401). (a) The light curve of GRD \# 06 high gain which contains the majority of net (burst) counts. (b) The RFD spectral fitting result. (c) The RFD spectral fitting result. The location credible region of (d) FIX, (e) RFD, and (f) APR localization. The captions are the same as Figure \ref{fig2a}. }
        \label{fig2_SGR1935h}
    \end{figure*}

    \begin{figure*}
        \centering
        \subfigure[]{\includegraphics[height=3.5cm]{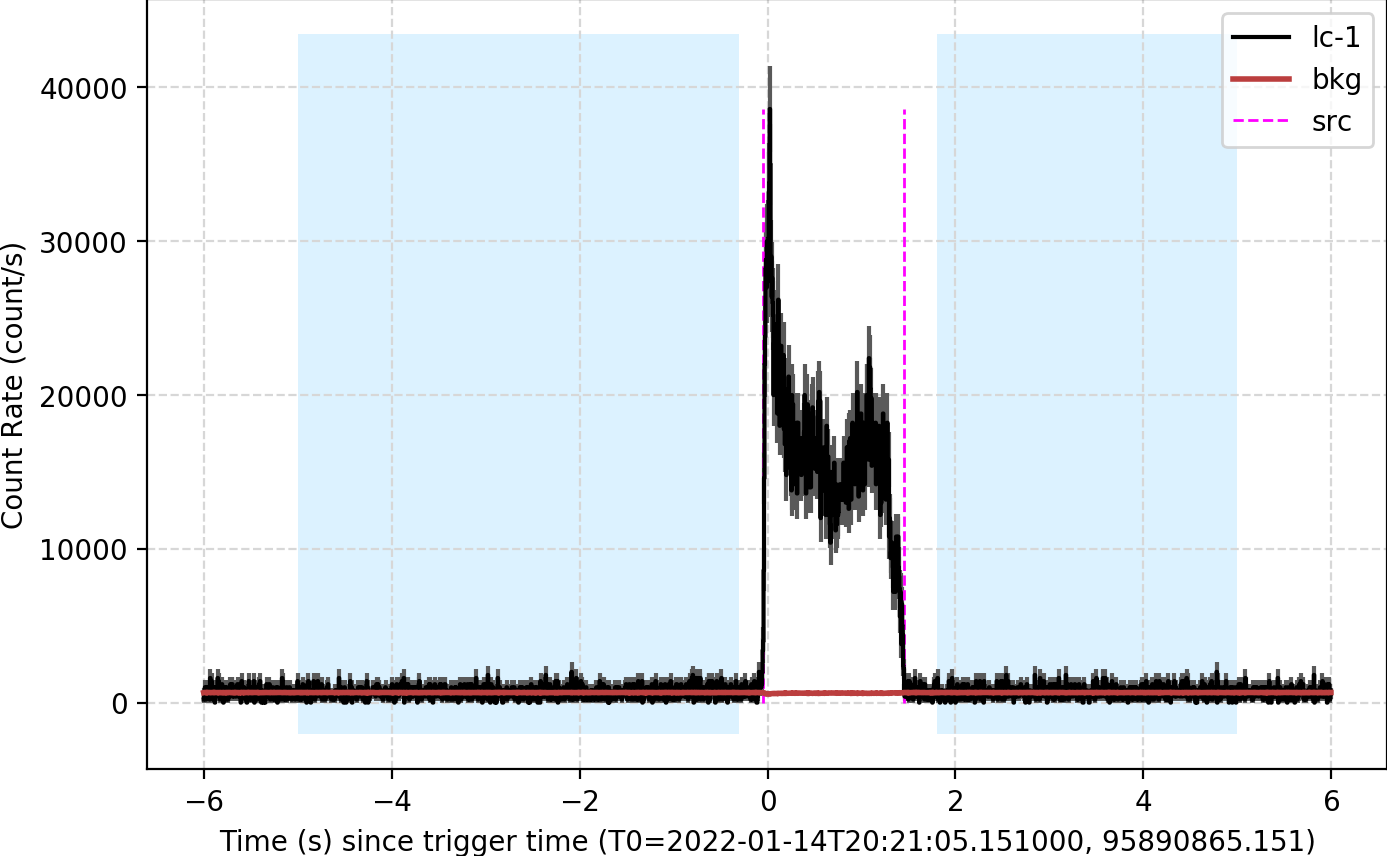}}
        \quad
        \subfigure[]{\includegraphics[height=3.5cm]{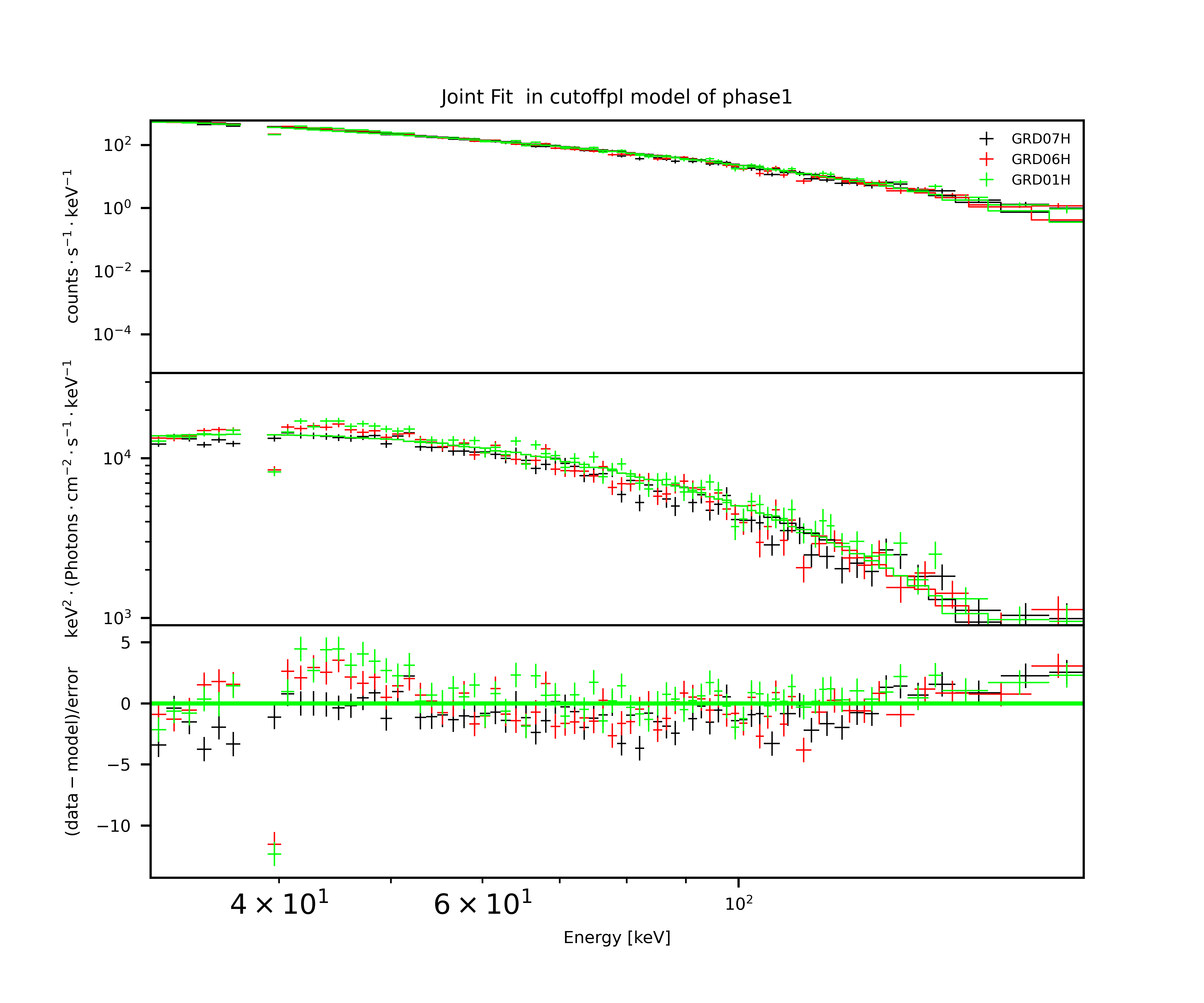}}
        \quad
        \subfigure[]{\includegraphics[height=3.5cm]{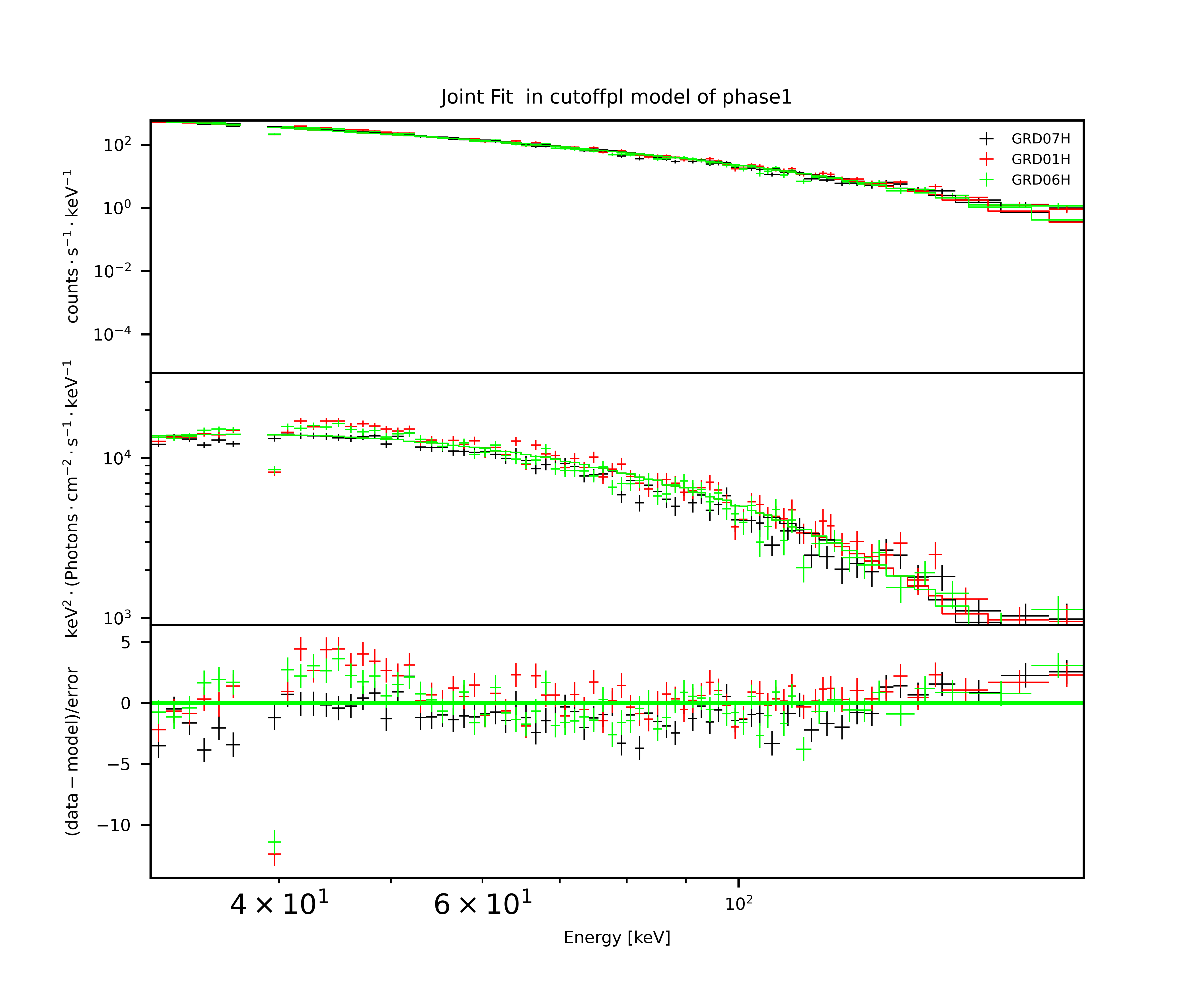}}
        \quad
        \\
        \subfigure[]{\includegraphics[width=4.5cm]{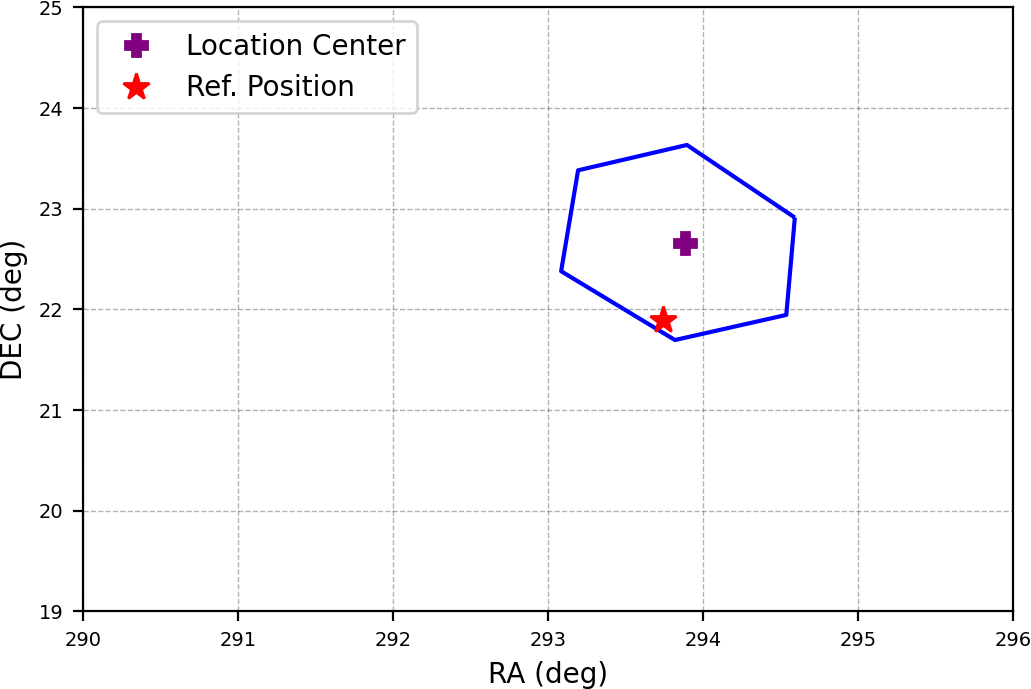}}
        \quad
        \subfigure[]{\includegraphics[width=4.5cm]{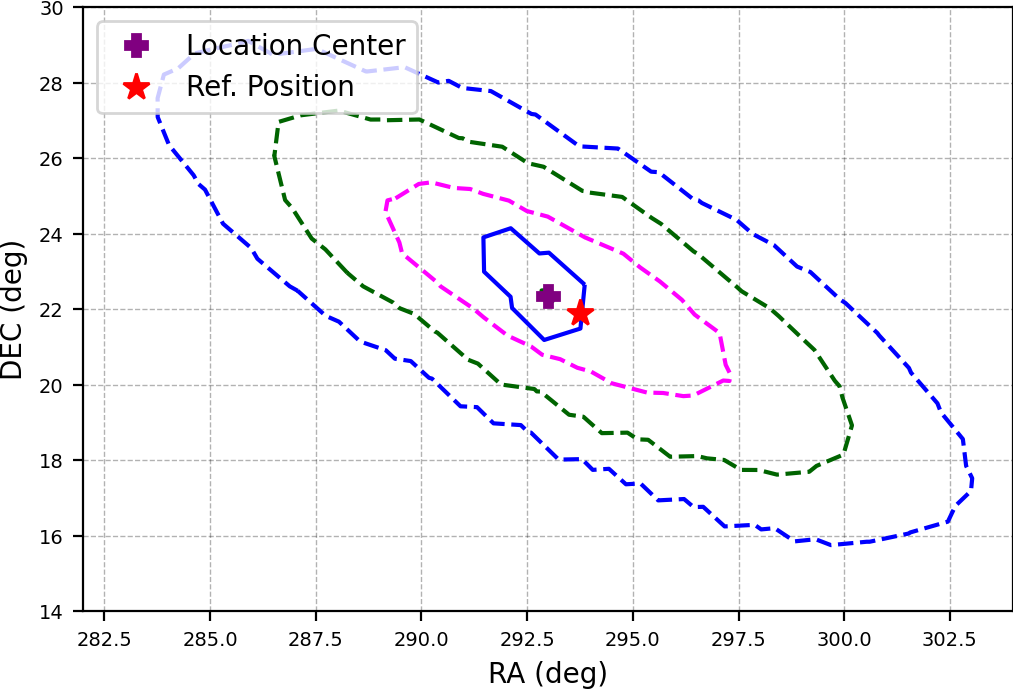}}
        \quad
        \caption{ GECAM localization results of SGR 1935+2154 (UT 2022-01-14T20-21-05.151). (a) The light curve of GRD \# 01 high gain which contains the majority of net (burst) counts. (b) The RFD spectral fitting result. (c) The RFD spectral fitting result. The location credible region of (d) FIX, (e) RFD, and (f) APR localization. The captions are the same as Figure \ref{fig2a}. }
        \label{fig2_SGR1935i}
    \end{figure*}

    \begin{figure*}
        \centering
        \subfigure[]{\includegraphics[height=3.5cm]{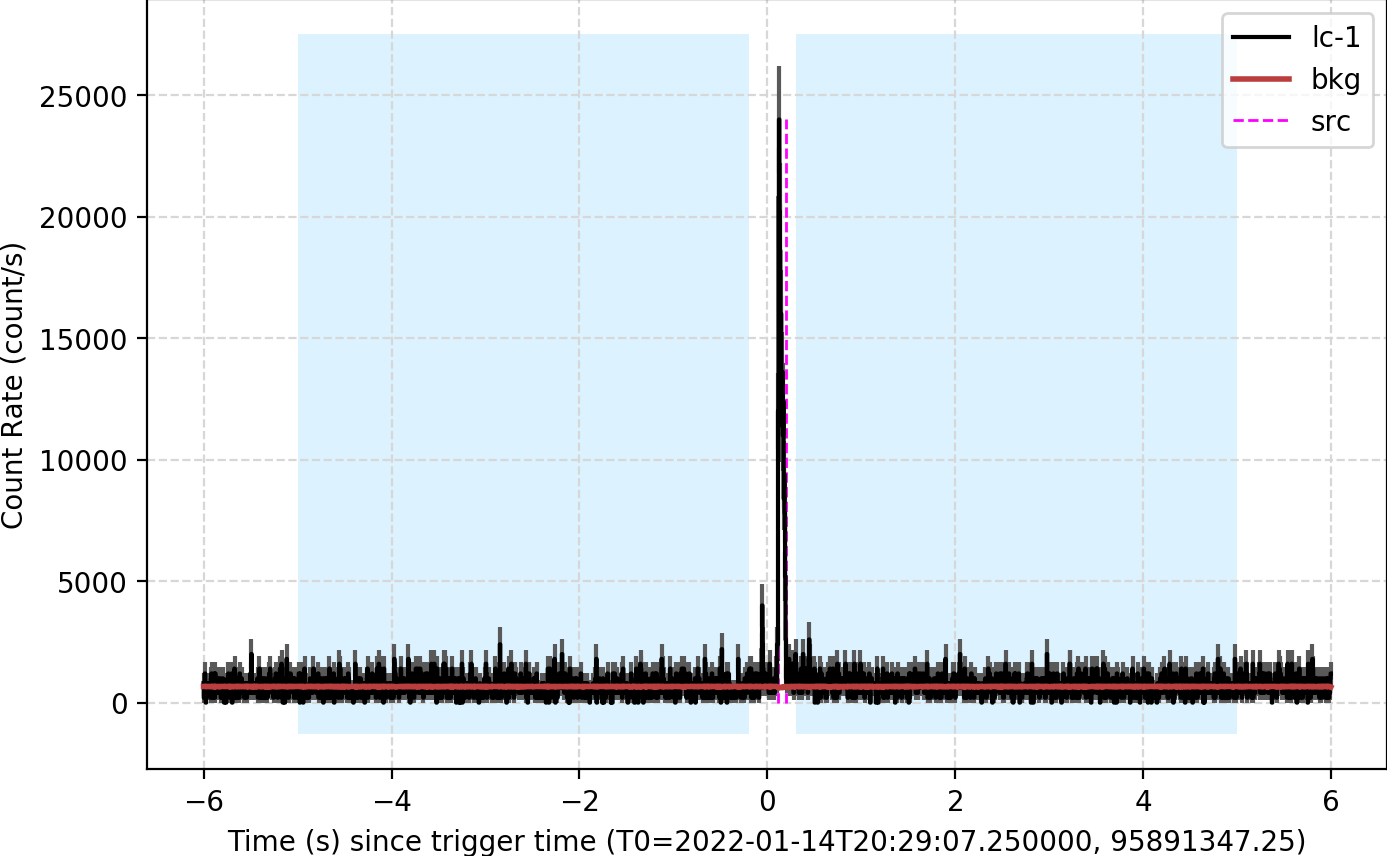}}
        \quad
        \subfigure[]{\includegraphics[height=3.5cm]{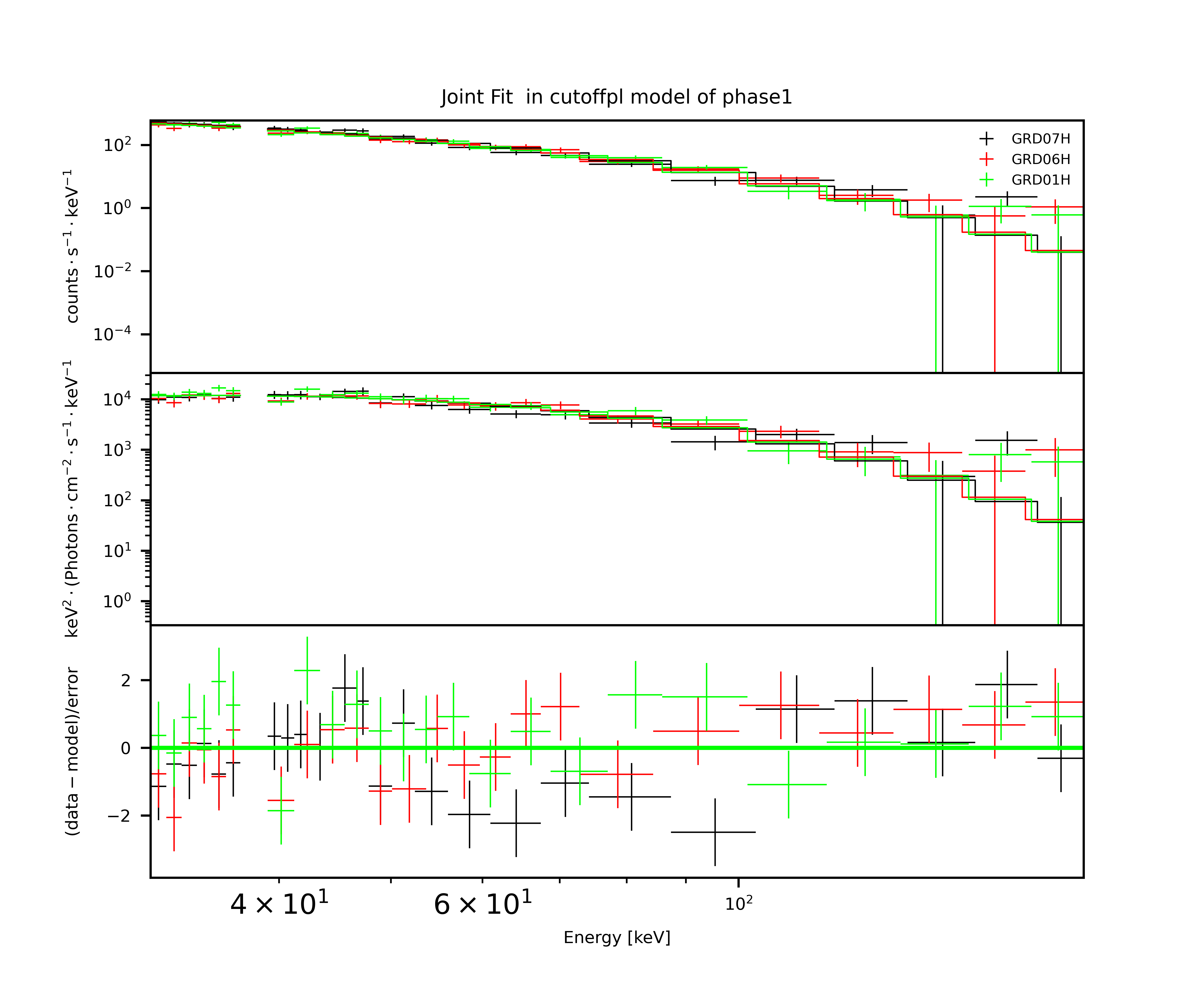}}
        \quad
        \subfigure[]{\includegraphics[height=3.5cm]{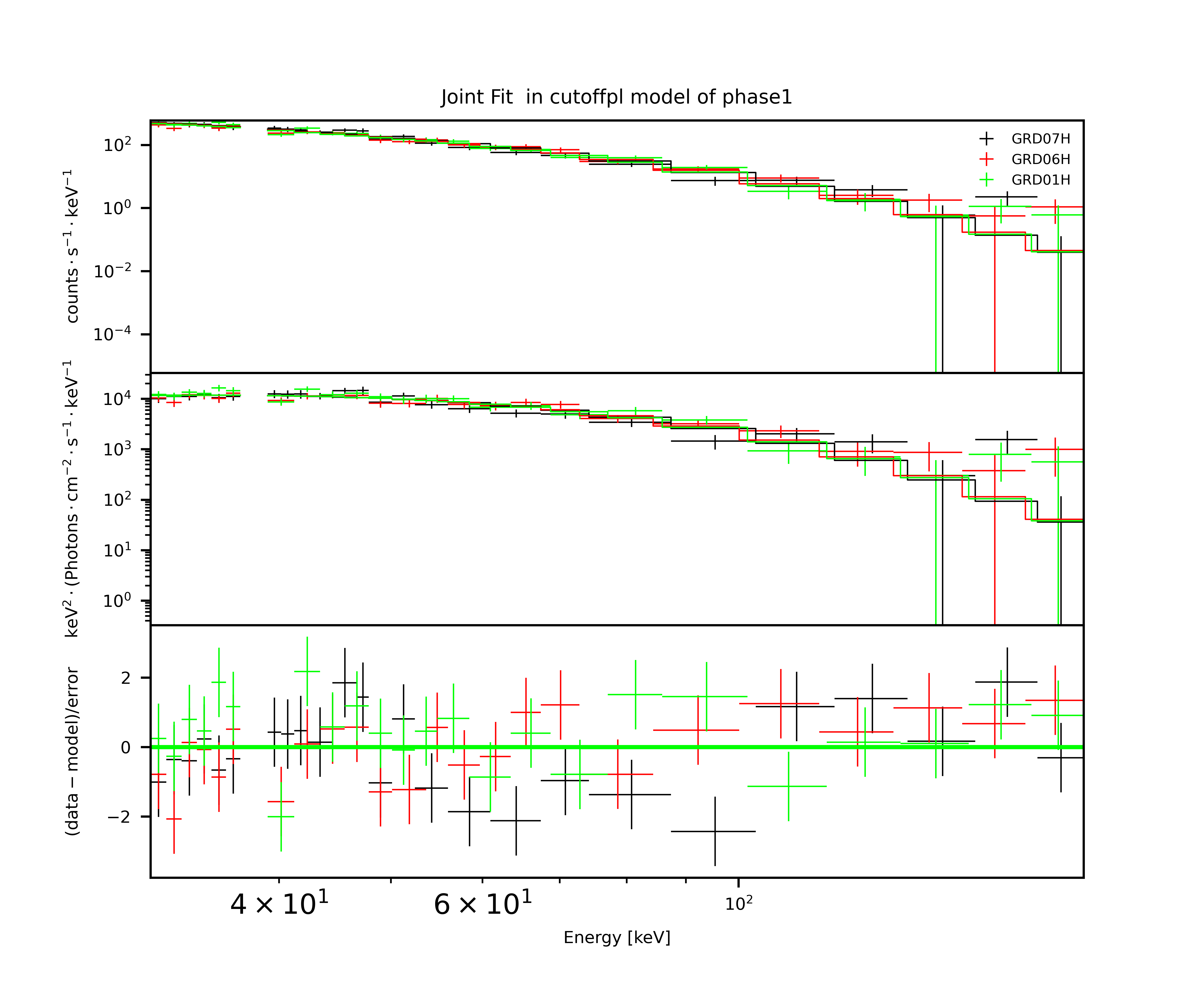}}
        \quad
        \\
        \subfigure[]{\includegraphics[width=4.5cm]{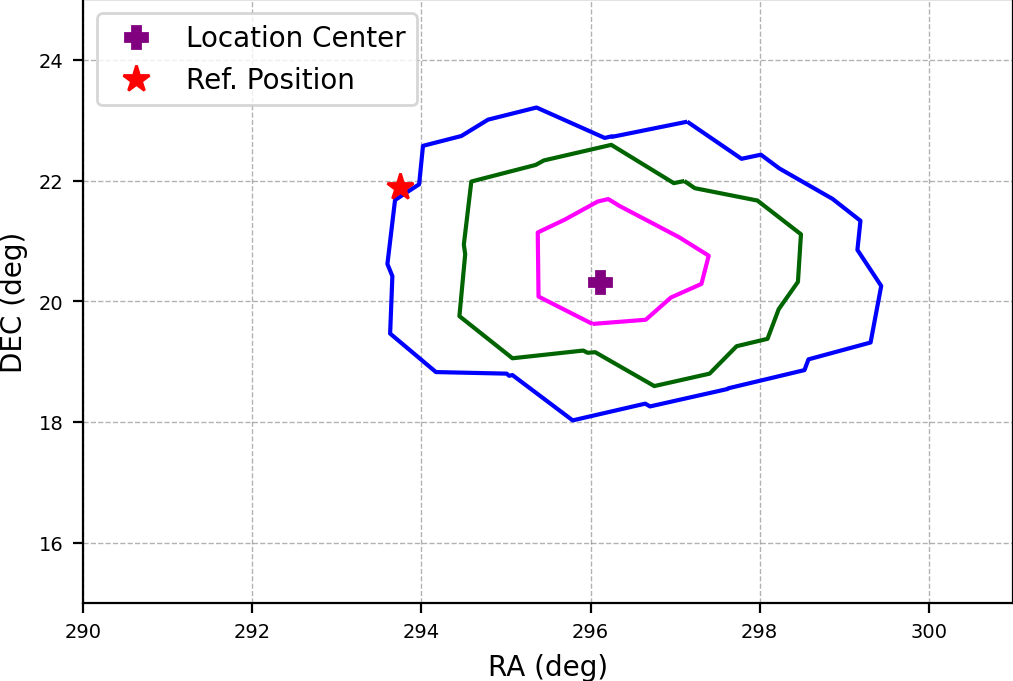}}
        \quad
        \subfigure[]{\includegraphics[width=4.5cm]{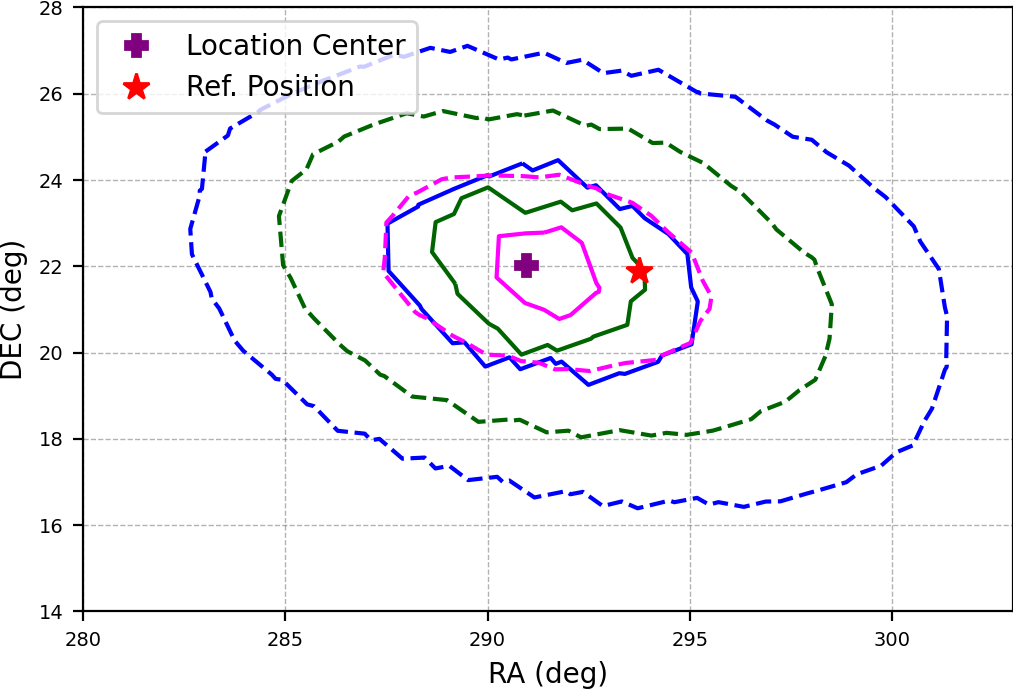}}
        \quad
        \subfigure[]{\includegraphics[width=4.5cm]{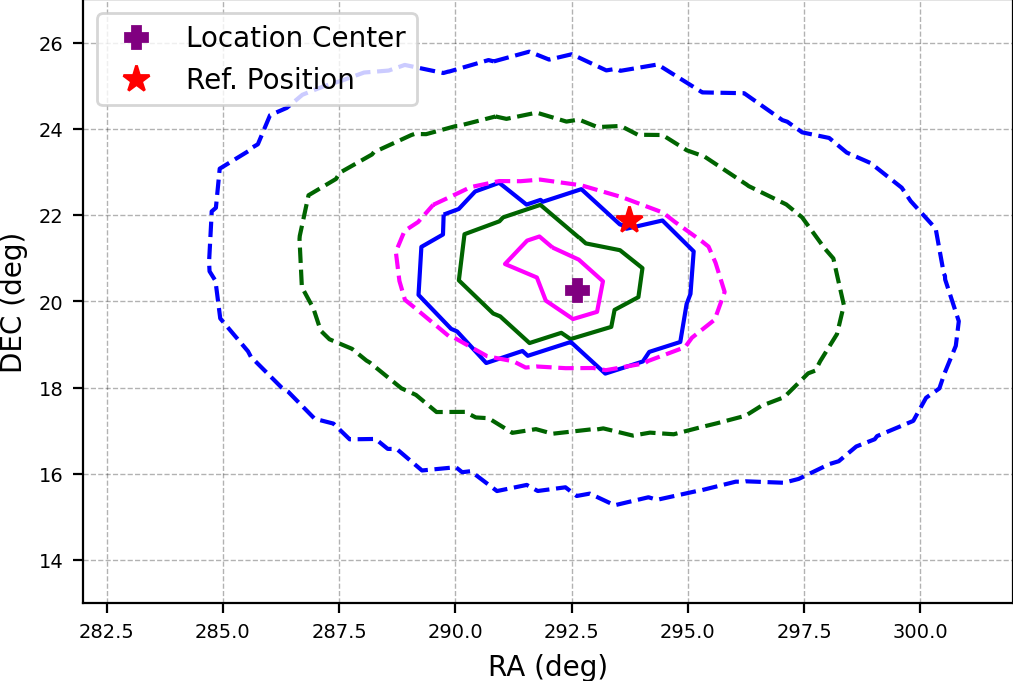}}
        \caption{ GECAM localization results of SGR 1935+2154 (UT 2022-01-14T20-29-07.250). (a) The light curve of GRD \# 01 high gain which contains the majority of net (burst) counts. (b) The RFD spectral fitting result. (c) The RFD spectral fitting result. The location credible region of (d) FIX, (e) RFD, and (f) APR localization. The captions are the same as Figure \ref{fig2a}. }
        \label{fig2_SGR1935j}
    \end{figure*}

    \begin{figure*}
        \centering
        \subfigure[]{\includegraphics[height=3.5cm]{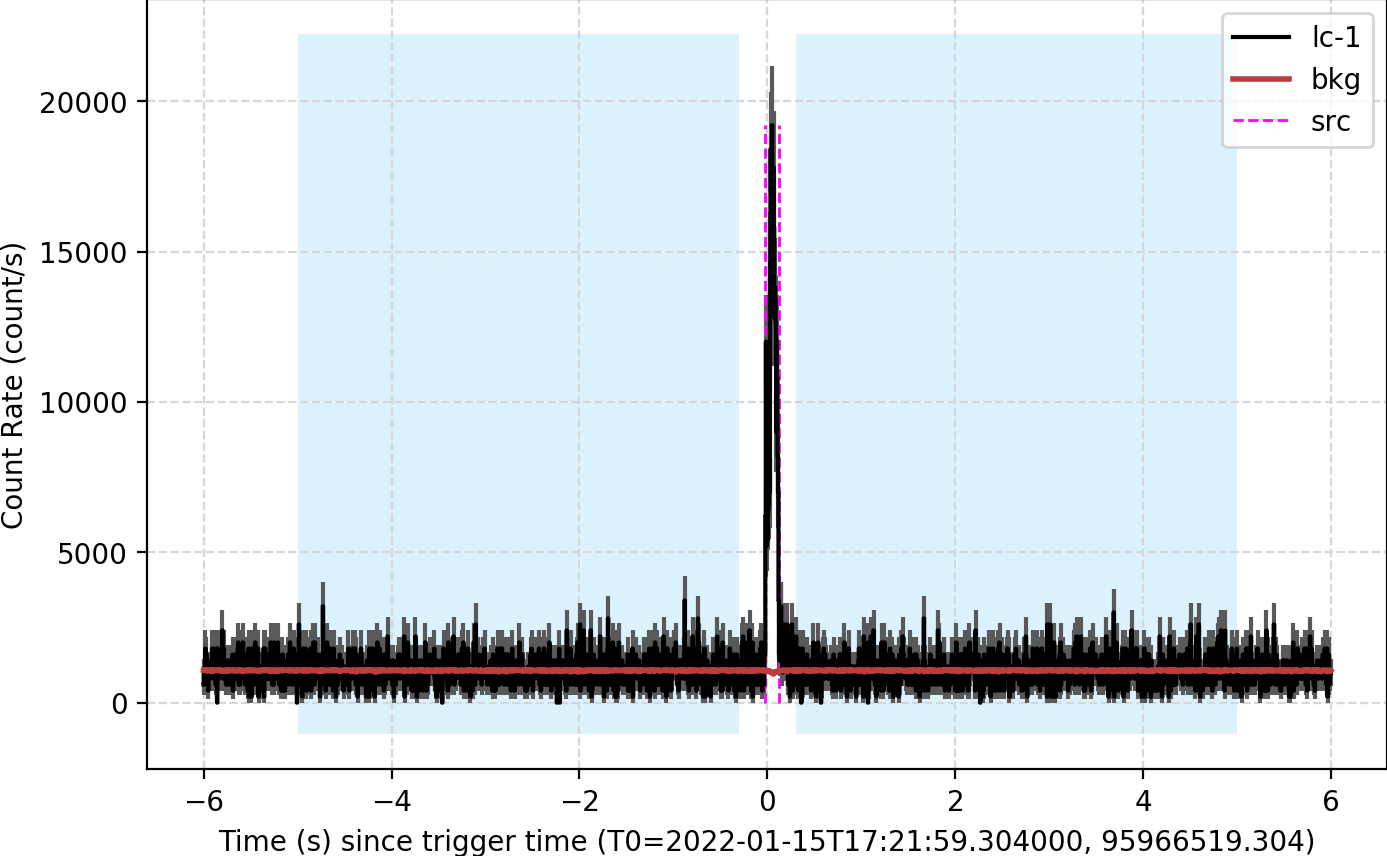}}
        \quad
        \subfigure[]{\includegraphics[height=3.5cm]{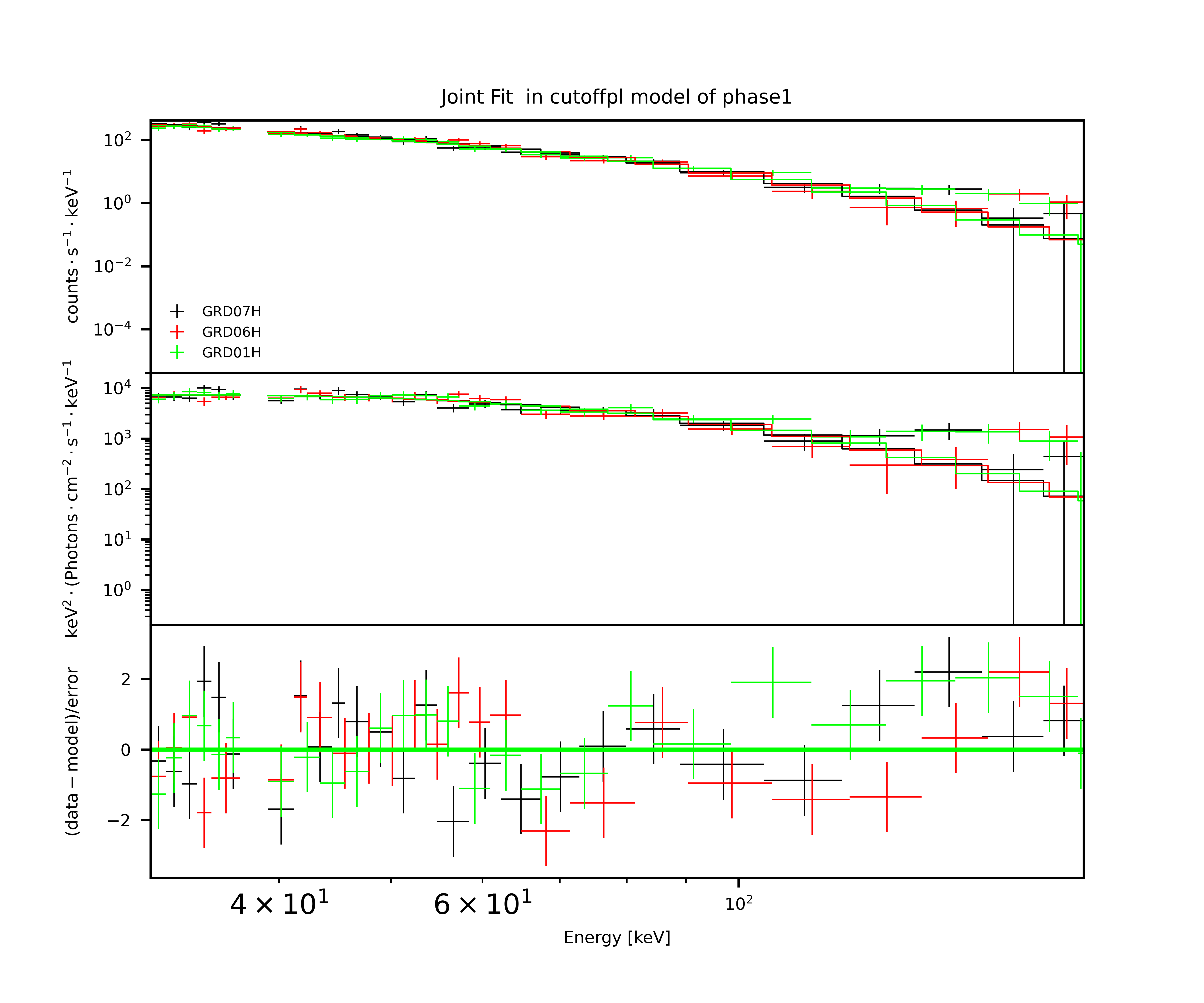}}
        \quad
        \subfigure[]{\includegraphics[height=3.5cm]{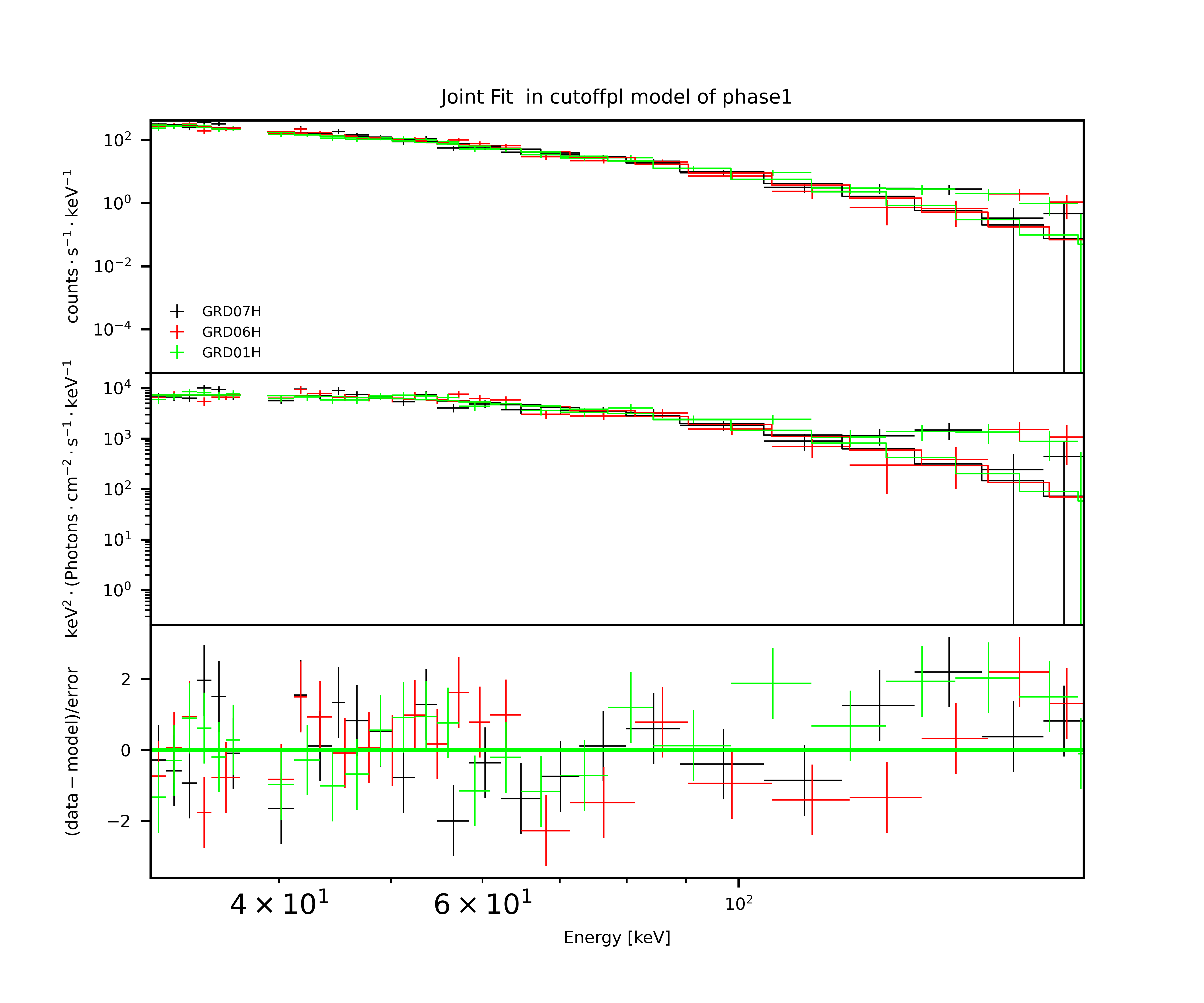}}
        \quad
        \\
        \subfigure[]{\includegraphics[width=4.5cm]{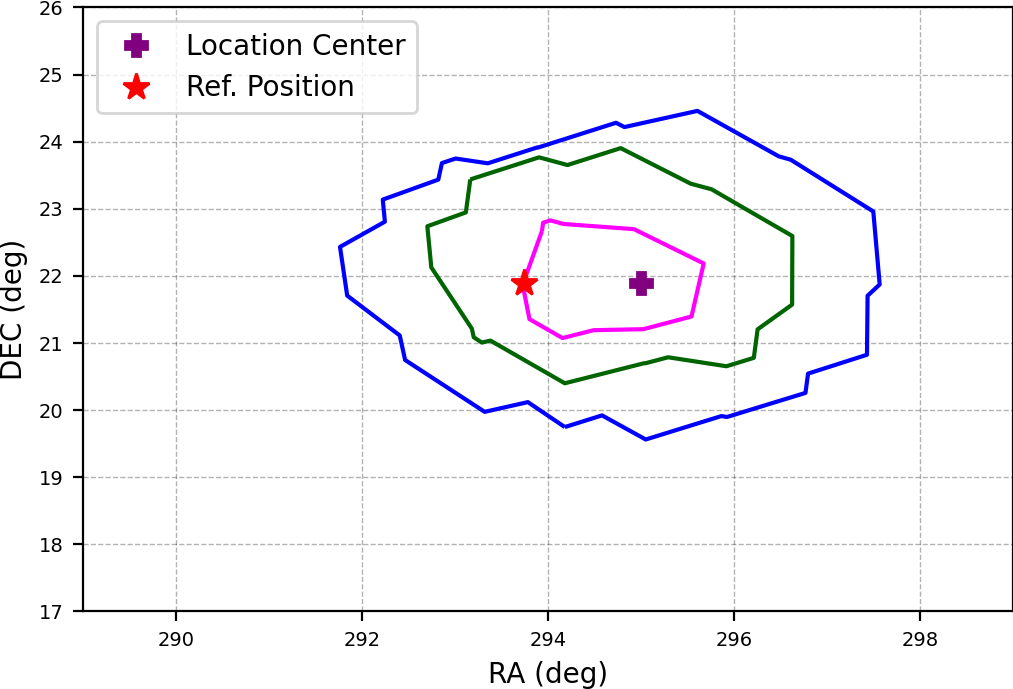}}
        \quad
        \subfigure[]{\includegraphics[width=4.5cm]{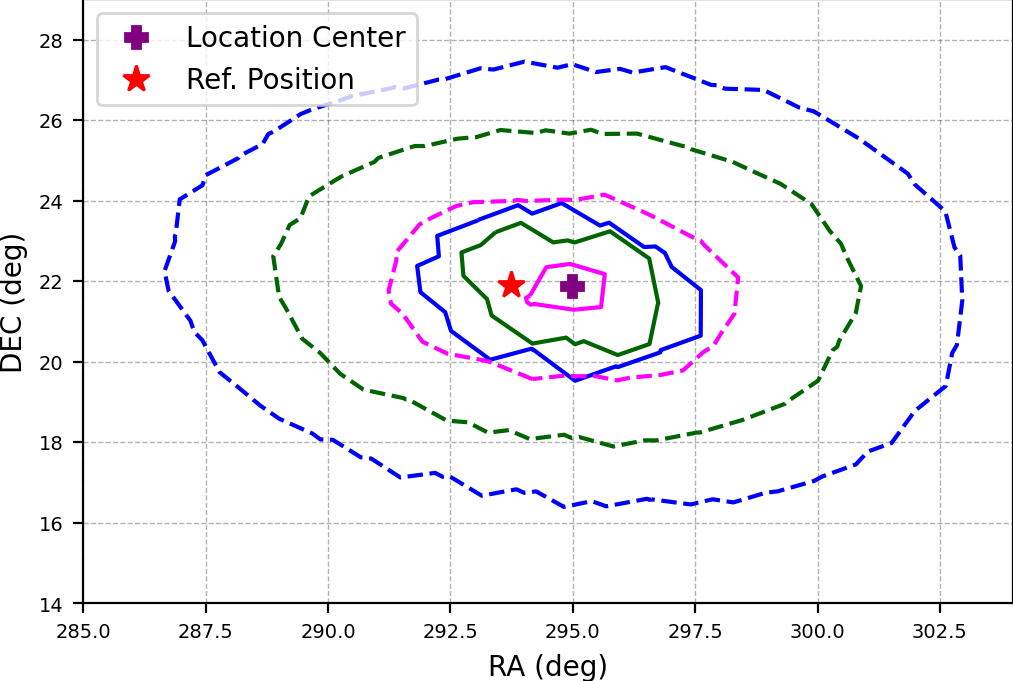}}
        \quad
        \subfigure[]{\includegraphics[width=4.5cm]{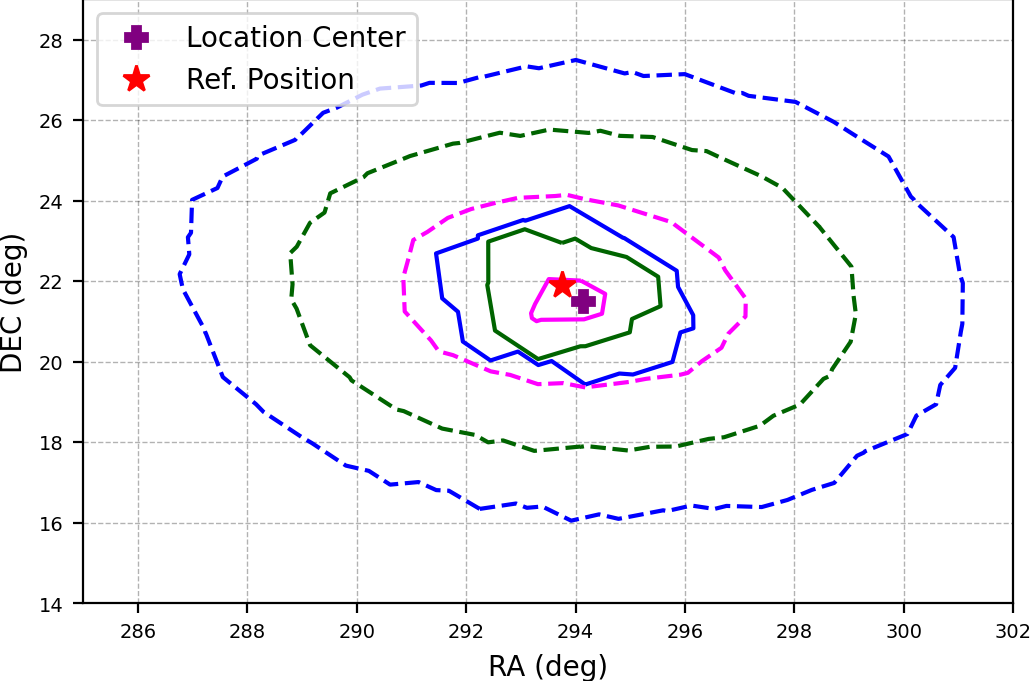}}
        \caption{ GECAM localization results of SGR 1935+2154 (UT 2022-01-15T17-21-59.304). (a) The light curve of GRD \# 07 high gain which contains the majority of net (burst) counts. (b) The RFD spectral fitting result. (c) The RFD spectral fitting result. The location credible region of (d) FIX, (e) RFD, and (f) APR localization. The captions are the same as Figure \ref{fig2a}. }
        \label{fig2_SGR1935k}
    \end{figure*}

    \begin{figure*}
        \centering
        \subfigure[]{\includegraphics[height=3.5cm]{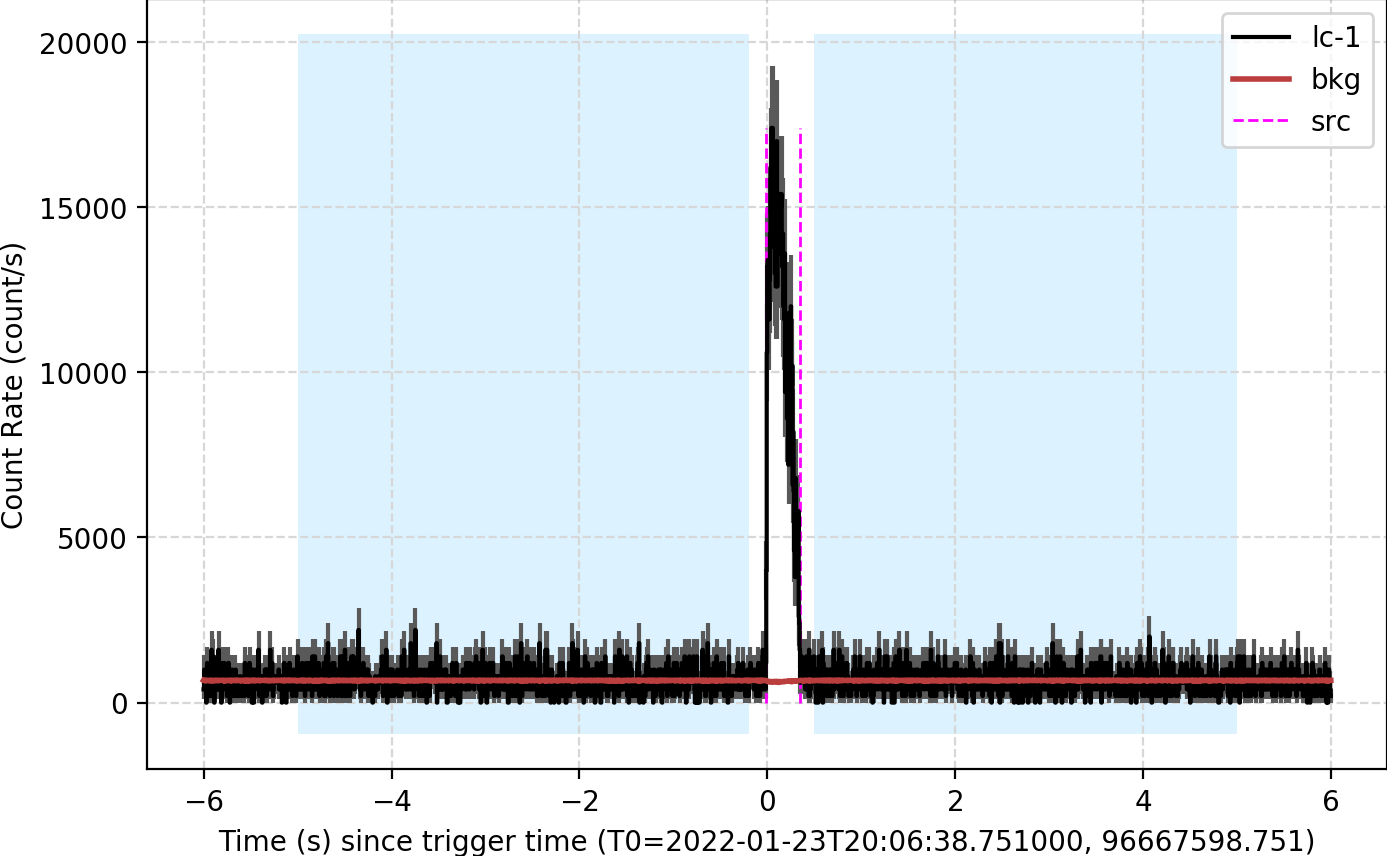}}
        \quad
        \subfigure[]{\includegraphics[height=3.5cm]{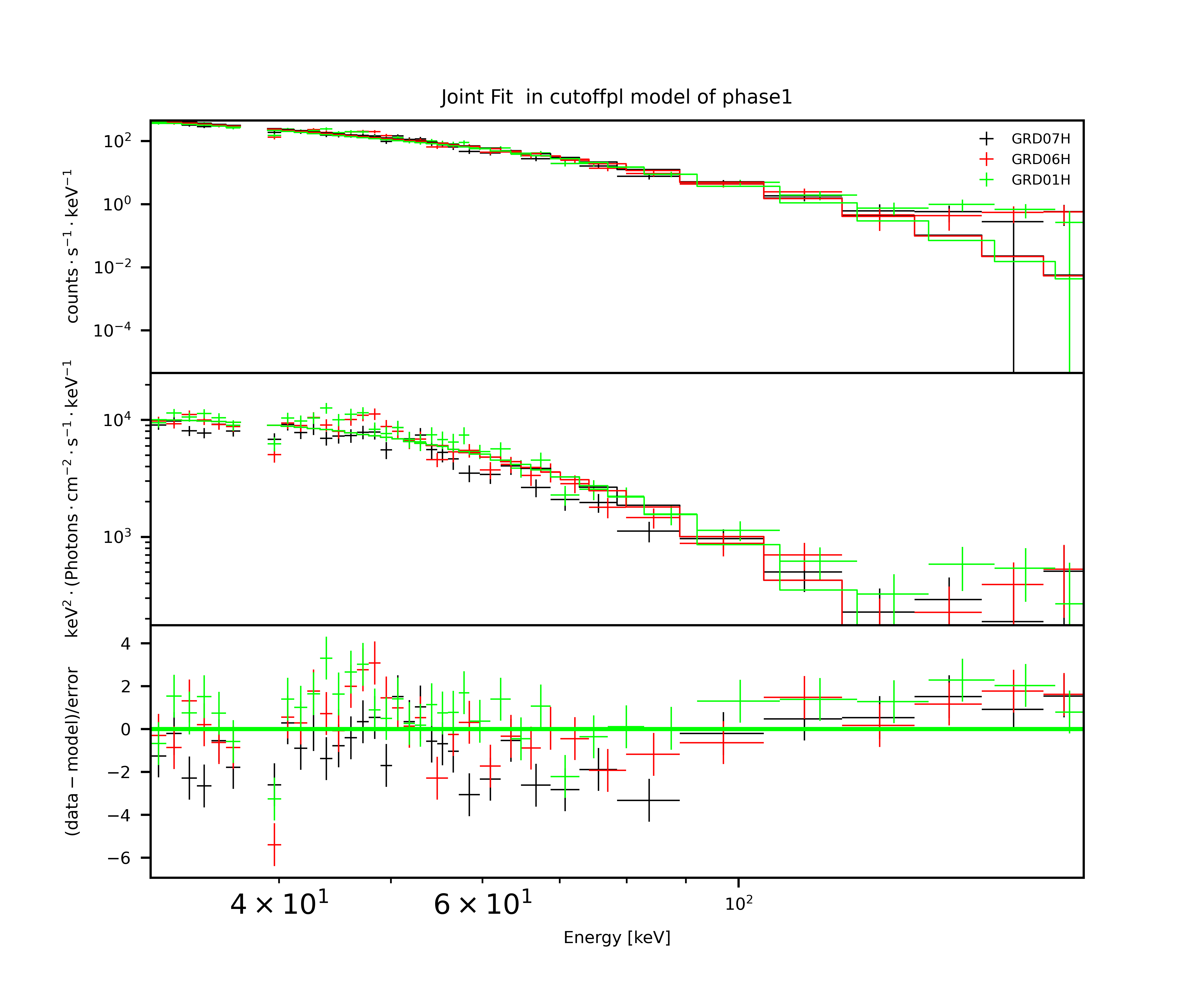}}
        \quad
        \subfigure[]{\includegraphics[height=3.5cm]{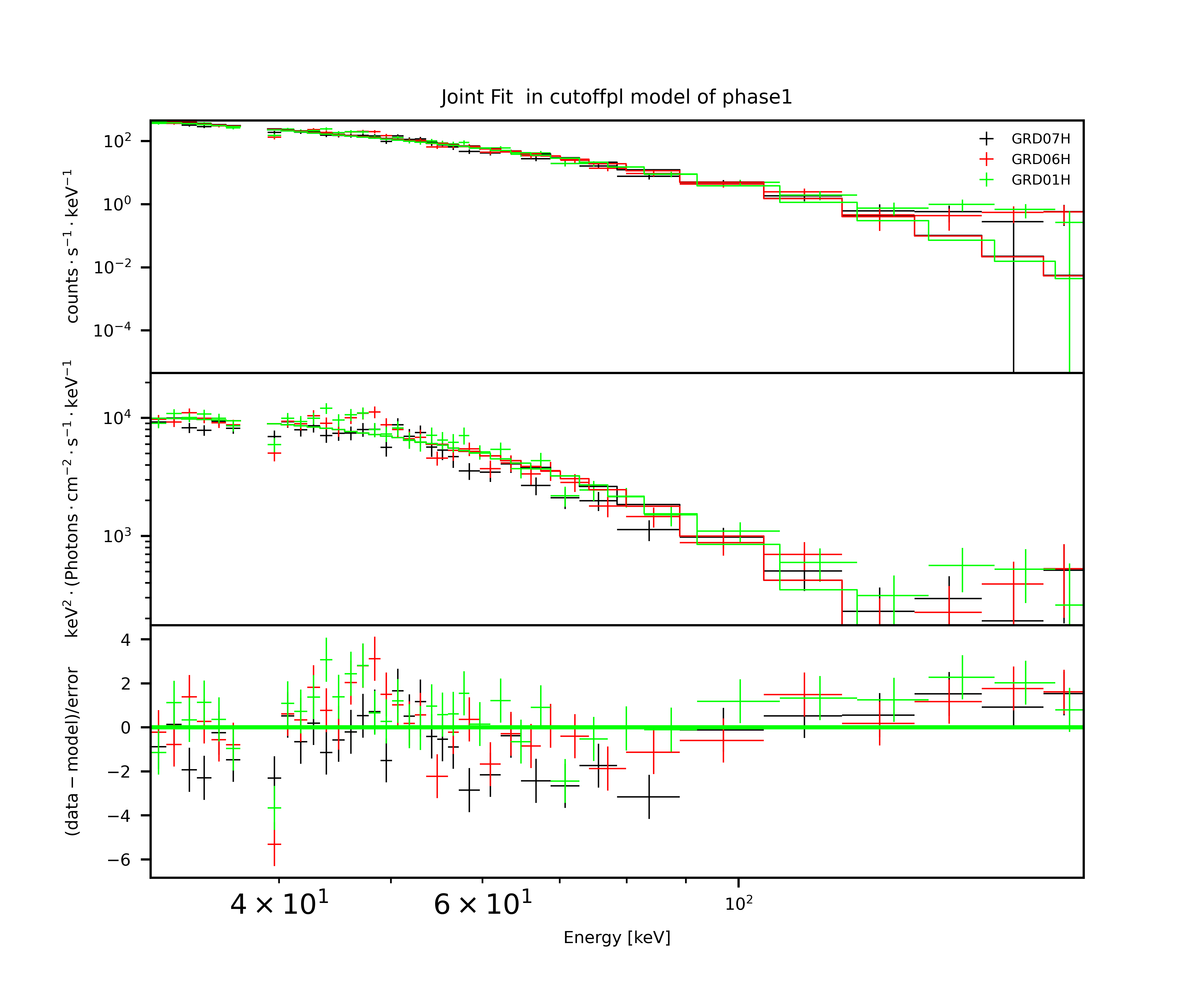}}
        \quad
        \\
        \subfigure[]{\includegraphics[width=4.5cm]{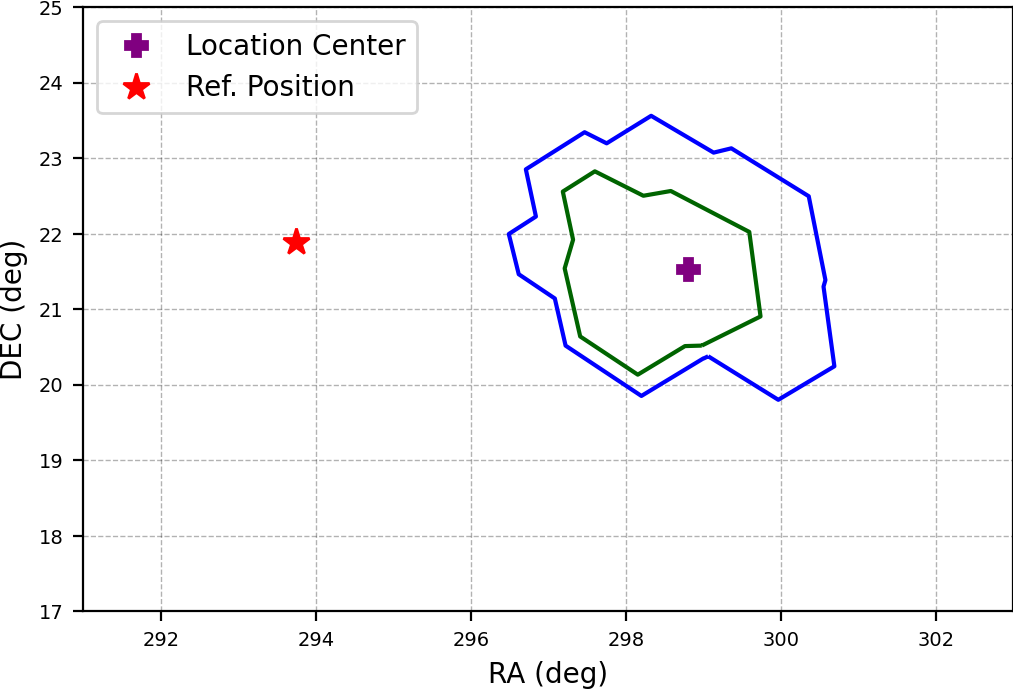}}
        \quad
        \quad
        \subfigure[]{\includegraphics[width=4.5cm]{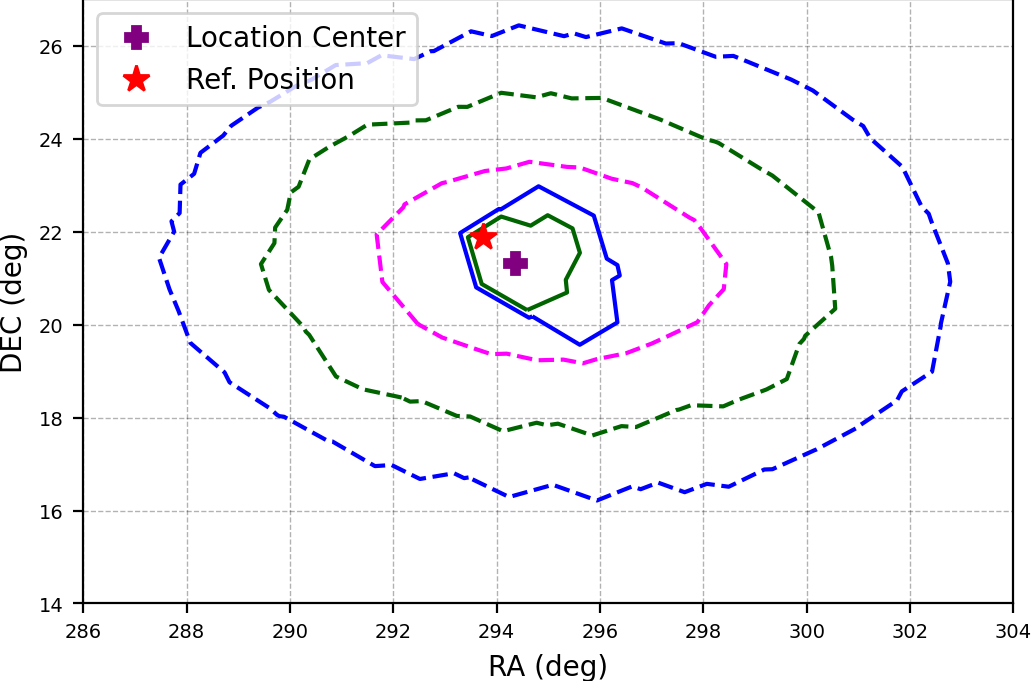}}
        \caption{ GECAM localization results of SGR 1935+2154 (UT 2022-01-23T20-06-38.751). (a) The light curve of GRD \# 06 high gain which contains the majority of net (burst) counts. (b) The RFD spectral fitting result. (c) The RFD spectral fitting result. The location credible region of (d) FIX, (e) RFD, and (f) APR localization. The captions are the same as Figure \ref{fig2a}. }
        \label{fig2_SGR1935l}
    \end{figure*}


\clearpage
\section{GECAM Localization Results for TGFs} \label{SECTION_LocRes_TGF}

    The GECAM localization result of 2 bright TGFs (One is shown in Figure \ref{fig4b}).

    \begin{figure*}
        \centering
        \subfigure[]{\includegraphics[width=9.0cm]{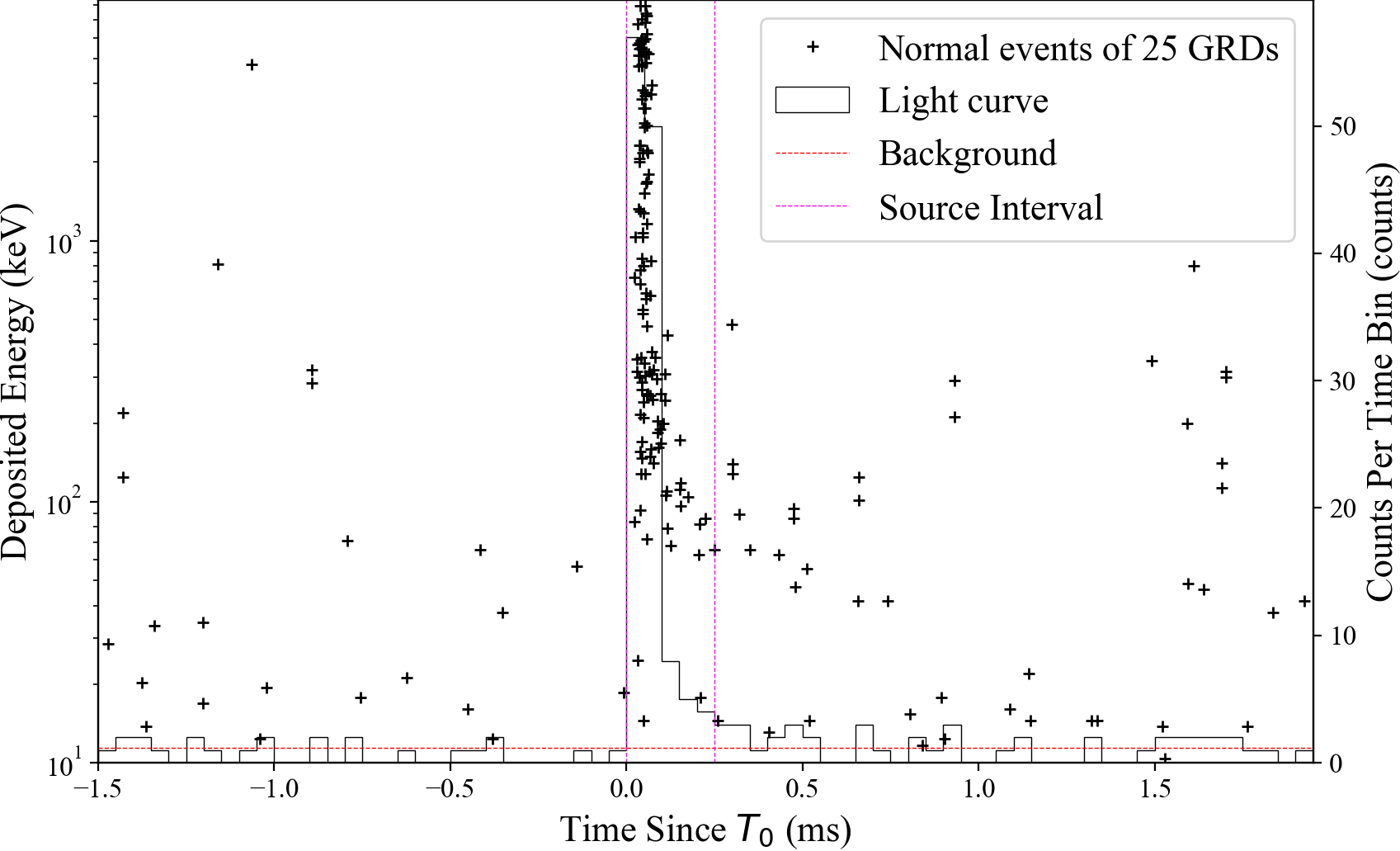}}
        \quad
        \subfigure[]{\includegraphics[width=9.0cm]{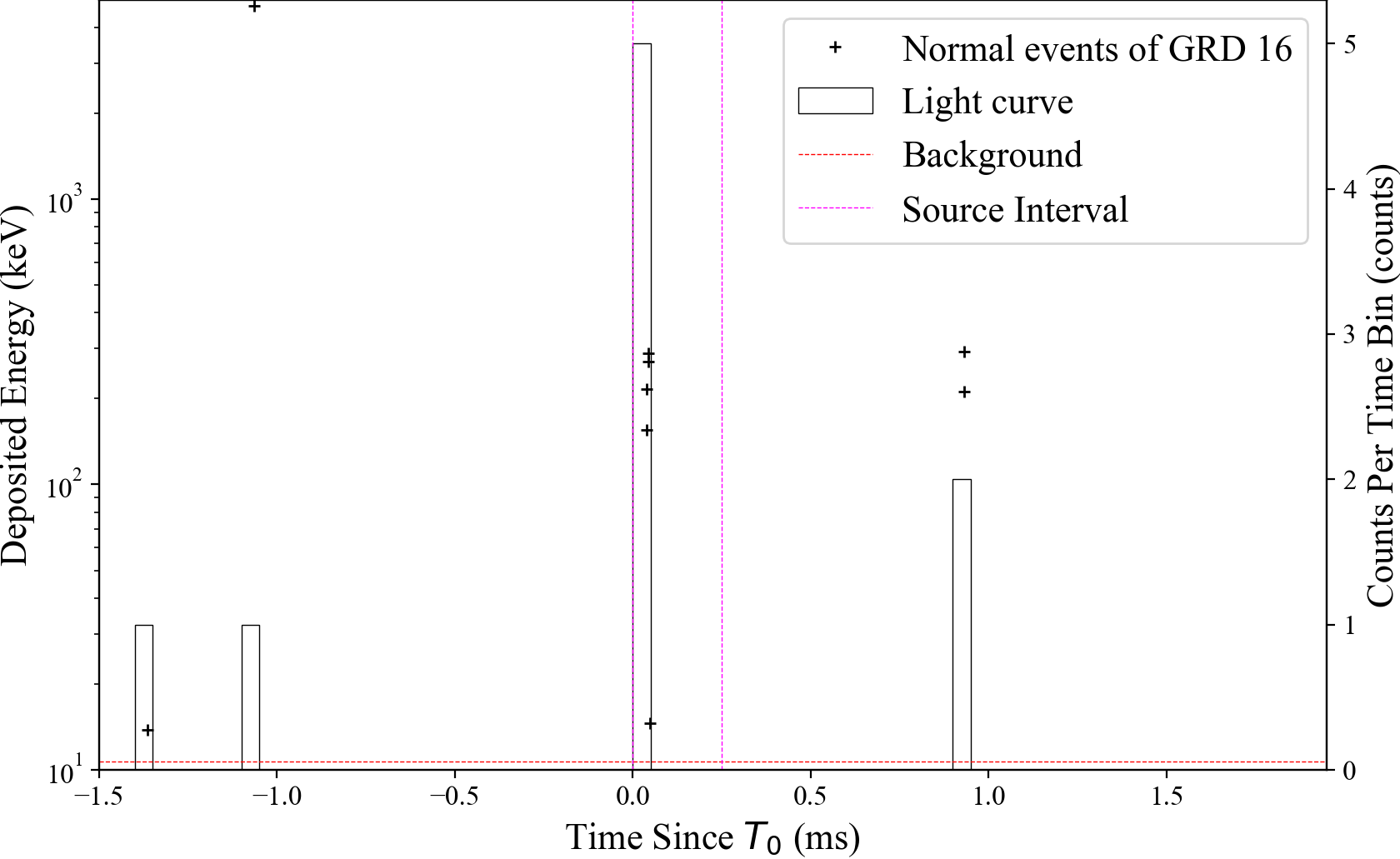}}
        \quad
        \subfigure[]{\includegraphics[width=9.0cm]{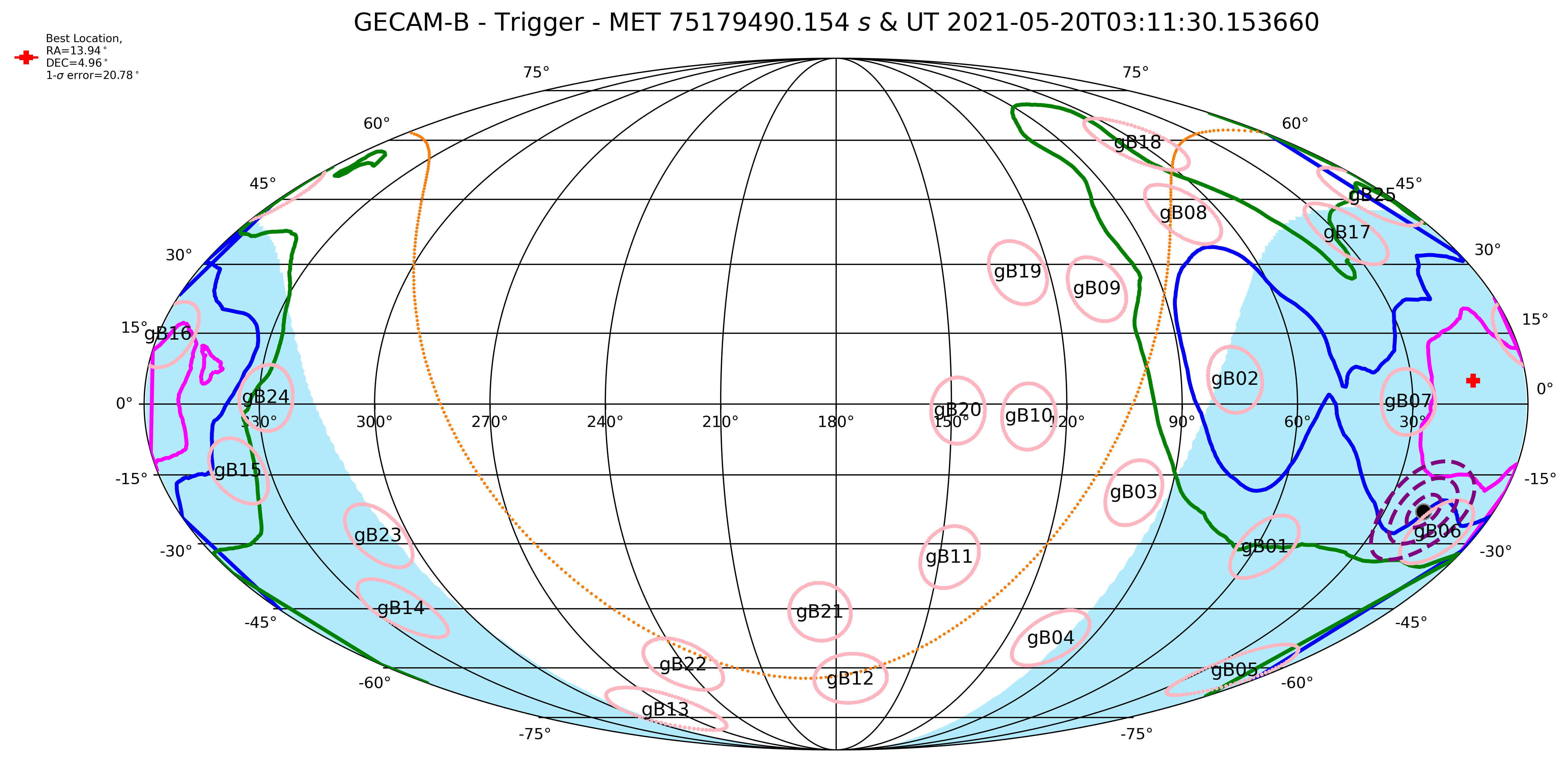}}
        \quad
        \caption{ GECAM localization results of TGF 210520. (a) The light curve and time-energy scatter plot of all 25 detectors. (b) The light curve and time-energy scatter plot of GRD \# 16 which contains the majority of net counts in the discovery bin. (c) The location sky map. The captions are the same as Figure \ref{fig4b}. }
        \label{fig4b_02}
    \end{figure*}

    \begin{figure*}
        \centering
        \subfigure[]{\includegraphics[width=9.0cm]{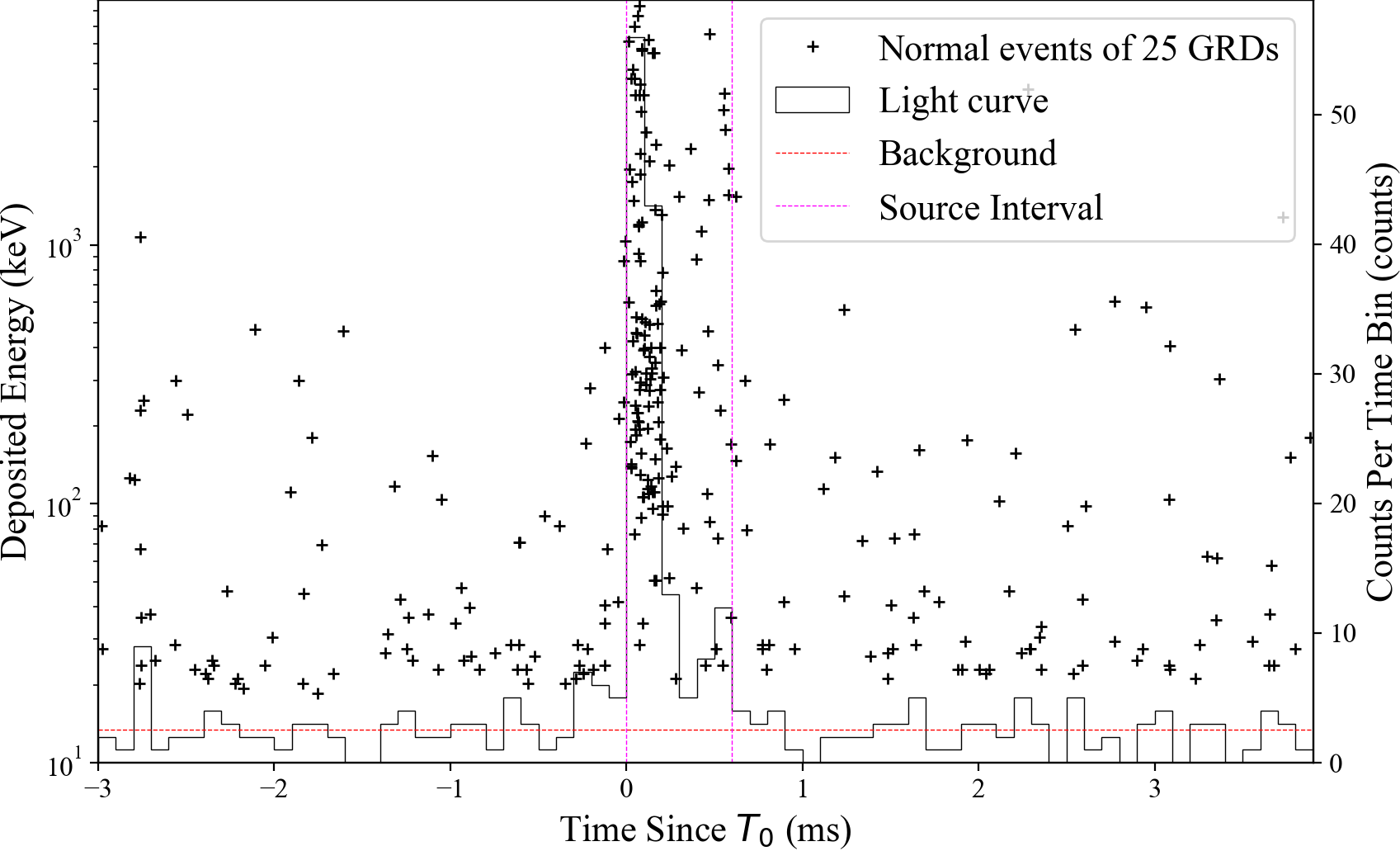}}
        \quad
        \subfigure[]{\includegraphics[width=9.0cm]{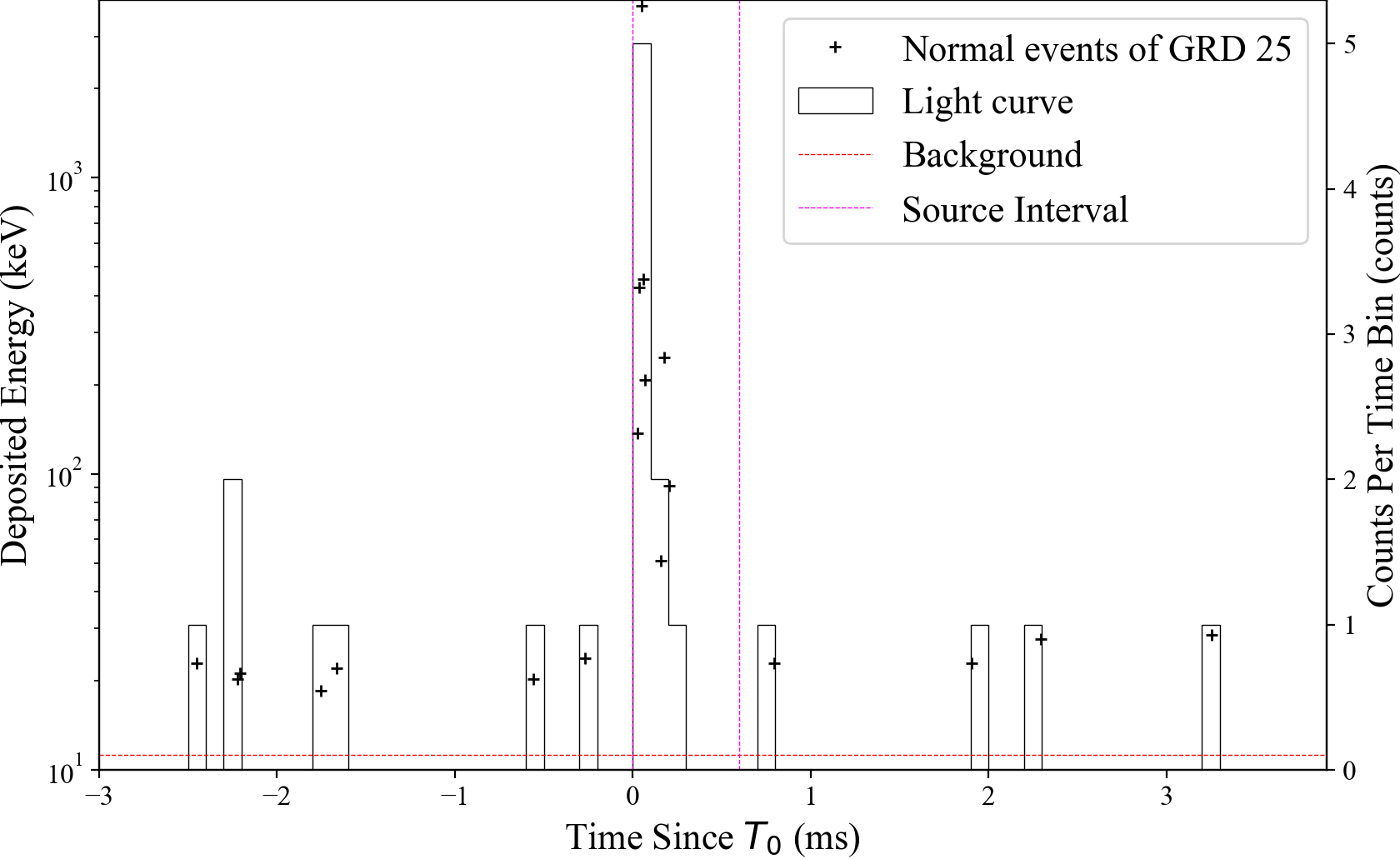}}
        \quad
        \subfigure[]{\includegraphics[width=9.0cm]{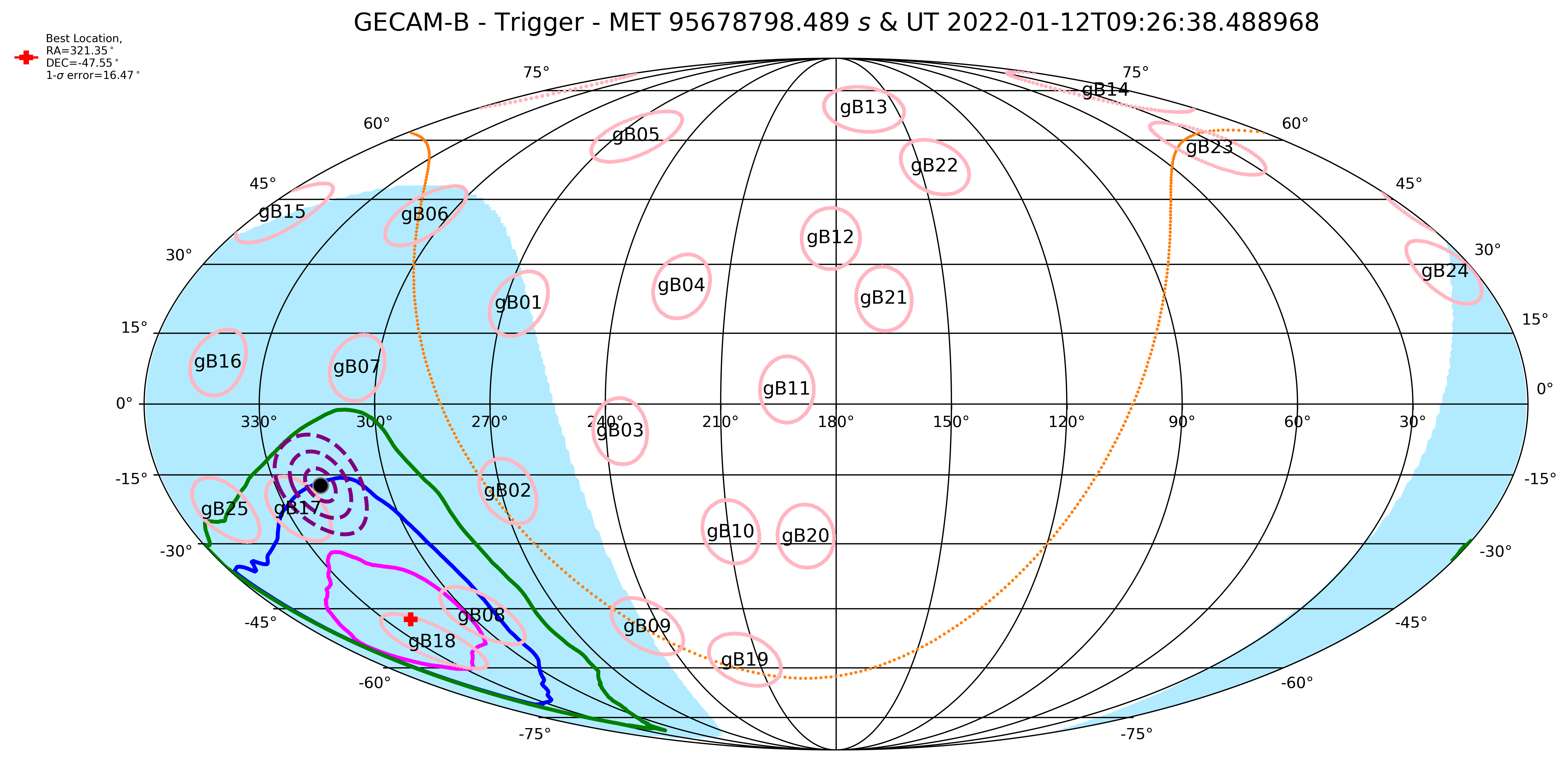}}
        \quad
        \caption{ GECAM localization results of TGF 220112. (a) The light curve and time-energy scatter plot of all 25 detectors. (b) The light curve and time-energy scatter plot of GRD \# 25 which contains the majority of net counts in the discovery bin. (c) The location sky map. The captions are the same as Figure \ref{fig4b}. }
        \label{fig4b_03}
    \end{figure*}

\clearpage 

\input{main.bbl}
\bibliographystyle{aasjournal}

\end{document}

%% file: authors.tex
\author{Yi Zhao}
\affiliation{Department of Astronomy, Beijing Normal University, Beijing 100875, China}
\affiliation{Key Laboratory of Particle Astrophysics, Institute of High Energy Physics, Chinese Academy of Sciences, Beijing 100049, China}
\author{Wang-Chen Xue}
\affiliation{Key Laboratory of Particle Astrophysics, Institute of High Energy Physics, Chinese Academy of Sciences, Beijing 100049, China}
\affiliation{University of Chinese Academy of Sciences, Beijing 100049, China}
\author{Shao-Lin Xiong}
\affiliation{Key Laboratory of Particle Astrophysics, Institute of High Energy Physics, Chinese Academy of Sciences, Beijing 100049, China}
\correspondingauthor{Shao-Lin Xiong}
\email{xiongsl@ihep.ac.cn}
\author{Yuan-Hao Wang}
\affiliation{Key Laboratory of Particle Astrophysics, Institute of High Energy Physics, Chinese Academy of Sciences, Beijing 100049, China}
\author{Jia-Cong Liu}
\affiliation{Key Laboratory of Particle Astrophysics, Institute of High Energy Physics, Chinese Academy of Sciences, Beijing 100049, China}
\affiliation{University of Chinese Academy of Sciences, Beijing 100049, China}
\author{Qi Luo}
\affiliation{Key Laboratory of Particle Astrophysics, Institute of High Energy Physics, Chinese Academy of Sciences, Beijing 100049, China}
\affiliation{University of Chinese Academy of Sciences, Beijing 100049, China}
%
%
\author{Yan-Qiu Zhang}
\affiliation{Key Laboratory of Particle Astrophysics, Institute of High Energy Physics, Chinese Academy of Sciences, Beijing 100049, China}
\affiliation{University of Chinese Academy of Sciences, Beijing 100049, China}
\author{Jian-Chao Sun}
\affiliation{Key Laboratory of Particle Astrophysics, Institute of High Energy Physics, Chinese Academy of Sciences, Beijing 100049, China}
\author{Xiao-Yun Zhao}
\affiliation{Key Laboratory of Particle Astrophysics, Institute of High Energy Physics, Chinese Academy of Sciences, Beijing 100049, China}
\author{Ce Cai}
\affiliation{College of Physics, Hebei Normal University, 20 South Erhuan Road, Shijiazhuang 050024, Hebei, China}
\affiliation{Key Laboratory of Particle Astrophysics, Institute of High Energy Physics, Chinese Academy of Sciences, Beijing 100049, China}
\affiliation{University of Chinese Academy of Sciences, Beijing 100049, China}
\author{Shuo Xiao}
\affiliation{Guizhou Provincial Key Laboratory of Radio Astronomy and Data Processing, Guizhou Normal University, Guiyang 550001, GuiZhou, China}
\affiliation{School of Physics and Electronic Science, Guizhou Normal University, Guiyang 550001, GuiZhou, China}
\affiliation{Key Laboratory of Particle Astrophysics, Institute of High Energy Physics, Chinese Academy of Sciences, Beijing 100049, China}
\affiliation{University of Chinese Academy of Sciences, Beijing 100049, China}
\author{Yue Huang}
\affiliation{Key Laboratory of Particle Astrophysics, Institute of High Energy Physics, Chinese Academy of Sciences, Beijing 100049, China}
\author{Xiao-Bo Li}
\affiliation{Key Laboratory of Particle Astrophysics, Institute of High Energy Physics, Chinese Academy of Sciences, Beijing 100049, China}
\author{Zhen Zhang}
\affiliation{Key Laboratory of Particle Astrophysics, Institute of High Energy Physics, Chinese Academy of Sciences, Beijing 100049, China}
\author{Jin-Yuan Liao}
\affiliation{Key Laboratory of Particle Astrophysics, Institute of High Energy Physics, Chinese Academy of Sciences, Beijing 100049, China}
\author{Sheng Yang}
\affiliation{Key Laboratory of Particle Astrophysics, Institute of High Energy Physics, Chinese Academy of Sciences, Beijing 100049, China}
\author{Rui Qiao}
\affiliation{Key Laboratory of Particle Astrophysics, Institute of High Energy Physics, Chinese Academy of Sciences, Beijing 100049, China}
\author{Dong-Ya Guo}
\affiliation{Key Laboratory of Particle Astrophysics, Institute of High Energy Physics, Chinese Academy of Sciences, Beijing 100049, China}
\author{Chao Zheng}
\affiliation{Key Laboratory of Particle Astrophysics, Institute of High Energy Physics, Chinese Academy of Sciences, Beijing 100049, China}
\affiliation{University of Chinese Academy of Sciences, Beijing 100049, China}
\author{Qi-Bin Yi}
\affiliation{Key Laboratory of Particle Astrophysics, Institute of High Energy Physics, Chinese Academy of Sciences, Beijing 100049, China}
\affiliation{School of Physics and Optoelectronics, Xiangtan University, Xiangtan 411105, Hunan, China}
\author{Sheng-Lun Xie}
\affiliation{Key Laboratory of Particle Astrophysics, Institute of High Energy Physics, Chinese Academy of Sciences, Beijing 100049, China}
\affiliation{Institute of Astrophysics, Central China Normal University, Wuhan 430079, China}
\author{Zhi-Wei Guo}
\affiliation{Key Laboratory of Particle Astrophysics, Institute of High Energy Physics, Chinese Academy of Sciences, Beijing 100049, China}
\affiliation{College of Physics Sciences Technology, Hebei University, No. 180 Wusi Dong Road, Lian Chi District, Baoding City, Hebei Province 071002, China}
\author{Chao-Yang Li}
\affiliation{Key Laboratory of Particle Astrophysics, Institute of High Energy Physics, Chinese Academy of Sciences, Beijing 100049, China}
\affiliation{Physics and Space Science College, China West Normal University, Nanchong 637002, China}
\author{Chen-Wei Wang}
\affiliation{Key Laboratory of Particle Astrophysics, Institute of High Energy Physics, Chinese Academy of Sciences, Beijing 100049, China}
\affiliation{University of Chinese Academy of Sciences, Beijing 100049, China}
\author{Wen-Jun Tan}
\affiliation{Key Laboratory of Particle Astrophysics, Institute of High Energy Physics, Chinese Academy of Sciences, Beijing 100049, China}
\affiliation{University of Chinese Academy of Sciences, Beijing 100049, China}
\author{Yue Wang}
\affiliation{Key Laboratory of Particle Astrophysics, Institute of High Energy Physics, Chinese Academy of Sciences, Beijing 100049, China}
\affiliation{University of Chinese Academy of Sciences, Beijing 100049, China}
\author{Wen-Xi Peng}
\affiliation{Key Laboratory of Particle Astrophysics, Institute of High Energy Physics, Chinese Academy of Sciences, Beijing 100049, China}
\author{Shi-Jie Zheng}
\affiliation{Key Laboratory of Particle Astrophysics, Institute of High Energy Physics, Chinese Academy of Sciences, Beijing 100049, China}
\author{Jian-Jian He}
\affiliation{Key Laboratory of Particle Astrophysics, Institute of High Energy Physics, Chinese Academy of Sciences, Beijing 100049, China}
\author{Ping Wang}
\affiliation{Key Laboratory of Particle Astrophysics, Institute of High Energy Physics, Chinese Academy of Sciences, Beijing 100049, China}
\author{Jin Wang}
\affiliation{Key Laboratory of Particle Astrophysics, Institute of High Energy Physics, Chinese Academy of Sciences, Beijing 100049, China}
\author{Xiang Ma}
\affiliation{Key Laboratory of Particle Astrophysics, Institute of High Energy Physics, Chinese Academy of Sciences, Beijing 100049, China}
\author{Xin-Ying Song}
\affiliation{Key Laboratory of Particle Astrophysics, Institute of High Energy Physics, Chinese Academy of Sciences, Beijing 100049, China}
\author{Hong-Mei Zhang}
\affiliation{Key Laboratory of Particle Astrophysics, Institute of High Energy Physics, Chinese Academy of Sciences, Beijing 100049, China}
\author{Bing Li}
\affiliation{Key Laboratory of Particle Astrophysics, Institute of High Energy Physics, Chinese Academy of Sciences, Beijing 100049, China}
\author{Peng Zhang}
\affiliation{College of Electronic and information Engineering, Tongji University, Shanghai 201804, China}
\affiliation{Key Laboratory of Particle Astrophysics, Institute of High Energy Physics, Chinese Academy of Sciences, Beijing 100049, China}
\author{Hong Wu}
\affiliation{Key Laboratory of Particle Astrophysics, Institute of High Energy Physics, Chinese Academy of Sciences, Beijing 100049, China}
\affiliation{School of Computing and Artificial Intelligence, Southwest Jiaotong University, Chengdu 611756, China}
\author{Yan-Qi Du}
\affiliation{Key Laboratory of Particle Astrophysics, Institute of High Energy Physics, Chinese Academy of Sciences, Beijing 100049, China}
\affiliation{School of Computing and Artificial Intelligence, Southwest Jiaotong University, Chengdu 611756, China}
\author{Jing Liang}
\affiliation{Key Laboratory of Particle Astrophysics, Institute of High Energy Physics, Chinese Academy of Sciences, Beijing 100049, China}
\affiliation{School of Computing and Artificial Intelligence, Southwest Jiaotong University, Chengdu 611756, China}
\author{Guo-Ying Zhao}
\affiliation{Key Laboratory of Particle Astrophysics, Institute of High Energy Physics, Chinese Academy of Sciences, Beijing 100049, China}
\affiliation{School of Physics and Optoelectronics, Xiangtan University, Xiangtan 411105, Hunan, China}
\author{Xin-Qiao Li}
\affiliation{Key Laboratory of Particle Astrophysics, Institute of High Energy Physics, Chinese Academy of Sciences, Beijing 100049, China}
\author{Xiang-Yang Wen}
\affiliation{Key Laboratory of Particle Astrophysics, Institute of High Energy Physics, Chinese Academy of Sciences, Beijing 100049, China}
\author{Zheng-Hua An}
\affiliation{Key Laboratory of Particle Astrophysics, Institute of High Energy Physics, Chinese Academy of Sciences, Beijing 100049, China}
\author{Xi-Lei Sun}
\affiliation{Key Laboratory of Particle Astrophysics, Institute of High Energy Physics, Chinese Academy of Sciences, Beijing 100049, China}
\author{Yan-Bing Xu}
\affiliation{Key Laboratory of Particle Astrophysics, Institute of High Energy Physics, Chinese Academy of Sciences, Beijing 100049, China}
\author{Fan Zhang}
\affiliation{Key Laboratory of Particle Astrophysics, Institute of High Energy Physics, Chinese Academy of Sciences, Beijing 100049, China}
\author{Da-Li Zhang}
\affiliation{Key Laboratory of Particle Astrophysics, Institute of High Energy Physics, Chinese Academy of Sciences, Beijing 100049, China}
\author{Ke Gong}
\affiliation{Key Laboratory of Particle Astrophysics, Institute of High Energy Physics, Chinese Academy of Sciences, Beijing 100049, China}
\affiliation{University of Chinese Academy of Sciences, Beijing 100049, China}
\author{Ya-Qing Liu}
\affiliation{Key Laboratory of Particle Astrophysics, Institute of High Energy Physics, Chinese Academy of Sciences, Beijing 100049, China}
\author{Xiao-Hua Liang}
\affiliation{Key Laboratory of Particle Astrophysics, Institute of High Energy Physics, Chinese Academy of Sciences, Beijing 100049, China}
\author{Xiao-Jing Liu}
\affiliation{Key Laboratory of Particle Astrophysics, Institute of High Energy Physics, Chinese Academy of Sciences, Beijing 100049, China}
\author{Min Gao}
\affiliation{Key Laboratory of Particle Astrophysics, Institute of High Energy Physics, Chinese Academy of Sciences, Beijing 100049, China}
\author{Jin-Zhou Wang}
\affiliation{Key Laboratory of Particle Astrophysics, Institute of High Energy Physics, Chinese Academy of Sciences, Beijing 100049, China}
\author{Li-Ming Song}
\affiliation{Key Laboratory of Particle Astrophysics, Institute of High Energy Physics, Chinese Academy of Sciences, Beijing 100049, China}
\author{Gang Chen}
\affiliation{Key Laboratory of Particle Astrophysics, Institute of High Energy Physics, Chinese Academy of Sciences, Beijing 100049, China}
\author{Ke-Ke Zhang}
\affiliation{Innovation Academy for Microsatellites of Chinese Academy of Sciences, Shanghai, 201304, China}
\author{Xing-Bo Han}
\affiliation{Innovation Academy for Microsatellites of Chinese Academy of Sciences, Shanghai, 201304, China}
\author{Hai-Yan Wu}
\affiliation{National Space Science Center, Chinese Academy of Sciences, Beijing 100190, China}
\author{Tai Hu}
\affiliation{National Space Science Center, Chinese Academy of Sciences, Beijing 100190, China}
\author{Hao Geng}
\affiliation{National Space Science Center, Chinese Academy of Sciences, Beijing 100190, China}
\author{Fang-Jun Lu}
\affiliation{Key Laboratory of Particle Astrophysics, Institute of High Energy Physics, Chinese Academy of Sciences, Beijing 100049, China}
\author{Shu Zhang}
\affiliation{Key Laboratory of Particle Astrophysics, Institute of High Energy Physics, Chinese Academy of Sciences, Beijing 100049, China}
\author{Shuang-Nan Zhang}
\affiliation{Key Laboratory of Particle Astrophysics, Institute of High Energy Physics, Chinese Academy of Sciences, Beijing 100049, China}
\affiliation{University of Chinese Academy of Sciences, Beijing 100049, China}
\author{Gao-Peng Lu}
\affiliation{School of Earth and Space Sciences, University of Science and Technology of China, Hefei 230026, China}
\author{Ming Zeng}
\affiliation{Key Laboratory of Particle and Radiation Imaging (Tsinghua University), Ministry of Education, Beijing 100084, China}
\affiliation{Department of Engineering Physics, Tsinghua University, Beijing 100084, China}
\author{Heng Yu}
\affiliation{Department of Astronomy, Beijing Normal University, Beijing 100875, China}